\documentclass[a4paper,UKenglish,cleveref, autoref, thm-restate]{lipics-v2021}

\hideLIPIcs  

\title{Simple qudit ZX~and ZH~calculi, via integrals}

\titlerunning{Simple qudit ZX~and ZH~calculi, via integrals} 

\author{Niel de Beaudrap}{University of Sussex, United Kingdom}{niel.debeaudrap@sussex.ac.uk}{https://orcid.org/
0000-0001-9549-5146}{}

\author{Richard D. P. East}{Haiqu}{rdp.east@gmail.com}{https://orcid.org/0000-0002-5818-483X}{}

\authorrunning{N.~de\;Beaudrap and R.~D.~P.~East} 

\Copyright{Niel~de\;Beaudrap and Richard\;D.\;P.\;East} 



\relatedversion{} 

\acknowledgements{Thanks to Patrick Roy, Titouan Carette, John van de Wetering, and Robert Booth for technical discussions, and to the anonymous referees for suggestions on presentation.}

\nolinenumbers 
\EventEditors{John Q. Open and Joan R. Access}
\EventNoEds{2}
\EventLongTitle{42nd Conference on Very Important Topics (CVIT 2016)}
\EventShortTitle{CVIT 2016}
\EventAcronym{CVIT}
\EventYear{2016}
\EventDate{December 24--27, 2016}
\EventLocation{Little Whinging, United Kingdom}
\EventLogo{}
\SeriesVolume{42}
\ArticleNo{23}

\usepackage[utf8]{inputenc}
\usepackage{mathtools}
\usepackage{amsthm,amsmath,amssymb}
\usepackage{mathrsfs}
\usepackage{hyperref}
\usepackage[hyphenbreaks]{breakurl}
\usepackage{stmaryrd}
\usepackage{graphicx}
\usepackage{scalefnt}
\usepackage{multicol}
\usepackage{array}
\usepackage[numbers,compress]{natbib}

\usepackage{tikz}
\usetikzlibrary{shapes,arrows,calc,positioning,fit}
\pgfdeclarelayer{background}
\pgfsetlayers{background,main}

\usepackage{tikz}
\usetikzlibrary{shapes,arrows,calc,positioning,fit}
\pgfdeclarelayer{background}
\pgfsetlayers{background,main}

\tikzstyle{every picture}=[baseline=-0.25em]

\newcommand\tikzfig[1]{%
\IfFileExists{./figures/#1.pdf}%
  {\raisebox{-0.25em}{$\begin{gathered}\includegraphics{./figures/#1.pdf}\end{gathered}$}}%
  {%
    \IfFileExists{./Figures-Source/#1.tikz}%
      {\input{./Figures-Source/#1.tikz}}%
      {%
	\IfFileExists{./#1.tikz}%
      	  {\input{./#1.tikz}}%
      	  {\tikz[baseline=-0.5em]{%
		\node[draw=red!50!black,font=\color{red!50!black},fill=red!10!white]
		{\textit{#1}};}}%
      }%
  }%
}
\newlength\vtikzfigoffset
\newcommand\vtikzfig[2][0ex]{%
\addtolength\vtikzfigoffset{#1}
\IfFileExists{./figures/#2.pdf}%
  {\raisebox{-0.25em}{$\begin{gathered}~\\[\vtikzfigoffset]\includegraphics{./figures/#2.pdf}\end{gathered}$}}%
  {%
    \IfFileExists{./Figures-Source/#2.tikz}%
      {\input{./Figures-Source/#2.tikz}}%
      {%
	\IfFileExists{./#2.tikz}%
      	  {\input{./#2.tikz}}%
      	  {\tikz[baseline=-0.5em]{%
		\node[draw=red!50!black,font=\color{red!50!black},fill=red!10!white]
		{\textit{#2}};}}%
      }%
  }%
}

\tikzstyle{Z dot}=
  [fill={rgb,255:red,216;green,248;blue,216}, inner sep=0mm, minimum size=2mm, draw=black, inner sep=2pt, shape=circle]
\tikzstyle{X dot}=
  [Z dot, fill={rgb,255:red,227;green,145;blue,145}]
\tikzstyle{white dot}=
  [Z dot, fill={rgb,255:red,255;green,255;blue,255}]
\tikzstyle{not dot}=
  [Z dot, fill={rgb,255:red,108;green,108;blue,108}]
\tikzstyle{gray dot}=
  [Z dot, fill={rgb,255:red,208;green,208;blue,208}]
\tikzstyle{Z phase dot}=
  [fill={rgb,255:red,216;green,248;blue,216},  minimum size=5mm, font={\footnotesize}, rounded corners=2mm, inner sep=0.2mm, outer sep=-2mm, scale=0.8,draw=black]
\tikzstyle{X phase dot}=
  [Z phase dot, fill={rgb,255:red,227;green,145;blue,145}]
\tikzstyle{H box}=
  [fill=yellow!50!white, draw=black, shape=rectangle, inner sep=0.6mm, minimum height=3mm, minimum width=3mm, font=\footnotesize]
\tikzstyle{small H box}=
  [fill=yellow!50!white, draw=black, shape=rectangle, inner sep=0.6mm, minimum width=2mm, minimum height=2mm, font=\footnotesize]
\tikzstyle{rfarr}=
	[draw, signal, fill=gray!10!white, signal to=east, signal from=west, inner sep=2pt, minimum height=8pt]
\tikzstyle{mini H box}=
  [fill=yellow!50!white, draw=black, shape=rectangle, inner sep=0.6mm, minimum height=1mm, minimum width=1mm, font=\footnotesize]

\newlength\defaultrulelabelwd
\newlength\rulelabelwd
\newlength\defaultrulediagramwd
\newlength\rulediagramwd
\newlength\defaultrulexnwd
\newlength\rulexnwd
\def\rewriterule[#1] #2{%
	\node (#1) at (0,0) {%
	\begin{minipage}{\rulediagramwd}\centering%
		#2%
	\end{minipage}%
	\begin{minipage}{\rulexnwd}\centering%
		$\xleftrightarrow{\textsf{(#1)}}$
	\end{minipage}%
	};
	\coordinate (anchor-next-diagram) at (#1.east);
}
\def\nextrewriterule[#1] #2{%
	\node [anchor=west] (#1) at (anchor-next-diagram) {%
	\begin{minipage}{\rulediagramwd}\centering%
		#2%
	\end{minipage}%
	\begin{minipage}{\rulexnwd}\centering%
		$\xleftrightarrow{\textsf{(#1)}}$
	\end{minipage}%
	};
	\coordinate (anchor-next-diagram) at (#1.east);
}
\def\rewritetarget #1{%
	\node [anchor=west] at (anchor-next-diagram) {%
	\begin{minipage}{\rulediagramwd}\centering%
		#1%
	\end{minipage}%
	};
}

\newcommand\Rule[2][]{%
	\noindent\textbf{#1\textsf{(#2)}:}%
}
\newcommand\Corollary[2][sf]{%
	\Rule[Corollary ]{\csname #1family\endcsname #2}%
}

\let\subset\subseteq
\let\oldepsilon\epsilon
\let\epsilon\varepsilon
\let\varepsilon\oldepsilon
\let\le\leqslant
\let\ge\geqslant
\let\vec\boldsymbol

\newcommand\e{{\mathrm e}}

\newcommand\D{{\mathbf D}}

\newcommand\N{{\mathbb N}}
\newcommand\Z{{\mathbb Z}}
\newcommand\R{{\mathbb R}}
\newcommand\C{{\mathbb C}}

\newcommand\cH{{\mathcal H}}
\newcommand\cD{{\mathcal D}}

\newcommand\s{\!\!\;\mathbin;\!\!\;}

\newcommand\uchi{
	\mathchoice
		{\raisebox{0.1875ex}{$\chi$}}%
		{\raisebox{0.1875ex}{$\chi$}}%
		{\raisebox{0.1875ex}{$\chi$}}%
		{\raisebox{0.1875ex}{$\chi$}}%
	}
	
\DeclareFontFamily{U}{wncy}{}
\DeclareFontShape{U}{wncy}{m}{n}{<->wncyr10}{}
\DeclareSymbolFont{mcy}{U}{wncy}{m}{n}
\DeclareMathSymbol{\Sh}{\mathord}{mcy}{"58}

\renewcommand\t[2][0.707]{\text{\normalsize\scalefont{#1}$\!\!\;#2\!\!\;$}}
\newcommand\tp{\t{\texttt+}}
\newcommand\tm{\t{\texttt-}}

\newcommand\tmp{\t{\texttt{\llap{\raisebox{0.35ex}{-}}{\raisebox{-0.15ex}{+}}}}}

\newcommand\sox[1]{^{\otimes #1}}
\newcommand\herm{^{\dagger}}
\newcommand\trans{^{\mathsf T}}

\newcommand\ket[1]{\left| #1 \right\rangle}
\newcommand\bra[1]{\left\langle #1 \right|}
\newcommand\bracket[2]{\left\langle #1 \!\!\:\left| #2 \right\rangle\right.}
\newcommand\kket[1]{{%
    \left|\left.\!\!\: #1\right\rangle\mspace{-4.25mu}\right\rangle}}
\newcommand\bbra[1]{{%
    \left\langle\mspace{-4.25mu}\left\langle #1\right.\!\right\rvert}}
\newcommand\bbracket[2]{{%
  	\left\langle \mspace{-4.25mu}\left\langle\left.\!\!\;
  		#1%
  	\:\!\right|\mspace{-8mu}\left|\:\!
  		#2%
  	\right.\!\!\:\right\rangle\mspace{-4.25mu}\right\rangle}}

\newcommand\sem[1]{%
  {[\mspace{-2.75mu}[ \smash{#1} ]\mspace{-2.75mu}]}%
}
\newcommand\bigsem[1]{%
  {\bigl[\mspace{-4.5mu}\bigl[ \smash{#1} \bigr]\mspace{-4.5mu}\bigr]}%
}
\newcommand\Bigsem[1]{%
  {\Bigl[\mspace{-5.5mu}\Bigl[\, \smash{#1} \,\Bigr]\mspace{-5.5mu}\Bigr]}%
}
\newcommand\biggsem[1]{%
  {\biggl[\mspace{-6.5mu}\biggl[\, \smash{#1} \,\biggr]\mspace{-6.5mu}\biggr]}%
}
\newcommand\Biggsem[1]{%
  {\Biggl[\mspace{-7.5mu}\Biggl[\, \smash{#1} \,\Biggr]\mspace{-7.5mu}\Biggr]}%
}
\newcommand\Sem[2]{%
  {\left[\mspace{-8.25mu}\left[
    \smash{#2} \mathclap{\begin{matrix} \\[#1] \end{matrix}}
  \right]\mspace{-8mu}\right]}%
}

\begin{document}

\maketitle

\begin{abstract}
	The ZX~calculus and ZH~calculus use diagrams to denote and compute properties of quantum operations, using `rewrite rules' to transform between diagrams which denote the same operator through a functorial \emph{semantic map}.
	Different semantic maps give rise to different rewrite systems, which may prove more convenient for different purposes.
	Using discrete measures, we describe semantic maps for ZX~and ZH~diagrams, well-suited to analyse unitary circuits and measurements on qudits of any fixed dimension ${D \!>\! 1}$ as a single `ZXH-calculus'.
	We demonstrate rewrite rules for the `stabiliser fragment' of the ZX~calculus and a `multicharacter fragment' of the ZH~calculus.
\end{abstract}

\section{Introduction}

The ZX~calculus~\cite{CD-2011,DP-2009,Backens-2015,PW-2016,BPW-2017,JPVW-2017,CK-2017,GW-2017,NW-2017,JPV-2017,NW-2018,Wang-2018,CW-2018,Vilmart-2019-minimal,Vilmart-2019-ToffoliH,JPV-2019,Wang-2019,Wang-2019-semirings,dB-2021,Wang-2020,vdW-2020,TYF-2021,Wang-2021,WYK-2022,JPV-2022,SH-2022,BC-2022,WY-2022,Majid-2022} and ZH~calculus~\cite{BK-2019,WW-2019,KWK-2019,BKMWW-2021,Roy-2022,BKK-2021,dB-2021} are systems using annotated graphs (`ZX~diagrams' and `ZH~diagrams'), to denote tensor networks for quantum computation, and other problems involving tensors over $\mathbb C^2$~\cite{BKK-2021,EWCG-2022,TM-2021,EMW-2021,GC-2022,LMW-2022}.
They include rewrite rules, to perform computations  on diagrams without recourse to exponentially large matrices.
Complicated procedures may involve diagrams of mounting complexity to analyse, but the ZX- and ZH-calculi often simplify the analysis of many-qubit procedures.
It is also increasingly common to consider versions of the ZX- and ZH-calculi for qudits~\cite{GW-2017,Wang-2018,Wang-2021,BC-2022,WY-2022,Majid-2022,Roy-2022,PWSYYC-2023}, which promise similar benefits for the analysis of procedures on qudits.

Most treatments of these calculi \cite{Backens-2015,PW-2016,BPW-2017,JPVW-2017,CK-2017,NW-2017,JPV-2017,NW-2018,Vilmart-2019-minimal,Vilmart-2019-ToffoliH,JPV-2019,Wang-2019,Wang-2019-semirings,dB-2021,Wang-2020,TYF-2021,Wang-2021,WYK-2022,JPV-2022,SH-2022,BC-2022,BK-2019,WW-2019,KWK-2019,BKMWW-2021,Roy-2022} are `scalar exact', \emph{i.e.},~they are equational theories which do not introduce changes by scalar factors.
Changes by scalar factors do not matter for some applications (\emph{e.g.},~testing equivalence of unitary transformations), but \emph{are} important for probabilistic processes (\emph{e.g.},~postselection) or to compute specific numerical values~\cite{LMW-2022}.
But `scalar exact' treatments often involve the frequent accumulation or deletion of \emph{scalar gadgets}: disconnected sub-diagrams which obliquely denote normalisation factors.
Presentations of these calculi which avoid such book-keeping, are simpler  for both instruction and practical use.
They may also allow a simple presentation of a unified rewrite system (a~`ZXH~calculus') incorporating the rules of each~\cite{dB-2021,EWCG-2022,EMW-2021}.

In previous work~\cite{dB-2021}, one of us addressed this issue of bookkeeping of scalars for ZX- and ZH-diagrams on qubits through a carefully constructed semantic map.
The result, described as `well-tempered' versions of these calculi, are scalar exact while avoiding the modifications of scalar gadgets for the most often-used rewrites.
However, while the rewrite rules of this `well-tempered' notation are simple, the notational convention itself (\emph{i.e.},~the semantics of the generators of the calculi) is slightly unwieldly. 
Furthermore, it left open how to address similar issues with scalars for versions of these calculi on qudits of dimension $D > 2$.

In this work, we consider how different normalisations of the ZX- and ZH-calculus may be expressed in a more uniform way, by representing operators on qudits (of any fixed dimension $D>1$) through the use of integrals with respect to a discrete measure.

For a finite set $S$, let $\#S$ denote its cardinality.
Consider a measure $\mu(S) = \#S \cdot \nu^2$ on subsets $S \subseteq \mathbb Z$, for $\nu > 0$ to be fixed later.
This is a `measure' on sets $S$ (see Section~\ref{sec:maths-prelims}) which allows us to define a formal notion of integration of functions $f: \mathbb Z \to \C$\;,\;~\\[-3.0ex]
\begin{equation}
\label{eqn:introducing-discrete-integral}
	\int\limits_{\mathclap{x \in S}}
		f(x) 
	\;:=\;
	\int\limits_{\mathclap{x \in S}}
		f(x) \; \mathrm d\mu(x)
	\;:=\;
		 \sum_{x \in S} f(x) \, \nu^2\,.
\end{equation}~\\[-1.5ex]
(For the sake of brevity, we often use the standard convention of suppressing the differential $\mathrm d\mu$, as on left-hand expression below, when the measure of integration is understood.)
Such integrals allow us to express sums with certain normalising factors more uniformly, by absorbing the factors into the measure $\mu$ by an appropriate choice of $\nu > 0$.
For a finite-dimensional Hilbert space $\mathcal H$ with standard basis $\ket{x} := \mathbf e_x$ for some index set $x \in \D$, we may define the (not-necessarily normalised) \emph{point-mass distributions} $\kket{x} = \nu^{-1} \ket{x} \in \cH$, and their adjoints $\bbra{x} = \kket{x}^\dagger$.
Then, if we have some `state-function' $\kket{f} := \int_{x \in \D} f(x) \;\kket{x}$ for an arbitrary function $f: \mathbb Z \to \mathbb C$, it is easy to show that~\\[-4.0ex]
\begin{equation}
\label{eqn:point-mass-integration}%
    \bbracket{z}{f}
\;:=\;
		\int\limits_{\mathclap{x \in \D}}
			 \bbracket{z}{x}  \; f(x)
\;=\;
		f(z),
\end{equation}~\\[-2.5ex]
similar to how Dirac measures are used with integration over $\R$.
While a similar result $\bracket{z}{f} = f(z)$ holds if we simply define $\ket{f} = \sum_{x \in D} f(x) \,\ket{x}$, couching this sort of analysis in terms of discrete integrals and point-mass functions $\kket{x}$ allows us to  accommodate scalar factors which may arise when manipulating expressions involving operators such as
$
    \sum_{x \in \D} \ket{x}\sox{n}\bra{x}\sox{m}
$
for $m,n > 1$.
This is an example of the sort of operator, for which book-keeping of scalar factors frequently arises in most versions of the ZX- or ZH-calculi.

By introducing the additional layer of abstraction, provided by discrete integrals and their accompanying point-mass functions $\kket{x}$, we  describe semantics for ZX- and ZH-diagrams which are simple, and which admits a system of rewrites which largely dispenses with the need for modifications to scalar gadgets in the diagrams.
This approach to notation, and the rewrites which we demonstrate, are applicable for generators representing operators on qudits of \emph{any} finite dimension, and enables the two calculi to be used interoperably as a single `ZXH-calculus'.
We present this approach in the hopes that it facilitates the development of practically useful extensions of these calculi beyond qubits.

\begin{proof}[Structure of this article.]
Section~\ref{sec:preliminaries} sets out number-theoretic preliminaries,  some background in string diagrams, and common approaches to defining ZX and ZH~calculi.
Section~\ref{sec:discrete-integrals} introduces discrete measures and integrals on $\D$, including what little measure theory we require, and considers the constraints that follow from a particular treatment of discrete Fourier transforms.
Section~\ref{sec:application-ZX-ZH} demonstrates how using such discrete integrals as the basis for a semantic map for ZX and ZH diagrams, leads to convenient representations of particular unitary operators and convenient rewrites for both the ZX and ZH calculi.
In the Appendices,
we outline a normal form for qudit ZH~diagrams for all $D>1$ (building on a similar construction for $D \!=\! 2$~\cite{BK-2019}), provide proofs of all the rewrites and the constraints on discrete measures through the treatment of the Fourier transform,
and remark on connections to related subjects such as Fourier analysis and Gauss sums.
\phantom{\qedhere}
\end{proof}

\begin{proof}[Related work.]
As we note above, there is recent and ongoing work~\cite{GW-2017,Wang-2018,Wang-2021,BC-2022,WY-2022,Majid-2022,Roy-2022,PWSYYC-2023,RWY-2023} on ZX, ZH, and related calculi on qudits of dimension $D>2$ (though often restricted to the case of $D$ an odd prime).
Our work is influenced in particular by Booth and Carette~\cite{BC-2022}, and  Roy~\cite{Roy-2022}, and we are aware of parallel work by collaborations involving these authors~\cite{RWY-2023,PBCWY-2023}.
Our work is distinguished in presenting convenient semantics for both ZX~and ZH~diagrams for arbitrary ${D\!>\!1}$, notably including the case where $D$ is composite or even.
\phantom{\qedhere}
\end{proof}

\section{Preliminaries}
\label{sec:preliminaries}

\subsection{Number-theoretic preliminaries}
\label{sec:maths-prelims}

Let $D>1$ be a fixed integer, and $\omega = \e^{2\pi i \!\!\;/\!\!\;D}$.
We assume basic familiarity with number theory, in particular with $\Z_D$, the integers modulo~$D$.
While it is common to associate $\Z_D$ with the set $\{0,1,\ldots,D\!-\!1\}$ of  non-negative `residues' of integers modulo $D$, one may associate them $\Z_D$ with any contiguous set of residues $\D = \{L,L\,{+}\,1,\ldots,U\,{-}\,1,U\}$ where $U-L+1 =D$.%
    \footnote{%
        The reader may wonder why we do not simply adopt the conventional choice of $\D = \{0,1,\ldots,D-1\}$.
        There are multiple reasons, the simplest of which being that allowing for the index to include negative integers may be useful for representing certain `quantum numbers' in application to physics.%
    }
We may then occasionally substitute $\Z_D$ for $\D$ when this is unlikely to cause confusion: this will most often occur in the context of expressions such as $\omega^{xy}$, which is well-defined modulo $D$ in each of the variables $x$ and $y$ (\emph{i.e.},~adding any multiple of $D$ to either $x$ or $y$ does not change the value of the expression).
In such an expression, while we may intend for one of $x$ or $y$ or both may be an element of $\Z_D$ in principle, they would in practise be interpreted as a representative integer $x,y \in \D \subseteq \Z$.

\subsection{String diagrams}
\label{sec:string-diagrams}

ZX- and ZH-diagrams are examples of \emph{string diagrams}, which can be described as diagrams composed of dots (or boxes) and wires, where the wires denote objects and the dots/boxes denote maps on those objects.

In the string diagrams which we consider in this article, diagrams are composed of dots or boxes, and wires.
These diagrams can be described as being a composition of `generators',
which typically consist of one (or zero) dots/boxes with some amount of meta-data, and any number (zero or more) directed wires, where the direction is usually represented by an orientation in the diagram.
(In this article, wires are oriented left-to-right, though they are also allowed to bend upwards or downwards.)
For any two diagrams $\cD_1$ and $\cD_2$, we may define composite diagrams $\cD_1 \!\!\;\otimes\!\!\; \cD_2$ and $\cD_1 \!\!\;\mathbin;\!\!\; \cD_2$, represented schematically by~\\[-3.25ex]
\begin{equation}
\begin{aligned}
	\begin{tikzpicture}
		\node (D) at (0,0) [draw=black, line width=.75pt,
				minimum width=1em, minimum height=9ex]
				{$\cD_1 \otimes \cD_2$};
		\foreach \dy in {-0.45,-0.35,-0.25,-0.15,0.15,0.25,0.35,0.45} {
			\draw ($(D.west) + (0,\dy)$) -- ++(-0.25,0);
			\draw ($(D.east) + (0,\dy)$) -- ++(0.25,0);
		}
	\end{tikzpicture}
\end{aligned}	
\;\;=\;\;
\begin{aligned}
	\begin{tikzpicture}
		\node (D1) at (0,.75) [draw=black, line width=.75pt,
				minimum width=1em, minimum height=4ex]
				{$\cD_1$};
		\node (D2) at (0,0) [draw=black, line width=.75pt,
				minimum width=1em, minimum height=4ex]
				{$\cD_2$};
		\foreach \n in {D1,D2} {%
			\foreach \dy in {-0.15,-0.05,0.05,0.15} {
			\draw ($(\n.west) + (0,\dy)$) -- ++(-0.25,0);
			\draw ($(\n.east) + (0,\dy)$) -- ++(0.25,0);
		}}
	\end{tikzpicture}
\end{aligned}\quad;
\qquad
\qquad
\begin{aligned}
	\begin{tikzpicture}
		\node (D) at (0,0) [draw=black, line width=.75pt,
				minimum width=1em, minimum height=5ex]
				{$\cD_1 \mathbin; \cD_2$};
		\foreach \dy in {-0.15,-0.05,0.05,0.15} {
			\draw ($(D.west) + (0,\dy)$) -- ++(-0.25,0);
			\draw ($(D.east) + (0,\dy)$) -- ++(0.25,0);
		}
	\end{tikzpicture}
\end{aligned}	
\;\;=\;\;
\begin{aligned}
	\begin{tikzpicture}
		\node (D1) at (0,0) [draw=black, line width=.75pt,
				minimum width=1em, minimum height=4ex]
				{$\cD_1$};
		\node (D2) at (1.125,0) [draw=black, line width=.75pt,
				minimum width=1em, minimum height=4ex]
				{$\cD_2$};
		\foreach \n in {D1,D2} {%
			\foreach \dy in {-0.15,-0.05,0.05,0.15} {
			\draw ($(\n.west) + (0,\dy)$) -- ++(-0.25,0);
			\draw ($(\n.east) + (0,\dy)$) -- ++(0.25,0);
		}}
	\end{tikzpicture}
\end{aligned}\quad,
\end{equation}~\\[-2.25ex]
which we call the `parallel' and `serial' composition of $\cD_1$ and $\cD_2$.
In the latter case we require that the number of output wires of $\cD_1$ (on the right of $\cD_1$) equal the number of input wires of $\cD_2$ (on the left of $\cD_2$), for the composition to be well-defined.

String diagrams may be used to denote maps in a monoidal category $\mathbf C$ (in which objects can be aggregated to form composite objects through a parallel product, which we denote by `$\otimes$').
This is done through a \emph{semantic map} $\sem{\,\cdot\,}$ which maps each generator to a map in $\mathbf C$.
This semantic map is defined to be consistent with respect to composition, in the sense that~\\[-3.75ex]
\begin{equation}
	\Bigsem{\cD_1 \otimes \cD_2} \;=\; \bigsem{\cD_1} \otimes \bigsem{\cD_2},
\qquad
\qquad	
	\Bigsem{\cD_1 \mathbin; \cD_2} \;=\; \bigsem{\cD_2} \circ \bigsem{\cD_1}.
\end{equation}~\\[-4.0ex]
Note the reversal of the order for sequential composition, which is just an artefact of the difference in orientation of diagrams (left-to-right), and the conventional right-to-left application order of functions that is common \emph{e.g.}~in quantum information theory. 

\subsection{Preliminary remarks on ZX and ZH diagrams}

ZX and ZH diagrams are string diagrams which denote multi-linear operators on some finite-dimensional vector space $\cH \cong \mathbb C^{\D}$, equipped with a standard basis $\lvert x \rangle$ for $x \in \D$ and functionals $\bra{x} = \ket{x}\herm$.
(For $\D$ a set of $D$ consecutive integers, we may use arithmetic expressions, such as $\ket{x+y}$, to index basis vectors; specifically in the labels of `kets' and `bras', such expressions can be understood to be evaluated mod~$D$.)
The parallel product in this case is the usual tensor product $\otimes$, and the sequential product is composition of operators.
For a generator $\cD$ with $m$ input wires and $n$ output wires, one assigns an operator $\sem{\cD}: \cH\sox{m} \to \cH\sox{n}$.
To represent string diagrams to represent maps in which some of the `parallel' operands are being permuted or unaffected, we also consider generators consisting only of wires.
We consider four such generators, to which we assign semantics as follows:~\\[-3.0ex]
\begin{small}
\begin{gather}
\label{eqn:stringGenerators}
\mspace{-36mu}
\begin{aligned}
  \bigsem{\input{./Figures-Source/id-wire.tikz}\,}
    &=\!\:
    \sum_{\mathclap{x \in \D}} 
    	\ket{x}\bra{x}
,&
    \mspace{6mu}
    \Bigsem{\!\!\:\input{./Figures-Source/swap.tikz}}
    &=\!\:
    \sum_{\mathclap{x,y \in \D}}
    	\ket{y,\!\!\:x\!\:}\bra{x,\!\!\:y\!\:},
&
    \Bigsem{\!\!\:\input{./Figures-Source/cup.tikz}}\!
    &=\;\!
    \sum_{\mathclap{x \in \D}} 
    	\ket{x,\!\!\:x\!\:},&
    \mspace{6mu}
    \Bigsem{\!\!\:\input{./Figures-Source/cap.tikz}}\!
    &=\;\!
    \sum_{\mathclap{x \in \D}} 
    	\bra{x,\!\!\:x\!\:}.
\end{aligned}
\mspace{-36mu}
\end{gather}~\\[-2.5ex]
\end{small}
%
%
ZX and ZH~diagrams are designed with different priorities, but have common features.
ZX~diagrams are effective for representing operations generated by single-qubit rotations and controlled-NOT gates; in most cases (excepting, \emph{e.g.}, Refs.~\cite{Wang-2020,BC-2022}) it rests on the unitary equivalence of two conjugate bases.
ZH~diagrams were developed to facilitate reasoning about quantum circuits over the Hadamard-Toffoli gate set~\cite{Shi-2003,Aharonov-2003}.
Both were originally defined so that the semantics is preserved by a change in the presentation of the underlying graph, which preserves the connectivity of the diagram~\cite{CD-2011,CW-2018}.

\begin{proof}[ZX Diagrams.]
We define the following ZX generators on qudits with state-space $\cH$,~\\[-4.25ex]
    \begin{equation}
    \label{eqn:ZXnodeFamilies}
    \begin{gathered}
    \qquad
      \input{./Figures-Source/ZX-green-phase-dot-arity.tikz}
    \;,\qquad
      \input{./Figures-Source/ZX-red-phase-dot-arity.tikz}
    \;,\qquad
      \input{./Figures-Source/ZX-H-plus-box.tikz}
    \;,\qquad
      \input{./Figures-Source/ZX-H-minus-box.tikz}
    \;\;,
    \end{gathered}
    \end{equation}~\\[-3.5ex]
where $m,n \in \N$, and for any function $\Theta: \Z \to \C$.
(Our approach of using functions mildly extends the approach of Wang~\cite{Wang-2021}, who prefers to parameterise the generators with vectors indexed from $1$.
For the constant function $\Theta(x) = 1$, we may omit the label $\Theta$ entirely.)
We call these generators `green dots', `red dots', `Hadamard plus boxes', and `Hadamard minus boxes'.
The usual approach to assigning semantics to ZX~generators is by considering the green and red dots to represent similar operations, subject to different (conjugate) choices of orthonormal basis, and a unitary `Hadamard' gate relating the two bases.
One defines a semantic map $\sem{\,\cdot\,}$ in which the (lighter-coloured)  `green' dots are mapped to an action on the basis $\ket{x}$, and the (darker-coloured) `red' are mapped to an action on the basis $\ket{\smash{\omega^x}}$, where $\ket{\smash{\omega^k}} = \smash{\tfrac{1}{\sqrt D} \sum_x \omega^{-kx} \ket{x}}$%
\label{discn:def-Fourier-basis}
for $k \in \D$ (and where again $\omega = \e^{2\pi i / D}$).%
	\footnote{%
		In the notation of Ref.~\cite{BC-2022}, we have $\ket{\smash{\omega^k}} = \ket{k\:\!{:} X}$, up to a relabeling of the basis elements of $\cH$.
	}
The conventional choice would be, for a green dot, to assign an interpretation such as $\smash{\sum_{x \in \D} \Theta(x) \ket{x}^{\!\otimes n}\!\bra{x}^{\!\!\;\otimes m}}$; and for a red dot, to assign the interpretation $\smash{\sum_{x \in \D} \Theta(x) \ket{\smash{\omega^x}}^{\!\otimes n}\!\bra{\smash{\omega^x}}^{\!\!\;\otimes m}}$.
Unfortunately, for $D>2$, such a conventional interpretation does not yield a `flexsymmetric'~\cite{Carette-2021} calculus, in effect because $\bra{\smash{\omega^a}}\trans = \ket{\smash{\omega^a}}^\ast = \ket{\smash{\omega^{-a}}}$.
In particular, the conventional approach described just above would mean that the equality~\\[-3.5ex]
\begin{equation}
	\label{eqn:red-deg-2-dots-flexsymmetric}
	\biggsem{\,%
	\begin{aligned}~\\[-2ex]
    \begin{tikzpicture}
      \node (X) at (0,0) [X dot, label=above:\small$\Theta$] {};
      \draw (X) -- ++(.625,0);
      \draw (X) -- ++(-.125,0) arc (270:90:.3125) -- ++(.75,0);
    \end{tikzpicture}
    \end{aligned}
    \,}
    \;=\;    
	\biggsem{\!\! %
	\begin{aligned}
    \begin{tikzpicture}
      \node (X) at (0,0) [X dot, label=left:\small$\Theta$] {};
      \draw (X) .. controls (0.125,0.3175) .. ++(.625,0.3125);
      \draw (X) .. controls (0.125,-0.3125) .. ++(.625,-0.3125);
    \end{tikzpicture}
    \end{aligned}
    \,}
    \;=\;
	\biggsem{\,%
	\begin{aligned}~\\[-4.5ex]
    \begin{tikzpicture}
      \node (X) at (0,0) [X dot, label=below:\small$\Theta$] {};
      \draw (X) -- ++(.625,0);
      \draw (X) -- ++(-.125,0) arc (90:270:.3125) -- ++(.75,0);
    \end{tikzpicture}
    \end{aligned}
    \,}
\end{equation}~\\[-1.25ex]
would not hold: the first diagram would denote $\sum_x \Theta(x) \ket{\smash{\omega^{-x}, \omega^x}}$, the second would denote $\sum_x \Theta(x) \ket{\smash{\omega^x, \omega^x}}$, and the third would denote $\sum_x \Theta(x)\ket{\smash{\omega^x, \omega^{-x}}}$. 
This represents a way in which such a calculus would fail to have the useful syntactic property that ``only the connectivity matters''~\cite{CD-2011,CW-2018}; and other inconveniences would also arise, which would make these diagrams more difficult to work with.
To avoid this problem, we endorse the convention adopted by Refs.~\cite{BC-2022,WY-2022} of involving a generator which is related to the green dot by \emph{different} unitary transformations on the inputs and outputs, but which differ only by a permutation.
We then interpret the generators of Eqn.~\eqref{eqn:ZXnodeFamilies} as operators using a model $\sem{\,\cdot\,}$ which satisfies~\\[-3.5ex]
  \begin{small}
  \begin{equation}{}
  \label{eqn:ZX-conventional-model}
  \begin{aligned}{}
  \Biggsem{\!\!\!\input{./Figures-Source/ZX-green-phase-dot-arity.tikz}\!\!\!}
  &\propto\!\;
  	\sum_{x \in \D}
	    \Theta(x) \, \ket{x}^{\!\otimes n}\!\bra{x}^{\!\!\;\otimes m}
\;,
%
%
&\quad\!\!
  \Bigsem{\!\!\: \input{./Figures-Source/ZX-H-plus-box.tikz} \!\!\:}
  &\propto\,
  	\mathop{\sum \sum}_{x,k \in \D}\;
  		\omega^{k x } \ket{x}\bra{k}
\\[0.75ex]
  \Biggsem{\!\!\!\input{./Figures-Source/ZX-red-phase-dot-arity.tikz}\!\!\!}
  &\propto\!\;
  	\sum_{k \in \D} \!\!\;
  	\Theta(k) \,
  	\ket{\smash{\omega^{-k}}}\sox{n} \!\bra{\smash{\:\!\omega^{k}}\;\!}\sox{m}
  	\;,
&\quad\!\!\!\!\!\!\!
  \Bigsem{\!\!\: \input{./Figures-Source/ZX-H-minus-box.tikz} \!\!\:}
  &\propto\,
  	\mathop{\sum \sum}_{x,k \in \D}\;
  		\omega^{- k x } \ket{x}\bra{k}
  \end{aligned}
  \end{equation}
  \end{small}~\\[-2.25ex]
so that the `Hadamard' plus and minus boxes are proportional to the quantum Fourier transform $\ket{\smash{\omega^{k}}} \mapsto \ket{k}$ (\emph{i.e.},~the inverse discrete Fourier transform), and its adjoint.
\phantom{\qedhere}
\end{proof}

\begin{proof}[ZH Diagrams.]
We define the following ZH generators on qudits with Hilbert space $\cH$,~\\[-3.75ex]
    \begin{equation}
    \label{eqn:ZHnodeFamilies}
    \begin{gathered}
    \qquad\;\;
      \input{./Figures-Source/ZH-white-dot-arity.tikz}
    ,\quad
      \input{./Figures-Source/ZH-H-phase-box-arity.tikz}
    ,\quad
      \input{./Figures-Source/ZH-gray-dot-arity.tikz}
    ,\quad
      \input{./Figures-Source/ZH-gen-not-dot.tikz}
    \;\;,
    \end{gathered}
    \end{equation}~\\[-3.25ex]
where $m,n \in \N$, $c \in \Z$, and for any function $\mathrm{A} : \Z \to \C$.
(If $\mathrm{A}(t) = \alpha^t$ for some $\alpha \in \C^\times$, we may write the scalar $\alpha$ in place of $\mathrm{A}$, consistent with the notation for ZH generators in Refs.~\cite{BK-2019,dB-2021}.
Following Roy~\cite{Roy-2022}, we later define a further short-hand notation for $\mathrm{A}(t) = \uchi_c(t) = \exp(2\pi i c t/D)$ with $c \in \Z$.)
We call these generators `white dots', `H-boxes', `gray dots', and `generalised-not dots'.%
	\,\footnote{%
		We follow Ref.~\cite{dB-2021} in considering the gray and not dots to be (primitive) generators, rather than gadgets or `derived generators', \emph{e.g.},~as in Refs.~\cite{BK-2019,Roy-2022}.
	}
We interpret the generators of Eqn.~\eqref{eqn:ZHnodeFamilies} as operators using a model $\sem{\,\cdot\,}$ which satisfies the following:~\\[-3.75ex]
\begin{small}%
\begin{equation}{}
\label{eqn:ZH-conventional-model}
\mspace{-18mu}
\begin{aligned}
  \Biggsem{\!\!\!\input{./Figures-Source/ZH-H-phase-box-arity.tikz}\!\!}
  &\propto\;
    \mathop{\sum \sum}_{%
      \mathclap{
        {x \in \D^m\!\!\!\;,\, y\in \D^n}
      }}
      \,
      \mathrm{A}(x_1 \!\cdot\!\cdot\!\cdot x_m y_1 \!\cdot\!\cdot\!\cdot y_n) \,
      \ket{y}\!\!\bra{x}
%
%
&
%
%
  \Bigsem{\input{./Figures-Source/ZH-gen-not-dot.tikz}}
  &\propto
  	\sum_{x \in \D} \ket{-c{-}x}\bra{x}
    \mspace{-26mu}
%
%
\\[.75ex]
%
%
  \Biggsem{\!\!\!\input{./Figures-Source/ZH-white-dot-arity.tikz}\!\!}
  &\propto
  	\sum_{x \in \D}
    \ket{x}^{\!\otimes n}\!\bra{x}^{\!\!\;\otimes m}
%
%
&\mspace{-80mu}
%
%
  \Biggsem{\!\!\!\input{./Figures-Source/ZH-gray-dot-arity.tikz}\!\!}
  &\propto\;
    \mathop{\sum \sum}_{%
      \mathclap{\substack{
        {x \in \D^m \!\!\:,}
        \,
        {y \in \D^n} \\[.5ex]
      	\sum\limits_h \!\!\; x_h + \sum\limits_k \!\!\; y_k \;\!\equiv\, 0
      }}}
      \;\;
      \ket{y}\!\!\bra{x} 
  	,
\\[-2ex]
\end{aligned}%
\mspace{-30mu}
\end{equation}%
\end{small}~\\[-1.0ex]
where for the gray dots, we constrain the indices $x \in \D^m$ and $y \in \D^n$, so that the sum of their entries is $0$ mod~$D$; and for the not-dots we interpret the index of the vector $\ket{-c-x}$ modulo $D$.
By contrast, note that for the H-boxes, we consider the products of the input and output labels $x_1, \ldots, x_m, y_1, \ldots, y_m$ as \emph{integers},%
	\footnote{%
		Note that our use of the index-set $x \in \D = \{L,L\,{+}\,1,\ldots,U\,{-}\,1,U\}$ means that the exponential function $t \mapsto \alpha^t$ for $\alpha = 0$ is not well-defined if $L < 0$.
		We may instead consider a function $\mathbf X_{\{0\}}: \Z \to \C$ given by $\mathbf X_{\{0\}}(t) = 1$ for $t = 0$, with $\mathbf X_{\{0\}}(t) = 0$ otherwise: this is substitution is adequate, \emph{e.g.},~for applications to counting complexity~\cite{BKK-2021,LMW-2022}.
	}
\emph{i.e.},~elements of $\mathbb Z$, whose product is the argument of $\mathrm A$ in the the expression $\mathrm A(x_1 \cdots y_m)$.
In particular, we fix the semantics so that~\\[-2.5ex]
\begin{equation}
  \label{eqn:ZH-scalar-box}
  \Bigsem{\input{./Figures-Source/ZH-scalar-box.tikz}}
  \;\,=\;\,\;
    \sum_{\mathclap{\text{(singleton)}}}
      \;
      \alpha^{\text{(empty product)}} \cdot 1
  \;=\;
    \alpha^1
  \;=\;
    \alpha,
\end{equation}~\\[-2.5ex]
again using the short-hand that $\alpha \in \mathbb C^\times$ stands for the function $\mathrm A(t) = \alpha^t$.
\phantom{\qedhere}
\end{proof}

\begin{proof}[Remarks on semantics, and rewriting systems.]
Eqns.~\eqref{eqn:ZX-conventional-model} and~\eqref{eqn:ZH-conventional-model}  describe not one semantic map $\sem{\,\cdot\,}$ for ZX diagrams or ZH diagrams, but rather the conventional approach (with minor elaborations) to choosing such semantic maps.
A specific semantic map determines which pairs of diagrams have the same semantics, and therefore which \emph{diagrammatic rewrites} are \emph{sound} (\emph{i.e.},~which local transformations one may perform to a diagram without changing its semantics).
We suggest that rewrite systems, in which the most commonly used diagrammatic rewrites can be expressed simply, are to be preferred over others.
However, this depends on obtaining a semantic map $\sem{\,\cdot\,}$ for which such a rewrite system is sound.
\begin{itemize}
\item 
    Some authors (\emph{e.g.},~\cite{CD-2011,CW-2018}) prefer to define semantics only up to proportionality, in which case Eqns.~\eqref{eqn:ZX-conventional-model} and~\eqref{eqn:ZH-conventional-model} suffice to determine when two diagrams are equivalent up to an neglected scalar factor.
    This has the virtue of simplicity, but does not provide the precision needed for all applications one might wish to consider for these calculi. 
\item
    Most `scalar exact' treatments of ZX and ZH fix a map $\sem{\,\cdot\,}$   by replacing the proportionalities in Eqns.~\eqref{eqn:ZX-conventional-model} and~\eqref{eqn:ZH-conventional-model} with equalities --- except for the `Hadamard' boxes of Eqn.~\eqref{eqn:ZX-conventional-model}, where a factor of $1/\sqrt D$ is used to yield unitary operators.
    However, the rewrites in those systems often involve book-keeping of auxiliary sub-diagrams (`scalar gadgets').
\item
    In the case $D=2$, Ref.~\cite{dB-2021} presents a different, unified semantic map $\sem{\,\cdot\,}_\nu$ for both ZX and ZH~diagrams, in order to support rewrites involving fewer scalar gadgets.
    However, the scalar factors involved in those semantics could be considered non-obvious, and does not provide insights into how one would achieve the same goal for arbitrary $D \ge 2$.
\end{itemize}
The aim of this work is to extend the results of Ref.~\cite{dB-2021}, providing a simple approach to fixing a semantic map $\sem{\,\cdot\,}$ for both ZX and ZH~diagrams for arbitrary $D \ge 2$, which supports a set of diagrammatic rewrites without much use of scalar gadgets.
\phantom{\qedhere}
\end{proof}

\section{Discrete integrals}
\label{sec:discrete-integrals}

Our main theoretical contribution is to demonstrate how discrete integrals provide a way to fix a semantic map for ZX and ZH~diagrams, with favourable properties.
In this section, we introduce discrete measures and discrete integrals independently of string diagrams, and consider the constraints on discrete measures obtained through a particular representation of discrete Fourier transforms.

\subsection{Introducing discrete measures and discrete integrals}
\label{sec:discrete-measures-ZD}

We begin by introducing more fully the concepts first described on page~\pageref{eqn:introducing-discrete-integral}.
For a set $\mathbf X$, let $\wp(\mathbf  X)$ be the power-set of $\mathbf  X$.
We may define a \emph{$\sigma$-algebra on $\mathbf  X$} to be a set $\Sigma \subset \wp(\mathbf  X)$ which contains $\mathbf  X$, which is closed under set complements ($S \in \Sigma \,\Longleftrightarrow\, \mathbf  X \!\!\;\setminus\!\!\; S \in \Sigma$), and which is closed under countable unions (if $S_1, S_2, \ldots \in \Sigma$, then $S_1 \cup S_2 \cup \cdots \in \Sigma$).
---
The purpose of defining $\Sigma$ is to allow the notion of a \emph{measure} $\mu: \Sigma \to \R \cup \{+\infty\}$ to be defined, where the sets $S \in \Sigma$ are the ones which have a well-defined measure.
Such a function $\mu$ is a measure, if and only if $\mu(\varnothing) = 0$, $\mu(S) \ge 0$ for all $S \in \Sigma$, and if~\\[-3.0ex]
\begin{equation}
	\mu\bigl(S_1 \cup S_2 \cup \cdots\bigr) = \mu(S_1) + \mu(S_2) + \cdots	
\end{equation}~\\[-3.5ex]
for any sequence of disjoint sets $S_j \in \Sigma$.
An example is the $\sigma$-algebra $\Sigma$ consisting of all countable unions of intervals over $\R$, with $\mu$ defined by assigning $\mu(J) = b{\!\;-\!\;}a$ to any interval $J \!\!\;=\!\!\; (a,b)$, $J \!\!\;=\!\!\; (a,b]$, $J \!\!\;=\!\!\; [a,b)$, or $J \!\!\;=\!\!\; [a,b]$ for $a \le b$.
A somewhat more exotic measure is the Dirac distribution $\mu_\delta$ on $\R$, for which $\mu_\delta(S) \!\!\;=\!\!\; 1$ if $0 \in S$, and $\mu_\delta(S) \!\!\;=\!\!\; 0$ otherwise.
(We remark on the Dirac distribution and related concepts in Appendix~\ref{apx:discrete-measures-R}.)
However, we are mainly interested in measures $\mu$ defined on subsets  $S \subseteq \D$, for which $\mu(S) \,\propto\, \# S$.

For the set $\D = \{L,L\,{+}\,1,\ldots,U\,{-}\,1,U\}$, consider the $\sigma$-algebra $\mathcal B = \wp(\D)$ consisting of all subsets of $\D$.
Define the measure ${\mu: \mathcal B \to \R}$ on this $\sigma$-algebra given by $\mu(S) \;=\; \# S \cdot \nu^2$, where $\nu > 0$ can in principle be chosen freely.
This presents $\D$ as a measure space, the purpose of which is to allow us to define (multi-)linear operators on $\cH$ as arising from integrals with respect to that measure.
For a function $f: \Z \to \C$, we may define a notion of integration of $f$ over a subset $S \subset \D$:~\\[-3.0ex]
\begin{equation}
\label{eqn:integral-over-ZD}
\begin{aligned}[b]
	\int\limits_{x \in S} \!\!f(x) \;	\mathrm d\mu(x)
\;&=\;
	\sum_{x \in S}	\, f(x) \, \mu(\{x\})
\;=\,
	 \sum_{x \in S} f(x) \,\nu^2
	 \;.
\end{aligned}
\end{equation}~\\[-2.25ex]
We may apply this notion of integration to operator-valued functions, as is typical for wave-functions in quantum mechanics.
For instance, one may define~\\[-3.5ex]
\begin{equation}
	\int\limits_{x \in S} \! f(x) \,\ket{x}\, \mathrm d\mu(x)
\;=\;
	\nu^2 \sum_{x \in S}\, f(x) \, \ket{x}.
\end{equation}~\\[-2.5ex]
In the usual approach to describing wave-functions over $\R$, one takes $\ket{x}$ to represent a point-mass distribution (\emph{i.e.},~not a vector $\vec v \in \C^\R$ for which $v_x = 1$), so that the equality~\\[-3.5ex]
\begin{equation}
	\bra{z} \Biggl[\;\; \int\limits_{\mathclap{x \in \R}} f(x) \!\;\ket{x} \, \mathrm dx \,\Biggr]
=\,
	\int\limits_{\mathclap{x \in \R}} f(x) \!\;\delta_z(x) \, \mathrm dx \,
\,=\,
	f(z),
\end{equation}~\\[-2.25ex]
holds.
Here $\delta_z(x)$ is a shifted Dirac distribution (see Appendix~\ref{apx:dirac-distributions} for more details).%
    \footnote{%
        While it is not necessary to understand our results, readers who are interested in connections between the integrals and measures presented here with integration over compact groups, may be interested in remarks which we make in Appendix~\ref{apx:continuous-models}.
    }
To avoid notational confusion, we prefer to reserve the symbol `$\ket{x}$' to represent a unit-norm standard basis vector in $\cH$ (\emph{i.e.},~a~vector $\vec v \in \cH$ such that $v_x = 1$), and introduce a symbol `$\kket{x}$' which denotes the vector $\kket{x}  \,=\, \tfrac{1}{\nu} \ket{x}$, specifically so that~\\[-3.5ex]
\begin{equation}{}
\mspace{-18mu}
\label{eqn:initial-attempt-point-mass-distribution}
\begin{aligned}[b]
	\bbra{z} \Biggl[\, \int\limits_{\;x \in \D} \!\!\!f(x) \;\kket{x} \; \mathrm d\mu(x) \:\!\Biggr]
\,&=\!\!
	\int\limits_{\;x \in \D} \!\!\!f(x) \;\bbracket{z}{x} \; \mathrm d\mu(x)
\;=\;
	\nu^2 \!\!\: \sum_{x \in \D}  f(x) \frac{\bracket{z}{x}}{\nu^2} 
\,=\;
	f(z)
	\,,
\end{aligned}
\end{equation}~\\[-2.0ex]
and also~\\[-3.25ex]
\begin{equation}{}
\mspace{-18mu}
\label{eqn:resolution-of-the-identity}
\begin{aligned}[b]
	\int\limits_{\;x \in \D} \!\!\!\kket{x}\bbra{x} \; \mathrm d\mu(x) 
\;&=\;
	\nu^2  \sum_{x \in \D} \! \frac{\ket{x}\bra{x}}{\nu^2}
\,=\;
	\sum_{x \in \D} \ket{x}\bra{x}
\;=\;
	\mathbf 1
	\,.
\end{aligned}
\end{equation}~\\[-2.25ex]%
The notation `$\kket{x}$' provides us the flexibility to consider which measures $\mu: \mathcal B \to \R$ are best suited for defining convenient semantics for ZX and ZH generators, while retaining the features provided by Dirac distributions over $\R$, and without constraining $\nu$.

For maps $U$ and $V$ described in this way, one may analyse compositions $UV$   in the same way that one would do if $U$ and $V$ were given by sums of operators: by using the expressions for $U$ and $V$ in terms of discrete integrals, manipulating expressions within the integrals, and well-judged use of algebraic identities such as Eqns.~\eqref{eqn:initial-attempt-point-mass-distribution} and~\eqref{eqn:resolution-of-the-identity}.
For instance, if we have~\\[-3.25ex]
\begin{align}
    U   \;&=\;   \mathop{\int\!\!\!\!\int}\limits_{\mathclap{x,y \in \D}} u_{x,y} \; 
                    \kket{x}\bbra{y}    
&
    V   \;&=\;   \mathop{\int\!\!\!\!\int}\limits_{\mathclap{w,z \in \D}} v_{w,z}  \; 
                    \kket{w}\bbra{z}    
\end{align}~\\[-2.25ex]
then we may express the composite operation $UV$ by~\\[-3.0ex]
\begin{equation}
\begin{aligned}[b]
        UV
   \;&=\;   
        \Biggl[\;\;\,
            \mathop{\int\!\!\!\!\int}\limits_{\mathclap{x,y \in \D}} u_{x,y} \;
            \kket{x}\bbra{y}
        \Biggr]    
        \Biggl[\;\;\,
            \mathop{\int\!\!\!\!\int}\limits_{\mathclap{w,z \in \D}} v_{w,z} \;  
            \kket{w}\bbra{z}
        \Biggr]
        \mspace{-160mu}
\\[-0.25ex]&=\;
        \mathop{\int\!\!\!\!\int\!\!\!\!\int\!\!\!\!\int}\limits_{\mathclap{\substack{w,x,y,z \in \D}}} u_{x,y} v_{w,z} \; 
        \kket{x}\bbracket{y}{w} \bbra{z}
        \mspace{-18mu}     
&
   \;&=\; 
        \mathop{\int\!\!\!\!\int}\limits_{\mathclap{x,z \in \D}} \;\,
            \Biggl(\;\;
                \mathop{\int\!\!\!\!\int}\limits_{\mathclap{w,y \in \D}}
                    u_{x,y} v_{w,z} \; \bbracket{y}{w}
            \Biggr)\,
        \kket{x} \bbra{z}     
\\&&&=\;   
        \mathop{\int\!\!\!\!\int}\limits_{\mathclap{x,z \in \D}} \;\;
            \Biggl(\;\;\,
                \int\limits_{\mathclap{y \in \D}}
                    u_{x,y} v_{y,z} 
            \Biggr)\,
        \kket{x} \bbra{z}     \;.
\\[-3ex]
\end{aligned}
\end{equation}%
Composition of such integrals generalises straightforwardly for tensors of any signature: examples can be seen in Appendix~\ref{apx:ZXH-phase-gadgets}.
(Appendix~\ref{apx:proofs-rewrites} makes heavy use of such analysis of compositions to prove the soundness of various diagrammatic rewrites of ZX and ZH~diagrams.)
The only distinction between this approach and one expressed directly in terms of summation, are the scalar factors which are subsumed in the integral notation and operators such as $\kket{x}\bbra{z}$, both of which are governed by the choice of measure $\mu$.

\subsection{Constraints on normalisation motivated by the Fourier transform}
\label{sec:normalisation-constraints-unitary-FT}

Having defined the discrete measure (and discrete integrals) over $\D$, and the corresponding point-mass distributions $\kket{x}$ to satisfy Eqn.~\eqref{eqn:point-mass-integration}, we may consider how this might influence our approach to analysis of complex-valued functions over $\D$ (or $\Z_D$, using a similar measure).

In analogy to a common representation%
	\footnote{%
		We emulate the presentation of the Fourier transform in terms of an oscillation frequency $k$ (including the minus sign in the exponent, which for historical reasons is absent in the definition of the quantum Fourier transform).
		The main difference between Eqn.~\eqref{eqn:FT-of-f} and the usual Fourier transform over $\R$ is the factor of $1\!\!\;/\!\!\;D$ in the exponent: this can be shown to arise from a representation of functions $f: \Z_D \to \C$ in terms of discrete distributions on $\R$
		(see Appendix~\ref{apx:FT-and-measure}).
	}
of the Fourier transform of functions on $\R$,
we may describe the (discrete) Fourier transform of a function $f: \Z_D \to \C$ by
~\\[-3.75ex]
\begin{equation}
\label{eqn:FT-of-f}
	\hat f(k)
\;=\;	
	\int\limits_{\;\mathclap{x \in \Z_D}}
		\e^{-2\pi i k x / D} \, f(x)\;\mathrm d\mu(x) .
\end{equation}~\\[-2.75ex]
In principle, the domain $\Z_D$ of $\hat f$ indexes a character $\chi_k(x) = \e^{-2\pi i k x / D}$ in the dual group $\widehat{\Z_D}$.
The dual group $\widehat{\Z_D}$ can itself be assigned a measure $\tilde \mu$ which is in principle \emph{independent} of $\mu$.
As $\Z_D$ is a finite abelian group, we use the fact that there is an isomorphism $\epsilon: \widehat{\Z_D} \to \Z_D$ to describe $\hat f$ as a function $\mathbb Z_D \to \mathbb C$.
The isomorphism $\epsilon$ induces a measure $\mu' = \tilde \mu \circ \epsilon^{-1}$ on $\Z_D$, which may differ from $\mu$ and which would be relevant to any integrals involving the argument of $\hat f$.%
    \footnote{%
        The precise relationship between $\mu$ and $\mu'$, corresponds to the question in physics of the choice of units for $x$ and $k$ as continuous variables ranging over $\R$.%
    }
--- Note that there are different conventions for normalising the Fourier transform (over $\R$ or $\Z_D$): one might consider modifying Eqn.~\eqref{eqn:FT-of-f} to include a non-trivial scalar factor on the right-hand side.
This is related to the questions of whether we take the Fourier transform $f \mapsto \hat f$ to preserve the $\ell_2$-norm $\lVert f \rVert_2 = \Bigl(\int_x \,\lvert f(x) \rvert^2 \, \mathrm d\mu(x)\Bigr)\big.^{\!\!\;1/2}$, and whether we take $\mu'$ to differ from $\mu$.
We simply adopt the convention of defining the Fourier transform of $f: \Z_D \to \C$ as in Eqn.~\eqref{eqn:FT-of-f}, and consider the constraints that this imposes on these other considerations.

In analogy to standard practise in physics, we may use $f$ to describe a `wave-function',%
	\footnote{%
		Note that $\kket{f}$ may not be a unit vector; whether this is the case depends on the values taken by $f$.
	}%
~\\[-3.0ex]
\begin{equation}
	\kket{f}
\;:=\;
	\int\limits_{\;\mathclap{x \in \Z_D}}
		f(x) \; \kket{x} \; \mathrm d\mu(x)\,.
\end{equation}~\\[-2.0ex]
A similar ``wave function'' for $\hat f$ would involve the measure $\mu'$, the measure on the argument of $\hat f$, which may in principle differ from $\mu$:~\\[-3.75ex]
\begin{equation}
\label{eqn:initial-ket-f-hat}
	\kket{\:\!\smash{\hat f}\;\!}
\;=\;	
	\int\limits_{\;\mathclap{k \in \Z_D}}
		\hat f(k) \, \kket{k} \; \mathrm d\mu'(k)\,,
\end{equation}~\\[-2.75ex]
integrating with respect to that different measure.
Taking $\mu' \ne \mu$ would imply that the functions $f: (\Z_D,\mu) \to \C$, defined on $\Z_D$ considered as a space with measure $\mu$, are strictly speaking not of the same type as their Fourier transforms $\hat f: (\Z_D,\mu') \to \C$ which are defined over a space with a different measure.
String diagrams representing the transformations of such functions would then require wires of more than one type.
While this is admissible in principle, 
we prefer to consider $f$ and $\hat f$ to have the same measure space $(\Z_D,\mu)$ for their domains, so that we may treat them using string diagrams with wires of a single type, as we do in the ZX and ZH calculi.
Identifying $\mu' = \mu$, we obtain 
\\[-3.75ex]
\begin{equation}
\label{eqn:ket-f-hat}
\begin{aligned}[b]
	\kket{\:\!\smash{\hat f}\;\!}
\;&=\,	
	\int\limits_{\;\mathclap{k \in \Z_D}}
		\hat f(k) \, \kket{k} \; \mathrm d\mu(k)
\;=\,
	\mathop{\int \!\!\!\! \int}\limits_{\;\mathclap{k,x \in \Z_D}}
		\e^{-2\pi i k x / \!\!\: D}  f(x) \;\kket{k} \;\mathrm d\mu(x) \; \mathrm d\mu(k)
		\,.
\end{aligned}
\end{equation}~\\[-2.75ex]
This motivates the definition of the discrete Fourier transform operator $F$ over $\Z_D$, as~\\[-3.75ex]
\begin{align}
\label{eqn:integral-FT}
	F
\;&=\;	
	\mathop{\int \!\!\!\! \int}\limits_{\;\mathclap{k,x \in \Z_D}}
		\e^{-2\pi i k x / D} \;\kket{k}\bbra{x} \;\mathrm d\mu(x) \; \mathrm d\mu(k)
	\;,
\end{align}~\\[-2.25ex]%
so that $\kket{\smash{\hat f}\:\!} = F\kket{f}$; this is the interpretation given to the `Hadamard minus box' in Eqn.~\eqref{eqn:idealised-ZX-integrals}.
We adopt the convention that $F$ is unitary, to allow it to directly represent a possible transformation of state-vectors over $\cH$.
This has the benefit that the inverse Fourier transform can be expressed similarly (now suppressing the differentials $\mathrm d\mu$, for brevity):~\\[-3.0ex]
\begin{equation}{}
\label{eqn:integral-inv-FT}
\mspace{-18mu}
		f(x)
	\,=\,
		\bbra{x} F\herm \kket{\smash{\hat f}\:\!}
	\,=
	\mathop{\int \!\!\!\! \int \!\!\!\! \int}\limits_{\;\mathclap{x,h,k \in \Z_D}}
		\e^{2\pi i k x / D} \,\kket{x}\bbra{k} \Bigl( \hat f(h) \, \kket{h} \Bigr)
    \,=
	\int\limits_{\;\mathclap{k \in \Z_D}}
		\e^{2\pi i k x / D} \; \hat f(x)  \;.
\mspace{-9mu}
\end{equation}~\\[-2.5ex]%
The definition of $F$ in Eqn.~\eqref{eqn:integral-FT} and the constraint that it should be unitary, imposes a constraint on the measure $\mu$ on $\Z_D$.
We first prove a routine Lemma (which will be of some use in the Appendices in simplifying iterated integrals):

\smallskip
\begin{lemma}
\label{lemma:exponential-integral}
	Let $\omega = \e^{2\pi i\!\!\;/\!\!\;D}$ and $E \in \D$.
	Then \smash{
	$
		\displaystyle\int\limits_{\mathclap{k \in \Z_D}}
			\omega^{Ek} \; \mathrm d\mu(k)
		\,=\,
		\bbracket{E}{0} \, D\nu^4.
	$}
\end{lemma}
\smallskip
\begin{proof}
	This holds by reduction to the usual exponential sum:
	\begin{equation*}
	\begin{aligned}[b]
		\int\limits_{\mathclap{k \in \D}}
	 		\e^{2\pi i Ek/\!\!\:D} \; \mathrm d\mu(k)
	\;=\;
		\nu^2 \! \sum_{k \in \D}
			\bigl(\omega^E\bigr)^k 
	\;&=\;
		\left\{
		\begin{aligned}
			\nu^2\cdot \omega^{E L_D} \cdot \text{\small$\dfrac{(\omega^E)^D-1}{\omega-1}$}\,
			,&~ ~	\text{if $\omega^{E} \ne 1$}
		\\[2ex]
			\nu^2 \cdot D
			,&~ ~	\text{if $\omega^{E} = 1$}
		\end{aligned}
		\right\}
	\\[1ex]&=\;
		\delta_{E,0} \; D\nu^2 
	\;=\;
		\bracket{E}{0} D\nu^2 
	\;=\;
		\bbracket{E}{0} \, D\nu^4	\;.
	\end{aligned}
	\qedhere
	\end{equation*}
\end{proof}
\noindent
We may apply this in the case of the Fourier transform as follows.
If $F$ as expressed in Eqn.~\eqref{eqn:integral-FT} is unitary, we have~\\[-3.25ex]
\begin{equation}
\label{eqn:constraint-on-N-via-FT}
\begin{aligned}[b]{}
\mspace{-18mu}
	\mathbf 1
\,=\:\!
	F\herm F
\;\!&=\,
	\Biggl[\;\;\,
		\mathop{\int \!\!\!\! \int}\limits_{%
			\mathclap{
			y,h \in \D
		}}
	 		\e^{2\pi i hy/\!\!\:D} \; \kket{y} \bbra{h} \;
			 	\mathrm d\mu(y) \; \mathrm d\mu(h)
	\Biggr]
	\Biggl[\;\;\,
		\mathop{\int \!\!\!\! \int }\limits_{%
			\mathclap{
			k,x \in \D
		}}
	 		\e^{-2\pi i kx/\!\!\:D} \; \kket{k} \bbra{x} \;
			 	\mathrm d\mu(k) \; \mathrm d\mu(x)
	\Biggr]
\\&=\;
	\mathop{\int \!\!\!\! \int \!\!\!\! \int \!\!\!\! \int}\limits_{%
	\mathclap{
	y,h,k,x \in \D
	}}
	 \e^{2\pi i (hy - kx)/\!\!\:D} \; \kket{y} \bbracket{h}{k} \bbra{x} \;
	 	\mathrm d\mu(y) \; \mathrm d\mu(h) \; \mathrm d\mu(k) \; \mathrm d\mu(x)
\\[0.75ex]&=\;
	\mathop{\int \!\!\!\! \int \!\!\!\! \int}\limits_{%
	\mathclap{
	y,k,x \in \D
	}}
	 \e^{2\pi i k(y - x)/\!\!\:D} \; \kket{y} \bbra{x} \;
	 	\mathrm d\mu(y) \; \mathrm d\mu(k) \; \mathrm d\mu(x)
\\[.5ex]&=\;
	\mathop{\int \!\!\!\! \int}\limits_{%
	\mathclap{
	y,x \in \D
	}} \;\;
	 \Biggl[\;\;\; \int\limits_{\mathclap{k \in \D}}
	 \e^{2\pi i k(y - x)/\!\!\:D} \; \mathrm d\mu(k)  \Biggr]
	 \; \kket{y} \bbra{x} \;
	 	\mathrm d\mu(y) \; \mathrm d\mu(x)
\\[.75ex]&=\;
	\mathop{\int \!\!\!\! \int}\limits_{%
	\mathclap{
	y,x \in \D
	}} \,
	 \Bigl[ D\nu^4 \cdot \bbracket{y}{x} \Bigr]
	 \, \kket{y} \bbra{x} \;
	 	\mathrm d\mu(y) \; \mathrm d\mu(x)
\\[.25ex]&=\;
	D\nu^4 \!
	\mathop{\int}\limits_{%
	\mathclap{
	x \in \D
	}} \kket{x} \bbra{x} \;
	 	 \mathrm d\mu(x)
\;=\;
	D\nu^4 \cdot \mathbf 1
	 \,.
\end{aligned}	
\mspace{-18mu}
\end{equation}~\\[-2ex]
This implies that $\nu = D^{-1/4}$
(or equivalently, $N = \mu(\Z_D) = D \nu^2 = \sqrt D$).

As there are multiple conventions for representing the discrete Fourier transform, one might wish to consider how adopting a different convention to Eqn.~\eqref{eqn:FT-of-f} affects constraints on the measure $\mu$; we consider this question in Appendix~\ref{apx:alternative-normalisations-FT}.

\section{Semantics for ZX- and ZH-diagrams using discrete interals}
\label{sec:application-ZX-ZH}

We present an approach to simply and systematically define semantic maps for ZX and ZH generators, which \textbf{(a)}~yields simple diagrams for unitary transformations of interest, \textbf{(b)}~admits scalar-exact diagrammatic rewrites involving few scalar gadgets, and \textbf{(c)}~allows the two notational systems to be used seamlessly together.

Our approach is to subsume all considerations of normalising factors into the measure of a discrete integral, and its accompanying point-mass functions, as indicated on page~\pageref{eqn:introducing-discrete-integral}.
Our use of integrals and discrete measures in this way is standard, if somewhat uncommon in quantum information theory: see Ref.~\cite{Schlingemann-2004,Majid-2022} for comparable examples.
Our intent is explicitly to draw attention to the freedom involved in the choice of measure, as a way forward to defining a semantic map $\sem{\,\cdot\,}$ for ZX and ZH~diagrams that has desirable features.


Defining a discrete integral on $\D$ with $\mu(\D) = \sqrt{D}$, as we do in the preceding Section, allows us to easily define semantic maps for ZX and ZH diagrams with a number of convenient properties.
Let~\\[-3.75ex]
\begin{equation}
\label{eqn:Fourier-basis}
		\kket{\smash{\omega^k}}
	\;=\;
		F \kket{k}
	\;=\;
		\int\limits_{\mathclap{x \in \D}} \omega^{-kx} \,\kket{x}	\;
\end{equation}~\\[-2.5ex]
be the non-normalised point-mass distributions analogous to the Fourier basis states $\ket{\smash{\omega^k}}$ introduced on page~\pageref{discn:def-Fourier-basis} (so that $\kket{\smash{\omega^k}}$ is an $\omega^k$-eigenvector of the cyclic shift operator $X$ given by $X \ket{a} = \ket{a{+}1}$).
We then define a semantic map $\sem{\,\cdot\,}$ on the ZX generators of Eqns.~\eqref{eqn:ZXnodeFamilies},~\\[-3.0ex]
\begin{small}%
\begin{equation}
\label{eqn:idealised-ZX-integrals}%
  \mspace{-36mu}
  \begin{aligned}{}
  \Biggsem{\!\!\!\input{./Figures-Source/ZX-green-phase-dot-arity.tikz}\!\!\!}
  \!\:&=
  	\int\limits_{\mathclap{x \in \D}} 
	    \Theta(x) \; \kket{x}^{\!\otimes n}\bbra{x}^{\!\!\;\otimes m} 
%
%
&\quad
%
%
  \Bigsem{\!\!\: \input{./Figures-Source/ZX-H-plus-box.tikz} \!\!\:}
  \!\:&=
  F\herm
  =
    \!\!\:
  	\mathop{\int \!\!\!\! \int}\limits_{\mathclap{x,k \in \D}}
  		\e^{2\pi i k x \!\!\;/\!\!\; D} \kket{k}\bbra{x} 
%
%
\\[.75ex]
%
%
  \Biggsem{\!\!\!\input{./Figures-Source/ZX-red-phase-dot-arity.tikz}\!\!\!}
  \!\:&=
  	\int\limits_{\mathclap{k \in \D}} 
	    \Theta(k) \; \kket{\smash{\omega^{-k}}}\sox{n} \bbra{\smash{\,\omega^{k}\,}}\sox{m}
%
&
%
%
  \Bigsem{\!\!\: \input{./Figures-Source/ZX-H-minus-box.tikz} \!\!\:}
  \!\:&=
  F
  =
  \!\!\:
  	\mathop{\int \!\!\!\! \int}\limits_{\mathclap{x,k \in \D}}
  		\e^{-2\pi i k x \!\!\;/\!\!\; D} \kket{k}\bbra{x}
\end{aligned}
  	\mspace{-100mu}
\end{equation}%
\end{small}~\\[-2.5ex]
and the ZH generators of Eqn.~\eqref{eqn:ZHnodeFamilies}:%
~\\[-3ex]
\begin{small}%
\begin{gather}
\label{eqn:idealised-ZH-integrals}%
\qquad\qquad\qquad
\begin{aligned}[b]{}
%
  \Biggsem{\!\!\!\input{./Figures-Source/ZH-H-phase-box-arity.tikz}\!\!}
  \;&=\,
  \mathop{\int \!\!\!\!\int}_{%
      \mathclap{
        {\substack{x \in \D^m \\ y\in \D^n}}
      }}
      \!\mathrm{A}(x_1 \!\cdot \!\cdot\! \cdot x_m y_1 \!\cdot \!\cdot\! \cdot  y_n)\;
      \kket{y}\bbra{x}
     \,,
%
\\[-.25ex]
%
 \Biggsem{\!\!\!\input{./Figures-Source/ZH-gray-dot-arity.tikz}\!\!}
 \;&=\,
  \mathop{\int \!\!\!\!\int}_{%
      \mathclap{
        {\substack{x \in \D^m \\ y\in \D^n}}
      }}
    \bigl\langle\!\!\bigl\langle
    \;
    	\smash{\textstyle \sum\limits_h x_h + \sum\limits_k y_k}
    \,
    \big\vert
    \,
    0
    \,
    \bigr\rangle\!\!\bigr\rangle
      \;\;
      \kket{y}\bbra{x}
  	\,,
%
\end{aligned}
\\[-1ex]
\notag\begin{aligned}
 \Biggsem{\!\!\!\input{./Figures-Source/ZH-white-dot-arity.tikz}\!\!}
 \:&=\;
  	\int\limits_{\mathclap{x \in \D}}
    \kket{x}^{\!\otimes n}\!\bbra{x}^{\!\!\;\otimes m} 
  	\,,
%
&\qquad
%
  \Bigsem{\input{./Figures-Source/ZH-gen-not-dot.tikz}}
  \;&=\;
  	\int\limits_{\mathclap{x \in \D}}
  		\kket{-c{-}x}\bbra{x} 
  	\,,
\end{aligned}
\end{gather}%
\end{small}~\\[-2.5ex]
These semantics are consistent with those set out in Eqns.~\eqref{eqn:ZX-conventional-model} and~\eqref{eqn:ZH-conventional-model}, replacing the sums and the vectors $\ket{x}$ with discrete integrals and the corresponding point-mass distributions $\kket{x}$, and substitute proportionality relations with equalities. 
The discrete integrals (and point-mass functions) serve to specify specific scalar factors for the proportionalities.

\begin{figure}[p]
\begin{align*}{}
\mspace{48mu}
		{\vtikzfig[-2.5ex]{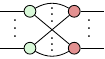}}
		\;&\longleftrightarrow\;
		{\vtikzfig[-3.5ex]{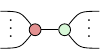}}
&
        \mspace{-48mu}
		{\vtikzfig[-3.5ex]{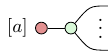}}
		\;&\longleftrightarrow\;
        {\vtikzfig[-2.5ex]{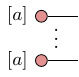}}
\\[-0.25ex]
\mspace{48mu}
		{\vtikzfig[-2.5ex]{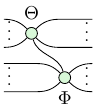}}
		\;\;&\longleftrightarrow\;\;
        {\vtikzfig[-2.5ex]{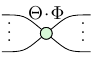}}
&
		{\vtikzfig[-2.5ex]{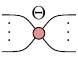}}
		\;&\longleftrightarrow\;
		{\vtikzfig[-2.5ex]{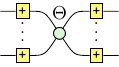}}
\\[-2.750ex]
\mspace{48mu}
		{\vtikzfig[-2.5ex]{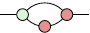}}
		\;\;&\longleftrightarrow\;\;
		{\vtikzfig[-2.5ex]{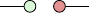}}
&
		{\vtikzfig[-2.5ex]{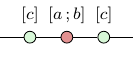}}
		\;&\longleftrightarrow\;
		{\vtikzfig[-2.5ex]{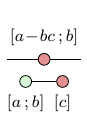}}
\\[-10.25ex]
\end{align*}
\begin{gather*}{}
		{\vtikzfig[-2.5ex]{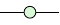}}
		\;\;\longleftrightarrow\;\;
		{\vtikzfig[-2.5ex]{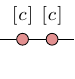}}
		\;\;\longleftrightarrow\;\;
        {\vtikzfig[-2.5ex]{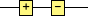}}
		\;\;\longleftrightarrow\;\;
        {\vtikzfig[-2.5ex]{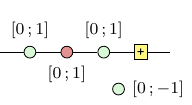}}
		\!\!\longleftrightarrow\;\;\;
        {\vtikzfig[-2.5ex]{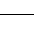}}
\\[-6.0ex]
\end{gather*}
\hrule
\smallskip
\caption{%
	\label{fig:eg-ZX-rewrites}
	A sample of the (scalar-exact) rewrites which are sound for ZX diagrams with semantics as in Eqn.~\eqref{eqn:idealised-ZX-integrals}, when $\nu = D^{-1/4}$.
	Throughout, we have $\Theta, \Phi: \Z \to \C$, and $a, b, c \in \Z$. 
	Node labels of the form ${[\:\! a\:\!]}$ or ${[\:\!a \s b\:\!]}$ stand respectively for the amplitude functions $x \mapsto \tau^{2ax}$ and $x \mapsto \tau^{2ax + bx^2}$, where $\tau = \exp(\pi i (D^2 {+} 1)/D)$. 
	A more complete list of rewrites, and proofs of their soundness, may be found in Appendix~\ref{apx:sound-ZX-rewrites} (page~\pageref{apx:sound-ZX-rewrites}).
}

\bigskip

\begin{gather*}
\qquad\quad
		{\vtikzfig[-3.5ex]{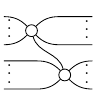}}
        \;\;\longleftrightarrow\;\;
        {\vtikzfig[-2.5ex]{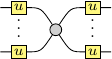}}
        \;\;\longleftrightarrow\;\;	
		{\vtikzfig[-2.5ex]{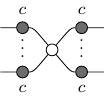}}
		\;\;\longleftrightarrow\;\;
		{\vtikzfig[-3.5ex]{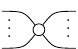}}
\\[-0.5ex]
\qquad\quad
		{\vtikzfig[-2.5ex]{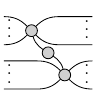}}
		\;\;\longleftrightarrow\;\;
		\;\;{\vtikzfig[-2.5ex]{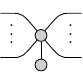}}\;\;	
		\;\;\longleftrightarrow\;\;
		{\vtikzfig[-2.5ex]{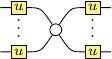}}
		\;\;\longleftrightarrow\;\;
		{\vtikzfig[-3.5ex]{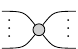}}
\\[1.0ex]
	\begin{aligned}
	    {\vtikzfig[-3.5ex]{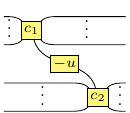}}
	    \longleftrightarrow\,
	    {\vtikzfig[-3.5ex]{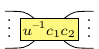}}
	&&
		\qquad
	&&
	   {\vtikzfig[-2.5ex]{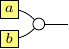}}
	   \;\longleftrightarrow\;
	   {\vtikzfig[-2.5ex]{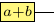}}
	&&
		\qquad
	&&
	   {\vtikzfig[-2.5ex]{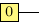}}
	   \;\longleftrightarrow\;
	   {\vtikzfig[-2.5ex]{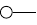}}
	\end{aligned}
\\[1.0ex]
\qquad\quad
	\begin{aligned}
        {\vtikzfig[-2.5ex]{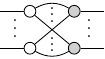}}
        \;\longleftrightarrow\;
        {\vtikzfig[-3.5ex]{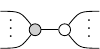}}
	&&\;\;\;
		\qquad
	&&
	   {\vtikzfig[-2.5ex]{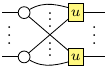}}
	   \;\longleftrightarrow\;
	   {\vtikzfig[-2.5ex]{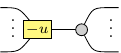}}
	\end{aligned}
\\[1.0ex]
\mspace{-24mu}
	\begin{aligned}
	   {\vtikzfig[-2.5ex]{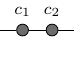}}
	   \;\longleftrightarrow\;
	   {\vtikzfig[-2.5ex]{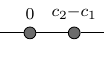}}
	&&
		\qquad
	&&
	   {\vtikzfig[-2.5ex]{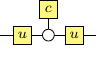}}
	   \;\longleftrightarrow\;
	   {\vtikzfig[-2.5ex]{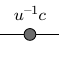}}
	&&
		\qquad
	&&
	   {\vtikzfig[-2.5ex]{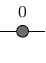}}
	   \;\longleftrightarrow\;
	   {\vtikzfig[-2.5ex]{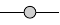}}
	\end{aligned}
\\[-4.25ex]
    \quad\qquad
		{\vtikzfig[-2.5ex]{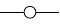}}
		\;\;\longleftrightarrow\;\;
		{\vtikzfig[-2.5ex]{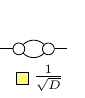}}
		\;\longleftrightarrow\;\;
		{\vtikzfig[-2.5ex]{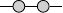}}
		\;\;\longleftrightarrow\;\;
		{\vtikzfig[-2.5ex]{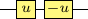}}
        \;\;\longleftrightarrow\;\;
        {\vtikzfig[-2.5ex]{id-wire}}
\\[-7.25ex]
\end{gather*}
\hrule
\smallskip
\caption{%
	\label{fig:eg-ZH-rewrites}
	A sample of the (scalar-exact) rewrites which are sound for ZH diagrams with semantics as in Eqn.~\eqref{eqn:idealised-ZH-integrals} when $\nu = D^{-1/4}$.
	H-boxes which are labeled inside with an integer parameter such as $c \in \Z$, indicate an amplitude of $\omega^c = \e^{2\pi i c/D}$; H-boxes labelled with `$\texttt+$' or `$\texttt-$' indicate $c = \pm 1$
	.
	Throughout, we have$a,b,c, c_1, c_2,u,v \in \Z$ (which may be evaluated modulo $D$), where in particular $u$ and $v$ are coprime to $D$.
    A more complete list of rewrites, and proofs of their soundness, may be found in Appendix~\ref{apx:sound-ZH-rewrites} (page~\pageref{apx:sound-ZH-rewrites}).
}
\end{figure}

These definitions are ones that we could choose to make, regardless of the measure $\mu$ that we consider for $\D$.
Regardless of the choice made for $\nu$, the above interpretations are certainly similar in their simplicity to the standard interpretations.
By taking $\nu = D^{-1/4}$ as suggested in the preceding section, 
we not only obtain rewrite systems involving very few scalar gadgets --- see Figs.~\ref{fig:eg-ZX-rewrites} and~\ref{fig:eg-ZH-rewrites} --- but also,
the most commonly considered states and unitary operations of qudit circuits  admit simple presentations using these semantics.
We may demonstrate this as follows.

\subsection{The stabiliser sub-theory of ZX for arbitrary $D>1$}
\label{sec:stabiliser-ZX}

We describe below a  \emph{stabiliser subtheory} of the ZX calculus, concerning ZX diagrams which suffice to represent stabiliser states~\cite{dB-2013} on systems of arbitrary dimension $D>1$.
These are characterised by ZX diagrams whose phase parameters are governed by restricted functions, for which arithmetic modulo $D$ plays a central role.

We begin by describing the stabiliser sub-theory of quantum circuits.
Following Ref.~\cite{dB-2013}, define the complex unit $\tau = \e^{\pi i(D^2 + 1)/D}$, which is relevant to the analysis of stabiliser circuits on qudits of dimension $D$.
The scalar $\tau$ is defined in such a way that $\tau^2 = \omega$, but also so that $\tau X\herm Z\herm$ is an operator of order $D$, where $X$ and $Z$ given by~\\[-3.75ex]
\begin{equation}
	X \ket{t}	\;=\;	\ket{t+1},
\qquad\qquad	
	Z \ket{t}	\;=\;	\omega^t \ket{t},
\end{equation}~\\[-4.0ex]
are the $D$-dimensional generalised Pauli operators.
(As always, arithmetic performed in the kets are evaluated modulo $D$.)
Choosing $\tau$ in this way makes it possible~\cite{dB-2013} to define a simple and uniform theory of unitary stabiliser circuits on qudits of dimension $D$, generated by the single-qudit operators\,%
	\footnote{%
		Despite the different convention we adopt for the labeling of the standard basis, the definitions below are equivalent to those of Ref.~\cite{dB-2013}: the relative phases $\tau^{\raisebox{-0.125ex}{$\scriptscriptstyle 2ax + bx^2$}}$ remain well-defined on substitution of values $x < 0$ with $D+x$, as $\tau^{\raisebox{-0.125ex}{$\scriptscriptstyle 2a(D+x) + b(D+x)^2$}} = \tau^{\raisebox{-0.125ex}{$\scriptscriptstyle 2aD + 2ax + bD^2 + 2bD + bx^2$}} =  \tau^{\raisebox{-0.125ex}{$\scriptscriptstyle 2ax + bx^2$}}$ (using the fact that $\tau^{\raisebox{-0.125ex}{$\scriptscriptstyle D^2$}} = \tau^{\raisebox{-0.125ex}{$\scriptscriptstyle 2D$}} = 1$ for both even and odd $D$).
	}%
~\\[-3.5ex]
\begin{equation}{}
\mspace{-24mu}
	S
\,=\!\!\:
	\int\limits_{\mathclap{x\in\D}} 
	\!
	\tau^{\,x^2} \;\kket{x}\bbra{x}
\;;\;
\mspace{24mu}
	F
\,=\!\!\:
	\mathop{\int\!\!\!\!\int}\limits_{\mathclap{k,x\in\D}} 
	\!
	\tau^{-2kx} \;\kket{k}\bbra{x}
\;;\;
\mspace{24mu}
	M_u
\,=\!\!\:
	\int\limits_{\mathclap{x\in\D}} \kket{ux}\bbra{x}  \;,
\end{equation}~\\[-2.75ex]
where in the case of $M_u$ we restrict to $u \in \Z$ which is relatively prime to $D$; and either one of the two-qudit operators~\\[-3.5ex]
\begin{align}
	\mathrm{CX}
\,&=
	\mathop{\int\!\!\!\!\int}\limits_{\mathclap{x,y\in\D}} \kket{x}\bbra{x} \otimes \kket{x{+}y}\bbra{y}
\;;
&
	\mathrm{CZ}
\,&=
	\mathop{\int\!\!\!\!\int}\limits_{\mathclap{x,y\in\D}} \tau^{2xy}\;\kket{x,y}\bbra{x,y} \;.
\end{align}~\\[-2.5ex]
Finally, the full range of stabiliser circuits also admit measurements in the standard basis (and bases which may be related to the standard basis by the above unitaries).

Booth and Carette~\cite{BC-2022} describe a version of the ZX calculus which is complete for this subtheory, for the special case of $D$ an odd prime.
Following them, we may describe how the semantics of Eqn.~\eqref{eqn:idealised-ZX-integrals} allows a simplification of these rewrites, extending them in most cases to arbitrary $D > 1$.
To this end, it will be helpful to use a slightly different notational convention to Booth and Carette~\cite{BC-2022}, we may easily denote these with ZX~diagrams using the semantics of Eqn.~\eqref{eqn:idealised-ZX-integrals}.
For $a,b \in \Z$, when parameterising a green or red dot, let $[a \s b]$ stand for the amplitude function $\Theta(x) = \tau^{2ax + bx^2}$, so that~\\[-3.75ex]
\begin{align}{}
	\biggsem{\!\!\!%
		\begin{aligned}{}
		~\\[-1.25ex]
		\begin{tikzpicture}[]
      		\node (z) at (0,0.375) [Z dot, label=left:\small{$[a \s b]$}] {};
      		\draw (z) -- ++(.5,0);
		\end{tikzpicture}	
		\end{aligned}\,}
	\,&=\,
		\int\limits_{\mathclap{x \in \D}}\!
			\tau^{\,2ax \,+\, bx^2} \,\kket{x}
\;;
&
	\biggsem{\!\!\!%
		\begin{aligned}{}
		~\\[-1.25ex]
		\begin{tikzpicture}[]
      		\node (x) at (0,0.375) [X dot, label=left:\small{$[a \s b]$}] {};
      		\draw (x) -- ++(.5,0);
		\end{tikzpicture}	
		\end{aligned}\,}
	\,&=\,
		\int\limits_{\mathclap{k \in \D}}\!
			\tau^{\,2ak \,+\, bk^2} \,\kket{\smash{\omega^{-k}}}
\;;
\end{align}~\\[-2.75ex]%
generalising these to dots with multiple edges (or with none) similarly to Ref.~\cite{BC-2022}.
When $b = 0$, we may abbreviate this function simply by $[a]$, so that we may represent the states $\kket{a}$ and $\kket{\omega^a}$ straightforwardly (albeit with the use of auxiliary red dots to represent an antipode operator, mapping $\kket{\omega^a} \mapsto \kket{\smash{\omega^{-a}}}$ and $\kket{a} \mapsto \kket{-a}$ for $a \in \Z_D$):~\\[-3.5ex]
\begin{subequations}{}%
\allowdisplaybreaks
\begin{align}
\mspace{-18mu}
	\biggsem{\!%
		\vtikzfig[-1ex]{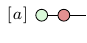}
	\,}
	\,&=\,
		\mathop{\int \!\!\!\!\int}\limits_{\mathclap{k,x \in \D}}\!
			\tau^{2ax} \,\kket{\smash{\omega^{-k}}}\bbracket{\smash{\omega^k}}{x}
	\;=\;
		\kket{\smash{\omega^a}}
\,;
\mspace{-12mu}
\\[0.25ex]
\mspace{-18mu}
	\biggsem{\!%
		\vtikzfig[-1ex]{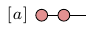}
	\,}
	\,&=\,
		\mathop{\int \!\!\!\! \int}\limits_{\mathclap{h,k \in \D}}\!
			\tau^{2ah} \,\kket{\smash{\omega^{-k}}}
							\bbracket{\smash{\omega^k}}{\smash{\omega^{-h}}}
	\;=\;
		\kket{a}
\,.
\mspace{-12mu}
\end{align}%
\end{subequations}~\\[-3.0ex]%
We may also easily represent the operators $Z$, and $X$ as $1 \to 1$ dots:~\\[-3.25ex]
\begin{equation}{}%
\begin{gathered}
\mspace{-18mu}
	\biggsem{\,
		\vtikzfig[-1ex]{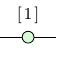}
	\,}
	\,=
		\int\limits_{\mathclap{x \in \D}}\!
			 	\tau^{2x}	\,
				\kket{x}\bbra{x}
	\,=\,
		Z
\,;
\qquad\qquad\qquad
\label{eqn:ZX-X-gadget}
	\biggsem{\,
		\vtikzfig[-1ex]{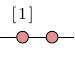}
	\,}
	\,=
		\int\limits_{\mathclap{h \in \D}}\!
			 	\tau^{2h}	\,
				\kket{\smash{\omega^{h}}}\bbra{\smash{\omega^{h}}}
	\,=\,
		X
\,.
\mspace{-12mu}
\end{gathered}
\end{equation}~\\[-3.0ex]%
Regarding the unitary stabiliser operators on qudits, we may express them without any phases, using multi-edges between green and red dots, or using Hadamard boxes:%
\begin{small}%
\begin{align}{}
\mspace{-20mu}
	\Biggsem{\!\:
		\vtikzfig[-1ex]{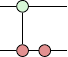}
	\!\:}
	&=
		\mathrm{CX}
,\;
&
	\Biggsem{\!\:%
		\begin{aligned}{}
		\begin{tikzpicture}[]
      		\node (c) at (0,0) [Z dot] {};
      		\draw (c) -- ++(.375,0);
      		\draw (c) -- ++(-.375,0);
      		\node (t) at ($(c) + (0,-.75)$) [Z dot] {};
      		\draw (t) -- ++(.375,0);
      		\draw (t) -- ++(-.375,0);
      		\draw (c) -- node [midway, small H box] {\tp} (t);
		\end{tikzpicture}	
		\end{aligned}\!\:}
	&=
		\mathrm{CZ}
,\;
&
	\biggsem{\,%
		\begin{aligned}{}
		\begin{tikzpicture}[scale=0.75]
      		\node  at (0,0.375) [Z dot, fill=none, draw=none, label=below:\phantom{\footnotesize{$[\;\!0 \s 1\;\!]$}}] {};
      		\node (x) at (0,0.375) [Z dot, label=above:\footnotesize{$[\;\! 0 \s 1\;\!]$}] {};
      		\draw (x) -- ++(.625,0);
      		\draw (x) -- ++(-.625,0);
		\end{tikzpicture}	
		\end{aligned}\,}
	&=
		S
,\;
&
	\Biggsem{\begin{aligned}
	\begin{tikzpicture}
		\node (Z) [Z dot] at (0,0) {};
		\draw (Z) -- ++(-0.375,0);
		\node (G) [X dot] at (1.125,0) {};
		\draw (G) -- ++(0.375,0) node [X dot] {} -- ++(0.375,0);
		\draw [out=70,in=110] (Z) to (G);
		\draw [out=45,in=135] (Z) to (G);
		\draw [out=-45,in=-135] (Z) to (G);
		\draw [out=-70,in=-110] (Z) to (G);
		\node at ($(Z)!0.375!(G) + (0,0.0875)$) {$\vdots$};
		\node at ($(Z)!0.5875!(G) + (0,0.)$) {$\left. \begin{matrix} \\[4ex] \end{matrix}\right\} \! u$};
	\end{tikzpicture}
	\end{aligned}}
	&=
		M_u
\,.
\mspace{-9mu}
\end{align}~\\[-1.75ex]%
\end{small}%
(The diagram shown for $M_u$ also generalises to operators $M_u = \int_x \kket{ux}\bbra{x}$ for $u$ not a multiplicative unit modulo $D$, though in that case the operator will not be invertible.)

The stabiliser subtheory of ZX may produce green or red dots \emph{of degree zero} with phase parameter $[a \s b]$ for some $a,b \in \Z$.
These may occur when evaluating the probability of measurement outcomes (\emph{e.g.},~in the standard basis) arising from a stabiliser qudit circuit.
As we show in Section~\ref{apx:sound-ZX-rewrites} (as a simple corollary of a more general fusion rule), we have~\\[-3.75ex]
\begin{equation}
	\begin{gathered}
		\Sem{11ex}{\vtikzfig[-2.5ex]{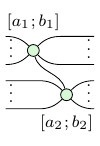}}
		\;=\;
		\Sem{9ex}{\vtikzfig[-0.5ex]{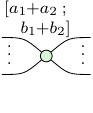}}
	\end{gathered}    
\end{equation}~\\[-3.0ex]
where each instance of ``\,$\vdots$\,'' denotes some number (zero or more)  of incident wires.
In the case where there are \emph{no} other dots connected to two green dots as above, the right-hand side would be an isolated dot denoting a scalar, for which we define the notation $\Gamma(a,b,D)$:~\\[-3.0ex]
\begin{equation}{}
\mspace{-18mu}
	\Bigsem{\!
		\begin{aligned}{}
		~\\[-1.25ex]
		\begin{tikzpicture}[]
      		\node (x) at (0,0.375) [Z dot, label=left:\footnotesize{$[a \s b]\!$}] {};
		\end{tikzpicture}	
		\end{aligned}\,}
	\;=\,
		\int\limits_{\mathclap{x \in \D}} \tau^{2ax + bx^2}
	=:\;
		\Gamma(a,b,D)
    \;.
\mspace{-18mu}
\end{equation}~\\[-2.5ex]
Evaluating such a discrete integral is connected with the subject of quadratic Gaussian sums, which we address in some detail in Appendix~\ref{apx:quadratic-Gaussian-integrals}.
As a result of the normalisation convention for our discrete integrals, it is possible to show (see Eqn.~\eqref{eqn:quadratic-Gauss-integral-formula} on page~\pageref{eqn:quadratic-Gauss-integral-formula}) that~\\[-3.75ex]
\begin{equation}
\label{eqn:quadratic-Gauss-integral}
	\Gamma(a,b,D)
\;:=\;
	\int\limits_{\mathclap{x \in \D}} \tau^{2ax + bx^2}	
	\,=\;
		\left\{
		\begin{aligned}
			\sqrt{t\,} \cdot \e^{i\gamma}		,
			&\quad	\text{if $t = \gcd(b,D)$ and $a$ is divisible by $t$};
		\\
			0,
			&\quad	\text{otherwise},									
		\end{aligned}
		\right.
\mspace{-18mu}
\end{equation}~\\[-2.5ex]
where $\gamma$ is a phase parameter described in more detail in Eqn.~\eqref{eqn:quadratic-Gauss-integral-formula}.
In particular, if $b$ is~a multiplicative unit modulo $D$, this represents a global phase factor.
(If we also have $a = 0$, then $\Gamma(a,b,D)$ is in fact a power of $\e^{\pi i /4}$.)
More generally, $ \Gamma(a,b,D)$ will either be $0$, or have magnitude $\sqrt t$, where $t = \gcd(b,D)$.

In this way, we obtain a diagrammatic language which is capable of expressing the rewrites similar to those described by Ref.~\cite{BC-2022}, while involving fewer scalar factors (see Figure~\ref{fig:ZX-rewrites} in Appendix~\ref{apx:sound-ZX-rewrites} for a more complete list of sound rewrites). 

\subsection{Multipliers and multicharacters in qudit ZH}
\label{sec:multicharacters}

It would be cumbersome to reason about stabiliser multiplication operators $M_u$ or iterated $\mathrm{CX}$ or $\mathrm{CZ}$ gates using parallel edges between dots.
Booth and Carette~\cite{BC-2022} describe how these may be denoted using gadgets called `multipliers', denoted
$\,\tikz
	\draw (0,0) -- node [midway, rfarr] {\t c} (1,0);
\,$
for $c \in \N$, which represent a limited form of scalable ZX~notation~\cite{CHP-2019,CDP-2021}.
Using discrete integrals and the semantics described in Eqn.~\eqref{eqn:idealised-ZX-integrals}, we would simply write~\\[-3.75ex]
\begin{equation}
	\Biggsem{\,\begin{aligned}
	\begin{tikzpicture}
		\node (a) [rfarr] at (0,0) {\t c};
		\draw (a) -- ++(-0.4375,0);
		\draw (a) -- ++(0.5,0);
	\end{tikzpicture}
	\end{aligned}}
\;=\;
	\Biggsem{\,\begin{aligned}
	\begin{tikzpicture}
		\node (Z) [Z dot] at (0,0) {};
		\draw (Z) -- ++(-0.375,0);
		\node (G) [X dot] at (1.125,0) {};
		\draw (G) -- ++(0.375,0) node [X dot] {} -- ++(0.375,0);
		\draw [out=70,in=110] (Z) to (G);
		\draw [out=45,in=135] (Z) to (G);
		\draw [out=-45,in=-135] (Z) to (G);
		\draw [out=-70,in=-110] (Z) to (G);
		\node at ($(Z)!0.375!(G) + (0,0.0875)$) {$\vdots$};
		\node at ($(Z)!0.5875!(G) + (0,0.)$) {$\left. \begin{matrix} \\[4ex] \end{matrix}\right\} \! c$};
	\end{tikzpicture}
	\end{aligned}}
\;=\;
	\int\limits_{\mathclap{x \in \D}} \kket{cx}\bbra{x}
\;.
\end{equation}~\\[-2.5ex]
Using these multipliers, Booth and Carette~\cite[p.\,24]{BC-2022} then define `Fourier boxes' $
	\begin{aligned}
	\begin{tikzpicture}
		\node (K) [H box] at (0,0) {\t{c}};
		\draw (K) -- ++(-0.5,0);
		\draw (K) -- ++(0.5,0);
	\end{tikzpicture}
	\end{aligned}
\,:=\,
	\begin{aligned}
	\begin{tikzpicture}
		\node (a) [rfarr] at (0,0) {\t c};
		\node (K) [small H box] at ($(a) + (0.625,0)$) {\tp};
		\draw (a) -- ++(-0.5,0);
		\draw (K) -- ++(0.375,0);
		\draw (a) -- (K);
	\end{tikzpicture}
	\end{aligned}
$ (using our notation for Hadamard boxes), 
whose interpretation coincides with the ones we assign using Eqn.~\eqref{eqn:idealised-ZH-integrals} to an H-box with an amplitude parameter $\omega^c$.
Using this as a primitive, and composing this with the inverse 
$\,\tikz
	\draw (0,0) -- node [midway, small H box] {\tm} (.75,0);
\,$
of the positive Hadamard box
$\,\tikz
	\draw (0,0) -- node [midway, small H box] {\tp} (.75,0);
\,$, we may directly describe multipliers instead as a ZH gadget, loosely following Roy~\cite{Roy-2022}:~\\[-3.75ex]
\begin{equation}
	\begin{aligned}
	\begin{tikzpicture}
		\node (K) [H box] at (0,0) {\t{c}};
		\node (h) [small H box] at ($(K) + (0.625,0)$) {\tm};
		\draw (K) -- ++(-0.5,0);
		\draw (h) -- ++(0.375,0);
		\draw (K) -- (h);
	\end{tikzpicture}
	\end{aligned}
\;\;=:\;\;
	\begin{aligned}
	\begin{tikzpicture}
		\node (a) [rfarr] at (0,0) {\t c};
		\draw (a) -- ++(-0.5,0);
		\draw (a) -- ++(0.5,0);
	\end{tikzpicture}
	\end{aligned}
\;\;.
\end{equation}~\\[-3.25ex]
On the left, we employ a short-hand for H-boxes with an amplitude parameter $\omega^c$.
This is short-hand for a character function $\uchi_c: \Z \to \C$ given by $\uchi_c(x) = \omega^{cx}$, which is well-defined modulo $D$, and which we may then regard as a character on $\Z_D$.
The function $\Z \times \Z \to \C$ given by $(x,y) \mapsto \uchi_c(xy)$ is a \emph{bicharacter}, which is also well-defined modulo $D$ on each of its arguments; and more generally we may consider \emph{multicharacters}, which are functions $\Z_D \times \cdots \times \Z_D \to \C$ given by $(x_1, \ldots, x_n) \mapsto \omega^{c x_1 \cdots x_n}$.
We may call H-boxes with any number of edges, and with amplitude parameter $\omega^c$ for some $c \in \Z_D$, a ($\Z_D$-)\emph{multicharacter box}.

We may use these ideas to define a \emph{multicharacter subtheory} of ZH, consisting of the subtheory in which the H-boxes are indexed by paramters $c \in \Z_D$ in this way.
Roy~\cite{Roy-2022} has substantially investigated this fragment of ZH, in odd prime dimension.
Our choice of semantic map allows us to reproduce many of the rewrites considered by Roy (see Fig.~\ref{fig:ZH-rewrites} on page~\pageref{fig:ZH-rewrites} in the Appendices), while making minor simplifications and extending them to arbitrary dimensions $D>1$.

We may use multiplier gadgets and multicharacter boxes to usefully describe unitary transformations, such as exponentiations of the qudit controlled-$X$ and controlled-$Z$ gates:~\\[-3.75ex]
\begin{align}{}
\mspace{-24mu}
	\Sem{8.5ex}{\,
		\vtikzfig[-2ex]{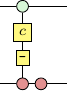}
	\,}
	\!\!\:&=\,
		\mathrm{CX}^c
	=
		\mathop{\int\!\!\!\!\int}\limits_{\mathclap{x,y \in \D}}
		  \!\!\:
			\kket{x,y\!\:{+}\!\:cx}\bbra{x,y}
\;,
\;\;
&
	\Sem{7ex}{\!\;
		\vtikzfig[-1.5ex]{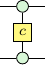}
	\!\;}
	\!\!\:&=\,
		\mathrm{CZ}^c
	=
		\mathop{\int\!\!\!\!\int}\limits_{\mathclap{x,y \in \D}}
		\!
			\omega^{cxy} \; \kket{x,y}\bbra{x,y}
\;.
\mspace{-18mu}
\end{align}~\\[-2.5ex]
These degree-2 multicharacter boxes are effectively a Fourier transform over an isomorphic presentation of $\Z_D$ in some cases.
This occurs in particular when $c = u \in \Z_D^\times$ is a multiplicative unit modulo $D$.
We can witness this by rewrites which are valid for H-boxes parameterised by units $u \in \Z_D^\times$, such as ones which relate the white dots and the gray dots among the ZH generators (similar remarks apply for the green and red ZX generators):~\\[-3.0ex]
\begin{align}
\begin{aligned}
		\Sem{6ex}{\vtikzfig[-2.5ex]{ZH-gray-w-unitH}}
		\;&=\;
		\Sem{6ex}{\vtikzfig[-3.5ex]{ZH-white-dot}}
\end{aligned}
&\quad
\text{and}
&&
\begin{aligned}
		\Sem{6ex}{\vtikzfig[-2.5ex]{ZH-white-w-unitH}}
	\;=\;
		\Sem{6ex}{\vtikzfig[-3.5ex]{ZH-gray-dot}}
\end{aligned}
\end{align}~\\[-2.5ex]
While there cannot be any perfect symmetry between the white and gray dots in general in the ZH calculus (as it involves the standard basis as a preferred basis), in this case a symmetry is recovered which one does not normally expect of presentations of the ZH calculusx.

We may also easily describe multi-qudit analogues of the qudit controlled-$X$ and controlled-$Z$ gates, using the fact that the H-boxes denote multi-characters.
For example:~\\[-3.75ex]
\begin{small}%
\begin{subequations}%
\begin{align}{}
\mspace{-48mu}
		\Sem{9ex}{\;
			\vtikzfig[-2.00ex]{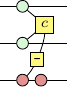}
		\;}
	&=\,
		\mathrm{CCX}^c
	=
		\mathop{\int\!\!\!\!\int\!\!\!\!\int}\limits_{\mathclap{x,y,z \in \D}}
			\!
			\kket{x,y,z\!\:{+}\!\:cxy}\bbra{x,y,z}
\,,
\\[-0.25ex]
		\Sem{9ex}{\,\;\vtikzfig[-1.75ex]{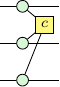}\,\;}
	&=\,
		\mathrm{CCZ}^c
	=
		\mathop{\int\!\!\!\!\int\!\!\!\!\int}\limits_{\mathclap{x,y,z \in \D}}
			\!
			\omega^{cxyz} \, \kket{x,y,z}\bbra{x,y,z}
\,.
\mspace{-36mu}
\end{align}%
\end{subequations}%
\end{small}~\\[-2.5ex]
While it would quickly become cumberson to represent each of the integrals in such an operation --- this being a motivation for diagrammatic calculi in general --- this demonstrates the genericity of the representation for these unitary transformations, and the relative lack of minor details to attend to in using them.
Finally, we note the quasi-spider property that H-boxes are known for in the qubit case and in (and also shown in a more complicated form for odd prime $D$ by Roy~\cite{Roy-2022}), which can also be shown for a pair of multicharacter boxes connected to a common H-box with parameter $u \in \Z_D^\times$:~\\[-3.75ex]
\begin{equation}
	\begin{aligned}
		\Sem{10ex}{\vtikzfig[-2.5ex]{ZH-H-complicated-fuse}}
		\;\;=\;\;
		\Sem{6ex}{\vtikzfig[-2.5ex]{ZH-H-complicated-phase-box}}
	\end{aligned}
\end{equation}~\\[-2.5ex]
We do not claim to have a complete multicharacter subtheory for ZH over arbitrary qudits, but many of the rewrites which one may show in this case (Fig.~\ref{fig:ZH-rewrites} on page~\pageref{fig:ZH-rewrites} of Appendix~\ref{apx:sound-ZH-rewrites}) can be specialised in a useful way to the multicharacter case.


\begin{figure}[t]
\begin{align*}
~\\[-7.0ex]
\mspace{-36mu}
	\begin{gathered}
		{\vtikzfig[-3.5ex]{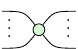}}
		\;\;\longleftrightarrow\;\;
		{\vtikzfig[-3.5ex]{ZH-white-dot}}
	\\[-0ex]
		{\vtikzfig[-3.5ex]{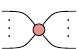}}
		\;\;\longleftrightarrow\;\;
		{\vtikzfig[-3.5ex]{ZH-gray-dot}}
	\\[1ex]
		{\vtikzfig[-3.5ex]{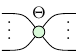}}
		\;\;\longleftrightarrow\;\;
        {\vtikzfig[-2.5ex]{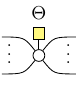}}
	\\[-3.75ex]
		{\vtikzfig[-2.5ex]{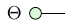}}
		\;\;\longleftrightarrow\;\;
		{\vtikzfig[-2.5ex]{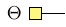}}
	\end{gathered}
\mspace{0mu}&&%
&&\mspace{0mu}
	\begin{gathered}
	~\\[-6ex]
		{\vtikzfig[-2.5ex]{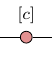}}
		\;\;\longleftrightarrow\;\;
		{\vtikzfig[-2.5ex]{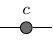}}
	\\[-3.5ex]
		{\vtikzfig[-2.5ex]{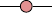}}
		\;\;\longleftrightarrow\;\;
		{\vtikzfig[-2.5ex]{ZH-gray-id}}
	\\[7ex]
		{\vtikzfig[-2.5ex]{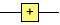}}
		\;\;\longleftrightarrow\;\;
		{\vtikzfig[-2.5ex]{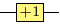}}
	\\[-.5ex]
		{\vtikzfig[-2.5ex]{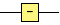}}
		\;\;\longleftrightarrow\;\;
		{\vtikzfig[-2.5ex]{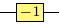}}
	\end{gathered}
\mspace{-36mu}&&%
\mspace{-18mu}&&
	\begin{gathered}
		{\vtikzfig[-2.5ex]{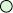}}
		\;\;\longleftrightarrow\;\;
		{\vtikzfig[-2.5ex]{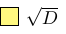}}
	\\[-2ex]
		{\vtikzfig[-2.5ex]{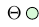}}
		\;\;\longleftrightarrow\;\;
		{\vtikzfig[-2.5ex]{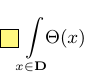}}
	\\[2.5ex]
		\;\;\;\;
		{\vtikzfig[-2.5ex]{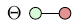}}
		\;\;\longleftrightarrow\;\;
		{\vtikzfig[-2.5ex]{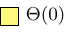}}
	\\[-0.5ex]
		{\vtikzfig[-2.5ex]{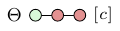}}
		\;\;\longleftrightarrow\;\;
		{\vtikzfig[-2.5ex]{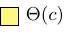}}
	~\\[0ex]
	\end{gathered}
    \mspace{-18mu}
\end{align*}~\\[-5.5ex]
\hrule
\smallskip
\caption{%
	\label{fig:ZXH-relations}
		Sound rewrites between the ZX~generators and the ZH~generators, subject to the semantics of Eqns.~\eqref{eqn:idealised-ZX-integrals} and~\eqref{eqn:idealised-ZH-integrals} in the case $\nu = D^{-1/4}$.
		The proofs of the soundness of these rewrites are shown in Appendix~\ref{apx:ZX-ZH-relations} (page~\pageref{apx:ZX-ZH-relations}).
	}
%
%
\end{figure}

\subsection{Compatibility and universality}

In addition to the semantics of 
Eqns.~\eqref{eqn:idealised-ZX-integrals} and~\eqref{eqn:idealised-ZH-integrals} yielding ZX and ZH~calculi which are each convenient in their own right, it also assigns the same semantics to certain ZX~generators and certain ZH~generators.
This is illustrated in Fig.~\ref{fig:ZXH-relations}.
This allows us to relate the two calculi to each other, to describe a `ZXH~calculus' which has the features of both.

It is not necessary to consider such a united calculus to be able to denote arbitrary operators: see Appendix~\ref{apx:beyond-subtheories} (and Section~\ref{sec:sketch-ZH-normal-form} in particular) for a sketch of a proof of universality of the ZH~diagrams, in terms of a `normal form'-like diagram which mirrors the construction of Ref.~\cite{BK-2019}.
However, while using both sets of generators may be redundant in principle, it should be expected to facilitate analysis, as the rewrite rules of each system effectively represents at least an important Lemma or Theorem of the other system.

We do not demonstrate any completeness results for these calculi; more rewrites may be necessary to prove completeness for arbitrary $D>1$, for each of these two calculi, even in the stabiliser and multicharacter fragments described above.

\section{
    Discussion}
\label{sec:summary}


As well as providing an approach to define ZX and ZH calculi with simple rewrite systems on qudits, this approach is a simpler, and apparently independent, way to reproduce%
	\footnote{%
		Strictly speaking, the calculi of Ref.~\cite{dB-2021} involve red and green dots with parameters $\theta \in \R$, H-boxes with parameters $\alpha \in \C$, only one type of Hadamard box instead of two, and a `nu~box' which is missing from the calculus presented here.
		We may bridge these differences using the short-hand described for phases / amplitudes to parameterise these nodes, identifying both Hadamard plus and minus boxes with the single Hadamard box of Ref.~\cite{dB-2021}, and replacing the nu-boxes with some suitable scalar gadgets (such as H-boxes parameterised by powers of $\nu = D^{-1/4}$).
	}
the `well-tempered' semantic map $\sem{\,\cdot\,}_\nu$ described in Ref.~\cite{dB-2021} for $D=2$.
In this way, Eqns.~\eqref{eqn:idealised-ZX-integrals} and~\eqref{eqn:idealised-ZH-integrals} provide a more intuitive definition of those semantics, and extend them to arbitrary $D>1$.
It is possible to show that this is essentially down to the constraints imposed on the representation of the discrete Fourier transform in Section~\ref{sec:normalisation-constraints-unitary-FT}.
Appendix~\ref{apx:proofs-rewrites-constrained-Ockhamic} describes \textbf{(a)}~the way that Eqns.~\eqref{eqn:idealised-ZX-integrals} and~\eqref{eqn:idealised-ZH-integrals} fail to constrain a generic `Ockhamic' interpretation of ZH~diagrams while fixing a specific `Ockhamic' interpretations of ZX~diagrams; and \textbf{(b)}~to what extent this approach to fixing semantics actually differs from the approach of Ref.~\cite{dB-2021}.

As well as discrete integrals, and amplitude functions $\Theta, \mathrm A: \Z \to \C$ in place of (vectors of) phases or amplitudes, we consider an index set for $\cH$ which is not simply $\{0,1,\ldots,D{-}1\}$, but instead $\{L,L\,{+}\,1,\ldots,U\,{-}\,1,U\}$ for some integers such that $U-L+1=D$.
One conventional choice is $L=0$ and $U=D-1$, but most of out results (in particular: all those to do with the stabiliser / multicharacter fragments of ZX or ZH) hold equally well with any such set of labels for the standard basis.
This less committal choice of index set demonstrates the flexibility of this system, which may prove useful for future applications (\emph{e.g.},~problems in physics where it may prove useful to consider negative index values).

We conclude with a highly speculative thought regarding discrete measures.
One constraint which we imposed on the measure $\mu$ on $\D$
--- interpreted as a measure on $\Z_D$ --- was that the Fourier transform should be interpretable as an involution $\C^{(\Z_D,\mu)} \to \C^{(\Z_D,\mu)}$ on functions on the measure space $(\Z_D,\mu)$, rather than a bijection $\C^{(\Z_D,\mu)} \to \C^{(\Z_D,\mu')}$ between functions on distinct measure spaces $(\Z_D,\mu)$ and $(\Z_D,\mu')$.
This may seem like a technical but necessary step; for a conventional presentation of ZX~diagrams, it \emph{is} necessary, if all of the wires are to have the same type.
However, many quantum algorithms have a structure in which some classical operation with a distinguished control register, where that control operates on a state which is conceived as being in the Fourier basis.
This structure is consistent with the control register having different `datatypes' at different stages of the algorithm.
Could it be more appropriate to make a distinction on logical qudits of each dimension $D$, between a `standard' type and a `Fourier' type (possibly among others), than to have just a single `type' for each $D$?
It would be interesting to consider what insights into the structure of quantum algorithms might arise by investigating along these lines; it is conceivable that this could give rise to new insights into structured quantum programming.

\bibliography{discrete-integrals}

\appendix
\titlerunning{Simple qudit ZX~and ZH~calculi, via integrals}
\authorrunning{N.~de\;Beaudrap and R.~D.~P.~East}


\section{Remarks on ZX and ZH beyond special subtheories}
\label{apx:beyond-subtheories}

In this section, we demonstrate the features of the semantic map of Eqns.~\eqref{eqn:idealised-ZX-integrals} and~\eqref{eqn:idealised-ZH-integrals} subject to $\nu = D^{-1/4}$.
This is to demonstrate how simple operators would be denoted using our semantics for the ZX and ZH generators, but also in part to demonstrate how those same operators can be denoted using discrete integrals.

\subsection{Multi-qudit phase operations in qudit ZXH}
\label{apx:ZXH-phase-gadgets}

The above construction is not special to multicharacters over $\Z_D$, and can also be used in conjunction with an arbitrary amplitude $\alpha \in \C^\times$, or indeed a more general function $\mathrm A: \Z \to \C$: for example,
\begin{small}%
\begin{align}{}
		\Sem{11ex}{\;%
		\begin{aligned}{}
		~\\[-3.5ex]
		\begin{tikzpicture}[]
      		\node (x) at (0,0) [white dot] {};
      		\draw (x) -- ++(1,0);
      		\draw (x) -- ++(-.375,0);
      		\node (w) at ($(x) + (0,.5875)$) [white dot] {};
      		\draw (w) -- ++(1,0);
      		\draw (w) -- ++(-.375,0);
      		\node (y) at ($(x) + (0,-.5875)$) [white dot] {};
      		\draw (y) -- ++(1,0);
      		\draw (y) -- ++(-.375,0);
      		\node (z) at ($(y) + (0,-.5875)$) [white dot] {};
      		\draw (z) -- ++(1,0);
      		\draw (z) -- ++(-.375,0);
      		\node [H box, label=right:\small$\!\!\:\mathrm{A}$] (p) at ($(x)!0.5!(y) + (.375,0)$) {};
      		\draw (w) -- (p);
      		\draw (x) -- (p);
      		\draw (y) -- (p);
      		\draw (z) -- (p);
		\end{tikzpicture}	
		\end{aligned}\;}
	\begin{split}
	&=
		\mathop{\int\!\!\!\!\int\!\!\!\!\int\!\!\!\!\int}\limits_{\mathclap{w,x,y,z \in \D}}\!
			\mathrm{A}(wxyz) \; \kket{w,\!\!\;x,\!\!\;y,\!\!\;z}\bbra{w,\!\!\;x,\!\!\;y,\!\!\;z}
	\\[.5ex]&\;\;\;\;{}=\;
		\mathop{\sum}\limits_{\mathclap{w,x,y,z \in \D}}\,
			\mathrm{A}(wxyz) \, \ket{w,\!\!\;x,\!\!\;y,\!\!\;z}\bra{w,\!\!\;x,\!\!\;y,\!\!\;z}
	\!\:.
	\end{split}
\end{align}%
\end{small}
This yields a unitary operator in all cases (\emph{i.e.},~for arbitrarily many operand qubits) if and only if $\lvert \mathrm A(t) \rvert = 1$ for all $t \in \Z$.
In particular, recall that the product of the variables of integration is evaluated in $\Z$.

For whatever index range $\D = \{L, L\,{+}\,1, \ldots, U\,{-}\,1, U\}$ one chooses, there is a simple involution $\neg: \D \to \D$, given by
\begin{equation}
	\neg x
	\::=\;
	\sigma - x\,,
\qquad
\text{where $\sigma = U + L$}.
\end{equation}
(For instance, one then has $\neg U = L$,\, $\neg L = U$,\, and $\neg (x+1) = \neg x - 1$ when $x,x+1 \in \D$.)
A `not-dot' which is parameterised by $-\sigma$ then has interesting properties which in the case $D = 2$ one observes for ZX red nodes with angular parameter $\pi$.
Using the syntactic sugar
\begin{equation}
    \smash{\,\tikz
	\draw (0,0) -- node [midway, not dot, label=above:\footnotesize$\neg$] {} (.75,0);
\, := \,
\,\tikz
	\draw (0,0) -- node [midway, not dot, label=above:\footnotesize$-\sigma$] {} (.75,0);
\,} \;,
\end{equation}
we may show certain quantum circuit equivalences, that hold without any scalar factor adjustment:
\begin{small}%
\allowdisplaybreaks
\begin{align}{}
\begin{aligned}[b]{}
\mspace{-39mu}
		\Sem{15ex}{\,%
		\begin{aligned}{}
		~\\[-4.5ex]
		\begin{tikzpicture}[]
      		\node (x) at (0,0) [white dot] {};
      		\draw (x) -- ++(1,0);
      		\draw (x) -- ++(-1,0);
      		\node (w) at ($(x) + (0,.5875)$) [white dot] {};
      		\draw (w) -- ++(1,0);
      		\draw (w) -- node [midway, not dot, label=above:\footnotesize$\neg$] {} ++(-1,0);
      		\node (y) at ($(x) + (0,-.5875)$) [white dot] {};
      		\draw (y) -- ++(1,0);
      		\draw (y) -- ++(-1,0);
      		\node (z) at ($(y) + (0,-.5875)$) [white dot] {};
      		\draw (z) -- ++(1,0);
      		\draw (z) -- ++(-1,0);
      		\node [H box, label=right:\small$\!\!\:\alpha$] (p) at ($(x)!0.5!(y) + (.375,0)$) {};
      		\draw (w) -- (p);
      		\draw (x) -- (p);
      		\draw (y) -- (p);
      		\draw (z) -- (p);
		\end{tikzpicture}	
		\end{aligned}\,}
	\!\;=\!\;
		\Sem{15ex}{\,%
		\begin{aligned}{}
		~\\[-4.5ex]
		\begin{tikzpicture}[]
      		\node (x) at (0,0) [white dot] {};
      		\draw (x) -- ++(1,0);
      		\draw (x) -- ++(-.5,0);
      		\node (w) at ($(x) + (0,.5875)$) [white dot] {};
      		\draw (w) -- node [midway, not dot, label=above:\footnotesize$\neg$] {} ++(1,0);
      		\draw (w) -- ++(-.5,0);
      		\node (y) at ($(x) + (0,-.5875)$) [white dot] {};
      		\draw (y) -- ++(1,0);
      		\draw (y) -- ++(-.5,0);
      		\node (z) at ($(y) + (0,-.5875)$) [white dot] {};
      		\draw (z) -- ++(1,0);
      		\draw (z) -- ++(-.5,0);
      		\node [H box, label=right:\small$\!\!\:\alpha$] (p) at ($(x)!0.5!(y) + (.375,0)$) {};
      		\draw (w) -- node [pos=0.3125, not dot, label=left:\footnotesize$\neg\!\!\!\;$] {} (p);
      		\draw (x) -- (p);
      		\draw (y) -- (p);
      		\draw (z) -- (p);
		\end{tikzpicture}	
		\end{aligned}\,}
	\,&=
		\mathop{\int\!\!\!\!\int\!\!\!\!\int\!\!\!\!\int}\limits_{\mathclap{w,x,y,z \in \D}}\!
			\alpha^{(\sigma - w)xyz} \; \kket{\neg w,x,y,z}\bbra{w,x,y,z}
	\\[-1.5ex]&=
		\mathop{\int\!\!\!\!\int\!\!\!\!\int\!\!\!\!\int}\limits_{\mathclap{w,x,y,z \in \D}}\!
			\alpha^{\sigma xyz} \alpha^{- wxyz} \; \kket{\neg w,x,y,z}\bbra{w,x,y,z}
	\\\,&=\!\;
			\Sem{15ex}{\,%
		\begin{aligned}{}
		~\\[-4.5ex]
		\begin{tikzpicture}[]
      		\node (x) at (0,0) [white dot] {};
      		\draw (x) -- ++(-.5,0);
      		\node (w) at ($(x) + (0,.5875)$) [white dot] {};
      		\draw (w) -- ++(-.5,0);
      		\node (y) at ($(x) + (0,-.5875)$) [white dot] {};
      		\draw (y) -- ++(-.5,0);
      		\node (z) at ($(y) + (0,-.5875)$) [white dot] {};
      		\draw (z) -- ++(-.5,0);
      		\node [H box, label=right:\small$\!\!\:\alpha^{-1}$] (p) at ($(x)!0.5!(y) + (.375,0)$) {};
      		\draw (w) -- (p);
      		\draw (x) -- (p);
      		\draw (y) -- (p);
      		\draw (z) -- (p);
			\coordinate (w') at ($(w) + (1.5,0)$);
			\node [white dot] (x') at ($(x) + (1.5,0)$) {};
			\node [white dot] (y') at ($(y) + (1.5,0)$) {};
			\node [white dot] (z') at ($(z) + (1.5,0)$) {};
      		\draw (w) -- (w');
      		\draw (x) -- (x');
      		\draw (y) -- (y');
      		\draw (z) -- (z');
      		\node [H box, label=right:\small$\!\!\:\alpha^{\sigma}$] (p') at ($(x')!0.5!(y') + (.375,0)$) {};
      		\draw (x') -- (p');
      		\draw (y') -- (p');
      		\draw (z') -- (p');
      		\draw (w') -- node [midway, not dot, label=above:\footnotesize$\neg$] {} ++(1,0);
      		\draw (x') -- ++(1,0);
      		\draw (y') -- ++(1,0);
      		\draw (z') -- ++(1,0);
		\end{tikzpicture}	
		\end{aligned}\,}
\!\;.
\mspace{-36mu}
\end{aligned}%
\end{align}%
\end{small}
As a consequence, for a green node with arbitrary phase parameter $\theta \in \R$ (representing an amplitude function $\Theta(x) = \e^{i\theta x}$), it is possible to show that
\begin{equation}
\label{eqn:not-dot-green-phase}
	\Sem{9ex}{\;
	\begin{aligned}
    \begin{tikzpicture}
      \node (Z) at (0,0) [Z dot, label=above:\small$\theta$] {};
      \node (h1) at (-.5,0.375) [not dot, label=above:\small$\neg$] {};
      \node (h2) at (-.5,-0.375) [not dot, label=below:\small$\neg$] {};
      \draw (Z) .. controls (-0.3175,0.375) .. (h1) -- ++(-0.375,0);
      \draw (Z) .. controls (-0.3175,-0.375) .. (h2) -- ++(-0.375,0);
      \node (dots) at ($(Z) + (-0.5,0.125)$) {\footnotesize$\mathbf\vdots$};
      \node (h1) at (.5,0.375) [not dot, label=above:\small$\neg$] {};
      \node (h2) at (.5,-0.375) [not dot, label=below:\small$\neg$] {};
      \draw (Z) .. controls (0.3175,0.375) .. (h1) -- ++(0.375,0);
      \draw (Z) .. controls (0.3175,-0.375) .. (h2) -- ++(0.375,0);
      \node (dots) at ($(Z) + (0.5,0.125)$) {\footnotesize$\mathbf\vdots$};
    \end{tikzpicture}
	\end{aligned}\; }
\;=\;
	\Sem{10ex}{\;
	\begin{aligned}
    \begin{tikzpicture}
      \node (Z) at (0,0) [Z dot, label=above:\small$-\theta$] {};
      \draw (Z) .. controls (-0.3175,0.375) .. ++(-0.75,.375);
      \draw (Z) .. controls (-0.3175,-0.375) .. ++(-0.75,-.375);
      \node (dots) at ($(Z) + (-0.5,0.125)$) {\footnotesize$\mathbf\vdots$};
      \draw (Z) .. controls (0.3175,0.375) ..  ++(0.75,.375);
      \draw (Z) .. controls (0.3175,-0.375) .. ++(0.75,-.375);
      \node (dots) at ($(Z) + (0.5,0.125)$) {\footnotesize$\mathbf\vdots$};
        \node (H) at ($(Z) + (-0.4375,0.75)$) [small H box, label=right:\small$\e^{i\theta\sigma}_{\big.}$] {};
        \node at ($(Z) + (-0.4375,-0.75)$) [small H box, fill=none, draw=none, label=right:\phantom{\small$\e^{i\theta\sigma}$}] {};
    \end{tikzpicture}
	\end{aligned}\;},
\end{equation}
generalising the situation in conventional presentations of the ZX~calculus for $D = 2$, in which red $\pi$-phase dots play the role of the $\neg\;\!$-dots.

Note that it is possible to derive the above equivalences entirely syntactically using the rewrites that we demonstrate in Fig.~\ref{fig:ZH-rewrites}, \ref{fig:ZXH-relations-redux}, and~\ref{fig:ZH-rewrites} --- specifically, in particular using the rewrites \textsf{(ZH-EC)}, \textsf{(ZXH-WH)}, \textsf{(ZXH-S)}, and \textsf{(ZX-NS)}.
We show this using semantics rather than diagrammatic rewrites, partly in order to \emph{demonstrate} the semantics, and partly in order to show --- for all that we must write several integrals to do so --- that it is \emph{feasible} to analyse these diagrams through their semantics.
Such analysis is in fact the means by which we will demonstrate the soundness of the diagrammatic rewrites in Sections~\ref{apx:sound-ZX-rewrites} and~\ref{apx:sound-ZH-rewrites}.


\subsection{Sketch of a qudit normal form for ZH}
\label{sec:sketch-ZH-normal-form}

In this Section, we outline a normal form for ZH~diagrams which, together with standard techniques for ZH~normal forms~\cite{BK-2019}, suffice to denote an arbitrary operator on $\cH$.

Unlike the rest of our work, these techniques are sensitive to the specific choice of the residues $\D = \{L,L\,{+}\,1, \ldots,U\,{-}\,1,U\}$, as they involve comparing products of elements of $\D$ to constants.
We consider two possible examples which we suppose to be the most likely to be useful in practise:

We may use general amplitude functions $\mathrm{A}: \Z \to \C$, to define more fine-grained operators than with scalar amplitude parameters.
For instance, let $\mathbf{X}{S}$ be the characteristic function of a set $S \subseteq \Z$ (so that $\mathbf{X}_{S}(t) = 0$ if $t \notin S$, and $\mathbf{X}_{S}(t) = 1$ otherwise), and let $\mathbf V_{S,\alpha}(t) = \alpha^{\mathbf{X}_{S}(t)}$.
Then 
\begin{align}{}
		\Sem{8ex}{\;%
		\begin{aligned}{}
		~\\[-3.5ex]
		\begin{tikzpicture}[]
      		\node (x) at (0,0) [white dot] {};
      		\draw (x) -- ++(1.5,0);
      		\draw (x) -- ++(-.375,0);
      		\node (y) at ($(x) + (0,-1)$) [white dot] {};
      		\draw (y) -- ++(1.5,0);
      		\draw (y) -- ++(-.375,0);
      		\node [H box, label=right:\small$\!\!\:\mathbf V_{\{1\},\alpha}$] (p) at ($(x)!0.5!(y) + (.375,0)$) {};
      		\draw (x) -- (p);
      		\draw (y) -- (p);
		\end{tikzpicture}	
		\end{aligned}\;}
	\,&=\,
		\mathop{\int\!\!\!\!\int}\limits_{\mathclap{x,y \in \D}}
			\mathbf V_{\{1\},\alpha}(xy) \; \kket{x,y}\bbra{x,y}
\end{align}
induces a phase (or a change in amplitude) factor of $\alpha$ on the states $\ket{x,y}$ for which $xy = 1$, and leaves all other states $\ket{x,y}$ unchanged.
If $1 \in \D$ (for instance, if $L = 0$ and $U = D-1$), then this includes the state $\ket{\texttt{1},\texttt{1}}$.
However, if $-1 \in \D$ as well as $+1 \in \D$  (for instance, is $L = -\lfloor \tfrac{1}{2}(D-1) \rfloor$ and $U = \lceil \tfrac{1}{2}(D-1) \rceil$ for $D > 2$), then both the states $\ket{\texttt{-1},\texttt{-1}}$ and $\ket{\texttt{+1},\texttt{+1}}$.%
Whichever choice of $\D$ we use, the ability to impose independently chosen factors for specific values of the product of the indices, provides us with the ability to manipulate individual coefficients on a relatively fine-grained level.
	
We may use similar gadgets as the basis of a normal form for qudit ZH diagrams.
For an arbitrary number $m \ge 0$ of input wires, we may consider a gadget of the form%
	\footnote{%
		Similar constructions may be possible, though the reader should beware that some promising ones may not work for all values of $D$ if $\D$ contains both positive and negative values.
		An example is the gadget\\
		\begin{minipage}[t]{0.8\textwidth}
			 illustrated on the right, taking $\boldsymbol \epsilon_\alpha(t) = \alpha$ for $t = 1$ and $\boldsymbol \epsilon_\alpha(t) = 1$ otherwise, relying on the fact that $1 - x^2 = \pm 1$ has no solutions over the integers except for $x = 0$, in which case $1 - x^2 = +1$.
			However, to be more precise, we must consider the conditions under which $\rho_D(-\rho_D(-1{-}x)) \!\:\cdot\!\: \rho_D(1{-}x) = \pm 1$ admits solutions for $x \in \D$, where $\rho_D : \Z \to \D$ maps integers to representatives mod $D$.
			For $D = 4$, $\D = \{-1,0,+1,+2\}$, and $x = 2$, we have $\rho_4(-\rho_4(-1{-}x)) \!\:\cdot\!\: \rho_4(1-x) = (-1) \!\cdot\! (-1) = +1$, so that this gadget does not suffice to single out $\ket{\texttt{+1}}$ on each input.
			Constructions which single out $\kket{x} = \kket{t}\sox{n}$ for a fixed $t \in \Z$ will likely fail for some single value of $D$, for similar reasons.
		\end{minipage}
		\hfill
		\begin{minipage}[t]{0.125\textwidth}\raggedleft
		\medskip
		
		$%
		\begin{aligned}[t]~\\[-6ex]
		\begin{tikzpicture}[]
      		\node (x) at (0,0) [white dot] {};
      		\draw (x) -- ++(-1,0);
      		\node (y) at ($(x) + (0,-1.75)$) [white dot] {};
      		\draw (y) -- ++(-1,0);
			\node (dots) at ($(x)!0.5!(y) + (-.75,0.09875)$) {$\vdots$};
			\node (dots) at ($(x)!0.5!(y) + (-.75,0.09875)$) {$\vdots$};
      		\node [Z dot] (p) at ($(x)!0.5!(y) + (.875,0)$) {};
      		\draw (x) .. controls ++(.375,-.125) and ++(-.125,.375) .. node [midway, not dot, label=above:\footnotesize$\;-1$] {} (p);
      		\draw (x) .. controls ++(.125,-.375) and ++(-.375,.125) .. node [pos=0.3125, not dot, label=left:\footnotesize$+1\!\!\;$] {}
      			node [pos=0.625, gray dot] {} (p);
      		\draw (y) .. controls ++(.375,.125) and ++(-.125,-.375) .. node [midway, not dot, label=below:\footnotesize$\;-1$] {} (p);
      		\draw (y) .. controls ++(.125,.375) and ++(-.375,-.125) .. node [pos=0.3125, not dot, label=left:\footnotesize$+1\!\!\;$] {} node [pos=0.625, gray dot] {} (p);
      		\node [H box, label=right:$\!\!\:\boldsymbol \epsilon_\alpha$] at (p) {};
		\end{tikzpicture}
		\end{aligned}
		$	
		\end{minipage}
		\hfill~
	}%
\begin{align}{}
\label{eqn:ZH-phase-only-on-Us-gadget}
		{%
		\begin{aligned}{}
		~\\[-3.5ex]
		\begin{tikzpicture}[]
      		\node (x) at (0,0) [white dot] {};
      		\draw (x) -- ++(-.75,0);
      		\node (y) at ($(x) + (0,-1.5)$) [white dot] {};
      		\draw (y) -- ++(-.75,0);
     		\node (brace) at ($(x)!0.5!(y) + (-1.25,0)$) {%
      				$m \left\{ \begin{matrix} \\[10ex] \end{matrix} \right.$
      			};
			\node (dots) at ($(x)!0.5!(y) + (-0.375,0.09875)$) {$\vdots$};
      		\node [Z dot] (p) at ($(x)!0.5!(y) + (.75,0)$) {};
      		\draw (x) .. controls ++(.375,-.125) and ++(-.125,.375) .. node [midway, not dot, label=above:\footnotesize$\;\;\;1{-}\sigma$] {} (p);
      		\draw (x) .. controls ++(.125,-.375) and ++(-.375,.125) .. (p);
      		\draw (y) .. controls ++(.375,.125) and ++(-.125,-.375) .. node [midway, not dot, label=below:\footnotesize$\;\;\;1{-}\sigma$] {} (p);
      		\draw (y) .. controls ++(.125,.375) and ++(-.375,-.125) .. (p);
      		\node [H box, label=right:$\mathbf M^{(2m)}_\alpha$] (p) at ($(x)!0.5!(y) + (.75,0)$) {};
		\end{tikzpicture}	
		\end{aligned}}
	\;,
	\qquad\qquad
	\text{where $
		\mathbf M^{(k)}_\alpha(t)
		\;=\;
			\begin{cases}
				\alpha,	&	\text{if $t = U_{\!D}^{\,k}$};
			\\
				1,		&	\text{otherwise}.	
			\end{cases}
	$}
\end{align}
We may describe the behaviour of this gadget as it acts on standard basis states.
Each input wire of this gadget admits a state $\kket{x_j}$, and copies it to produce a pair $\kket{x_j, x_j}$.
One copy is then acted on with a $(1{-}\sigma)$-not-dot, yielding a state ${\kket{x_j,\, \rho_D(\sigma{-}1{-}x_j)}} = {\kket{x_j,\, \rho_D(\neg x \:\!{-}\:\!1)}}$, where $\rho_D : \Z \to \D$ is a map which reduces each integer modulo $D$ to a representative in $\D$. 
(The map $\rho_D$ is implicit in many of the transformations in which one might prefer to evaluate arithmetic modulo $D$: it arises here because we must map the associate the expression $\sigma{-}1{-}x_j$\,, which we might prefer to implicitly evaluate modulo $D$ in the basis labels, to an explicit element of $\Z$.)
The $\mathbf M^{(2m)}_\alpha$ box then maps $\kket{x} \mapsto \alpha$ if and only if
\begin{equation}
\label{eqn:normal-form-gadget-constraint}
	\prod_{j=1}^n	\Bigl( x_j \cdot \rho_D(\neg x - 1)  \Bigr)
\;=\;
	U_{\!\!\:D}^{\,2m},
\end{equation}
and maps $\kket{x} \mapsto +1$ otherwise.
For $x \in \Z^m$, Eqn.~\eqref{eqn:normal-form-gadget-constraint} is satisfied only if ${x_j \cdot \rho_D(\neg x - 1)} {{}= \pm U_{\!\!\:D}^{\,2}}$ for each $x_j$ individually, as $U_D \in \D$ is the element with the largest absolute value.
\begin{itemize}
\item
	Note that $\rho_D(\neg x - 1) \,\ne\, \neg x - 1$ if and only if $\neg x = L_D$, which is to say precisely when $x = U_D$: in this case, we have $x \cdot \rho_D(\neg x - 1) = U_{\!\!\:D}^{\,2}$.

\item
	Otherwise, for $x < U_D$, we have $x \cdot \rho_D(\neg x - 1) = x \cdot (\neg x - 1) = \sigma x - x - x^2$.
	For $D$ even, we have $\sigma x - x - x^2 = -x^2$; for $D$ odd, we instead have  $\sigma x - x - x^2 = -x^2 = -x(x+1)$.
	The absolute values of these expressions are bounded strictly below $U_{\!\!\;D}^{\,2}$ in either case.
\end{itemize}
Then Eqn.~\eqref{eqn:normal-form-gadget-constraint} is satisfied if and only if $x_j = U_D$ for each $1 \le j \le m$.
The action of the gadget in Eqn.~\eqref{eqn:ZH-phase-only-on-Us-gadget} is then to map $\kket{U_{\!\!\:D}}\sox{m} \mapsto \alpha$ and $\kket{x} \mapsto +1$ for all other $x \in \D^m$. 

Using gadgets of this form, we may express any operator $\Omega: \cH\sox{m} \to \cH\sox{n}$ in terms of its coefficients, in a similar manner to the normal form for ZH diagrams in $D=2$ presented by Backens and Kissinger~\cite{BK-2019}.
That is, we may take any operator, transform it into a vector using cup operators, use white dots to make copies of each qudit in the standard basis (one copy for every coefficient of the operator), and then apply not-dots and gadgets of the form in Eqn.~\eqref{eqn:ZH-phase-only-on-Us-gadget} to fix the value of each coefficient $\alpha_{x,y} = \bbra{x} \,\Omega\, \kket{y} = \nu^{-m-n} \, \Omega_{\!\;x,y}$ for $x \in \D^m$ and $y \in \D^n$. 


We do not expect that the rewrite rules that we have set out, for ZX~or ZH~diagrams, suffice to transform arbitrary diagrams to such a normal form.
We leave open the problem of demonstrating rewrites to transform arbitrary qudit ZH~diagrams into a form such as the one we have sketched here, or a similar one.
We speculate that simple rewrite rules can be found, which might allow the gadget of Eqn.~\eqref{eqn:ZH-phase-only-on-Us-gadget} to be expressed using only phase-amplitude gadgets.

\subsection{Rewrites and relations for the ZX and ZH calculi}
\label{apx:proofs-rewrites}

Having considered semantic maps for ZX- and ZH-diagrams as above, we now briefly survey what rewrites we may show to be sound for these semantics.

We make no attempt to achieve complete calculi for these semantics: we expect that doing so will be somewhat involved, though perhaps they can be informed by recent work~\cite{PWSYYC-2023} on a complete ZXW-calculus in arbitrary dimensions $D > 1$.
We also do not aim to achieve a minimal set of rewrite rules, though we will in some instances point out when some rules may be derived from some others.
Our objective is to demonstrate the soundness of scalar-exact versions of familiar rewrite rules, motivated by rules from complete versions of the ZX- and ZH-calculi~\cite{dB-2021,BKMWW-2021} in the special case $D = 2$, by ZX~rewrite rules presented for odd $D>1$ presented by Booth and Carette~\cite{BC-2022}, and by ZH-rewrite rules for $D>1$ presented by Roy~\cite{Roy-2022}.
We aim also to describe versions of these rules, which are both simple and versatile enough that their usefulness may be plausible despite the absence of a completeness proof.


Our approach is concrete, consisting of the direct manipulation of discrete integrals.
We hope to demonstrate that, in the large majority of cases, this is not difficult to do.
In the following, we will model what we consider to be a reasonable approach to using discrete integrals in practise.%
    \footnote{%
        For example: we do not need to explicitly write $\mathrm d\mu(x)$ or $\mathrm d\mu(y)$ when integrating over $x,y \in \D$, nor $\mathrm d\mu^m(x)$ when integrating over $x \in \D^m$, and so forth.
        This is consistent with occasional practise in the mathematics literature (see, \emph{e.g.},~Ref.~\cite{Majid-2022}).
        One may regard the integral symbol itself as an operator on a given expression, with the measure understood from context (\emph{e.g.},~the variable of integration).
    }

Note also that, in order to shorten our calculations, we occasionally omit expressions
$\int_{y \in \D} \bbracket{x}{y}\, \Phi(y)$
that might arise due to $x$ and $y$ being indices for two different nodes, corresponding to an edge incident to both: we may implicitly simplify such integrals to $\Phi(x)$\,.
Similarly, we may simplify expressions ${\int\!\!\!\!\:\int}_{\!x\in \D^m,y \in \D} \bbracket{f(x_1,\ldots,x_m)}{y}\,\Phi(y) \longrightarrow \int_{\!\!\;x\in \D^m} \Phi(f(x_1,\ldots,x_m))$.
On occasion, when we wish to emphasise the role of $x_1, x_2, \ldots \in \D$ as representatives of the integers modulo $D$, we may write these same integrals  as $\int_{\!\!\;x\in \Z_D^m} \Phi(f(x_1,\ldots,x_m))$ rather than $\int_{\!\!\;x\in \D^m} \Phi(f(x_1,\ldots,x_m))$; however, we do so only when $\Phi$ is well-defined modulo $D$ in each of its arguments.
(We routinely interpret arithmetic expressions $E$ modulo $D$, when they are used to indicate point-mass functions $\kket{E}$; though $E \in \Z$ may not represent an element of $\D$ outside of that context.)
We also make liberal use of Lemma~\ref{lemma:exponential-integral}, which (except where we assume $\nu = D^{-1/4}$) will give rise to frequent occurrences of the scalar expression $D\nu^4$.

Section~\ref{apx:sound-ZH-rewrites} presents ZH~rewrites and proofs of their soundness, for the semantic map of Eqn.~\ref{eqn:idealised-ZH-integrals}; Section~\ref{apx:ZX-ZH-relations} (page~\pageref{apx:ZX-ZH-relations}) presents the relationships between the semantics of ZH generators, with those of the ZX generators when we fix their semantics as in~Eqn.~\ref{eqn:idealised-ZH-integrals}; and Section~\ref{apx:sound-ZX-rewrites} (page~\pageref{apx:sound-ZX-rewrites}) presents ZX~rewrites and proof of their soundness for those semantics.

\subsubsection{Sound ZH rewrites}
\label{apx:sound-ZH-rewrites}

\begin{figure}[p]
\begin{gather*}
~\\[-13.5ex]
\mspace{-75mu}
	\begin{tikzpicture}
	\setlength\rulediagramwd{4.25em}
		\rewriterule		[ZH-WI]	 	{\vtikzfig{ZH-white-id}}
		\nextrewriterule	[ZH-WQS]	{\vtikzfig{ZH-white-special-w-invSqrtD}}
		\nextrewriterule	[ZH-AI]		{\vtikzfig{ZH-2antipode}}
	\setlength\rulediagramwd{4.5em}
		\nextrewriterule	[ZH-HI]		{\vtikzfig{ZH-H-id}}
	\setlength\rulediagramwd{3em}
		\rewritetarget 					{\vtikzfig{id-wire}}
	\end{tikzpicture}
\\[-5.25ex]
\mspace{-75mu}
	\setlength\rulediagramwd{6em}
	\begin{tikzpicture}
		\rewriterule		[ZH-WF]	 	{\vtikzfig[-1ex]{ZH-white-fuse}}
		\coordinate (anchor-next-diagram) at ($(anchor-next-diagram) + (-.125,0)$);
		\nextrewriterule	[ZH-GWC]	{\vtikzfig{ZH-gray-w-unitH}}
		\coordinate (anchor-next-diagram) at ($(anchor-next-diagram) + (-.125,0)$);
		\setlength\rulediagramwd{6.5em}
		\nextrewriterule	[ZH-WNS]	{\vtikzfig{ZH-not-dot-symm}}
	\setlength\rulediagramwd{4.5em}
		\rewritetarget 					{\vtikzfig[-1ex]{ZH-white-dot}}
	\end{tikzpicture}
\\[-3.5ex]
\mspace{-75mu}
	\setlength\rulediagramwd{6em}
	\begin{tikzpicture}
		\rewriterule		[ZH-GF]		{\vtikzfig[-1ex]{ZH-gray-fuse}}
		\coordinate (anchor-next-diagram) at ($(anchor-next-diagram) + (-.125,0)$);
		\nextrewriterule	[ZH-GL]		{\vtikzfig{ZH-gray-dot-w-lollipop}}
		\coordinate (anchor-next-diagram) at ($(anchor-next-diagram) + (-.125,0)$);
		\setlength\rulediagramwd{6.5em}
		\nextrewriterule	[ZH-WGC]	{\vtikzfig{ZH-white-w-unitH}}
	\setlength\rulediagramwd{4.5em}
		\rewritetarget 					{\vtikzfig[-1ex]{ZH-gray-dot}}
	\end{tikzpicture}
\\[-3.5ex]
\mspace{-75mu}
	\begin{aligned}
	\begin{aligned}
	\begin{tikzpicture}
	\setlength\rulediagramwd{7em}
		\rewriterule		[ZH-MEH]		{\vtikzfig[.25ex]{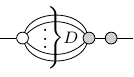}}
	\setlength\rulediagramwd{4.5em}
		\nextrewriterule	[ZH-A]	 	{\vtikzfig[.5ex]{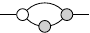}}
		\coordinate (anchor-next-diagram) at ($(anchor-next-diagram) + (-.125,0)$);
	\setlength\rulediagramwd{3em}
		\rewritetarget 					{\vtikzfig{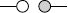}}
	\end{tikzpicture}
	\end{aligned}
	&&
		\Bigg\vert
	\mspace{-9mu}&&
	\begin{aligned}
	\begin{tikzpicture}
		\rewriterule		[ZH-WGB]	{\vtikzfig{ZH-bialg-white-gray}}
		\coordinate (anchor-next-diagram) at ($(anchor-next-diagram) + (-.125,0)$);
		\rewritetarget 					{\vtikzfig[-1ex]{ZH-bott-white-gray}}
	\end{tikzpicture}
	\end{aligned}
	\end{aligned}
\\[-6.5ex]
\mspace{-75mu}
	\begin{aligned}
	\begin{aligned}
	\begin{tikzpicture}
	\setlength\rulediagramwd{5.25em}	
		\rewriterule		[ZH-HM]		{\vtikzfig[.5ex]{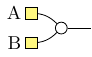}}
		\coordinate (anchor-next-diagram) at ($(anchor-next-diagram) + (-.125,0)$);
	\setlength\rulediagramwd{3.5em}	
		\rewritetarget 					{\!\!\!\!\!\!\vtikzfig[1.5ex]{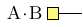}}
	\end{tikzpicture}
	\end{aligned}
	&&
		\Bigg\vert
	\mspace{-9mu}&&
	\begin{aligned}
	\begin{tikzpicture}
	\setlength\rulediagramwd{2.5em}	
		\rewriterule		[ZH-HU]		{\vtikzfig{ZH-H0-prep}}
	\setlength\rulediagramwd{1.5em}	
		\coordinate (anchor-next-diagram) at ($(anchor-next-diagram) + (-.125,0)$);
		\rewritetarget 					{\vtikzfig{ZH-white-prep}\!\!\!\!}
	\end{tikzpicture}
	\end{aligned}
	&&
		\Bigg\vert
	\mspace{-9mu}&&
	\begin{aligned}
	\begin{tikzpicture}
	\setlength\rulediagramwd{6em}	
		\rewriterule		[ZH-EC]		{\vtikzfig[-.5ex]{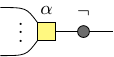}}
	\setlength\rulediagramwd{5.5em}	
		\coordinate (anchor-next-diagram) at ($(anchor-next-diagram) + (-.125,0)$);
		\rewritetarget 					{\vtikzfig[1.5ex]{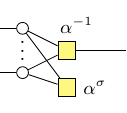}}
	\end{tikzpicture}
	\end{aligned}
	\end{aligned}
\\[-6.0ex]
\mspace{-75mu}
	\begin{aligned}
	\begin{aligned}
	\begin{tikzpicture}
	\setlength\rulediagramwd{6.5em}	
		\rewriterule		[ZH-MF]		{\vtikzfig[-1ex]{ZH-H-complicated-fuse}}
		\coordinate (anchor-next-diagram) at ($(anchor-next-diagram) + (-.125,0)$);
	\setlength\rulediagramwd{4.25em}	
		\rewritetarget 					{\vtikzfig[-1ex]{ZH-H-complicated-phase-box}}
	\end{tikzpicture}
	\end{aligned}
	&&
		\Bigg\vert
	\mspace{-18mu}&&
	\begin{aligned}
	\begin{tikzpicture}
	\setlength\rulediagramwd{4.5em}	
		\rewriterule		[ZH-MCA]	{\!\!\!\!\vtikzfig[1.5ex]{ZH-multichar-add}}
	\setlength\rulediagramwd{3em}	
		\rewritetarget 					{\!\!\!\!\vtikzfig[1ex]{ZH-multichar-prep}\!\!\!}
	\end{tikzpicture}
	\end{aligned}
	\mspace{-18mu}
	&&
		\Bigg\vert
	\mspace{-18mu}&&
	\begin{aligned}
	\begin{tikzpicture}
	\setlength\rulediagramwd{4.5em}	
		\rewriterule		[ZH-UM]		{\vtikzfig{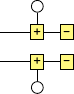}}
		\coordinate (anchor-next-diagram) at ($(anchor-next-diagram) + (-.125,0)$);
	\setlength\rulediagramwd{3.5em}	
		\rewritetarget 					{\vtikzfig{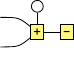}}
	\end{tikzpicture}
	\end{aligned}
	\end{aligned}
\\[-3.75ex]
\mspace{-75mu}
	\mspace{-9mu}
	\begin{aligned}
	\begin{aligned}
	\begin{tikzpicture}
	\setlength\rulediagramwd{10em}
	\setlength\rulexnwd{3em}
		\rewriterule		[ZH-O]		{\vtikzfig[.25ex]{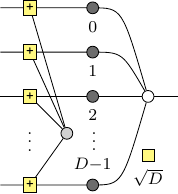}}
		\rewritetarget 					{\vtikzfig{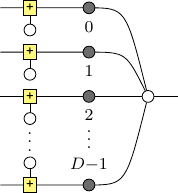}}
	\end{tikzpicture}
	\end{aligned}
	&&
		\Bigg\vert
	&&
	\begin{aligned}
	\begin{tikzpicture}
	\setlength\rulediagramwd{7.5em}	
		\rewriterule		[ZH-HWB]	{\vtikzfig{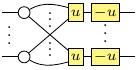}}
	\setlength\rulediagramwd{6.5em}	
		\rewritetarget 					{\!\!\!\!\vtikzfig[1ex]{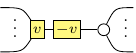}\!\!\!}
	\end{tikzpicture}
	\\[-1.5ex]
	\begin{tikzpicture}
		\rewriterule		[ZH-HMB]		{\vtikzfig{ZH-bialg-alt-white-H}}
		\rewritetarget 						{\vtikzfig{ZH-bott-alt-H-gray}}
	\end{tikzpicture}
	\\[-3.5ex]
	\begin{tikzpicture}
	\setlength\rulediagramwd{5.5em}	
		\rewriterule		[ZH-ME]			{\vtikzfig{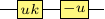}}
		\coordinate (anchor-next-diagram) at ($(anchor-next-diagram) + (-.125,0)$);
	\setlength\rulediagramwd{6.5em}	
		\rewritetarget 						{\vtikzfig{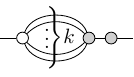}}
	\end{tikzpicture}
	\end{aligned}
	\end{aligned}
\\[-5.25ex]
\mspace{-75mu}
	\begin{aligned}
	\begin{aligned}
	\begin{tikzpicture}
	\setlength\rulediagramwd{4.5em}
		\rewriterule		[ZH-ND]		{\vtikzfig[.25ex]{ZH-not1-not2}}
	\setlength\rulediagramwd{5em}
		\rewritetarget 					{\vtikzfig{ZH-not-transport}}
	\end{tikzpicture}
	\end{aligned}
	&&
		\bigg\vert
	\mspace{-9mu}&&
	\begin{aligned}
	\begin{tikzpicture}
	\setlength\rulediagramwd{5.5em}	
		\rewriterule		[ZH-NH]			{\vtikzfig{ZH-gen-not-gadget}}
	\setlength\rulediagramwd{3em}
		\rewritetarget 						{\vtikzfig{ZH-var-not-dot}}
	\end{tikzpicture}
	\end{aligned}
	&&
		\bigg\vert
	\mspace{-9mu}&&
	\begin{aligned}
	\begin{tikzpicture}
	\setlength\rulediagramwd{3em}
		\rewriterule		[ZH-NA]	 	{\vtikzfig[.25ex]{ZH-not-antipode}}
		\coordinate (anchor-next-diagram) at ($(anchor-next-diagram) + (-.125,0)$);
		\rewritetarget 					{\vtikzfig{ZH-gray-id}}
	\end{tikzpicture}
	\end{aligned}
	\end{aligned}
\\[-8.75ex]
\end{gather*}
\hrule
\smallskip
\caption{%
	\label{fig:ZH-rewrites}
	Various scalar-exact rewrites (including axioms and corollaries) on ZH diagrams, which are sound for the semantics described in Eqn.~\eqref{eqn:idealised-ZH-integrals} subject to $\nu = D^{-1/4}$.
	Chains of rewrites \smash{$\Big.\cD_j \!\xleftrightarrow{\!\!\:\textsf{(x)}\!\:} \cdots \leftrightarrow \cD_{\!\!\;f}$} are intended to indicate that \smash{$\cD_j \!\xleftrightarrow{\!\!\:\textsf{(x)}\!\:}\! \cD_{\!\!\;f}$} is either an axiom or notable corollary.
	Throughout, we have $k \in \N$, $a,b,c, c_1, c_2 \in \Z$ (which may be evaluated modulo $D$); $u,v \in \Z_D^\times$\,; $\mathrm A, \mathrm B: \Z \to \C$; and $\alpha \in \C^\times$.
	H-boxes which are labeled inside with an integer parameter such as $c \in \Z$, indicate an amplitude of $\omega^c = \e^{2\pi i c/D}$; H-boxes labelled with $\texttt+$ or $\texttt-$ indicate $c = \pm 1$ (see Figure~\ref{fig:ZXH-relations}).
	A not dot labeled with $\neg$ indicates a dimension-dependent parameter $-\sigma \in \Z$, where $\sigma = 0$ for $D$ odd and $\sigma = 1$ for $D$ even; more generally, not-dots may be parameterised by $c \in \Z$ for the sake of convenience and reduced modulo $D$ to an element of $\D$.
}
\end{figure}

Figure~\ref{fig:ZH-rewrites} presents a list of all of the rewrites of ZH diagrams which we consider, which are sound for the semantic map defined in Eqn.~\eqref{eqn:idealised-ZH-integrals} subject to $\nu = D^{-1/4}$.
In this section, we demonstrate the soundness of these rewrites.

We first consider rewrites for the ZH~generators, which are sound for a semantic map as consdered in Section~\ref{apx:constraining-Ockhamic-ZH}, \emph{without} fixing the value of $\nu$.
This provides us with multiple versions of the ZH~calculus, parameterised by $\nu$ as well as by $D$, which may be suitable for different purposes.
For instance, following our observation at the end of Section~\ref{apx:constraining-Ockhamic-ZH}, it is likely that a measure in which $\nu = 1$ (corresponding to the original presentation of Backens and Kissinger~\cite{BK-2019} in the special case $D = 2$) may be most appropriate
to analyse counting problems; whereas for the analysis of unitary transformations and interoperation with the ZX~calculus, our analysis suggests that $\nu = D^{-1/4}$ would be more suitable (as well as giving rise to simpler rewrites overall).

In the following, we make frequent use of the `multicharacter' notation for H-boxes, in which an H-box inscribed with $c \in \Z$ denotes that the amplitude $\omega^c$ is intended.
Note that this is different from the original annotation convention set out for the ZH~calculus in Ref.~\cite{BK-2019} and related works such as Ref.~\cite{BKK-2021}, in which amplitudes $\alpha \in \C$ are written in the box.
(In particular, in our work, an H-box with $0 \in \Z$ inscribed indicates the constant amplitude function $\mathrm A(t) = \omega^{0\cdot t} = 1^t = 1$, rather than $\mathrm A(t) = 0^t$, which is not a function on $\D$ for $D > 2$.)
As $\mathrm A(t) = \omega^t$ is well-defined modulo $D$, we may abuse notation slightly and parameterise such `multicharacter' boxes by $c \in \Z_D$, or conversely interpret $c \in \Z$ as though it were an element of $\Z_D$.
In particular, for $u \in \Z_D^\times$, we will make frequent use of the generator $
\tikz \draw (0,0) -- node [midway, H box] {\t u} ++(.75,0);
$
which denotes the operator $\int_{x,y} \omega^{uxy} \, \kket{y}\bbra{x}$.
We also occasionally make use of the short-hand 
$
\tikz \draw (0,0) -- node [midway, small H box] {\tp} ++(.75,0);
$
and
$
\tikz \draw (0,0) -- node [midway, small H box] {\tm} ++(.75,0);
$
for $u = \pm 1$ in particular.

\medskip
\nopagebreak
\smallskip
\Rule{ZH-WF} White dot fusion.
\\[-3ex]
\begin{gather*}
	\Sem{8ex}{\vtikzfig[-3ex]{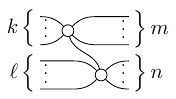}}
\;\;=\;\;
	\Biggsem{\vtikzfig[-3ex]{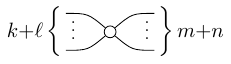}}
\end{gather*}
\bigskip
\allowdisplaybreaks
\noindent
\textbf{Proof.}~\\[-12ex]
\begin{align*}
	\qquad&\mathop{\int \!\!\!\! \int}\limits_{\mathclap{
		x,y \in \D
	}}
	\Bigl(
		\mathbf 1^{\otimes n} \otimes \kket{y}^{\otimes n} \bbra{y}^{\otimes \ell+1}
	\Bigr)
	\Bigl(
		\kket{x}^{\otimes m+1} \bbra{x}^{\otimes k} \otimes \mathbf 1^{\otimes m}
	\Bigr)
\\[1ex]&=\;
	\mathop{\int \!\!\!\! \int}\limits_{\mathclap{
		x,y \in \D
	}}
		\kket{x}^{\otimes m} \bbra{x}^{\otimes k} \,\otimes\; \bbracket{y}{x} \;\otimes\; \kket{y}^{\otimes n} \bbra{y}^{\otimes \ell}
\\[1ex]
\;&=\;
	\int\limits_{\mathclap{
		x \in \D
	}}
		\kket{x}^{\otimes m} \bbra{x}^{\otimes k} \,\otimes\; \kket{x}^{\otimes n} \bbra{x}^{\otimes \ell}
\\[1ex]&=\;
	\int\limits_{\mathclap{
		x \in \D
	}}
		\kket{x}^{\otimes m+n} \bbra{x}^{\otimes k+\ell} 	\;.
\tag*{\qed}
\end{align*}

\medskip
\hrule
\nopagebreak
\smallskip

\Rule{ZH-WQS} White quasi-special fusion.
\\[-3ex]
\begin{gather*}
	\Biggsem{\;\vtikzfig[-5ex]{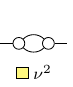}}
\;\;=\;\;
	\Bigsem{\vtikzfig[-2ex]{id-wire}}
\end{gather*}

\medskip
\allowdisplaybreaks
%
\noindent
\textbf{Proof.}~\\[-8.375ex]
\begin{align*}
	\mspace{48mu}
	\nu^2 \!
	\mathop{\int \!\!\!\! \int}\limits_{\mathclap{
		x,y \in \D
	}}
	\kket{y} \bbracket{y,y}{x,x} \bbra{x}
\,&=\,
	\nu^2 \!
	\mathop{\int \!\!\!\! \int}\limits_{\mathclap{
		x,y \in \D
	}}
	\kket{y} \Bigl(\bbracket{y}{x} \otimes \bbracket{y}{x} \Bigr) \bbra{x}
\\[1ex]&=
	\int\limits_{\mathclap{
		x \in \D
	}}
	\!
	\nu^2\,
	 \bbracket{x}{x}\;
	\kket{x} \bbra{x}
\;=
	\int\limits_{\mathclap{
		x \in \D
	}}
	\kket{x} \bbra{x}	\;.
	\mspace{-36mu}
\tag*{\qed}
\end{align*}

\noindent
This result essentially follows from the presence of a redundant factor of $\bbracket{y}{x}$ in the integral, which does not further constrain the variables of integration but does contribute a factor of $\bbracket{x}{x} = \nu^{-2}$.

\medskip
\vfill
\hrule
\nopagebreak
\smallskip

\Rule{ZH-WNS}  Not-dot symmetry.
\\[-3ex]
\begin{gather*}
	\Sem{7ex}{\begin{aligned}
    \begin{tikzpicture}
      \node (Z) at (0,0) [white dot] {};
      \node (h1) at (-.5,0.375) [not dot, label=above:\footnotesize$c$] {};
      \node (h2) at (-.5,-0.375) [not dot, label=below:\footnotesize$c$] {};
      \draw (Z) .. controls (-0.3175,0.375) .. (h1) -- ++(-0.375,0);
      \draw (Z) .. controls (-0.3175,-0.375) .. (h2) -- ++(-0.375,0);
      \node (dots) at ($(Z) + (-0.5,0.125)$) {\footnotesize$\mathbf\vdots$};
      \node (h1) at (.5,0.375) [not dot, label=above:\footnotesize$c$] {};
      \node (h2) at (.5,-0.375) [not dot, label=below:\footnotesize$c$] {};
      \draw (Z) .. controls (0.3175,0.375) .. (h1) -- ++(0.375,0);
      \draw (Z) .. controls (0.3175,-0.375) .. (h2) -- ++(0.375,0);
      \node (dots) at ($(Z) + (0.5,0.125)$) {\footnotesize$\mathbf\vdots$};
    \end{tikzpicture}
	\end{aligned}}
\;\;=\;\;
	\biggsem{\vtikzfig[-3ex]{ZH-white-dot}}
\end{gather*}

\medskip

\noindent
\textbf{Proof.}~\\[-9.5ex]
\begin{align*}
	\mspace{48mu}
	\mathop{\int\!\!\!\!\int\!\!\!\!\int}_{\mathclap{\substack{
		x \in \Z_D^m \\ y \in \Z_D^n \\ z \in \Z_D
	}}}
		\Biggl[ \bigotimes_{j=1}^n \kket{-c{-}y_j}\bbra{y_j} \Biggr]
		\Biggl[ \kket{z}\sox{n}\bbra{z}\sox{m} \Biggr]
		\Biggl[ \bigotimes_{h=1}^m \kket{x_h}\bbra{-c{-}x_h} \Biggr]
		&
\\[-4.5ex]{}=\,
	\mathop{\int}_{\mathclap{\substack{
		z \in \Z_D
	}}}
	\;\;
		\Biggl[ \bigotimes_{j=1}^n \kket{-c{-}z} \Biggr]
		\Biggl[ \bigotimes_{h=1}^m \bbra{-c{-}z} \Biggr]
\,&=\,
	\mathop{\int}_{\mathclap{\substack{
		z \in \Z_D
	}}}
		\kket{z}\bbra{z}	\;.	
\tag*{\qed}
\end{align*}~\\[-2ex]
\textbf{N.B.}~We expect this rewrite can be derived from the others:
the special case of $c=0$ is not difficult to show using \textsf{(ZH-WGB)} and \textsf{(ZH-NA)}.

\medskip
\vfill
\hrule
\smallskip

\Rule{ZH-WGB}  White-gray bialgebra.
\\[-3ex]
\begin{gather*}
	\Biggsem{\;\vtikzfig[-2ex]{ZH-bialg-white-gray}}
\;\;=\;\;
	\biggsem{\vtikzfig[-3ex]{ZH-bott-white-gray}}
\end{gather*}
\medskip

\noindent
\textbf{Proof.}~\\[-9.5ex]
\begin{align*}
	\mspace{90mu}&\mspace{-48mu}
	\mathop{\int \!\!\!\! \int}\limits_{\mathclap{\substack{
		x \in \D^m 
	\\[.25ex]
		z \in \D^n
	}}}
	\;
	\Biggl(
		\bigotimes_{j=1}^n \bbracket{\smash{z_j + \textstyle \sum\limits_{h=1}^m \!x_h}\, }{0\big.}
	\Biggr)
	\,\kket{z}\bbra{x}
\\[-2.5ex]&=\;
	\mathop{\int \!\!\!\! \int}\limits_{\mathclap{\substack{
		x \in \D^m 
	\\[.25ex]
		z \in \D^n
	}}}
	\;
	\Biggl(
		\prod_{j=2}^n \bracket{z_j}{z_1}
	\Biggr)
	\bbracket{\smash{z_1 + \textstyle \sum\limits_{h=1}^m \!x_h}\, }{0\big.}
	\,\Biggl(\bigotimes_{j=1}^n \kket{z_j}\Biggr)\bbra{x}
\\[.125ex]&=\;
	\mathop{\int \!\!\!\! \int}\limits_{\mathclap{\substack{
		x \in \D^m 
	\\[.25ex]
		z_1 \in \D
	}}}
	\bbracket{\smash{z_1 + \textstyle \sum\limits_{h=1}^m \!x_h}\, }{0\big.}
	\,\Biggl(\bigotimes_{j=1}^n \kket{z_1}\Biggr)\bbra{x}
\\[.75ex]&=\;
	\mathop{\int \!\!\!\! \int \!\!\!\! \int}\limits_{\mathclap{\substack{
		x \in \D^m 
	\\[.25ex]
		y,z \in \D
	}}}
	\bbracket{z}{y}
	\bbracket{\smash{y + \textstyle \sum\limits_{h=1}^m \!x_h}\, }{0\big.}
	\;
	\,\kket{z}\sox{n}\bbra{x}
\\[.5ex]&=\;
	\Biggl[\;\;\;
		\int\limits_{\mathclap{
			y \in \D
		}}
		\kket{z}\sox{n}\bbra{z}
	\Biggr]
	\Biggl[\;\;\;\;\,
		\int\limits_{\mathclap{\substack{
			x \in \D^m \\[.25ex] y \in \D
		}}}
		\bbracket{\smash{y + \textstyle \sum\limits_{h=1}^m \!x_h}\, }{0\big.}
		\,\kket{y}\bbra{x}
	\Biggr].
\tag*{\qed}
\end{align*}
\vspace*{-3.5ex}

\medskip
%
\hrule
\nopagebreak
\smallskip
\smallskip

\Rule{ZH-WGC}  White-gray colour change.
\\[-4ex]
\begin{gather*}
	\Biggsem{\vtikzfig[-1ex]{ZH-white-w-unitH}}
\;\;=\;\;
	\Sem{5ex}{\;\vtikzfig[-2ex]{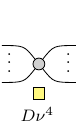}\;}
\end{gather*}

\bgroup
\bigskip
\bigskip
\allowdisplaybreaks
%
\noindent
\textbf{Proof.}~\\[-10ex]\nopagebreak
\begin{align*}
	\mspace{150mu}&\mspace{-90mu}
	\mathop{\int \!\!\!\!  \int \!\!\!\! \int}\limits_{\mathclap{\substack{
		x \in \D^m 
	\\[.25ex]
		y \in \D
	\\[.25ex]
		z \in \D^n
	}}}
		\Biggl(\bigotimes_{j=1}^m \omega^{u\:\!x_{\!\!\;j}\!\: y} \!\Biggr)
		\Biggl(\bigotimes_{k=1}^n \omega^{u\:\!y \!\:z_k} \!\Biggr)
		\kket{z}\bbra{x}
\\[-5ex]&=\;
	\mathop{\int \!\!\!\!   \int}\limits_{\mathclap{\substack{
		x \in \D^m 
	\\[.25ex]
		z \in \D^n
	}}}
	\;
	\Biggl(\;\;\;\;
		\int\limits_{\mathclap{
			y_j \in \D
		}}
		\omega^{uy\bigl(\smash{\sum\limits_j x_j + \sum\limits_k z_k}\bigr)}
	\Biggr)
		\kket{z}\bbra{x}
\\[2ex]&=\;
	\mathop{\int \!\!\!\!   \int}\limits_{\mathclap{\substack{
		x \in \D^m 
	\\[.25ex]
		z \in \D^n
	}}}
	\!
	D\nu^4\,
	\bbracket{\,\big.\smash{%
		\textstyle \sum\limits_j x_j + \sum\limits_k z_k}
    \,\big.}{0} \;
		\kket{z}\bbra{x}
	\;.
\tag*{\qed}
\end{align*}
\egroup

\medskip
\Corollary[rm]{``Double-H''}~\\[-10.5ex]
\begin{equation*}
\qquad\qquad\qquad\qquad\qquad\qquad\quad
\begin{aligned}
	\begin{tikzpicture}
		\node at (0,0) (h1) [small H box] {\t{u}};
		\draw (h1) -- ++(-0.375,0);
		\node at ($(h1) + (0.5,0)$) (h2) [small H box] {\t{u}};
		\draw (h2) -- ++(0.375,0);
		\draw (h1) -- (h2) ;
	\end{tikzpicture}
\end{aligned}
\;\longleftrightarrow\;
\begin{aligned}
	\begin{tikzpicture}
		\node at (0,0) (h1) [small H box] {\t{u}};
		\draw (h1) -- ++(-0.375,0);
		\node at ($(h1) + (0.375,0)$) (w1) [white dot] {};
		\node at ($(w1) + (0.375,0)$) (h2) [small H box] {\t{u}};
		\draw (h2) -- ++(0.375,0);
		\draw (h1) -- (w1) -- (h2);
	\end{tikzpicture}
\end{aligned}
\;\longleftrightarrow\;
\begin{aligned}
	\begin{tikzpicture}
		\node at (0,0) (g1) [gray dot] {};
		\draw (g1) -- ++(-0.625,0);
		\draw (g1) -- ++(0.625,0);
		\node at ($(g1) + (0,-.375)$) [small H box, label=below:$\footnotesize\mathclap{D\nu^4}$] {};
		\node at ($(g1) + (0,.375)$) [small H box, draw=none, fill=none,label=above:$\phantom{\footnotesize\mathclap{D\nu^4}}$] {};
	\end{tikzpicture}
\end{aligned}	
\end{equation*}

\Corollary{ZH-GWC} Gray-white colour change ---
Using the white-gray colour change rules, the double-H rule, and the `antipode symmetry' rule, we may show:~\\[-5.5ex]
\begin{equation*}{}
\mspace{-9mu}
\begin{aligned}
  		\input{./Figures-Source/ZH-gray-w-unitH-arity.tikz}	
  	\end{aligned}
    \;\leftrightarrow\;
 	\begin{aligned}
		\input{./Figures-Source/ZH-white-w-unitH2-and-scalar-arity.tikz}	
	\end{aligned}
    \;\leftrightarrow\;
 	\begin{aligned}
		\input{./Figures-Source/ZH-white-w-gray-scalar-arity.tikz}	
	\end{aligned}
    \;\leftrightarrow\;
 	\begin{aligned}
		\input{./Figures-Source/ZH-white-scalar-arity.tikz}	
	\end{aligned}	
\end{equation*}

\Corollary{ZH-GL} Gray lollipop.
\begin{equation*}
\begin{aligned}
    \begin{tikzpicture}
      \node (Z) at (0,0) [gray dot] {};
      \draw (Z) .. controls (-0.3175,0.375) .. (-.5,0.375) -- ++(-0.1875,0);
      \draw (Z) .. controls (-0.3175,-0.375) .. (-.5,-0.375) -- ++(-0.1875,0);
      \node (dots) at ($(Z) + (-0.5,0.125)$) {\footnotesize$\mathbf\vdots$};
      \draw (Z) .. controls (0.3175,0.375) .. (.5,0.375) -- ++(0.1875,0);
      \draw (Z) .. controls (0.3175,-0.375) .. (.5,-0.375) -- ++(0.1875,0);
      \node (dots) at ($(Z) + (0.5,0.125)$) {\footnotesize$\mathbf\vdots$};
      \node (w) at (0,.5) [gray dot, fill=none, draw=none] {};
      \node (w) at (0,-.5) [gray dot] {};
      \draw (Z) -- (w);
    \end{tikzpicture}
\end{aligned}
\;\longleftrightarrow\;
\begin{aligned}
    \begin{tikzpicture}
      \node (Z) at (0,0) [white dot] {};
      \node (h1) at (-.625,0.375) [mini H box] {\tp};
      \node (h2) at (-.625,-0.375) [mini H box] {\tp};
      \draw (Z) .. controls (-0.3175,0.375) .. (h1) -- ++(-0.25,0);
      \draw (Z) .. controls (-0.3175,-0.375) .. (h2) -- ++(-0.25,0);
      \node (dots) at ($(Z) + (-0.625,0.09375)$) {\footnotesize$\mathbf\vdots$};
      \node (h1) at (.625,0.375) [mini H box] {\tp};
      \node (h2) at (.625,-0.375) [mini H box] {\tp};
      \draw (Z) .. controls (0.3175,0.375) .. (h1) -- ++(0.25,0);
      \draw (Z) .. controls (0.3175,-0.375) .. (h2) -- ++(0.25,0);
      \node (dots) at ($(Z) + (0.625,0.09375)$) {\footnotesize$\mathbf\vdots$};
      \node (w) at (0,.75) [gray dot, fill=none, draw=none] {};
      \node (w) at (0,-.75) [gray dot] {};
      \draw (Z) -- (w) node [pos=0.5875, mini H box] {\tp};
      \node at (0,0.5) [small H box, label=above:\footnotesize$(D\nu^4)^{-1}$] {};
      \node at (0,-0.5) [small H box, fill=none, draw=none, label=below:\footnotesize$\phantom{(D\nu^4)^{-1}}$] {};
    \end{tikzpicture}
\end{aligned}
\;\longleftrightarrow\;
\begin{aligned}
    \begin{tikzpicture}
      \node (Z) at (0,0) [white dot] {};
      \node (h1) at (-.625,0.375) [mini H box] {\tp};
      \node (h2) at (-.625,-0.375) [mini H box] {\tp};
      \draw (Z) .. controls (-0.3175,0.375) .. (h1) -- ++(-0.25,0);
      \draw (Z) .. controls (-0.3175,-0.375) .. (h2) -- ++(-0.25,0);
      \node (dots) at ($(Z) + (-0.625,0.09375)$) {\footnotesize$\mathbf\vdots$};
      \node (h1) at (.625,0.375) [mini H box] {\tp};
      \node (h2) at (.625,-0.375) [mini H box] {\tp};
      \draw (Z) .. controls (0.3175,0.375) .. (h1) -- ++(0.25,0);
      \draw (Z) .. controls (0.3175,-0.375) .. (h2) -- ++(0.25,0);
      \node (dots) at ($(Z) + (0.625,0.09375)$) {\footnotesize$\mathbf\vdots$};
      \node (w) at (0,.5) [white dot, fill=none, draw=none] {};
      \node (w) at (0,-.5) [white dot] {};
      \draw (Z) -- (w) {};
      \node at (0,0.5) [small H box, label=above:\footnotesize$(D\nu^4)^{-1}$] {};
      \node at (0,-0.5) [small H box, fill=none, draw=none, label=below:\footnotesize$\phantom{(D\nu^4)^{-1}}$] {};
    \end{tikzpicture}
\end{aligned}
\;\longleftrightarrow\;
\begin{aligned}
    \begin{tikzpicture}
      \node (Z) at (0,0) [white dot] {};
      \node (h1) at (-.625,0.375) [mini H box] {\tp};
      \node (h2) at (-.625,-0.375) [mini H box] {\tp};
      \draw (Z) .. controls (-0.3175,0.375) .. (h1) -- ++(-0.25,0);
      \draw (Z) .. controls (-0.3175,-0.375) .. (h2) -- ++(-0.25,0);
      \node (dots) at ($(Z) + (-0.625,0.09375)$) {\footnotesize$\mathbf\vdots$};
      \node (h1) at (.625,0.375) [mini H box] {\tp};
      \node (h2) at (.625,-0.375) [mini H box] {\tp};
      \draw (Z) .. controls (0.3175,0.375) .. (h1) -- ++(0.25,0);
      \draw (Z) .. controls (0.3175,-0.375) .. (h2) -- ++(0.25,0);
      \node (dots) at ($(Z) + (0.625,0.09375)$) {\footnotesize$\mathbf\vdots$};
      \node at (0,0.5) [small H box, label=above:\footnotesize$(D\nu^4)^{-1}$] {};
      \node at (0,-0.5) [small H box, fill=none, draw=none, label=below:\footnotesize$\phantom{(D\nu^4)^{-1}}$] {};
    \end{tikzpicture}
\end{aligned}
\;\longleftrightarrow\;
\begin{aligned}
    \begin{tikzpicture}
      \node (Z) at (0,0) [gray dot] {};
      \draw (Z) .. controls (-0.3175,0.375) .. ++(-.625,0.375);
      \draw (Z) .. controls (-0.3175,-0.375) .. ++(-.625,-0.375);
      \node (dots) at ($(Z) + (-0.5,0.125)$) {\footnotesize$\mathbf\vdots$};
      \draw (Z) .. controls (0.3175,0.375) .. ++(.625,0.375);
      \draw (Z) .. controls (0.3175,-0.375) .. ++(.625,-0.375);
      \node (dots) at ($(Z) + (0.5,0.125)$) {\footnotesize$\mathbf\vdots$};
    \end{tikzpicture}
\end{aligned}	
\end{equation*}

\Corollary{ZH-GF} Gray dot fusion.~\\[-2.5ex]
\begin{equation*}
\begin{aligned}[b]
	\begin{aligned}\input{./Figures-Source/ZH-gray-fuse-arity.tikz}\end{aligned}
\;\;\longleftrightarrow{}&\!
	\begin{aligned}\input{./Figures-Source/ZH-white-w-H-gray-fuse-arity.tikz}\end{aligned}
\!\!\!\longleftrightarrow\!
	\begin{aligned}\input{./Figures-Source/ZH-white-w-H-fuse-arity.tikz}\end{aligned}
\\[-2.5ex]\;\longleftrightarrow{}&\;\;
	\begin{aligned}\input{./Figures-Source/ZH-white-w-H-sum-arity.tikz}\end{aligned}
\;\;\longleftrightarrow\;\;\;
	\begin{aligned}\input{./Figures-Source/ZH-gray-sum-arity.tikz}\end{aligned}
\end{aligned}
\end{equation*}

\smallskip
\Corollary{ZH-A} Antipode rule.\\[-2.5ex]
\begin{equation*}
\begin{aligned}[b]
	\begin{aligned}
	\begin{tikzpicture}
		\node (Z) [white dot] at (0,0) {};
		\draw (Z) -- ++(-0.375,0);
		\node (G) [gray dot] at (.75,0) {};
		\draw (G) -- ++(0.375,0);
		\draw [out=45,in=135] (Z) to (G);
		\draw [out=-45,in=-135] (Z) to node [midway, gray dot] {} (G);
	\end{tikzpicture}
	\end{aligned}
	\;\;\leftrightarrow\;\;
	\begin{aligned}
	\begin{tikzpicture}
		\node (Z) [white dot] at (0,0) {};
		\node (Z') [white dot, fill=none, draw=none] at ($(Z) + (0,0.375)$) {};
		\node (Z') [white dot] at ($(Z) + (0,-0.375)$) {};
		\draw (Z) -- ++(-0.375,0);
		\node (G) [gray dot] at (.5,0) {};
		\node (G') [gray dot] at ($(G) + (0,-.375)$) {};
		\draw (G) -- ++(0.375,0);
		\draw (Z) to (G);
		\draw (Z) to (G');
		\draw (Z') to (G);
		\draw (Z') to (G');
	\end{tikzpicture}
	\end{aligned}
	\;\;\leftrightarrow\;\;
	\begin{aligned}
	\begin{tikzpicture}
		\node (Z) [white dot] at (0,0) {};
		\node (Z') [white dot, fill=none, draw=none] at ($(Z) + (0,0.375)$) {};
		\node (Z') [white dot] at ($(Z) + (0,-0.375)$) {};
		\draw (Z) -- ++(-0.625,0);
		\draw (Z') -- ++(-0.3125,0) node [white dot] {};
		\node (G) [gray dot] at (.5,0) {};
		\node (G') [gray dot] at ($(G) + (0,-.375)$) {};
		\draw (G) -- ++(0.625,0);
		\draw (G') -- ++(0.3125,0) node [gray dot] {};
		\draw (Z) to (G);
		\draw (Z) to (G');
		\draw (Z') to (G);
		\draw (Z') to (G');
	\end{tikzpicture}
	\end{aligned}
	\;\;\leftrightarrow\;\;
	\begin{aligned}
	\begin{tikzpicture}
		\node (Z) [white dot, fill=none, draw=none] at (0,0) {};
		\node (Z') [white dot, fill=none, draw=none] at ($(Z) + (0,0.375)$) {};
		\node (Z') [white dot, fill=none, draw=none] at ($(Z) + (0,-0.375)$) {};
		\node (G) [gray dot] at ($(Z)!0.5!(Z')$) {};
		\draw [out=135,in=0] (G) to ($(G) + (-0.625,0.1875)$);
		\draw [out=225,in=0] (G) to ($(G) + (-0.3125,-0.1875)$) node [white dot] {};
		\node (G'') [gray dot, draw=none, fill=none] at (.5,0) {};
		\node (G') [gray dot, draw=none, fill=none] at ($(G'') + (0,-.375)$) {};
		\node (Z) [white dot] at ($(G')!0.5!(G'')$) {};
		\draw [out=45,in=180] (Z) to ($(Z) + (0.625,0.1875)$) -- ++(0.25,0);
		\draw [out=-45,in=180] (Z) to ($(Z) + (0.3125,-0.1875)$) node [gray dot] {};
		\draw (Z) to (G);
	\end{tikzpicture}
	\end{aligned}
	\;\;\leftrightarrow\;\;
	\begin{aligned}
	\begin{tikzpicture}
		\node (Z) [white dot, fill=none, draw=none] at (0,0) {};
		\node (Z') [white dot, fill=none, draw=none] at ($(Z) + (0,0.375)$) {};
		\node (Z') [white dot, fill=none, draw=none] at ($(Z) + (0,-0.375)$) {};
		\node (G) [white dot] at ($(Z)!0.5!(Z')$) {};
		\draw [out=135,in=0] (G) to ($(G) + (-0.625,0.1875)$);
		\node (G'') [gray dot, draw=none, fill=none] at (.75,0) {};
		\node (G') [gray dot, draw=none, fill=none] at ($(G'') + (0,-.375)$) {};
		\node (Z) [white dot] at ($(G')!0.5!(G'')$) {};
		\draw (Z) -- ++(-0.3125,0) node [white dot] {};
		\draw [out=45,in=180] (Z) to ($(Z) + (0.625,0.1875)$) -- ++(0.25,0);
		\draw [out=-45,in=180] (Z) to ($(Z) + (0.3125,-0.1875)$) node [gray dot] {};
	\end{tikzpicture}
	\end{aligned}
	\;\;\leftrightarrow\;\;
	\begin{aligned}
	\begin{tikzpicture}
		\node (Z) [white dot] at (0,0) {};
		\draw (Z) -- ++(-0.5,0);
		\node (G) [gray dot] at (.5,0) {};
		\draw (G) -- ++(0.5,0);
	\end{tikzpicture}
	\end{aligned}
\end{aligned}
\end{equation*}

\smallskip
\hrule
\nopagebreak
\smallskip
\smallskip

\Rule{ZH-EC} 
Exponent complement.
\\[-3ex]
\begin{gather*}
	\Biggsem{\,\vtikzfig[-3ex]{ZH-exponent-compl}\,}
\;\;=\;\;
	\Sem{8ex}{\;\vtikzfig[-3ex]{ZH-exponent-compl-gadget}\;}
\end{gather*}

\noindent
\textbf{Proof.}~\\[-8ex]
\begin{align*}
	\mspace{60mu}&
		\mathop{\int \!\!\!\! \int \!\!\!\! \int }\limits_{\mathclap{\substack{
		x \in \D^m
	\\[.5ex]
		y,z \in \D
	}}}
		\alpha^{\Pi(x)\!\:y} \;
		\kket{\neg z} \bbracket{z}{y} \bbra{x}
\;=\;
	\mathop{\int \!\!\!\! \int \!\!\!\! \int }\limits_{\mathclap{\substack{
		x \in \D^m
	\\[.5ex]
		y,z \in \D
	}}}
		\alpha^{\Pi(x)\!\:y} \;
		\kket{z} \bbracket{\neg z}{y} \bbra{x}
		\mspace{-36mu}
\\[1ex]&=\;
	\mathop{\int \!\!\!\! \int \!\!\!\! \int }\limits_{\mathclap{\substack{
		x \in \D^m
	\\[.5ex]
		y,z \in \D
	}}}
		\alpha^{\Pi(x)\!\:y} \;
		\kket{z} \bbracket{\sigma {-} z}{y} \bbra{x}
\;\;=\;
	\mathop{\int \!\!\!\! \int }\limits_{\mathclap{\substack{
		x \in \D^m
	\\[.5ex]
		z \in \D
	}}}
		\alpha^{\Pi(x)\!\:(\sigma - z)} \;
		\kket{z} \bbra{x}
		\mspace{-36mu}
\\[1ex]&=\;
	\mathop{\int \!\!\!\! \int }\limits_{\mathclap{\substack{
		x \in \D^m
	\\[.5ex]
		z \in \D
	}}}
		\bigl(\, \alpha^{\sigma} \,\bigr)^{\!\Pi(x)} \cdot
		\bigl(\, \alpha^{-1} \,\bigr)^{\!\Pi(x) z} \;
		\kket{z} \bbra{x}
	\;.
\tag*{\qed}
\end{align*}

\medskip
\hrule
\nopagebreak
\smallskip
\smallskip

\Rule{ZH-MF} Multicharacter fusion.
~\\[-3ex]
\begin{gather*}
	\Sem{10ex}{\;\vtikzfig[-1ex]{ZH-H-complicated-fuse}}
\;\;=\;\;
	\Sem{6ex}{\;\vtikzfig[-2ex]{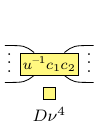}\;}
\end{gather*}

\medskip
\noindent
We use the notation $\Pi(x)$, in effect representing a function ${\Pi : \Z_D^m \to \Z_D}$ which we define by $\Pi(x) = x_1 x_2 \cdots x_m$.
As a mild abuse of notation, we allow $\Pi$ to act on arguments of various sizes, \emph{e.g.},~$x \in \Z_D^m$ and $y \in \Z_D^n$.

\smallskip
\bigskip
\allowdisplaybreaks
\noindent
\textbf{Proof.}~\\[-9ex]
\begin{align*}
	\mspace{90mu}&\mspace{-12mu}
	\mathop{\int \!\!\!\! \int \!\!\!\! \int \!\!\!\! \int \!\!\!\! \int \!\!\!\! \int }\limits_{\mathclap{\substack{
		r \in \Z_D^k\!\!\;,\, s \in \Z_D^\ell
	\\[.5ex]
		x \in \Z_D^m \!\!\;,\, y \in \Z_D^n
	\\[.5ex]
		t,z \in \Z_D
	}}}
		\omega^{c_1 \!\: \Pi(r)\!\:t\,\Pi(x)} \;
		\omega^{-utz} \,
		\omega^{c_2 \!\:\Pi(s)\!\:z\!\;\Pi(y)} \;
		\kket{x,y}\bbra{r,s}
\\[.5ex]&=
	\mathop{\int \!\!\!\! \int \!\!\!\! \int \!\!\!\! \int \!\!\!\! \int }\limits_{\mathclap{\substack{
		r \in \Z_D^k\!\!\;,\, s \in \Z_D^\ell
	\\[.5ex]
		x \in \Z_D^m \!\!\;,\, y \in \Z_D^n
	\\[.5ex]
		t \in \Z_D
	}}}
	\omega^{c_1\Pi(r)\!\:t\,\Pi(x)} \;
	\Biggl(\;\;\;
		\int\limits_{\mathclap{
			z \in \Z_D
		}}
		\omega^{z\bigl(-ut + v_2\:\!\Pi(s)\,\Pi(y)\bigr)}
	\Biggr)\,
		\kket{x,y}\bbra{r,s}
\\[1.5ex]&=
	\mathop{\int \!\!\!\! \int \!\!\!\! \int \!\!\!\! \int \!\!\!\! \int }\limits_{\mathclap{\substack{
		r \in \Z_D^k\!\!\;,\, s \in \Z_D^\ell
	\\[.5ex]
		x \in \Z_D^m \!\!\;,\, y \in \Z_D^n
	\\[.5ex]
		t \in \Z_D
	}}}
	D\nu^4 \cdot \bbracket{-ut + c_2\:\!\Pi(s)\,\Pi(y)\big.}{0} \,
	\cdot \omega^{c_1\Pi(r)\!\:t\,\Pi(x)} \;
		\kket{x,y}\bbra{r,s}
	\mspace{-30mu}
\\[1.25ex]&=\,
	D\nu^4 \!\mathop{\int \!\!\!\! \int \!\!\!\! \int \!\!\!\! \int }\limits_{\mathclap{\substack{
		r \in \Z_D^k\!\!\;,\, s \in \Z_D^\ell
	\\[.5ex]
		x \in \Z_D^m \!\!\;,\, y \in \Z_D^n
	}}}
		\omega^{u^{-1}c_1c_2 \cdot \Pi(r)\,\Pi(s)\,\Pi(x)\,\Pi(y)} \;
		\kket{x,y}\bbra{r,s}
	\;.
\tag*{\qed}
\end{align*}~\\[-4ex]

\Corollary[rm]{``Multicharacter negation''} ~~
$
\begin{aligned}
	\begin{tikzpicture}
		\node at (0,0) (h1) [H box] {\t c};
		\draw (h1.north west) .. controls ++(-0.1875,0.25) .. ++(-0.375,0.25);
		\draw (h1.south west) .. controls ++(-0.1875,-0.25) .. ++(-0.375,-0.25);
		\node at ($(h1) + (-0.375,0.0875)$) {$\vdots$};
		\node at ($(h1) + (0.5,0)$) (g) [gray dot] {};
		\draw (g) -- ++(0.375,0);
		\draw (h1) -- (g) ;
	\end{tikzpicture}
\end{aligned}
\;\longleftrightarrow\;
\begin{aligned}
	\begin{tikzpicture}
		\node at (0,0) (h1) [H box] {\t c};
		\draw (h1.north west) .. controls ++(-0.1875,0.25) .. ++(-0.375,0.25);
		\draw (h1.south west) .. controls ++(-0.1875,-0.25) .. ++(-0.375,-0.25);
		\node at ($(h1) + (-0.4125,0.0875)$) {$\vdots$};
		\node at ($(h1) + (0.5,0)$) (h2) [small H box] {\tp};
		\node at ($(h2) + (0.5,0)$) (h3) [small H box] {\tp};
		\draw (h3) -- ++(0.375,0);
		\draw (h1) -- (h2) -- (h3) ;
		\node at ($(h2) + (0,-.375)$) [small H box, label=below:\footnotesize$\mathclap{(D\nu^4)^{-1}}$] {};
		\node at ($(h2) + (0,.375)$) [small H box, draw=none, fill=none,label=above:\footnotesize$\phantom{\mathclap{D\nu^4)^{-1}}}$] {};
	\end{tikzpicture}
\end{aligned}
\;\longleftrightarrow\;
\begin{aligned}
	\begin{tikzpicture}
		\node at (0,0) (h1) [H box] {\t {-c}};
		\draw (h1.north west) .. controls ++(-0.1875,0.25) .. ++(-0.375,0.25);
		\draw (h1.south west) .. controls ++(-0.1875,-0.25) .. ++(-0.375,-0.25);
		\node at ($(h1) + (-0.4125,0.0875)$) {$\vdots$};
		\draw (h1) -- ++(0.625,0);
		\draw (h1) ;
	\end{tikzpicture}
\end{aligned}
$

\medskip
\noindent
This corollary is also in principle a special case of the ``mixed bialgebra rule'' \textsf{(ZH-HMB)}.
We note it here as a frequently-used special case involving the relationship between multicharacters and the degree-2 gray dot, which may be construed as actions on single qudits.

\medskip
\hrule
\nopagebreak
\smallskip

\Rule{ZH-HM}  H-box multiplication.
\\[-3ex]
\begin{gather*}
	\biggsem{\vtikzfig[-3ex]{ZH-mult}}
\;=\;
	\Bigsem{\vtikzfig[-2.25ex]{ZH-H-prod-prep}\;}
\end{gather*}
\smallskip

\noindent
\textbf{Proof.}~\\[-9.5ex]
\begin{align*}
	\mspace{120mu}&\mspace{-80mu}
	\mathop{\int \!\!\!\! \int \!\!\!\! \int}\limits_{\mathclap{
		x,y,z \in \D 
	}}
		\mathrm{A}(x) \, \mathrm{B}(y) \; \kket{z}\bbracket{z,z}{x,y}
\\[-1.5ex]&=\;
	\int\limits_{\mathclap{
		z \in \D 
	}}
		\mathrm{A}(z) \;\! \mathrm{B}(z)\; \kket{z}
\;=\;
	\int\limits_{\mathclap{
		z \in \D 
	}}
		\bigl[\mathrm{A}\cdot \mathrm{B}\bigr](z) \; \kket{z}
\;.
\tag*{\qed}
\end{align*}~\\[-1ex]

\Corollary{ZH-MCA} Multicharacter addition. ~~
    $\begin{aligned}\\[-2.5ex]\input{./Figures-Source/ZH-multichar-add.tikz}\end{aligned}
    \;\longleftrightarrow\;
    \begin{aligned}\\[-2.5ex]\input{./Figures-Source/ZH-multichar-prep.tikz}\end{aligned}$
    
\smallskip
\hrule
\smallskip

\Rule{ZH-HU} H-box unit.
\\[-3ex]
\begin{gather*}
	\Bigsem{\,\input{./Figures-Source/ZH-H0-prep.tikz}\,} \;=\;
	\Bigsem{\;\input{./Figures-Source/ZH-white-prep.tikz}\;}
\end{gather*}
\bigskip
\noindent
\textbf{Proof.}
$
\begin{aligned}[b]
\begin{aligned}[t]
	\int\limits_{\mathclap{
		x \in \D 
	}}
		\omega^{0 \;\!\cdot x} \; \kket{x}
\;&=\;
	\int\limits_{\mathclap{
		x \in \D 
	}}
		\kket{x}	\;.
\end{aligned}
\end{aligned}
$
\hfill
$\qed$

\vspace*{-1ex}
\hrule
\nopagebreak
\smallskip

\Rule{ZH-NH} \\
Generalised-not dots:
\\[-3ex]
\begin{gather*}
	\biggsem{\;\vtikzfig[0.25ex]{ZH-gen-not-gadget}}
\;\;=\;\;
	\Biggsem{\;\vtikzfig[-3ex]{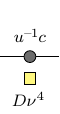}\;}
\end{gather*}
\medskip\medskip

\noindent
\textbf{Proof.}~\\[-10ex]
\begin{align*}
	\mspace{150mu}&\mspace{-72mu}
	\mathop{\int \!\!\!\!  \int \!\!\!\! \int}\limits_{\mathclap{\substack{
		x,y,z \in \D 
	}}}
		\omega^{uxy} \omega^{cy} \omega^{uyz} \kket{z}\bbra{x}
\\[-2ex]&=\;
	\mathop{\int \!\!\!\! \int}\limits_{\mathclap{\substack{
		x,z \in \D 
	}}}
	\,
	\Biggl(\;\;\;
	\int\limits_{\mathclap{
		y \in \D 
	}}
		\omega^{uy(z\,+\,x\,+\,u^{-1}c)}
	\Biggr)\,
		 \kket{z}\bbra{x}
\\[2ex]&=\;
	\mathop{\int \!\!\!\! \int}\limits_{\mathclap{\substack{
		x,z \in \D 
	}}}
	\,
	D\nu^4 \, \bbracket{z}{-u^{-1}c-x} \;
		 \kket{z}\bbra{x}
\\[1ex]&=\;
	D\nu^4 	\int\limits_{\mathclap{
		x \in \D 
	}}
		 \kket{-u^{-1}c-x}\bbra{x}.
\tag*{\qed}
\end{align*}

\vspace*{-4ex}
\Corollary{ZH-NA} Not-antipode. ~
$
\begin{aligned}[t]
\input{./Figures-Source/ZH-not-antipode.tikz}
\;&\longleftrightarrow\;
\begin{aligned}
	\begin{tikzpicture}
		\node at (0,0) (h1) [small H box] {\tp};
		\draw (h1) -- ++(-0.375,0);
		\node at ($(h1) + (0.375,0)$) (w1) [white dot] {};
		\node at ($(w1) + (0.375,0)$) (h2) [small H box] {\tp};
		\draw (h2) -- ++(0.375,0);
		\draw (h1) -- (w1) -- (h2);
		\node at ($(w1) + (0,0.375)$) (ha) [small H box] {\t 0};
		\node at ($(w1) + (0,-0.375)$) [small H box, draw=none, fill=none] {\phantom{\t c}};
		\draw (w1) -- (ha);
		\node at ($(w1) + (0,-.5)$) [small H box, label=below:\footnotesize$(D\nu^4)^{-1}$] {};
		\node at ($(w1) + (0,.5)$) [small H box, draw=none, fill=none, label=above:\footnotesize$\phantom{(D\nu^4)^{-1}}$] {};
	\end{tikzpicture}
\end{aligned}
\;\longleftrightarrow\;
\begin{aligned}
	\begin{tikzpicture}
		\node at (0,0) (h1) [small H box] {\tp};
		\draw (h1) -- ++(-0.375,0);
		\node at ($(h1) + (0.375,0)$) (w1) [white dot] {};
		\node at ($(w1) + (0.375,0)$) (h2) [small H box] {\tp};
		\draw (h2) -- ++(0.375,0);
		\draw (h1) -- (w1) -- (h2);
		\node at ($(w1) + (0,0.375)$) (ha) [white dot] {};
		\node at ($(w1) + (0,-0.375)$) [white dot, draw=none, fill=none] {};
		\draw (w1) -- (ha);
		\node at ($(w1) + (0,-.5)$) [small H box, label=below:\footnotesize$(D\nu^4)^{-1}$] {};
		\node at ($(w1) + (0,.5)$) [small H box, draw=none, fill=none, label=above:\footnotesize$\phantom{(D\nu^4)^{-1}}$] {};
	\end{tikzpicture}
\end{aligned}
\;\longleftrightarrow\;
\begin{aligned}
	\begin{tikzpicture}
		\node at (0,0) (h1) [small H box] {\tp};
		\draw (h1) -- ++(-0.375,0);
		\node at ($(h1) + (0.375,0)$) (w1) [white dot] {};
		\node at ($(w1) + (0.375,0)$) (h2) [small H box] {\tp};
		\draw (h2) -- ++(0.375,0);
		\draw (h1) -- (w1) -- (h2);
		\node at ($(w1) + (0,-.5)$) [small H box, label=below:\footnotesize$(D\nu^4)^{-1}$] {};
		\node at ($(w1) + (0,.5)$) [small H box, draw=none, fill=none, label=above:\footnotesize$\phantom{(D\nu^4)^{-1}}$] {};
	\end{tikzpicture}
\end{aligned}
\;\longleftrightarrow\;
\begin{aligned}
	\begin{tikzpicture}
		\node (g) [gray dot] {};
		\draw (g) -- ++(-0.5,0);
		\draw (g) -- ++(0.5,0);
	\end{tikzpicture}
\end{aligned}
\end{aligned}
$

\medskip
\Corollary{ZH-ND} Not-displacement.
~\\[-5ex]
\begin{align*}
\input{./Figures-Source/ZH-not1-not2.tikz}
\;&\longleftrightarrow\;
\begin{aligned}
	\begin{tikzpicture}
		\node at (0,0) (h1) [small H box] {\tp};
		\draw (h1) -- ++(-0.375,0);
		\node at ($(h1) + (0.375,0)$) (w1) [white dot] {};
		\node at ($(w1) + (0.375,0)$) (h2) [small H box] {\tp};
		\node at ($(h2) + (0.375,0)$) (h3) [small H box] {\tp};
		\node at ($(h3) + (0.375,0)$) (w2) [white dot] {};
		\node at ($(w2) + (0.375,0)$) (h4) [small H box] {\tp};
		\draw (h4) -- ++(0.375,0);
		\draw (h1) -- (w1) -- (h2) -- (h3) -- (w2) -- (h4);
		\node at ($(w1) + (0,0.375)$) (ha) [small H box] {\t{c_1}};
		\node at ($(w2) + (0,0.375)$) (hb) [small H box] {\t{c_2}};
		\node at ($(w2) + (0,-0.375)$) [small H box, draw=none, fill=none] {\phantom{\t c}};
		\draw (w1) -- (ha);
		\draw (w2) -- (hb);
		\node at ($(h2)!0.5!(h3) + (0,-.5)$) [small H box, label=below:\footnotesize$(D\nu^4)^{-2}$] {};
		\node at ($(h2)!0.5!(h3) + (0,.5)$) [small H box, draw=none, fill=none, label=above:\footnotesize$\phantom{(D\nu^4)^{-2}}$] {};
	\end{tikzpicture}
\end{aligned}
\;\longleftrightarrow\;
\begin{aligned}
	\begin{tikzpicture}
		\node at (0,0) (h1) [small H box] {\tp};
		\draw (h1) -- ++(-0.375,0);
		\node at ($(h1) + (0.375,0)$) (w1) [white dot] {};
		\node at ($(w1) + (0.375,0)$) (g) [gray dot] {};
		\node at ($(g) + (0.375,0)$) (w2) [white dot] {};
		\node at ($(w2) + (0.375,0)$) (h4) [small H box] {\tp};
		\draw (h4) -- ++(0.375,0);
		\draw (h1) -- (w1) -- (g) -- (w2) -- (h4);
		\node at ($(w1) + (0,0.375)$) (ha) [small H box] {\t{c_1}};
		\node at ($(w2) + (0,0.375)$) (hb) [small H box] {\t{c_2}};
		\node at ($(w2) + (0,-0.375)$) [small H box, draw=none, fill=none] {\phantom{\t c}};
		\draw (w1) -- (ha);
		\draw (w2) -- (hb);
		\node at ($(g) + (0,-.5)$) [small H box, label=below:\footnotesize$(D\nu^4)^{-1}$] {};
		\node at ($(g) + (0,.5)$) [small H box, draw=none, fill=none, label=above:\footnotesize$\phantom{(D\nu^4)^{-1}}$] {};
	\end{tikzpicture}
\end{aligned}
\;\longleftrightarrow\;
\begin{aligned}
	\begin{tikzpicture}
		\node at (0,0) (h1) [small H box] {\tp};
		\draw (h1) -- ++(-0.375,0);
		\node at ($(g1) + (0.375,0)$) (g) [gray dot] {};
		\node at ($(g) + (0.375,0)$) (w1) [white dot] {};
		\node at ($(w1) + (0.5,0)$) (w2) [white dot] {};
		\node at ($(w2) + (0.375,0)$) (h4) [small H box] {\tp};
		\draw (h4) -- ++(0.375,0);
		\draw (h1) --  (g) -- (w1) -- (w2) -- (h4);
		\node at ($(w1) + (0,0.375)$) (g') [gray dot] {};
		\node at ($(g') + (0,0.375)$) (ha) [small H box] {\t{c_1}};
		\node at ($(w2) + (0,0.375)$) (hb) [small H box] {\t{c_2}};
		\node at ($(w2) + (0,-0.375)$) [small H box, draw=none, fill=none] {\phantom{\t c}};
		\draw (w2) -- (hb);
		\draw (w1) -- (g') -- (ha);
		\node at ($(w1) + (0,-.5)$) [small H box, label=below:\footnotesize$(D\nu^4)^{-1}$] {};
		\node at ($(w1) + (0,.5)$) [small H box, draw=none, fill=none, label=above:\footnotesize$\phantom{(D\nu^4)^{-1}}$] {};
	\end{tikzpicture}
\end{aligned}
\;\longleftrightarrow\;
\begin{aligned}
	\begin{tikzpicture}
		\node at (0,0) (h1) [small H box] {\tp};
		\draw (h1) -- ++(-0.375,0);
		\node at ($(h1) + (0.5,0)$) (g) [gray dot] {};
		\node at ($(g) + (0.5,0)$) (w1) [white dot] {};
		\node at ($(w1) + (0.5,0)$) (h4) [small H box] {\tp};
		\draw (h4) -- ++(0.375,0);
		\draw (h1) -- (g) -- (w1)  -- (h4);
		\node at ($(w1) + (-0.375,0.4125)$) (ha) [small H box] {\t{-c_1}};
		\node at ($(w1) + (0.375,0.4125)$) (hb) [small H box] {\t{\;\!\mathrlap{\phantom -}c_2\;\!}};
		\node at ($(w1) + (0,-0.375)$) [small H box, draw=none, fill=none] {\phantom{\t c}};
		\draw (w1) -- (ha);
		\draw (w1) -- (hb);
		\node at ($(g)!0.75!(w1) + (0,-.5)$) [small H box, label=below:\footnotesize$(D\nu^4)^{-1}$] {};
		\node at ($(g)!0.75!(w1) + (0,.5)$) [small H box, draw=none, fill=none, label=above:\footnotesize$\phantom{(D\nu^4)^{-1}}$] {};
	\end{tikzpicture}
\end{aligned}
\\[-1ex]&\longleftrightarrow\;
\begin{aligned}
	\begin{tikzpicture}
		\node at (0,0) (h1) [small H box] {\tp};
		\draw (h1) -- ++(-0.375,0);
		\node at ($(h1) + (0.4125,0)$) (h2) [small H box] {\tp};
		\node at ($(h2) + (0.4125,0)$) (h3) [small H box] {\tp};
		\node at ($(h3) + (0.4125,0)$) (w1) [white dot] {};
		\node at ($(w1) + (0.4125,0)$) (h4) [small H box] {\tp};
		\draw (h4) -- ++(0.375,0);
		\draw (h1) --  (h2) -- (h3) -- (w1) -- (h4);
		\node at ($(w1) + (-0,0.4125)$) (h') [small H box] {\t{\,c_2 \!\!\;-\!\!\; c_1\,}};
		\node at ($(w1) + (0,-0.375)$) [small H box, draw=none, fill=none] {\phantom{\t c}};
		\draw (w1) -- (h');
		\node at ($(h2) + (0,-.5)$) [small H box, label=below:\footnotesize$(D\nu^4)^{-2}$] {};
		\node at ($(h2) + (0,.5)$) [small H box, draw=none, fill=none, label=above:\footnotesize$\phantom{(D\nu^4)^{-2}}$] {};
	\end{tikzpicture}
\end{aligned}
\;\longleftrightarrow\;
\begin{aligned}
	\begin{tikzpicture}
		\node at (0,0) [not dot, draw=none, fill=none, label=below:\footnotesize\phantom{$c_1{-}c_2$}] {};
		\node at (0,0) (n) [not dot, label=above:\footnotesize$c_2{\!\!\;-}c_1$] {};
		\draw (n) -- ++(0.5,0);
		\node at ($(n) + (-0.625,0)$) (g) [gray dot] {};
		\draw (g) -- ++(-0.5,0);
		\draw (n) -- (g);
	\end{tikzpicture}
\end{aligned}
\;\longleftrightarrow\;
\input{./Figures-Source/ZH-not-transport.tikz}
\end{align*}

\hrule
\medskip

\noindent
\textbf{\textsf{(ZH-WI)},\! \textsf{(ZH-AI)},\! \textsf{(ZH-HI)},\! \textmd{\itshape etc.}} ---
Identity rules:

\medskip
$
    \bigsem{\,\begin{aligned}
	\begin{tikzpicture}
		\node at (0,0) (w) [white dot] {};
		\draw (w) -- ++(-0.375,0);
		\draw (w) -- ++(0.375,0);
	\end{tikzpicture}
	\end{aligned}
	\,}
=
    \bigsem{\,\begin{aligned}
	\begin{tikzpicture}
		\node at (0,0) (g1) [gray dot] {};
		\draw (g1) -- ++(-0.3125,0);
		\node at ($(g1) + (0.4125,0)$) (g2) [gray dot] {};
		\draw (g2) -- ++(0.3125,0);
		\draw (g1) -- (g2);
	\end{tikzpicture}
\end{aligned}
\,}
=
    \Bigsem{\begin{aligned}
	\begin{tikzpicture}
		\node at (0,0) (g1) [not dot, draw=none, fill=none, label=below:$\phantom c$] {};
		\node at (0,0) (g1) [not dot, label=above:\footnotesize$c$] {};
		\draw (g1) -- ++(-0.3125,0);
		\node at ($(g1) + (0.4125,0)$) (g2) [not dot, label=above:\footnotesize$c$] {};
		\draw (g2) -- ++(0.3125,0);
		\draw (g1) -- (g2);
	\end{tikzpicture}
\end{aligned}}
=
    \Sem{10ex}{\begin{aligned}
	\begin{tikzpicture}
		\node at (0,0) (h1) [small H box] {\t{\;\!\mathrlap{\phantom-}u\;\!}};
		\draw (h1) -- ++(-0.4375,0);
		\node at ($(h1) + (0.5375,0)$) (h2) [small H box] {\t{-u}};
		\draw (h2) -- ++(0.5,0);
		\draw (h1) -- (h2);
		\node at ($(h1)!0.5!(h2) + (0,-.375)$) [small H box, label=below:\footnotesize$(D\nu^4)^{-1}$] {};
		\node at ($(h1)!0.5!(h2) + (0,.375)$) [small H box, draw=none, fill=none,label=above:\footnotesize$\phantom{(D\nu^4)^{-1}}$] {};
	\end{tikzpicture}
\end{aligned}}
=
    \Sem{10ex}{\,\begin{aligned}
	\begin{tikzpicture}
		\node at (0,0) (h1) [small H box] {\t{\mathrlap{\phantom-}u}};
		\draw (h1) -- ++(-0.375,0);
		\node at ($(h1) + (0.375,0)$) (h2) [small H box] {\t{\mathrlap{\phantom-}u}};
		\node at ($(h2) + (0.375,0)$) (h3) [small H box] {\t{\mathrlap{\phantom-}u}};
		\node at ($(h3) + (0.375,0)$) (h4) [small H box] {\t{\mathrlap{\phantom-}u}};
		\node at ($(h2)!0.5!(h3) + (0,-.375)$) [small H box, label=below:\footnotesize$(D\nu^4)^{-2}$] {};
		\node at ($(h2)!0.5!(h3) + (0,.375)$) [small H box, draw=none, fill=none,label=above:\footnotesize$\phantom{(D\nu^4)^{-2}}$] {};
		\draw (h4) -- ++(0.375,0);
		\draw (h1) -- (h2) -- (h3) -- (h4);
	\end{tikzpicture}
\end{aligned}\,}
=
    \bigsem{\,\begin{aligned}\\[-3ex] \input{./Figures-Source/id-wire.tikz}\end{aligned}\,}
$\!\!

\smallskip
\bigskip
\noindent
\textbf{Proof.}  The white-dot identity is trivial: its semantics is the resolution of the identity, Eqn.~\eqref{eqn:resolution-of-the-identity}.
The generalised not-dot identity rule can be proven from \textsf{(ZH-NS)}:
\begin{equation}
\begin{aligned}
	\begin{tikzpicture}
		\node at (0,0) (g1) [not dot, draw=none, fill=none, label=below:$\phantom c$] {};
		\node at (0,0) (g1) [not dot, label=above:\footnotesize$c$] {};
		\draw (g1) -- ++(-0.375,0);
		\node at ($(g1) + (0.5,0)$) (g2) [not dot, label=above:\footnotesize$c$] {};
		\draw (g2) -- ++(0.375,0);
		\draw (g1) -- (g2);
	\end{tikzpicture}
\end{aligned}
\;\longleftrightarrow\;
\begin{aligned}
	\begin{tikzpicture}
		\node at (0,0) (g1) [not dot, draw=none, fill=none, label=below:$\phantom 0$] {};
		\node at (0,0) (g1) [not dot, label=above:\footnotesize$c$] {};
		\draw (g1) -- ++(-0.375,0);
		\node at ($(g1) + (0.5,0)$) (w) [white dot] {};
		\node at ($(w) + (0.5,0)$) (g2) [not dot, label=above:\footnotesize$c$] {};
		\draw (g2) -- ++(0.375,0);
		\draw (g1) -- (w) -- (g2);
	\end{tikzpicture}
\end{aligned}
\;\longleftrightarrow\;
\begin{aligned}
	\begin{tikzpicture}
		\node at (0,0) (w) [white dot] {};
		\draw (w) -- ++(-0.5,0);
		\draw (w) -- ++(0.5,0);
	\end{tikzpicture}
\end{aligned}
\;\longleftrightarrow\;
\begin{aligned}
	\begin{tikzpicture}
		\draw (-0.5,0) -- (.5,0);
	\end{tikzpicture}
\end{aligned}
\end{equation}
The gray-dot identity may be derived using this and \textsf{(ZH-NA)}:
\begin{equation}
\begin{aligned}
	\begin{tikzpicture}
		\node at (0,0) (g1) [gray dot] {};
		\draw (g1) -- ++(-0.375,0);
		\node at ($(g1) + (0.5,0)$) (g2) [gray dot] {};
		\draw (g2) -- ++(0.375,0);
		\draw (g1) -- (g2);
	\end{tikzpicture}
\end{aligned}
\;\longleftrightarrow\;
\begin{aligned}
	\begin{tikzpicture}
		\node at (0,0) [not dot,  draw=none, fill=none, label=below:\phantom{\footnotesize$0$}] {};
		\node at (0,0) (g1) [not dot, label=above:\footnotesize$0$] {};
		\draw (g1) -- ++(-0.375,0);
		\node at ($(g1) + (0.5,0)$) (g2) [not dot, label=above:\footnotesize$0$] {};
		\draw (g2) -- ++(0.375,0);
		\draw (g1) -- (g2);
	\end{tikzpicture}
\end{aligned}
\;\longleftrightarrow\;
\begin{aligned}
	\begin{tikzpicture}
		\draw (-0.5,0) -- (.5,0);
	\end{tikzpicture}
\end{aligned}
\end{equation}
Finally, the two H-box identity rules can be proven using \textsf{(ZH-MF)} for $c_1 = c_2 = 1$, the `double-H-box' rule, and the newly demonstrated gray-dot identity rule (simplifying scalar boxes immediately):~\\[-7ex]
\begin{equation}
\begin{aligned}
	\begin{tikzpicture}
		\node at (0,0) (h1) [small H box] {\t{\;\!\mathrlap{\phantom-}u\;\!}};
		\draw (h1) -- ++(-0.4375,0);
		\node at ($(h1) + (0.5375,0)$) (h2) [small H box] {\t{-u}};
		\draw (h2) -- ++(0.5,0);
		\draw (h1) -- (h2);
		\node at ($(h1)!0.5!(h2) + (0,-.375)$) [small H box, label=below:\footnotesize$(D\nu^4)^{-1}$] {};
		\node at ($(h1)!0.5!(h2) + (0,.375)$) [small H box, draw=none, fill=none,label=above:\footnotesize$\phantom{(D\nu^4)^{-1}}$] {};
	\end{tikzpicture}
\end{aligned}
\;\longleftrightarrow\;
\begin{aligned}
	\begin{tikzpicture}
		\node at (0,0) (h1) [small H box] {\t{\mathrlap{\phantom-}u}};
		\draw (h1) -- ++(-0.375,0);
		\node at ($(h1) + (0.375,0)$) (h2) [small H box] {\t{\mathrlap{\phantom-}u}};
		\node at ($(h2) + (0.375,0)$) (h3) [small H box] {\t{\mathrlap{\phantom-}u}};
		\node at ($(h3) + (0.375,0)$) (h4) [small H box] {\t{\mathrlap{\phantom-}u}};
		\node at ($(h2)!0.5!(h3) + (0,-.375)$) [small H box, label=below:\footnotesize$(D\nu^4)^{-2}$] {};
		\node at ($(h2)!0.5!(h3) + (0,.375)$) [small H box, draw=none, fill=none,label=above:\footnotesize$\phantom{(D\nu^4)^{-2}}$] {};
		\draw (h4) -- ++(0.375,0);
		\draw (h1) -- (h2) -- (h3) -- (h4);
	\end{tikzpicture}
\end{aligned}
\;\longleftrightarrow\;
\begin{aligned}
	\begin{tikzpicture}
		\node at (0,0) (g1) [gray dot] {};
		\draw (g1) -- ++(-0.375,0);
		\node at ($(g1) + (0.5,0)$) (g2) [gray dot] {};
		\draw (g2) -- ++(0.375,0);
		\draw (g1) -- (g2);
	\end{tikzpicture}
\end{aligned}
\;\longleftrightarrow\;
\begin{aligned}
	\begin{tikzpicture}
		\draw (-0.5,0) -- (.5,0);
	\end{tikzpicture}
\end{aligned}
\end{equation}

\medskip
\hrule
\nopagebreak
\smallskip
\smallskip

\Rule{ZH-HWB} 
H-box \& white-dot bialgebra. 
\\[-3ex]
\begin{gather*}
	\Sem{9ex}{\,\vtikzfig{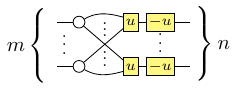}}
\;\;=\;\;
	\Sem{7ex}{\;\vtikzfig[-2ex]{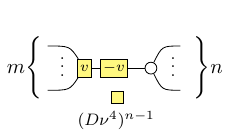}\;}
\end{gather*}

\medskip
\noindent
We use the notation $\Pi(x)$, in effect representing a function ${\Pi : \Z_D^m \to \Z_D}$ which we define by $\Pi(x) = x_1 x_2 \cdots x_m$.
As a mild abuse of notation, we allow $\Pi$ to act on arguments of various sizes, \emph{e.g.},~$x \in \Z_D^m$ and $y \in \Z_D^n$.

\bigskip
\medskip
\noindent
\textbf{Proof.}~\\[-10ex]
\begin{align*}
	\mspace{96mu}&\mspace{-48mu}%
	\mathop{\int \!\!\!\! \int \!\!\!\! \int}\limits_{\mathclap{\substack{
		x \in \D^m 
	\\[.25ex]
		y,z \in \D^n
	}}}
	\Biggl(
		\bigotimes_{j=1}^n 
			\omega^{-uy_j z_j} \omega^{u\Pi(x)\,y_j} \!
	\Biggr)
	\,\kket{z}\bbra{x}
\\[-2ex]&=\;
	\mathop{\int \!\!\!\! \int}\limits_{\mathclap{\substack{
		x \in \D^m 
	\\[.25ex]
		z \in \D^n
	}}}
	\;
	\Biggl(
		\prod_{j=1}^n 
		\mathop{\text{\Large$\int$}}\limits_{\mathclap{
			y_j \in \D
		}}
		\!\!
		\omega^{uy_j \bigl(\Pi(x) - z_j\bigr)} \!
	\Biggr)
	\,\Biggl(\bigotimes_{j=1}^n \kket{z_j}\Biggr)\bbra{x}
\\[1ex]&=\;
	\mathop{\int \!\!\!\! \int}\limits_{\mathclap{\substack{
		x \in \D^m 
	\\[.25ex]
		z \in \D^n
	}}}
	\;
	\Biggl(
		\prod_{j=1}^n 
		D\nu^4 \bbracket{z_j}{\big.\Pi(x)}
	\Biggr)
	\,\Biggl(\bigotimes_{j=1}^n \kket{z_j}\Biggr)\bbra{x}
\\[1ex]&=\;
	(D\nu^4)^{n-1} \mathop{\int \!\!\!\! \int}\limits_{\mathclap{\substack{
		x \in \D^m 
	\\[.25ex]
		z_1 \in \D
	}}}
		D\nu^4 \bbracket{z_1}{\big.\Pi(x)}
	\,\Biggl(\bigotimes_{j=1}^n \kket{z_1}\Biggr)\bbra{x}
\\[1ex]&=\;
	(D\nu^4)^{n-1} \mathop{\int \!\!\!\! \int}\limits_{\mathclap{\substack{
		x \in \D^m 
	\\[.25ex]
		z \in \D
	}}}
	\;
	\Biggl(\;\;\;\;
		\int\limits_{\mathclap{
			y \in \D
		}}
		\omega^{vy \bigl(\Pi(x)-z\bigr)} \!
	\Biggr)
	\,\kket{z}\sox{n}\bbra{x}
\\[1ex]&=\;
	(D\nu^4)^{n-1} 
	\Biggl[\;\;\;
	\int\limits_{\mathclap{
		z \in \D
	}}
	\kket{z}\sox{n}\bbra{z}
	\Biggr]
	\Biggl[\;\;\;
	\mathop{\int \!\!\!\! \int}\limits_{\mathclap{\substack{
		x \in \D^m 
	\\[.25ex]
		w,y \in \D
	}}}
	 \omega^{-vyw} \,\omega^{vy \,\Pi(x)}
	\,\kket{w}\bbra{x}
	\Biggr].
	\mspace{-18mu}
\tag*{\qed}
\end{align*}

\Corollary{ZH-HMB} H-box mixed bialgebra --- the rule described as `BA2' in Refs.~\cite{BK-2019,BKMWW-2021} can be generalised to $D>1$ (in a way equivalent to but different from that presented by Roy~\cite{Roy-2022}), using the above bialgebra rule in combination with \textsf{(ZH-HI)}, \textsf{(ZH-MF)}, and \textsf{(ZH-WGC)}:

\vspace*{-5ex}
$
\begin{aligned}[b]
\begin{aligned}[t]
\begin{aligned}\\[-3ex] \input{./Figures-Source/ZH-bialg-alt-white-H-arity.tikz}\end{aligned}
\;&\to\;
  \begin{aligned}
    \begin{tikzpicture}
      \node (H) at (.375,.375) [small H box] {\t{\mathrlap{\phantom-}u}};
      \node (H') at (.375,-.375) [small H box] {\t{\mathrlap{\phantom-}u}};
      \draw (H) -- ++(0.625,0) node [small H box] (H'') {\t{-u}}
      			-- ++(0.625,0) node [small H box] (H'''') {\t{\mathrlap{\phantom-}u}}
      			-- ++(0.5,0);
      \draw (H') -- ++(0.625,0) node [small H box] (H''') {\t{-u}}
      			-- ++(0.625,0) node [small H box] (H''''') {\t{\mathrlap{\phantom-}u}}
      			-- ++(0.5,0);
      \node (Z) at (-.5,.375) [white dot] {};
      \node (Z') at (-.5,-.375) [white dot] {};
      \draw (Z) -- ++(-0.375,0);
      \draw (Z') -- ++(-0.375,0);
      \draw [out=30,in=150] (Z) to ($(H.west) + (0,0.05)$);
      \draw [out=-30,in=-150] (Z') to ($(H'.west) + (0,-0.05)$);
      \draw (Z) to (H');
      \draw (H) to (Z');
      \node (dots) at ($(Z)!0.5!(Z') + (-0.25,0.09375)$) {\footnotesize$\mathbf\vdots$};
      \node [anchor=east] at ($(dots)+(-0.125,-0.125)$) {$m \left\{\begin{matrix}\\[4ex]\end{matrix}\right.$};
      \node (dots) at ($(Z)!0.5!(H) + (0,-0.125)$) {\mbox{\footnotesize\,.\llap{\raisebox{.875ex}.}\llap{\raisebox{1.75ex}.}\,}};
      \node (dots) at ($(Z')!0.5!(H') + (0,0.125)$) {\mbox{\footnotesize\,.\llap{\raisebox{.875ex}.}\llap{\raisebox{1.75ex}.}\,}};
      \node (dots) at ($(H'')!0.5!(H''') + (0,0.09375)$) {\footnotesize$\mathbf\vdots$};
      \node [anchor=west] at ($(dots)+(0.875,-0.125)$) {$\left.\,\begin{matrix}\\[4ex]\end{matrix}\right\} n$};
      \node (eta) at ($(H')!0.5!(Z') + (0,-0.5)$)
        [small H box, label=below:\footnotesize\!$(D\nu^4)^{-n}$] {};
      \node (eta) at ($(H)!0.5!(Z) + (0,0.5)$)
        [small H box, draw=none, fill=none, label=above:\footnotesize\!$\phantom{(D\nu^4)^{-n}}$] {};
    \end{tikzpicture}
  \end{aligned}
\;\to\;
  \begin{aligned}
    \begin{tikzpicture}
      \node (H) at (0,0) [small H box] {\t{\mathrlap{\phantom-}u}};
      \node (Z) at (1.125,0) [white dot] {};
      \draw (H) -- ++(0.5625,0) node (H') [small H box] {\t{-u}} -- (Z);
      \draw (H) .. controls ++(-0.3125,0.375) .. ++(-.625,0.375);
      \draw (H) .. controls ++(-0.3125,-0.375) .. ++(-.625,-0.375);
      \node (dots) at ($(H) + (-0.4125,0.125)$) {\footnotesize$\mathbf\vdots$};
      \node [anchor=east] at ($(dots)+(-0.0625,-0.125)$) {$m \left\{\begin{matrix}\\[4ex]\end{matrix}\right.$};
      \draw (Z) .. controls ++(0.1875,0.375) .. ++(.5,0.375) -- ++(0.125,0) node (H'') [small H box] {\t{\mathrlap{\phantom-}u}} -- ++(0.4125,0);
      \draw (Z) .. controls ++(0.1875,-0.375) .. ++(.5,-0.375) -- ++(0.125,0) node (H''') [small H box] {\t{\mathrlap{\phantom-}u}} -- ++(0.4125,0);;
      \node (dots) at ($(Z) + (0.625,0.09175)$) {\footnotesize$\mathbf\vdots$};
      \node [anchor=west] at ($(dots)+(0.25,-0.125)$) {$\left.\,\begin{matrix}\\[4ex]\end{matrix}\right\} n$};
      \node (H'') at ($(H') + (0,-0.5)$)
        [small H box, label=below:\footnotesize\!$(D\nu^4)^{-1}$] {};
      \node (H'') at ($(H') + (0,0.5)$)
        [small H box, draw=none, fill=none, label=above:\footnotesize\!$\phantom{(D\nu^4)^{-1}}$] {};
    \end{tikzpicture}
  \end{aligned}
\\[-2.5ex]\;&\to\;
  \begin{aligned}
    \begin{tikzpicture}
      \node (H) at (0,0) [small H box] {\t{\mathrlap{\phantom-}u}};
      \draw (H) -- ++(0.5,0) node [small H box] {\t{\mathrlap{\phantom-}u}}
      -- ++(0.5,0) node [small H box] {\t{\mathrlap{\phantom-}u}}
      -- ++(0.5,0) node [small H box] {\t{\mathrlap{\phantom-}u}}
      -- ++(0.5,0) node (Z) [white dot] {};
      \draw (H) .. controls ++(-0.3125,0.375) .. ++(-.625,0.375);
      \draw (H) .. controls ++(-0.3125,-0.375) .. ++(-.625,-0.375);
      \node (dots) at ($(H) + (-0.4125,0.125)$) {\footnotesize$\mathbf\vdots$};
      \node [anchor=east] at ($(dots)+(-0.0625,-0.125)$) {$m \left\{\begin{matrix}\\[4ex]\end{matrix}\right.$};
      \draw (Z) .. controls ++(0.1875,0.375) .. ++(.5,0.375) -- ++(0.125,0) node (H'') [small H box] {\t{\mathrlap{\phantom-}u}} -- ++(0.4125,0);
      \draw (Z) .. controls ++(0.1875,-0.375) .. ++(.5,-0.375) -- ++(0.125,0) node (H''') [small H box] {\t{\mathrlap{\phantom-}u}} -- ++(0.4125,0);;
      \node (dots) at ($(Z) + (0.625,0.09175)$) {\footnotesize$\mathbf\vdots$};
      \node [anchor=west] at ($(dots)+(0.25,-0.125)$) {$\left.\,\begin{matrix}\\[4ex]\end{matrix}\right\} n$};
      \node (H'') at ($(H)!0.5!(Z) + (0,-0.5)$)
        [small H box, label=below:\footnotesize\!$(D\nu^4)^{-2}$] {};
      \node (H'') at ($(H)!0.5!(Z) + (0,0.5)$)
        [small H box, draw=none, fill=none, label=above:\footnotesize\!$\phantom{(D\nu^4)^{-2}}$] {};
    \end{tikzpicture}
  \end{aligned}
\;\to\;
  \begin{aligned}
    \begin{tikzpicture}
      \node (H) at (0,0) [small H box] {\t{-u}};
      \draw (H) -- ++(.75,0) node (Z) [gray dot] {};
      \draw (H) .. controls ++(-0.3125,0.375) .. ++(-.625,0.375);
      \draw (H) .. controls ++(-0.3125,-0.375) .. ++(-.625,-0.375);
      \node (dots) at ($(H) + (-0.4125,0.125)$) {\footnotesize$\mathbf\vdots$};
      \node [anchor=east] at ($(dots)+(-0.0625,-0.125)$) {$m \left\{\begin{matrix}\\[4ex]\end{matrix}\right.$};
      \draw (Z) .. controls ++(0.1875,0.375) .. ++(.5,0.375) -- ++(0.125,0);
      \draw (Z) .. controls ++(0.1875,-0.375) .. ++(.5,-0.375) -- ++(0.125,0);
      \node (dots) at ($(Z) + (0.5,0.125)$) {\footnotesize$\mathbf\vdots$};
      \node [anchor=west] at ($(dots)+(0,-0.125)$) {$\left.\,\begin{matrix}\\[4ex]\end{matrix}\right\} n$};
    \end{tikzpicture}
  \end{aligned}
\end{aligned}
\end{aligned}
$

\smallskip
\hrule
\nopagebreak
\smallskip

\Rule{ZH-ME} 
White-gray multiplication.
\\[-3ex]
\begin{gather*}
	\Sem{9ex}{\,\begin{aligned}
	\begin{tikzpicture}
		\node (Z) [white dot] at (0,0) {};
		\draw (Z) -- ++(-0.375,0);
		\node (G) [gray dot] at (1.5,0) {};
		\draw (G) -- ++(0.5,0) node [gray dot] {} -- ++(0.5,0);
		\draw [out=70,in=110] (Z) to (G);
		\draw [out=45,in=135] (Z) to (G);
		\draw [out=-45,in=-135] (Z) to (G);
		\draw [out=-70,in=-110] (Z) to (G);
		\node at ($(Z)!0.4125!(G) + (0,0.0875)$) {$\vdots$};
		\node at ($(Z)!0.625!(G) + (0,0.)$) {$\left. \begin{matrix} \\[6ex] \end{matrix}\right\} \! k$};
		\node at (1.625,.5) [small H box, draw=none, fill=none, label=right:\phantom{\footnotesize$D\nu^4$}] {};
		\node at (1.625,-.5) [small H box, label=right:\footnotesize$D\nu^4$] {};
	\end{tikzpicture}
	\end{aligned}}
\;\;=\;\;
	\Sem{7ex}{\;\vtikzfig{ZH-multiplier}\;}
\end{gather*}

\medskip

\noindent
\textbf{Proof.}~\\[-8.75ex]
\begin{align*}{}
	\mspace{96mu}&\mspace{-48mu}
	D \nu^4 \mathop{\int \!\!\!\! \int \!\!\!\! \int}\limits_{\mathclap{\substack{
		x,z \in \D
	\\[1.75ex]
		y \in \D^{k{+}1}
	}}}
	\kket{z}\;
	\bbracket{z + y_{k{+}1}}{0}\;
    \bbracket{\big.\,
    	\smash{\textstyle \sum\limits_h y_h}
    \,}{0}
    \,
	\Biggl(
		\bigotimes_{j=1}^k 
			\bbracket{y_j}{x}
	\Biggr)\,
	\bbra{x}
	\mspace{-72mu}
\\[-2.375ex]&=\;
	\mathop{\int \!\!\!\! \int}\limits_{\mathclap{
		x,z \in \D
	}}
	\kket{z}\!\;
    \Bigl( D \nu^4 \, \bbracket{
    	\smash{\textstyle k x - z}
    }{0}
   	\Bigr)
	\bbra{x}
\\[.75ex]&=\;
	\mathop{\int \!\!\!\!  \int}\limits_{\mathclap{
		x,z \in \D
	}}
	\kket{z}\!\;
    \Biggl(\;\;\; \int\limits_{\mathclap{t \in \D}}
    	\omega^{ut(kx-z)}
   	\Biggr)
	\bbra{x}
\;=\;
	\mathop{\int \!\!\!\!  \int \!\!\!\!  \int \!\!\!\!  \int}\limits_{\mathclap{
		x,s,t,z \in \D
	}}
	\Bigl(\omega^{-usz} \, \kket{z}\bbra{s}\Bigr)
    \Bigl(
    	\omega^{uktx}\,
    	\kket{t}\bbra{x}
   	\Bigr)
	\;.
\mspace{-36mu}
\tag*{\qed}
\end{align*}

\Corollary{ZH-MEH} Multiedge Hopf Law ---
We may prove the following using the rules  \textsf{(ZH-AI)}, \textsf{(ZH-ME)}, \mbox{\textsf{(ZH-MF)}}, \textsf{(ZH-HU)}, \textsf{(ZH-WGC)}, \textsf{(ZH-GL)}, and \textsf{(ZH-HMB)}:~\\[-2ex]
\begin{equation*}
\begin{aligned}
	\begin{aligned}
	\begin{tikzpicture}
		\node (Z) [white dot] at (0,0) {};
		\draw (Z) -- ++(-0.375,0);
		\node (G) [gray dot] at (1.5,0) {};
		\draw (G) -- ++(0.5,0);
		\draw [out=70,in=110] (Z) to (G);
		\draw [out=45,in=135] (Z) to (G);
		\draw [out=-45,in=-135] (Z) to (G);
		\draw [out=-70,in=-110] (Z) to (G);
		\node at ($(Z)!0.4125!(G) + (0,0.0875)$) {$\vdots$};
		\node at ($(Z)!0.625!(G) + (0,-0.)$) {$\left. \begin{matrix} \\[6ex] \end{matrix}\right\} \! \raisebox{-0.125ex}{$D$}$};
		\node at (1.5,.625) [small H box, draw=none, fill=none, label=right:\phantom{\footnotesize$D\nu^4$}] {};
		\node at (1.5,-.625) [small H box, label=right:\footnotesize$D\nu^4$] {};
	\end{tikzpicture}
	\end{aligned}
	\!\!\!\leftrightarrow\;
	\begin{aligned}
	\begin{tikzpicture}
		\node (Z) [white dot] at (0,0) {};
		\draw (Z) -- ++(-0.375,0);
		\node (G) [gray dot] at (1.5,0) {};
		\draw (G) -- ++(0.4375,0) node [gray dot] {}
					-- ++(0.4375,0) node [gray dot] {} -- ++(0.4375,0);
		\draw [out=70,in=110] (Z) to (G);
		\draw [out=45,in=135] (Z) to (G);
		\draw [out=-45,in=-135] (Z) to (G);
		\draw [out=-70,in=-110] (Z) to (G);
		\node at ($(Z)!0.4125!(G) + (0,0.0875)$) {$\vdots$};
		\node at ($(Z)!0.625!(G) + (0,-0.)$) {$\left. \begin{matrix} \\[6ex] \end{matrix}\right\} \! \raisebox{-0.125ex}{$D$}$};
		\node at (1.625,.5) [small H box, draw=none, fill=none, label=right:\phantom{\footnotesize$D\nu^4$}] {};
		\node at (1.625,-.5) [small H box, label=right:\footnotesize$D\nu^4$] {};
	\end{tikzpicture}
	\end{aligned}
	\;&\leftrightarrow\;
	\begin{aligned}
	\begin{tikzpicture}
		\node (K) [H box] at (0,0) {\t D};
		\draw (K) -- ++(-0.5,0);
		\draw (K) -- ++(0.5,0) node [small H box] {\tm} -- ++(0.5,0) node [gray dot] {} -- ++(0.5,0);
	\end{tikzpicture}
	\end{aligned}
	\;\longleftrightarrow\;
	\begin{aligned}
	\begin{tikzpicture}
		\node (K) [H box] at (0,0) {\t 0};
		\draw (K) -- ++(-0.5,0);
		\draw (K) -- ++(0.5,0) node [small H box] {\tm}
					 -- ++(0.5,0) node [small H box] {\tp}
					 -- ++(0.5,0) node [small H box] {\tp} --  ++(0.5,0);
		\node at (.75,-.5) [small H box, draw=none, fill=none, label=below:\footnotesize\phantom{$\mathrlap{(D\nu^4)^{-1}}$\quad}] {};
		\node at (.75,.5) [small H box, label=above:\footnotesize$\mathrlap{(D\nu^4)^{-1}}\quad$] {};
	\end{tikzpicture}
	\end{aligned}
	\;\leftrightarrow\;
	\begin{aligned}
	\begin{tikzpicture}
		\node (K) [small H box] at (0,0) {\tp};
		\node at (0,-.875) [small H box, fill=none, draw=none] {\phantom{\t 0}};
		\draw (K) -- ++(0,0.4375) node [small H box] {\tp}
				-- ++(0,0.4375) node [small H box] {\t 0};
		\draw (K) -- ++(-0.5,0);
		\draw (K) -- ++(0.5,0) node [small H box] {\tp} -- ++(0.5,0);
		\node at (.5,-.5) [small H box, draw=none, fill=none, label=below:\footnotesize\phantom{$\mathrlap{(D\nu^4)^{-1}}$\quad}] {};
		\node at (.5,.5) [small H box, label=above:\footnotesize$\mathrlap{(D\nu^4)^{-1}}\quad$] {};
	\end{tikzpicture}
	\end{aligned}
	\\[-3ex]&\leftrightarrow\;
	\begin{aligned}
	\begin{tikzpicture}
		\node (K) [small H box] at (0,0) {\tm};
		\node at (0,-.875) [white dot, fill=none, draw=none] {};
		\draw (K) -- ++(0,0.4375) node [small H box] {\tp}
				-- ++(0,0.4375) node [white dot] {};
		\draw (K) -- ++(-0.5,0);
		\draw (K) -- ++(0.5,0) node [small H box] {\tp} -- ++(0.5,0);
		\node at (.5,-.5) [small H box, draw=none, fill=none, label=below:\footnotesize\phantom{$\mathrlap{(D\nu^4)^{-1}}$\quad}] {};
		\node at (.5,.5) [small H box, label=above:\footnotesize$\mathrlap{(D\nu^4)^{-1}}\quad$] {};
	\end{tikzpicture}
	\end{aligned}
	\;\leftrightarrow\;
	\begin{aligned}
	\begin{tikzpicture}
		\node (K) [small H box] at (0,0) {\tm};
		\node at (0,-0.5) [gray dot, fill=none, draw=none] {};
		\draw (K) -- ++(0,0.5) node [gray dot] {};
		\draw (K) -- ++(-0.5,0);
		\draw (K)
			-- ++(0.5,0) node [small H box] {\tp}
			-- ++(0.5,0);
		\node at (.5,-.5) [small H box, draw=none, fill=none, label=below:\footnotesize\phantom{$\mathrlap{(D\nu^4)^{-1}}$\quad}] {};
		\node at (.5,.5) [small H box, label=above:\footnotesize$\mathrlap{(D\nu^4)^{-1}}\quad$] {};
	\end{tikzpicture}
	\end{aligned}
	\;\leftrightarrow\;
	\begin{aligned}
	\begin{tikzpicture}
		\node (Z) [white dot] at (-0.1875,0) {};
		\draw (Z) -- ++(-0.375,0);
		\node (Z') [white dot] at (0.1875,0) {};
		\draw (Z') -- ++(0.4375,0) node [small H box] {\tp} -- ++(0.5,0);
		\node at ($(Z') + (0,-.5)$) [small H box, draw=none, fill=none, label=below:\footnotesize\phantom{$\mathrlap{(D\nu^4)^{-1}}$\quad}] {};
		\node at ($(Z') + (0,.5)$) [small H box, label=above:\footnotesize$\mathrlap{(D\nu^4)^{-1}}\quad$] {};
	\end{tikzpicture}
	\end{aligned}
	\;\leftrightarrow\;
	\begin{aligned}
	\begin{tikzpicture}
		\node (Z) [white dot] at (-0.1875,0) {};
		\draw (Z) -- ++(-0.375,0);
		\node (Z') [gray dot] at (0.1875,0) {};
		\draw (Z') -- ++(0.375,0);
	\end{tikzpicture}
	\end{aligned}
\end{aligned}
\end{equation*}
\vspace*{-5ex}

\hrule
\nopagebreak
\medskip

\noindent\textbf{``Zero product Lemma''.}
~\\[-3.5ex]
\begin{gather*}{}
\mspace{-18mu}
	\Sem{22ex}{\,\begin{aligned}
	\\[-4ex]
	\begin{tikzpicture}
		\coordinate (x0) at (0,0);
		\coordinate (x1) at (0,-.8);
		\coordinate (x2) at (0,-1.6);
		\coordinate (xdots) at (0,-2.4);
		\node at ($(xdots) + (0.875,0.085)$) {\footnotesize$\vdots$};
		\node at ($(xdots) + (2,0.1875)$) {\footnotesize$\vdots$};
		\coordinate (xf) at (0,-3.2);
		\node (w) [white dot] at ($(x2) + (2.75,0)$) {};
		\draw (w) -- ++(0.5,0);
		\node (g) [white dot] at ($(x2) + (1.375,-0.375)$) {};
		\foreach \n/\a/\s/\p in {x0/0/-/below,x1/1/-/below,x2/2/-/below,xf/{D{-}1}//above} {
			\draw (\n) -- ++(0.5,0) node [small H box] (h\n) {\tp}
				-- ++(1.5,0) node [not dot, label=\p:\footnotesize$\a$] {}
				.. controls ++(.625,0) .. (w);
			\node (h\n') [small H box] at ($(h\n) + (0.375,\s0.375)$) {\tm};
			\draw (h\n) .. controls ++(0,\s0.375) .. (h\n');
			\draw (h\n') .. controls ++(0.375,0) .. (g);
		}
	\end{tikzpicture}	
	\end{aligned}}
\;=\;
	\Sem{22ex}{\;\begin{aligned}
	\\[-4ex]
	\begin{tikzpicture}
		\coordinate (x0) at (0,0);
		\coordinate (x1) at (0,-.8);
		\coordinate (x2) at (0,-1.6);
		\coordinate (xdots) at (0,-2.4);
		\node at ($(xdots) + (0.875,0.085)$) {\footnotesize$\vdots$};
		\node at ($(xdots) + (2,0.1875)$) {\footnotesize$\vdots$};
		\coordinate (xf) at (0,-3.2);
		\node (w) [white dot] at ($(x2) + (2.75,0)$) {};
		\draw (w) -- ++(0.5,0);
		\foreach \n/\a/\s/\p in {x0/0/-/below,x1/1/-/below,x2/2/-/below,xf/{D{-}1}//above} {
			\draw (\n) -- ++(0.5,0) node [small H box] (h\n) {\tp}
				-- ++(1.5,0) node [not dot, label=\p:\footnotesize$\a$] {}
				.. controls ++(.625,0) .. (w);
			\node (h\n') [small H box] at ($(h\n) + (0.375,\s0.375)$) {\tm};
			\draw (h\n) .. controls ++(0,\s0.375) .. (h\n')
				-- ++ (0.5,0) node [gray dot] {};
		}
		\node at ($(w) + (0.25,-1)$) [small H box, label=below:\footnotesize$\nu^2$] {};
	\end{tikzpicture}	
	\end{aligned}\;}
\mspace{-18mu}
\end{gather*}

\noindent
Following Roy~\cite[p.\,36]{Roy-2022}, this Lemma represents the fact that for $x, a_0, a_1, a_2, \ldots \in \Z_D$ such that ${a_0 (0-x)} = {a_1(1-x)} = {a_2 (2-x)} = \cdots$, then in particular ${a_j (j{-}x)} = 0$ for $j {\;\!=\;\!} x$, so that in fact ${a_k(k{\;\!-\;\!}x)} = 0$ for all $k \in \Z_D$.
--- In the seventh line of the proof below, we introduce a redundant factor of $\bbracket{0}{z_{x}}$.
This allows us to re-incorporate the terms in the first set of parentheses into the product over $j \in \Z_D$.
As $\bbracket{0}{z_{x}}$ introduces a factor of $\tfrac{1}{\nu^2}$, we must multiply by $\nu^2$ to avoid changing the overall normalisation.
This is what introduces the factor of $\nu^2$ in this rewrite.

\bgroup
\allowdisplaybreaks
\bigskip
\noindent
\textbf{Proof.}~\\[-10ex]
\begin{align*}{}
	\mspace{144mu}&\mspace{-84mu}
	\mathop{\int \!\!\!\! \int \!\!\!\! \int \!\!\!\!  \int \!\!\!\! \int}\limits_{\mathclap{\substack{%
		a,y,z \in \Z_D^D \\[.5ex] u,x \in \Z_D
	}}}\;
	\Biggl( \bigotimes_{j=1}^D \, \omega^{a_j y_j (j-x)} \, \omega^{-y_j z_j} \bbracket{u}{z_j} \! \Biggr)
		\kket{x}\bbra{a}
\mspace{-90mu}
\\[-4.5ex]
&=\;
\mathop{\int \!\!\!\! \int \!\!\!\! \int \!\!\!\! \int}\limits_{\mathclap{\substack{%
		a,z \in \Z_D^D \\[.5ex] u,x \in \Z_D
	}}}
	\;
	\prod_{j \in \Z_D} \,
	\Biggl(\;\;\;\;
		\int\limits_{\mathclap{y_j \in \Z_D}}  \omega^{y_j\bigl(a_j (j-x)\,-\,z_j\bigr)} \bbracket{u}{z_j} \! \Biggr)
		\kket{x}\bbra{a}
\mspace{-90mu}
\\[.5ex]
&=\;
\mathop{\int \!\!\!\! \int \!\!\!\! \int \!\!\!\! \int}\limits_{\mathclap{\substack{%
		a,z \in \Z_D^D \\[.5ex] u,x \in \Z_D
	}}}
	\;
	\Biggl(
		\prod_{j \in \Z_D} 
		D \nu^4 \; \bbracket{z_j}{a_j (j-x)} \; \bbracket{u}{z_j} \Biggr)
		\kket{x}\bbra{a}
\mspace{-90mu}
\\[.5ex]
&=\;
\begin{aligned}[t]
	&\mathop{\int \!\!\!\! \int \!\!\!\! \int \!\!\!\! \int}\limits_{\mathclap{\substack{%
		a,z \in \Z_D^D \\[.5ex] u,x \in \Z_D
	}}}
	\;
	\Biggl(
		D \nu^4 \; \bbracket{z_{x}}{a_j \cdot 0} \; \bbracket{u}{z_{x}}
	\Biggr)
	\times{}\\[-5.5ex]&\mspace{80mu}
	\Biggl(
		\prod_{\substack{j \in \Z_D \\ j \ne x}}
		D \nu^4 \; \bbracket{z_j}{a_j (j-x)} \; \bbracket{u}{z_j}
	\Biggr)
		\kket{x}\bbra{a}
\end{aligned}
\mspace{-90mu}
\\[-.75ex]
&=\;
\begin{aligned}[t]
	&\mathop{\int \!\!\!\! \int \!\!\!\! \int \!\!\!\! \int}\limits_{\mathclap{\substack{%
		a,z \in \Z_D^D \\[.5ex] u,x \in \Z_D
	}}}
	\;
	\Biggl(
		D \nu^4 \; \bbracket{z_{x}}{a_j \cdot 0} \; \bbracket{u}{0}
	\Biggr)
	\times{}\\[-5.5ex]&\mspace{80mu}
	\Biggl(
		\prod_{\substack{j \in \Z_D \\ j \ne x}}
		D \nu^4 \; \bbracket{z_j}{a_j (j-x)} \; \bbracket{u}{z_j}
	\Biggr)
		\kket{x}\bbra{a}
\end{aligned}
\mspace{-90mu}
\\[-.75ex]
{}\;&=\;
\begin{aligned}[t]
	&\mathop{\int \!\!\!\! \int \!\!\!\! \int}\limits_{\mathclap{\substack{%
		a,z \in \Z_D^D \\[.5ex] x \in \Z_D
	}}}
	\;
	\Biggl(
		D \nu^4 \; \bbracket{z_{x}}{a_j \cdot 0} 
	\Biggr)
	\times{}\\[-5.5ex]&\mspace{80mu}
	\Biggl(
		\prod_{\substack{j \in \Z_D \\ j \ne -x}}
		D \nu^4 \; \bbracket{z_j}{a_j (j-x)} \; \bbracket{0}{z_j}
	\Biggr)
		\kket{x}\bbra{a}
\end{aligned}
\mspace{-72mu}
\\[-.75ex]
&=\;
\begin{aligned}[t]
	&\mathop{\int \!\!\!\! \int \!\!\!\! \int}\limits_{\mathclap{\substack{%
		a,z \in \Z_D^D \\[.5ex] x \in \Z_D
	}}}
	\nu^2 \Biggl(
		D \nu^4 \; \bbracket{z_{x}}{a_j \cdot 0} \cdot \bbracket{0}{z_{x}}
	\Biggr)
	\times{}\\[-5.5ex]&\mspace{80mu}
	\Biggl(
		\prod_{\substack{j \in \Z_D \\ j \ne x}}
		D \nu^4 \; \bbracket{z_j}{a_j (j-x)} \; \bbracket{0}{z_j}
	\Biggr)
		\kket{x}\bbra{a}
\end{aligned}
\mspace{-72mu}
\\[-0.5ex]
&=\;
\nu^2
\mathop{\int \!\!\!\! \int \!\!\!\! \int}\limits_{\mathclap{\substack{%
		a,z \in \Z_D^D \\[.5ex] x \in \Z_D
	}}}
	\;
	\Biggl(
		\prod_{j \in \Z_D}
		D \nu^4 \; \bbracket{z_j}{a_j (j-x)} \; \bbracket{0}{z_j}
	\Biggr)
		\kket{x}\bbra{a}
\mspace{-72mu}
\\ &=\;
\nu^2
\mathop{\int \!\!\!\! \int \!\!\!\! \int \!\!\!\! \int}\limits_{\mathclap{\substack{%
		a,z \in \Z_D^D \\[.5ex] u,x \in \Z_D
	}}}
	\;
	\prod_{j \in \Z_D} \,
	\Biggl(\;\;\;\;
		\int\limits_{\mathclap{y_j \in \Z_D}}  \omega^{y_j\bigl(a_j (j-x)\,-\,z_j\bigr)} \bbracket{0}{z_j} \! \Biggr)
		\kket{x}\bbra{a}
\mspace{-72mu}
\\[.5ex]
&=\;
\nu^2
	\mathop{\int \!\!\!\! \int \!\!\!\! \int \!\!\!\!  \int \!\!\!\! \int}\limits_{\mathclap{\substack{%
		a,y,z \in \Z_D^D \\[.5ex] u,x \in \Z_D
	}}}\;
	\Biggl( \bigotimes_{j=1}^D \, \omega^{a_j y_j (j-x)} \, \omega^{-y_j z_j} \bbracket{0}{z_j} \! \Biggr)
		\kket{x}\bbra{a}
	\,.
\mspace{-72mu}
\tag*{\qed}
\\[-5ex]
\end{align*}
\egroup

\Corollary{ZH-O} `Ortho' ---
Using the above equality, together with \textsf{(ZH-WGC)} and the the colour change rule and the ``multicharacter negation'' rule, we may obtain: ~\\[1ex]
\begin{equation*}
\begin{aligned}
\\[-8ex]
\begin{tikzpicture}
		\coordinate (x0) at (0,0);
		\coordinate (x1) at (0,-0.75);
		\coordinate (x2) at (0,-1.5);
		\coordinate (xdots) at (0,-2.25);
		\node at ($(xdots) + (0.5,0.085)$) {\footnotesize$\vdots$};
		\node at ($(xdots) + (1.5875,0.09375)$) {\footnotesize$\vdots$};
		\coordinate (xf) at (0,-3.0);
		\node (w) [white dot] at ($(x2) + (2.5,0)$) {};
		\draw (w) -- ++(0.5,0);
		\node (g) [gray dot] at ($(x2) + (1.125,-0.625)$) {};
		\foreach \n/\a/\p in {x0/0/below,x1/1/below,x2/2/below,xf/{D{-}1}/above} {
			\draw (\n) -- ++(0.5,0) node [small H box] (h\n) {\tp}
				-- ++(1.0625,0) node [not dot, label=\p:\footnotesize$\a$] {}
				.. controls ++(.5,0) .. (w);
			\draw (h\n) -- (g);
		}
		\node at ($(w) + (0,-1)$) [small H box, label=below:\footnotesize$D\nu^2$] {};
\end{tikzpicture}	
\end{aligned}
\;\;\leftrightarrow\;\;
\begin{aligned}
\\[-5ex]
\begin{tikzpicture}
		\coordinate (x0) at (0,0);
		\coordinate (x1) at (0,-.8);
		\coordinate (x2) at (0,-1.6);
		\coordinate (xdots) at (0,-2.4);
		\node at ($(xdots) + (0.875,0.085)$) {\footnotesize$\vdots$};
		\node at ($(xdots) + (2,0.09375)$) {\footnotesize$\vdots$};
		\coordinate (xf) at (0,-3.2);
		\node (w) [white dot] at ($(x2) + (3,0)$) {};
		\draw (w) -- ++(0.5,0);
		\node (g) [white dot] at ($(x2) + (1.625,-0.375)$) {};
		\foreach \n/\a/\s/\p in {x0/0/-/below,x1/1/-/below,x2/2/-/below,xf/{D{-}1}//above} {
			\draw (\n) -- ++(0.5,0) node [small H box] (h\n) {\tp}
				-- ++(1.5,0) node [not dot, label=\p:\footnotesize$\a$] {}
				.. controls ++(.375,0) .. (w);
			\node (h\n') [small H box] at ($(h\n) + (0.375,\s0.375)$) {\tm};
			\draw (h\n) .. controls ++(0,\s0.375) .. (h\n');
			\draw (h\n') .. controls ++(0.375,0) .. (g);
		}
		\node at ($(w) + (0,-1)$) [small H box, label=below:\footnotesize$\nu^{-2}$] {};
\end{tikzpicture}	
\end{aligned}
\;\leftrightarrow\;\;
\begin{aligned}
\\[-5ex]
\begin{tikzpicture}
		\coordinate (x0) at (0,0);
		\coordinate (x1) at (0,-.8);
		\coordinate (x2) at (0,-1.6);
		\coordinate (xdots) at (0,-2.4);
		\node at ($(xdots) + (0.875,0.085)$) {\footnotesize$\vdots$};
		\node at ($(xdots) + (2,0.1875)$) {\footnotesize$\vdots$};
		\coordinate (xf) at (0,-3.2);
		\node (w) [white dot] at ($(x2) + (3,0)$) {};
		\draw (w) -- ++(0.5,0);
		\foreach \n/\a/\s/\p in {x0/0/-/below,x1/1/-/below,x2/2/-/below,xf/{D{-}1}//above} {
			\draw (\n) -- ++(0.5,0) node [small H box] (h\n) {\tp}
				-- ++(1.5,0) node [not dot, label=\p:\footnotesize$\a$] {}
				.. controls ++(.375,0) .. (w);
			\node (h\n') [small H box] at ($(h\n) + (0.375,\s0.375)$) {\tm};
			\draw (h\n) .. controls ++(0,\s0.375) .. (h\n')
				-- ++ (0.5,0) node [gray dot] {};
		}
\end{tikzpicture}	
\end{aligned}
\;\leftrightarrow\;\;
\begin{aligned}
\\[-5ex]
\begin{tikzpicture}
		\coordinate (x0) at (0,0);
		\coordinate (x1) at (0,-0.75);
		\coordinate (x2) at (0,-1.5);
		\coordinate (xdots) at (0,-2.25);
		\node at ($(xdots) + (0.5,0.085)$) {\footnotesize$\vdots$};
		\node at ($(xdots) + (1.5,0.125)$) {\footnotesize$\vdots$};
		\coordinate (xf) at (0,-3.0);
		\node (w) [white dot] at ($(x2) + (2.5,0)$) {};
		\draw (w) -- ++(0.5,0);
		\foreach \n/\a/\s/\p in {x0/0/-/below,x1/1/-/below,x2/2/-/below,xf/{D{-}1}//above} {
			\draw (\n) -- ++(0.5,0) node [small H box] (h\n) {\tp}
				-- ++(1,0) node [not dot, label=\p:\footnotesize$\a$] {}
				.. controls ++(.625,0) .. (w);
			\draw (h\n) -- ++(0,\s0.375) node [white dot] {};
		}
\end{tikzpicture}	
\end{aligned}
\end{equation*}

\hrule
\nopagebreak
\smallskip
\smallskip

\Rule{ZH-UM} 
Unit multiplication.
\nopagebreak
\\[-3ex]
\begin{gather*}{}
	\Sem{9ex}{\;
	\begin{aligned}
	\begin{tikzpicture}
    	\node (H) at (0,0.25) [small H box] {\tp};
      	\draw (H) -- ++(-.625,0);
      	\draw (H) -- ++(0.5,0) node (H') [small H box] {\tm};
		\node (Z) at ($(H) + (0,0.4375)$) [white dot] {};
		\draw (H) -- (Z);
      	\node (H) at (0,-0.25) [small H box] {\tp};
      	\draw (H) -- ++(-.625,0);
      	\draw (H) -- ++(0.5,0) node (H') [small H box] {\tm};
		\node (Z) at ($(H) + (0,-0.4375)$) [white dot] {};
		\draw (H) -- (Z);
	\end{tikzpicture}	
	\end{aligned}\;\;}
\;\;=\;\;
	\Biggsem{\,
	\begin{aligned}
	\begin{tikzpicture}
    	\node (H) at (0,0) [small H box] {\tp};
      	\draw (H) .. controls ++(-0.3125,0.25) .. ++(-.625,0.25);
      	\draw (H) .. controls ++(-0.3125,-0.25) .. ++(-.625,-0.25);
      	\draw (H) 
				-- ++(0.5,0) node (H') [small H box] {\tm};
		\node at (0,-0.4375) [white dot, fill=none, draw=none] {};
		\node (Z) at ($(H) + (0,0.4375)$) [white dot] {};
		\draw (H) -- (Z);
		\node (eta) at ($(H') + (0.5875,0)$) [small H box, label=right:\footnotesize$\!\!\;D\nu^4$] {};
	\end{tikzpicture}	
	\end{aligned}\!\!}
\end{gather*}

\medskip
\noindent
On the third-to-last line of the proof, we use the fact that the expression is non-zero only for components in which $r \in \Z_D$, two times.
First, we use the fact that only the term where $s = r^{-1}$ contributes to the integral in that case.
Then, we perform the substitution $u = ry$, so that $sy = u$.
The particular value of $s$ then does not play a role in the integrated expression, so that we may simply eliminate $s$ as a variable of integration.

\bigskip
\noindent
\textbf{Proof.}~\\[-8.5ex]
\begin{align*}
	\mspace{140mu}&\mspace{-90mu}
	\mathop{\int \!\!\!\! \int \!\!\!\! \int \!\!\!\! \int \!\!\!\! \int \!\!\!\! \int}\limits_{\mathclap{r,s,t,x,y,z \in \Z_D}} 
		\omega^{rst} \, \omega^{-t} \, \omega^{xyz} \, \omega^{-z} \; \bbra{r,x}
\\[-1ex]&=\;
	\mathop{\int \!\!\!\! \int \!\!\!\! \int \!\!\!\! \int}\limits_{\mathclap{r,s,x,y \in \Z_D}}
	\;\;\Biggl(\;\;\;
		\int\limits_{\mathclap{z \in \Z_D}}
			\omega^{t(rs-1)}
	\Biggr)
	\Biggl(\;\;\;
		\int\limits_{\mathclap{z \in \Z_D}}
			\omega^{z(xy-1)}
	\Biggr) \; \bbra{r,x}
\mspace{-72mu}
\\[1ex]&=\;
	\mathop{\int \!\!\!\! \int \!\!\!\! \int \!\!\!\! \int}\limits_{\mathclap{r,s,x,y \in \Z_D}}
	\Bigl( D\nu^4 \, \bbracket{rs}{1} \Bigr)
	\Bigl( D\nu^4 \, \bbracket{xy}{1} \Bigr) \,
	\bbra{r,x}
\\[2ex]&=\;
	(D\nu^4)^2
	\mathop{\int \!\!\!\! \int \!\!\!\! \int \!\!\!\! \int}\limits_{\mathclap{r,s,x,y \in \Z_D}}
	\bbracket{rs}{1} \, \bbracket{rs \cdot xy}{1} \,
	\bbra{r,x}
\\[1ex]&=\;
	(D\nu^4)^2
	\mathop{\int \!\!\!\! \int \!\!\!\! \int \!\!\!\! \int}\limits_{\mathclap{r,s,x,y \in \Z_D}}
	\bbracket{rs}{1} \, \bbracket{rx \cdot sy}{1} \,
	\bbra{r,x}
\\[2ex]\mspace{-36mu}
\;&=\;
	D\nu^4
	\mathop{\int \!\!\!\! \int \!\!\!\! \int}\limits_{\mathclap{r,x,u \in \Z_D}}
	\Bigl( D\nu^4 \bbracket{rx \cdot u}{1} \Bigr) \,
	\bbra{r,x}
\\[1ex]&=\;
	D\nu^4
	\mathop{\int \!\!\!\!  \int \!\!\!\! \int}\limits_{\mathclap{r,x,u \in \Z_D}}
	\;\;\Biggl(\;\;\;
		\int\limits_{\mathclap{z \in \Z_D}}
			\omega^{z(rxu-1)}
	\Biggr)
	\bbra{r,x}
\\[2ex]&=\;
	D\nu^4
	\mathop{\int \!\!\!\! \int \!\!\!\! \int \!\!\!\! \int}\limits_{\mathclap{r,x,u,z \in \Z_D}}
	\omega^{rxu} \, \omega^{-z} \,
	\bbra{r,x}
	\;.
\tag*{\qed}
\end{align*}	

\noindent
This rule corresponds to Ref.~\cite[Lemma~2.28]{BKMWW-2021}, which in effect relies on the fact that for $x,y \in \{0,1\}$\,, $xy = 1$ if and only if $x=y=1$.
While this observation does not hold in $\Z_D$ for any $D > 2$, we may suitably generalise it by noting that for $x,y \in \Z_D$, $xy$ is a multiplicative unit if and only if both $x$ and $y$ are; and that $x,y$ are multiplicative units if and only if there are scalars $a,b \in \Z_D$ such that $ax = 1$ and $by = 1$.
Thus, for an input $x \in \Z_D$, we use a small gadget which tests whether there is any $a \in \Z_D$ for which $ax = 1$, in place of a small gadget which tests for $x = 1$.

\medskip
\hrule
\smallskip

\bigskip
\noindent
\textit{(End of ZH rewrites.)}
\bigskip

In the above, many rewrites involve scalars which are simply integer powers of $D\nu^4$.
In the case that $\nu = D^{-1/4}$, these all may be simplified by eliminating any scalar boxes which amount to multiplications by $+1$.
The rules presented in Figure~\ref{fig:ZH-rewrites} demonstrate rewrites which are sound subject to this value of $\nu$.
\vfill
\newpage

\subsubsection{ZX--ZH relations}
\label{apx:ZX-ZH-relations}

\begin{figure}[t]
\begin{align*}
~\\[-10.75ex]
\mspace{-48mu}
	\begin{gathered}
		\begin{tikzpicture}
		\setlength\rulediagramwd{4.25em}
		\rewriterule		[ZXH-GW]	{\vtikzfig{ZX-green-dot}\;\;\;}
		\setlength\rulediagramwd{3.75em}	
		\rewritetarget 					{\vtikzfig{ZH-white-dot}}
		\end{tikzpicture}		
	\\[-4ex]
		\begin{tikzpicture}
		\setlength\rulediagramwd{4.25em}	
		\rewriterule		[ZXH-RG]	{\vtikzfig{ZX-red-dot}}
		\setlength\rulediagramwd{3.75em}	
		\rewritetarget 					{\vtikzfig{ZH-gray-dot}}
		\end{tikzpicture}
	\\[-2.5ex]
		\begin{tikzpicture}
		\setlength\rulediagramwd{4.25em}	
		\rewriterule		[ZXH-GP]	{\vtikzfig[-1ex]{ZX-green-phase-dot}}
		\setlength\rulediagramwd{3.75em}	
		\rewritetarget 					{\vtikzfig{ZH-white-H-gadget}}
		\end{tikzpicture}
	\\[-7.25ex]
		\begin{tikzpicture}
		\setlength\rulediagramwd{4.25em}	
		\rewriterule		[ZXH-WH]	{\vtikzfig{ZX-green-fn-lollipop}}
		\setlength\rulediagramwd{3.75em}	
		\rewritetarget 					{\vtikzfig{ZH-H-fn-lollipop}}
		\end{tikzpicture}
	\end{gathered}
\mspace{-36mu}&&%
\mspace{-24mu}&&
	\begin{gathered}
		\begin{tikzpicture}
		\setlength\rulediagramwd{4.25em}	
		\rewriterule		[ZXH-RN]		{\vtikzfig{ZX-red-c-dot}}
		\setlength\rulediagramwd{3.75em}	
		\rewritetarget 					{\vtikzfig{ZH-gen-not-dot}}
		\end{tikzpicture}
	\\[-5.5ex]
		\begin{tikzpicture}
		\setlength\rulediagramwd{4.25em}	
		\rewriterule		[ZXH-RA]	{\vtikzfig{ZX-red-phase-free-dot}}
		\setlength\rulediagramwd{3.75em}	
		\rewritetarget 					{\vtikzfig{ZH-gray-id}}
		\end{tikzpicture}
	\\[2ex]
		\begin{tikzpicture}
		\setlength\rulediagramwd{4em}	
		\rewriterule		[ZXH-HP]		{\vtikzfig{ZX-H-plus-box}}
		\setlength\rulediagramwd{3.5em}	
		\rewritetarget 					{\vtikzfig{ZH-H-plus1-box}}
		\end{tikzpicture}
	\\[-2.5ex]
		\begin{tikzpicture}
		\setlength\rulediagramwd{4em}	
		\rewriterule		[ZXH-HM]		{\vtikzfig{ZX-H-minus-box}}
		\setlength\rulediagramwd{3.5em}	
		\rewritetarget 					{\vtikzfig{ZH-H-minus1-box}}
		\end{tikzpicture}
	\end{gathered}
\mspace{-36mu}&&%
\mspace{-24mu}&&
	\setlength\rulexnwd{5em}	
	\begin{gathered}
		\begin{tikzpicture}
		\setlength\rulediagramwd{1.5em}	
		\rewriterule		[ZXH-GH0]	{\vtikzfig{ZX-green-phase-free-deg0-dot}}
		\setlength\rulediagramwd{0em}	
		\rewritetarget 					{\vtikzfig{ZH-sqrtD-scalar-gadget}}
		\end{tikzpicture}
	\\[-5.5ex]
		\begin{tikzpicture}
		\setlength\rulediagramwd{2.5em}	
		\rewriterule		[ZXH-GH]	{\vtikzfig{ZX-green-Theta-deg0-dot}}
		\setlength\rulediagramwd{1em}	
		\rewritetarget 					{\vtikzfig{ZH-Theta-integral-gadget}}
		\end{tikzpicture}
	\\[-1ex]
		\begin{tikzpicture}
		\setlength\rulediagramwd{5.25em}	
		\rewriterule		[ZXH-S0]	{\vtikzfig[-0ex]{ZX-Theta-0-phase-gadget}}
		\setlength\rulediagramwd{1.75em}	
		\coordinate (anchor-next-diagram) at ($(anchor-next-diagram) + (-.25,0)$);
		\rewritetarget 					{\vtikzfig[-0ex]{ZH-Theta-0-phase-gadget}}
		\end{tikzpicture}
	\\[-4ex]
		\begin{tikzpicture}
		\setlength\rulediagramwd{5.25em}	
		\rewriterule		[ZXH-S]	{\vtikzfig{ZX-Theta-c-phase-gadget}}
		\coordinate (anchor-next-diagram) at ($(anchor-next-diagram) + (-.25,0)$);
		\setlength\rulediagramwd{1.5em}	
		\rewritetarget 					{\vtikzfig{ZH-Theta-c-phase-gadget}}
		\end{tikzpicture}
	~\\[-2ex]
	\end{gathered}
\end{align*}~\\[-5.5ex]
\hrule
\smallskip
\caption{%
	\label{fig:ZXH-relations-redux}
		Sound rewrites between the ZX~generators and the ZH~generators, subject to the semantics of Eqns.~\eqref{eqn:idealised-ZX-integrals} and~\eqref{eqn:idealised-ZH-integrals} in the case $\nu = D^{-1/4}$.
	}
\end{figure}

Figure~\ref{fig:ZXH-relations} presents some equivalences between ZX~and ZH~diagrams, which hold between the ZX and ZH generators in the common semantic map that we have constructed.
This will allow us to take advantage of these relationships to streamline the proving of certain ZX rewrites, which from their definitions would be somewhat more laborious to carry out than similar ZH rewrites, owing to the emphasis placed by the ZH calculus on the standard basis.

\bigskip\medskip
\hrule
\nopagebreak

\vspace*{-1ex}
\begin{tabbing}
\hspace*{23em} \= \hspace*{6em} \= \hspace*{8em} \= \kill
\Rule{ZXH-GW} Green and white dots.
	\> \parbox{6em}{~\hfill$
    	\Sem{4ex}{\;\!\begin{aligned}\\[-4ex] \input{./Figures-Source/ZX-green-dot.tikz}\end{aligned}\;\!}
    	$}
    \> $=
    	\Sem{4ex}{\;\!\begin{aligned}\\[-4ex] \input{./Figures-Source/ZH-white-dot.tikz}\end{aligned}\;\!}
		$
	\>
		(\textbf{Proof:} trivial.)
\\[2ex] 
\Rule{ZXH-WH} Degree-1 green phase dots and H-boxes.
	\>
		\parbox{6em}{~\hfill$
    	\Bigsem{\!\!\:\begin{aligned}
    	\begin{tikzpicture}
    		\node (z) [Z dot, label=left:\small$\Theta$] {};
    		\draw (z) -- ++(0.5,0);
    	\end{tikzpicture}
    	\end{aligned}\!\:}
    	$}
    \>  $=
	    \Bigsem{\!\!\:\begin{aligned}
    	\begin{tikzpicture}
    		\node (z) [small H box, label=left:\small$\Theta$] {};
    		\draw (z) -- ++(0.5,0);
    	\end{tikzpicture}
    	\end{aligned}\!\:}
		$
	\>
		(\textbf{Proof:} trivial.)
\\[4ex]
\Corollary{ZXH-GP} Green phase dots in ZH --- \\[-5.25ex]
we may show this using  \textsf{(ZX-GF)} to be shown below, 
	\>
		\parbox{6em}{~\hfill$
    	\begin{aligned}\\[-4ex] \input{./Figures-Source/ZX-green-phase-dot.tikz}\end{aligned}
    	$}
    \>  $\leftrightarrow
	    \begin{aligned}\\[-3ex] \input{./Figures-Source/ZH-white-H-gadget.tikz}\end{aligned}
		$
\\[-5.25ex]
\textsf{(ZXH-GW)}, and \textsf{(ZXH-GP)}.
\end{tabbing}

\vfill
\hrule
\nopagebreak
\smallskip

\noindent
\textbf{\textsf{(ZXH-HP)}, \textsf{(ZXH-HM)}:} Hadamard boxes and multicharacters --- these equalities serve to motivate our choice, to denote $1 \to 1$ multicharacter H-boxes with parameters $c = \pm 1$, by the same symbol as the quantum Fourier transform (which is to say, the inverse Fourier transform, $F\herm$):~\\[-3ex]
\begin{align}{}
\mspace{-18mu}
	\int\limits_{\mathclap{ x,y \in \D}} \omega^{xy} \; \kket{y}\bbra{x}
\;&=\,
	\int\limits_{\mathclap{ x,y \in \D}} \e^{2\pi i xy / D} \; \kket{y}\bbra{x}
\,=\,
	F\herm	;
\\
	\int\limits_{\mathclap{ x,y \in \D}} \omega^{-xy} \; \kket{y}\bbra{x}
\;&=\,
	\int\limits_{\mathclap{ x,y \in \D}} \e^{-2\pi i xy / D} \; \kket{y}\bbra{x}
\,=\,
	F 	.
\mspace{-36mu}
\end{align}

\hrule
\nopagebreak
\medskip

\Rule{ZXH-RG} Red and gray dots.
$
    \Sem{4ex}{\,\begin{aligned}\\[-4ex] \input{./Figures-Source/ZX-red-dot.tikz}\end{aligned}\,}
=
    \Sem{4ex}{\,\begin{aligned}\\[-4ex] \input{./Figures-Source/ZH-gray-dot.tikz}\end{aligned}\,}
$

\bigskip\noindent
\textbf{Proof:} We proceed by demonstrating the red-green colour change rule for ZX, which suffices to relate the gray and phase-free red nodes. through the equivalence of white and phase-free green nodes.
We take advantage of the fact that $\ket{\omega^x} = F\herm \ket{x}$.~\\[-2ex]
\begin{equation}{}
\begin{aligned}[b]
	(F\herm)\sox{n}
	\Biggl[\;\;\,\,
		\int\limits_{\mathclap{
			x \in \D
		}}\!
			\Theta(x)\; \kket{x}\sox{n}\bbra{x}\sox{m}
	\Biggr]
	(F\herm)\sox{m}
\,&=\,
		\int\limits_{\mathclap{
			x \in \D
		}}
			\Theta(x)\; \Bigl(F\herm\kket{x}\Bigr)^{\!\otimes n}\Bigl(\bbra{x}F\herm\Bigr)^{\!\otimes m}
\\[1ex]&=\,
		\int\limits_{\mathclap{
			x \in \D
		}}
			\Theta(x)\;
			\kket{\smash{\omega}^{-x}}\sox{n}
			\bbra{\smash{\;\!\omega^{x}\;\!}}\sox{m}
	\;.
\mspace{-36mu}
\end{aligned}
\end{equation}~\\[-2ex]
From this it follows that
\begin{equation}
	\Biggsem{\;\begin{aligned}
		\input{./Figures-Source/ZX-red-phase-dot.tikz}
	\end{aligned}\;}
\;=\;
		\Sem{6ex}{\begin{aligned}
  		\input{./Figures-Source/ZX-green-phase-w-H.tikz}	
  	\end{aligned}}
\end{equation}
so that in particular, when $\Theta(x) = 1$, we have~\\[-1ex]
\begin{equation}
	\Biggsem{\;\begin{aligned}
		\input{./Figures-Source/ZX-red-dot.tikz}
	\end{aligned}\;}
\;=\;
	\Sem{6ex}{\begin{aligned}
  		\input{./Figures-Source/ZX-green-phasefree-w-H.tikz}	
  	\end{aligned}}
\;=\;
	\Sem{6ex}{\begin{aligned}
  		\input{./Figures-Source/ZH-white-w-H.tikz}	
  	\end{aligned}}
\;=\;
	\Sem{6ex}{\begin{aligned}
  		\input{./Figures-Source/ZH-gray-dot.tikz}	
  	\end{aligned}}
\end{equation}

\medskip
\hrule
\nopagebreak
\smallskip

\Rule{ZXH-RN} Red and generalised not-dots.
$
    \Sem{4ex}{\,\begin{aligned}
    	\begin{tikzpicture}
    		\node at (0,0) [X dot, draw=none, fill=none, label=below:\phantom{\footnotesize{$[\,c\,]$}}] {};
    		\node (x) at (0,0) [X dot, label=above:\footnotesize{$[c]$}] {};
    		\draw (x) -- ++(-0.5,0);
    		\draw (x) -- ++(0.5,0);
    	\end{tikzpicture}
    \end{aligned}\,}
=
    \Sem{4ex}{\,\begin{aligned}\input{./Figures-Source/ZH-gen-not-dot.tikz}\end{aligned}\,}
$

\bigskip
\bigskip
\noindent
\textbf{Proof.} Using the amplitude function $\mathrm{A}_c(t) = \tau^{2ct} = \omega^{ct}$, which is denoted by $[\!\:c\!\:]$ in our stabiliser shorthand for ZX green and red dots, we have:
\begin{equation}
    \Sem{4ex}{\,\begin{aligned}
    	\begin{tikzpicture}
    		\node at (0,0) [X dot, draw=none, fill=none, label=below:\phantom{\footnotesize{$[c]$}}] {};
    		\node (x) at (0,0) [X dot, label=above:\footnotesize{$[\:\!c\:\!]$}] {};
    		\draw (x) -- ++(-0.5,0);
    		\draw (x) -- ++(0.5,0);
    	\end{tikzpicture}
    \end{aligned}\,}
\;=\;
    \Sem{4ex}{\,\begin{aligned}
    	\begin{tikzpicture}
    		\node at (0,0) [Z dot, draw=none, fill=none, label=below:\phantom{\footnotesize{$[c]$}}] {};
    		\node (h1) at (0,0) [mini H box] {\tp};
    		\node (z) at ($(h1) + (0.5,0)$)
    			[Z dot, label=above:\footnotesize{$[\!\:c\!\:]$}] {};
    		\node (h2) at ($(z) + (0.5,0)$) [mini H box] {\tp};
    		\draw (h1) -- ++(-0.5,0);
    		\draw (h2) -- ++(0.5,0);
    		\draw (h1) -- (z) -- (h2);
    	\end{tikzpicture}
    \end{aligned}\,}
\;=\;
    \Sem{9ex}{\,\begin{aligned}
    	\begin{tikzpicture}
    		\node (h1) at (0,0) [mini H box] {\tp};
    		\node (w) at ($(h1) + (0.5,0)$) [white dot] {};
    		\node (h2) at ($(z) + (0.5,0)$) [mini H box] {\tp};
    		\draw (h1) -- ++(-0.5,0);
    		\draw (h2) -- ++(0.5,0);
    		\draw (h1) -- (z) -- (h2);
    		\node at ($(w) + (0,-0.375)$) [small H box, draw=none, fill=none, label=below:\phantom{\small$\mathrm{A}_c$}] {};
    		\node (h') at ($(w) + (0,0.375)$) [small H box, label=above:\small$\mathrm{A}_c$] {};
    		\draw (z) -- (h');
    	\end{tikzpicture}
    \end{aligned}\,}
\;=\;
    \Sem{6ex}{\,\begin{aligned}
    	\begin{tikzpicture}
    		\node (h1) at (0,0) [mini H box] {\tp};
    		\node (w) at ($(h1) + (0.5,0)$) [white dot] {};
    		\node (h2) at ($(z) + (0.5,0)$) [mini H box] {\tp};
    		\draw (h1) -- ++(-0.5,0);
    		\draw (h2) -- ++(0.5,0);
    		\draw (h1) -- (z) -- (h2);
    		\node at ($(w) + (0,-0.4375)$) [small H box, draw=none, fill=none] {\footnotesize\phantom{$-c$}};
    		\node (h') at ($(w) + (0,0.4375)$) [H box]  {\footnotesize{$c$}};
    		\draw (z) -- (h');
    	\end{tikzpicture}
    \end{aligned}\,}
\;=\;
    \Sem{4ex}{\,\begin{aligned}\input{./Figures-Source/ZH-gen-not-dot.tikz}\end{aligned}\,}
 \end{equation}

\hrule
\nopagebreak
\smallskip

\noindent
\textbf{\textsf{(ZXH-S0)}, \textsf{(ZXH-S)}, \textsf{(ZXH-GH0)}, \textsf{(ZXH-GH)}:} Phase scalar ZX gadgets and degree-0 H-boxes.
\begin{equation}
\begin{aligned}
    \Bigsem{\!\!
    	\vtikzfig[-2.75ex]{ZX-Theta-0-phase-gadget}
    \,}
&=
    \Bigsem{\,\,
    	\vtikzfig[-2.75ex]{ZH-Theta-0-phase-gadget}
    \!}
&\qquad\qquad &&	
    \bigsem{\;\!
    	\vtikzfig[-2.75ex]{ZX-green-phase-free-deg0-dot}
    \;}
&=
    \Bigsem{\;
    	\vtikzfig[-2.75ex]{ZH-sqrtD-scalar-gadget}
    }
\\[1ex]	
    \biggsem{\!\!
    	\vtikzfig[-2.5ex]{ZX-Theta-c-phase-gadget}
    \!} 
=
    \biggsem{\!\!
    	\vtikzfig[-2.5ex]{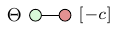}
    \!} 
&=
    \Bigsem{\,\,
    	\vtikzfig[-2.5ex]{ZH-Theta-c-phase-gadget}
    \!}
&	&&
    \Bigsem{\!\!
    	\vtikzfig[-2.75ex]{ZX-green-Theta-deg0-dot}
    \;}
&=
    \Sem{4ex}{\,\;
        	\vtikzfig[-2.75ex]{ZH-Theta-integral-gadget}
    \!}
\end{aligned}	
\end{equation}

\noindent
(\textbf{Proof}: trivial, with the first line being a special case of the second in each set.)

\medskip
\hrule
\medskip

\bigskip
\noindent
\textit{(End of ZX-ZH relations.)}
\vfill
\newpage

\begin{figure}[!h]
\begin{gather*}
~\\[-10.75ex]
\mspace{-48mu}
		\begin{tikzpicture}
		\setlength\rulediagramwd{3.5em}	
		\rewriterule		[ZX-GI]	{\vtikzfig[-0ex]{ZX-green-id}}
		\setlength\rulediagramwd{4.5em}	
		\nextrewriterule	[ZX-RI]	{\vtikzfig[-0ex]{ZX-2c-red-dots}}
		\setlength\rulediagramwd{5.0em}	
		\nextrewriterule	[ZX-HI]	{\vtikzfig[-0ex]{ZX-H-id}}
		\setlength\rulediagramwd{3em}	
		\rewritetarget 				{\vtikzfig[-0ex]{id-wire}}
	\end{tikzpicture}		
\\[-14.0ex]
\end{gather*}
\begin{align*}{}
\mspace{-96mu}
	\setlength\rulediagramwd{4.875em}	
	\begin{gathered}
	\begin{tikzpicture}
		\rewriterule		[ZX-GF]	{\vtikzfig[-0ex]{ZX-green-fn-phase-fuse}}
		\setlength\rulediagramwd{4em}	
		\coordinate (anchor-next-diagram) at ($(anchor-next-diagram) + (-.25,0)$);
		\rewritetarget 				{\vtikzfig[-0ex]{ZX-green-fn-phase-sum}}
	\end{tikzpicture}		
	\end{gathered}
\mspace{-48mu}
&&
&&
	\setlength\rulediagramwd{5em}	
	\begin{gathered}
	\begin{tikzpicture}
		\rewriterule		[ZX-GFP]	{\vtikzfig{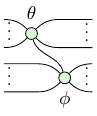}}
		\setlength\rulediagramwd{4em}	
		\coordinate (anchor-next-diagram) at ($(anchor-next-diagram) + (-.1875,0)$);
		\rewritetarget 					{\vtikzfig[-0ex]{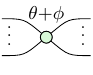}}
	\end{tikzpicture}		
	\end{gathered}
&&
&&
\mspace{-96mu}
	\setlength\rulediagramwd{5em}	
	\begin{gathered}
	\begin{tikzpicture}
		\rewriterule		[ZX-GFS]	{\vtikzfig[-0ex]{ZX-green-stab-fuse}}
		\setlength\rulediagramwd{4em}	
		\coordinate (anchor-next-diagram) at ($(anchor-next-diagram) + (-.1875,0)$);
		\rewritetarget 					{\vtikzfig{ZX-green-stab-sum}}
	\end{tikzpicture}		
	\end{gathered}
\mspace{-48mu}
\\[-10.5ex]
\end{align*}%
\begin{align*}
\mspace{-90mu}
	\begin{gathered}
	\setlength\rulediagramwd{7em}	
	\begin{tikzpicture}
	\setlength\rulediagramwd{6em}	
		\rewriterule		[ZX-RGC]	{\vtikzfig[-1ex]{ZX-red-phase-dot}}
	\setlength\rulediagramwd{4.25em}	
		\rewritetarget 					{\vtikzfig[-1ex]{ZX-green-phase-w-H}}
	\end{tikzpicture}
\\[-1ex]
	\begin{tikzpicture}
	\setlength\rulediagramwd{6em}	
		\rewriterule		[ZX-RGB]		{\vtikzfig[-1ex]{ZX-bialg-many}}
	\setlength\rulediagramwd{4.25em}	
		\rewritetarget 					{\vtikzfig[-1ex]{ZX-bott-many}}
	\end{tikzpicture}
\\[-3ex]
	\begin{tikzpicture}
	\setlength\rulediagramwd{6em}	
		\rewriterule		[ZX-CPY]	{\hspace*{-2ex}\vtikzfig{ZX-red-copy}\hspace*{1ex}}
	\setlength\rulediagramwd{3.5em}	
		\rewritetarget 					{\hspace*{-1ex}\vtikzfig{ZX-red-copies}\hspace*{1ex}}
	\end{tikzpicture}
\\[-5ex]
	\begin{tikzpicture}
	\setlength\rulediagramwd{5.25em}	
		\rewriterule		[ZX-NS]		{\vtikzfig{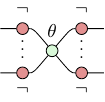}}
	\setlength\rulediagramwd{4em}	
		\rewritetarget 					{\hspace*{-2ex}\vtikzfig{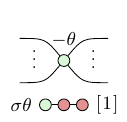}}
	\end{tikzpicture}
\\[-5.5ex]
	\begin{tikzpicture}
	\setlength\rulediagramwd{8em}	
		\rewriterule		[ZX-RS]		{\vtikzfig[-0ex]{ZX-conjugate-stab-dot}}
		\coordinate (anchor-next-diagram) at ($(anchor-next-diagram) + (-.625,0)$);
		\rewritetarget 					{\vtikzfig[.5ex]{ZX-shear-gadget}}
	\end{tikzpicture}
	\mspace{-18mu}
	\end{gathered}
&& &&
	\setlength\rulediagramwd{10em}	
	\begin{gathered}
	\begin{tikzpicture}
	\setlength\rulediagramwd{5.25em}	
		\rewriterule		[ZX-Z]		{\vtikzfig[-3ex]{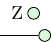}}
	\setlength\rulediagramwd{4em}	
		\rewritetarget 					{\hspace*{-2ex}\vtikzfig[-3ex]{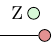}}
	\end{tikzpicture}
\\[-0.0ex]
	\mspace{-18mu}
	\setlength\rulediagramwd{3.25em}	
	\begin{tikzpicture}
		\rewriterule		[ZX-ZCP]		{\vtikzfig[-0ex]{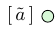}}
		\coordinate (anchor-next-diagram) at ($(anchor-next-diagram) + (-.25,0)$);
		\setlength\rulediagramwd{5.25em}	
		\nextrewriterule	[ZX-ZSP]		{\vtikzfig[-0ex]{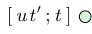}}
		\coordinate (anchor-next-diagram) at ($(anchor-next-diagram) + (-.4375,0)$);
		\setlength\rulediagramwd{3.25em}	
		\rewritetarget 					{\vtikzfig[-0ex]{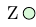}}
	\end{tikzpicture}
\\[-2.5ex]
	\begin{tikzpicture}
		\setlength\rulediagramwd{9em}	
		\rewriterule		[ZX-MH]	{\hspace*{-2ex}\vtikzfig{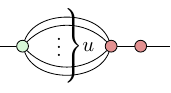}\hspace*{1ex}}
		\coordinate (anchor-next-diagram) at ($(anchor-next-diagram) + (.25,0)$);
		\rewritetarget 					{\hspace*{-1ex}\vtikzfig{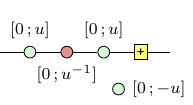}\hspace*{1ex}}
	\end{tikzpicture}
	\mspace{-36mu}
\\[-2.5ex]
	\begin{tikzpicture}
		\rewriterule		[ZX-ME]		{\vtikzfig{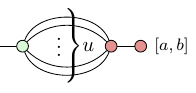}}
		\setlength\rulediagramwd{8.5em}	
		\rewritetarget 					{\hspace*{-2ex}\vtikzfig{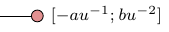}}
	\end{tikzpicture}
\\[-2.5ex]
\mspace{-90mu}
	\setlength\rulediagramwd{7.5em}	
	\begin{gathered}
	\begin{tikzpicture}
		\rewriterule		[ZX-MEH]		{\vtikzfig{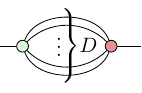}}
		\setlength\rulediagramwd{4.4375em}	
		\nextrewriterule	[ZX-A]		{\vtikzfig[1ex]{ZX-antipode}}
		\setlength\rulediagramwd{3.25em}	
		\coordinate (anchor-next-diagram) at ($(anchor-next-diagram) + (-.125,0)$);
		\rewritetarget 					{\vtikzfig[1ex]{ZX-dc}}
	\end{tikzpicture}
	\end{gathered}
	\end{gathered}
\\[-11.0ex]
\end{align*}
\begin{gather*}
\setlength\rulediagramwd{5em}	
\begin{tikzpicture}
	\rewriterule		[ZX-PU]	{\vtikzfig[-0ex]{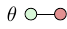}}
	\setlength\rulediagramwd{6em}	
	\coordinate (anchor-next-diagram) at ($(anchor-next-diagram) + (-.25,0)$);
	\nextrewriterule	[ZX-SU]	{\vtikzfig[-0ex]{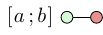}}
	\setlength\rulediagramwd{5.75em}	
	\coordinate (anchor-next-diagram) at ($(anchor-next-diagram) + (-.25,0)$);
	\nextrewriterule	[ZX-GU]	{\vtikzfig[-1ex]{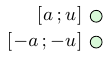}}
	\setlength\rulediagramwd{2.5em}	
	\rewritetarget 				{\vtikzfig[-0.0ex]{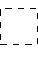}}
\end{tikzpicture}		
\\[-7.5ex]
\end{gather*}
\hrule
\smallskip
\caption{%
	\label{fig:ZX-rewrites}
	Various scalar-exact rewrites (including axioms and corollaries) on ZX diagrams, which are sound for the semantics described in Eqn.~\eqref{eqn:idealised-ZX-integrals} subject to $\nu = D^{-1/4}$.
	Chains of rewrites \smash{$\Big.\cD_j \!\xleftrightarrow{\!\!\:\textsf{(x)}\!\:} \cdots \leftrightarrow \cD_{\!\!\;f}$} are intended to indicate that \smash{$\cD_j \!\xleftrightarrow{\!\!\:\textsf{(x)}\!\:}\! \cD_{\!\!\;f}$} is either an axiom or notable corollary.
	Throughout, we have $\theta, \phi \in \R$, and $\Theta, \Phi: \Z \to \C$, and $a, a_1, a_2, b, b_1, b_2, c \in \Z$.
	We define the constant $0$ function, $\mathrm Z: \Z \to \{ 0 \}$.
	We let $1 < t < D$ be a divisor of $D$, $1 < t' < t$ be a divisor of $t$ (and thus also of $D$), $u \in \N$ an integer which has no common factors $t>1$ with $D$ (sometimes interpreted as an element $u \in \Z_D^\times$), and $\tilde a \in \Z$ an integer which is not a multiple of $D$.
	Many of the rules involve green dots or red dots parameterised by a label ${[\:\! a\:\!]}$ or ${[\:\!a \s b\:\!]}$ for $a,b \in \Z$\;\!: these stand respectively for the amplitude functions $x \mapsto \tau^{-2ax}$ and $x \mapsto \tau^{2ax + bx^2}$, where $\tau = \exp(\pi i (D^2 {+} 1)/D)$. 
	An annotation $\neg$ on a red or green dot indicates a dimension-dependent parameter ${[\!\:\sigma \!\:]}$, where $\sigma = 0$ for $D$ odd and $\sigma = 1$ for $D$ even.
}
\end{figure}

\subsubsection{Sound ZX rewrites}
\label{apx:sound-ZX-rewrites}

Figure~\ref{fig:ZX-rewrites} presents a list of all of the rewrites of ZX diagrams which we consider, which are sound for the semantic map defined in Eqn.~\eqref{eqn:idealised-ZX-integrals} subject to $\nu = D^{-1/4}$.
In this section, we demonstrate the soundness of these rewrites.
(We will occasionally simplify our proofs using the fact that $F\herm \kket{x} = \ket{\smash{\omega^x}}$.)

We are guided in no small part by the rule-set that Booth and~Carette~\cite{BC-2022} set out for odd dimensions $D > 1$ --- with some differences due to \textbf{(a)}~our definition of red dots, and \textbf{(b)}~the different convention that we use for describing stabiliser phases.
For the latter, we broadly follow Ref.~\cite{dB-2013} in taking ${[\!\:a \s b\!\:]}$ to denote a phase-function  $x \mapsto \tau^{2ax + bx^2}$ for $\tau = \e^{\pi i (D^2 + 1)/D}$ (using ${[\!\:a\:\!]}$ as shorthand when $b=0$), rather than $x \mapsto \omega^{[(D{+}1)/2][ax + bx^2]} = \tau^{ax + bx^2}$ as in Ref.~\cite{BC-2022}.%
	\footnote{%
		While Ref.~\cite{BC-2022} does not appear to make explicit use of the 
		conventions of Ref.~\cite{dB-2013}, we may show that $\tau = \omega^{(D{+}1)/2}$, by observing that $\tau \cdot \omega^{-(D+1)/2} = \exp\bigl(\pi i (D^2 - D)/D\bigr) = (-1)^{D-1} = +1$.
	}
The motivation for the choice of $2ax + bx^2$ in the exponent in place of $ax + bx^2$, is to obtain a theory which describes stabiliser phases also when $D$ is even, and also for which~\\[-2ex]
\begin{align}
	\biggsem{\,
		\vtikzfig[-1ex]{ZX-Z}
	\,}
	&=\,
		Z
\,,
&\;\
	\biggsem{\,
		\vtikzfig[-1ex]{ZX-X}
	\,}
	&=\,
		X
\,,
\\\;\;
	\biggsem{\!%
		\vtikzfig[-1ex]{ZX-kket-a}
	\,}
	&=\,
		\kket{a}
\,,
&\;\;
	\biggsem{\!%
		\vtikzfig[-1ex]{ZX-kket-omega-a}
	\,}
	&=\,
	\kket{\smash{\omega^a}}
.
\end{align}~\\[-1ex]
We note also that the proof techniques of Booth and Carette~\cite{BC-2022} may  be preferable to the ones demonstrated below in many cases, at least in the case of $D$ an odd prime.
This constraint not only is the special case considered in Ref.~\cite{BC-2022}, but in particular implies that $\Z_D$ is a field, in which every $u \in \Z_D \setminus \{ 0 \}$ is a multiplicative unit.
This simplifies the situation with arithmetic substantially, and in particular allows Ref.~\cite{BC-2022} to obtain simple normal forms.
It is unclear to us how much of those techniques extend to the more general case of $D > 1$ arbitrary.
Our proof techniques are rather more direct and concrete: the aim of which is to  demonstrate the effectiveness (in principle) of analysing operators through discrete integrals, and also to reduce the amount of technical machinery on which our results depend.

\hfill

\medskip
\hrule
\nopagebreak
\bigskip

\noindent
\textbf{\textsf{(ZX-GI)},\! \textsf{(ZX-RI)},\! \textsf{(ZX-HI)},\! \textmd{\itshape etc.}} ---
Identity rules:

$
    \bigsem{\,\begin{aligned}\\[-3ex] \input{./Figures-Source/ZX-green-id.tikz}\end{aligned}\,}
=
    \Biggsem{\begin{aligned}
	\begin{tikzpicture}
		\node at (0,0) (g1) [X dot, draw=none, fill=none, label=below:\small$\phantom {[c]}$] {};
		\node at (0,0) (g1) [X dot, label=above:\small${[\!\:c\!\:]}$] {};
		\draw (g1) -- ++(-0.375,0);
		\node at ($(g1) + (0.5,0)$) (g2) [X dot, label=above:\small${[\!\:c\!\:]}$] {};
		\draw (g2) -- ++(0.375,0);
		\draw (g1) -- (g2);
	\end{tikzpicture}
	\end{aligned}}
=
    \bigsem{\begin{aligned}
	\begin{tikzpicture}
		\node at (0,0) (h1) [small H box] {\tp};
		\draw (h1) -- ++(-0.4375,0);
		\node at ($(h1) + (0.5375,0)$) (h2) [small H box] {\tm};
		\draw (h2) -- ++(0.5,0);
		\draw (h1) -- (h2);
	\end{tikzpicture}
	\end{aligned}}
=
    \bigsem{\,\begin{aligned}
	\begin{tikzpicture}
		\node at (0,0) (h1) [small H box] {\tp};
		\draw (h1) -- ++(-0.375,0);
		\node at ($(h1) + (0.375,0)$) (h2) [small H box] {\tp};
		\node at ($(h2) + (0.375,0)$) (h3) [small H box] {\tp};
		\node at ($(h3) + (0.375,0)$) (h4) [small H box] {\tp};
		\draw (h4) -- ++(0.375,0);
		\draw (h1) -- (h2) -- (h3) -- (h4);
	\end{tikzpicture}
\end{aligned}\,}
=
    \bigsem{\,\begin{aligned}\\[-3ex] \input{./Figures-Source/id-wire.tikz}\end{aligned}\,}
$

\bigskip
\noindent
\textbf{Proof.} These all follow from the relations between the generators described here, and corresponding ZH generators as described in Section~\ref{apx:ZX-ZH-relations}.

\medskip
\hrule
\nopagebreak
\smallskip
\Rule{ZX-GF} Green dot fusion.
\\[-3ex]
\begin{gather*}
	\Sem{8.5ex}{\input{./Figures-Source/ZX-green-fn-phase-fuse.tikz}}
\;\;=\;\;
	\Sem{5.5ex}{\input{./Figures-Source/ZX-green-fn-phase-sum.tikz}}
\end{gather*}

\bigskip\bigskip
%
\noindent
\textbf{Proof.}~\\[-9.25ex]
\begin{align*}
	\mspace{60mu}&\mathop{\int \!\!\!\! \int}\limits_{\mathclap{
		x,y \in \Z_D
	}}
	\Bigl(
		\mathbf 1^{\otimes n} \otimes \Phi(y)\; \kket{y}^{\otimes n} \bbra{y}^{\otimes \ell+1}
	\Bigr)
	\Bigl(
		\Theta(x) \;\kket{x}^{\otimes m+1} \bbra{x}^{\otimes k} \otimes \mathbf 1^{\otimes m}
	\Bigr)
	\mspace{-66mu}
\\[1ex]&=\;
	\mathop{\int \!\!\!\! \int}\limits_{\mathclap{
		x,y \in \Z_D
	}}
		\Theta(x) \, \Phi(y) \; \kket{x}^{\otimes m} \bbra{x}^{\otimes k} \,\otimes\; \bbracket{y}{x} \;\otimes\; \kket{y}^{\otimes n} \bbra{y}^{\otimes \ell}
\\[1ex]&=\;
	\int\limits_{\mathclap{
		x \in \Z_D
	}}
		\Theta(x) \, \Phi(x) \;  \kket{x}^{\otimes m} \bbra{x}^{\otimes k} \,\otimes\; \kket{x}^{\otimes n} \bbra{x}^{\otimes \ell}
\\[1ex]&=\;
	\int\limits_{\mathclap{
		x \in \Z_D
	}}
		\bigl[\Theta \!\cdot\! \Phi\bigr](x) \; \kket{x}^{\otimes m+n} \bbra{x}^{\otimes k+\ell}
	\;.
\tag*{\qed}
\end{align*}

\noindent
In addition to the proof provided here, this could easily be demonstrated using white dot fusion and H-box multiplication, using the correspondence to ZH~generators as described in Section~\ref{apx:ZX-ZH-relations}.

\medskip
\Corollary{ZX-GFP)\textrm, (ZX-GFS} Green phase fusion rules --- these are two special cases, presenting familiar versions of the green dot fusion rule, with additive phases. ~\\[-4ex]
\begin{align*}
	\begin{aligned}\input{./Figures-Source/ZX-green-phase-fuse.tikz}\end{aligned}
    \;\;&\longleftrightarrow\;\;
	\begin{aligned}\input{./Figures-Source/ZX-green-phase-sum.tikz}\end{aligned}
	& \Bigg/&&
	\begin{aligned}\input{./Figures-Source/ZX-green-stab-fuse.tikz}\end{aligned}
    \;\;&\longleftrightarrow\;\;
	\begin{aligned}\input{./Figures-Source/ZX-green-stab-sum.tikz}\end{aligned}	
\end{align*}

\medskip
\hrule
\nopagebreak
\smallskip

\Rule{ZX-RGC} 
 Red-green colour change.
 \\[-3ex]
\begin{gather*}
		\Sem{6ex}{\begin{aligned}
  		\input{./Figures-Source/ZX-green-phase-w-H.tikz}	
  	\end{aligned}}
\;\;=\;\;
	\Biggsem{\;\begin{aligned}
		\input{./Figures-Source/ZX-red-phase-dot.tikz}
	\end{aligned}\;}
\end{gather*}

\bigskip
%
\noindent
\textbf{Proof.}~\\[-9.5ex]
\begin{align*}{}
	\mspace{108mu}&\mspace{-60mu}
	(F\herm)\sox{n}
	\Biggl[\;\;\;\;
		\int\limits_{\mathclap{
			x \in \Z_D
		}}
			\Theta(x)\; \kket{x}\sox{n}\bbra{x}\sox{m}
	\Biggr]
	(F\herm)\sox{m}
\\[1ex]&=\;
		\int\limits_{\mathclap{
			x \in \Z_D
		}}
			\Theta(x)\; \Bigl(F\herm\kket{x}\Bigr)^{\!\otimes n}\Bigl(\bbra{x}F\herm\Bigr)^{\!\otimes m}
\;=\;
		\int\limits_{\mathclap{
			x \in \Z_D
		}}
			\Theta(x)\;
			\kket{\smash{\omega}^{-x}}\sox{n}
			\bbra{\smash{\;\!\omega^{x}\;\!}}\sox{m}
	\;.
\mspace{-36mu}
\tag*{\qed}
\end{align*}

\hrule
\nopagebreak
\smallskip

\Rule{ZX-CPY} Copy rule.
\\[-3ex]
\begin{gather*}
		\Sem{5ex}{\!\vtikzfig[-3ex]{ZX-red-copy}}
\;\;=\;\;
	\Sem{5ex}{\vtikzfig[-2ex]{ZX-red-copies}}
\end{gather*}

\bigskip
%
\noindent
\textbf{Proof.}~\\[-9.25ex]
\begin{align*}{}
\mspace{60mu}
	&%
	\mathop{\int \!\!\!\! \int}\limits_{\mathclap{
		k,y \in \Z_D
	}}
	\Bigl(
		\kket{y}\sox{n} \bbra{y}
	\Bigr)
	\Bigl(
		\tau^{2ak} \,\kket{\smash{\omega^{-k}}}
	\Bigr)
	\mspace{-66mu}
\\[1ex]&=\;
	\mathop{\int \!\!\!\! \int}\limits_{\mathclap{
		x,y,k \in \Z_D
	}}
	\Bigl(
		\kket{y}\sox{n} \bbra{y}
	\Bigr)
	\Bigl(
		\omega^{ak} \, \omega^{kx} \,\kket{x}
	\Bigr)
\\[1ex]&=\,
	\int\limits_{\mathclap{
		x \in \Z_D
	}}
		\kket{x}\sox{n} 
	\Bigl(\;\;\,
		\int\limits_{\mathclap{
			k \in \Z_D
		}}
		\omega^{k(x+a)} \!
	\Bigr)
\,=\,
	\int\limits_{\mathclap{
		x \in \Z_D
	}}
		\kket{x}\sox{n} 
		\;\bbracket{x}{-a}
\,=\,
	\kket{-a}\sox{n}
	\!\!\:.
\mspace{-36mu}
\tag*{\qed}
\end{align*}

\hrule

\nopagebreak
\smallskip

\Rule{ZX-PU}
Unit phase angle.
\\[-3ex]
\begin{gather*}
	\Bigsem{%
	\vtikzfig[-2.5ex]{ZX-phase-gadget} 
	\;\,}
\;=\;
	\Biggsem{\;\;\vtikzfig[-1.5ex]{empty}\;\;}
\end{gather*}

\bigskip
\noindent
\textbf{Proof.}~
$\int\limits_{\mathclap{
		x \in \D
	}}
	\e^{i\theta x}\; \bbracket{0}{x}
\;=\;
	\e^{0}
\;=\;
	1
	\;.
\hfill\qed$

\medskip
\vfill
\hrule

\nopagebreak
\smallskip

\Rule{ZX-SU} 
Unit stabiliser scalar.
\\[-3ex]
\begin{gather*}
	\biggsem{%
	\vtikzfig[-2.5ex]{ZX-stab-scalar-unit} 
	\;\,}
\;=\;
	\Biggsem{\;\;\vtikzfig[-1.5ex]{empty}\;\;}
\end{gather*}

\medskip
\noindent
\textbf{Proof.}~
$\int\limits_{\mathclap{
		x \in \D
	}}
	\tau^{2ax+bx^2}\; \bbracket{0}{x}
\;=\;
	\tau^0
\;=\;
	1
	\;.
\hfill\qed
$

\medskip
\hrule

\nopagebreak
\smallskip

\Rule{ZX-GU} 
Gaussian integral unit.
\\[-3.5ex]
\begin{gather*}
	\Biggsem{%
	\vtikzfig[-2ex]{ZX-conjugate-quadratic-dots} 
	\;\,}
\;=\;
	\Biggsem{\;\;\vtikzfig[-1.5ex]{empty}\;\;}
\end{gather*}

\bigskip
\noindent
\textbf{Proof.}~\\[-9.5ex]
\begin{align*}
	\mspace{48mu}&
	\Biggl[\;\;\;\;\int\limits_{\mathclap{
		x \in \D
	}}
	\tau^{-2ax-ux^2}
	\Biggr]
	\Biggl[\;\;\;\;\int\limits_{\mathclap{
		y \in \D
	}}
	\tau^{2ay+uy^2}
	\Biggr]
\\&=\;
	\Gamma(a,u,D)^\ast \, \Gamma(a,u,D)
\;=\;
	\bigl\lvert \Gamma(a,u,D) \bigr\rvert^2
\;=\;
	1
	\;.
\tag*{\qed}
\end{align*}

\medskip
\hrule
\nopagebreak
\smallskip

\Rule{ZX-Z} Zero scalar rule.
~\\[-3.5ex]
\begin{gather*}
	\Biggsem{%
	\vtikzfig[-3ex]{ZX-zero-and-green-lollipop} 
	\;\,}
\;=\;
	\Biggsem{%
	\vtikzfig[-3ex]{ZX-zero-and-red-lollipop} 
	\;\,}
\end{gather*}

\bigskip
\noindent
\textbf{Proof.}~\\[-9.25ex]
\begin{align*}
\mspace{90mu}
	\int\limits_{\mathclap{
			x \in \D
		}}\!
	\mathrm Z(x)	\,
	\bbra{x}
\;=
	\int\limits_{\mathclap{
			x \in \D
		}}\!
	0\cdot
	\bbra{x}
\;=\;
	\mathbf 0\herm
\,=
	\int\limits_{\mathclap{
			k \in \D
		}}\!
	0\cdot
	\bbra{\smash{\omega^k}}
\;=
	\int\limits_{\mathclap{
			k \in \D
		}}\!
	\mathrm Z(k)\,
	\bbra{\smash{\omega^k}}
\,.\;\;
\mspace{-60mu}
\tag*{\qed}
\end{align*}

\medskip

\hrule
\nopagebreak
\smallskip

\Rule{ZX-ZSP} 
Zero stabiliser phase.
~\\[-3ex]
\begin{gather*}
	\Bigsem{%
	\vtikzfig[-2ex]{ZX-awkward-quadratic-scalar-dot} 
	\;\,}
\;=\;
	\Bigsem{%
	\vtikzfig[-2ex]{ZX-zero-scalar-dot} 
	\;\,}
\end{gather*}

\bigskip
\noindent
\textbf{Proof.}~\\[-8.25ex]
\begin{align*}
\mspace{60mu}
	\int\limits_{\mathclap{
			x \in \D
		}}\!
	\tau^{2 u t'\!\!\: x + t x^2}
\;&=\;
	\Gamma(u \;\! t', t , D)
\\[-3ex]&=\;
	t \cdot \Gamma(u, t/t', D/t')
\;=\;
	0
\;=
	\int\limits_{\mathclap{
			x \in \D
		}}\!
	\mathrm Z(x)
\,.\;\;
\mspace{-60mu}
\tag*{\qed}
\end{align*}

\medskip

\hrule
\nopagebreak
\smallskip

\Rule{ZX-ZCP} 
Zero clock phase.
~\\[-3ex]
\begin{gather*}
	\Bigsem{%
	\vtikzfig[-2.5ex]{ZX-nonzero-stab-phase-dot} 
	\;\,}
\;=\;
	\Bigsem{%
	\vtikzfig[-2.5ex]{ZX-zero-scalar-dot} 
	\;\,}
\end{gather*}

\noindent
\textbf{Proof.} \qquad $
	\int\limits_{\mathclap{
			x \in \D
		}}\!
	\tau^{2 \tilde a  x}
\;=\;
	\bbracket{\tilde a}{0}
\;=\;
	0
\;=
	\int\limits_{\mathclap{
			x \in \D
		}}\!
	\mathrm Z(x)
\,.\;\;
\hfill
\qed
$
\bigskip

\noindent
This result could likely be subsumed as a special case of \textsf{(ZX-ZSP)}, if it were suitably modified to accommodate the case of a stabiliser phase ${[\:\! a, b \!\:]}$ in which $b$ is a multiple of $D$ (for $D$ odd) or of $2D$ (for $D$ even), and in which $a$ is not a multiple of $D$.

\medskip
\hrule

\nopagebreak
\smallskip

\Rule{ZX-RGB} Red-green bialgebra. $
	\Biggsem{\;\vtikzfig[-2ex]{ZX-bialg-many}}
	\biggsem{\vtikzfig[-3ex]{ZX-bott-many}}
$

\noindent
\textbf{Proof.} This follows from the correspondence between green and white dots, and between red and gray dots, shown in Section~\ref{apx:ZX-ZH-relations}.
\bigskip

\hrule
\nopagebreak
\smallskip

\Rule{ZX-NS} Not-dot symmetry.
$
	\Sem{8ex}{\vtikzfig[-1ex]{ZX-green-phase-nots}}
\;=\;
	\Sem{9ex}{\vtikzfig[-2ex]{ZX-green-negated-phase-gadget}}
$

\noindent
\textbf{Proof.}~
We may show this using \textsf{(ZXH-GP)}, \textsf{(ZXH-RN)}, and \textsf{(ZH-NS)}:~\\[-3ex]
\begin{equation}
	\begin{aligned}
    \begin{tikzpicture}
      \node (Z) at (0,0) [Z dot, label=above:\small$\theta$] {};
      \node (h1) at (-.5,0.375) [X dot, label=above:\small$\neg$] {};
      \node (h2) at (-.5,-0.375) [X dot, label=below:\small$\neg$] {};
      \draw (Z) .. controls (-0.3175,0.375) .. (h1) -- ++(-0.375,0);
      \draw (Z) .. controls (-0.3175,-0.375) .. (h2) -- ++(-0.375,0);
      \node (dots) at ($(Z) + (-0.5,0.125)$) {\footnotesize$\mathbf\vdots$};
      \node (h1) at (.5,0.375) [X dot, label=above:\small$\neg$] {};
      \node (h2) at (.5,-0.375) [X dot, label=below:\small$\neg$] {};
      \draw (Z) .. controls (0.3175,0.375) .. (h1) -- ++(0.375,0);
      \draw (Z) .. controls (0.3175,-0.375) .. (h2) -- ++(0.375,0);
      \node (dots) at ($(Z) + (0.5,0.125)$) {\footnotesize$\mathbf\vdots$};
    \end{tikzpicture}
	\end{aligned}
\;\leftrightarrow\;
\begin{aligned}
    \begin{tikzpicture}
      \node (Z) at (0,0) [white dot] {};
      \node (h1) at (-.5,0.375) [not dot, label=above:\small$\neg$] {};
      \node (h2) at (-.5,-0.375) [not dot, label=below:\small$\neg$] {};
      \draw (Z) .. controls (-0.3175,0.375) .. (h1) -- ++(-0.375,0);
      \draw (Z) .. controls (-0.3175,-0.375) .. (h2) -- ++(-0.375,0);
      \node (dots) at ($(Z) + (-0.5,0.125)$) {\footnotesize$\mathbf\vdots$};
      \node (h1) at (.5,0.375) [not dot, label=above:\small$\neg$] {};
      \node (h2) at (.5,-0.375) [not dot, label=below:\small$\neg$] {};
      \draw (Z) .. controls (0.3175,0.375) .. (h1) -- ++(0.375,0);
      \draw (Z) .. controls (0.3175,-0.375) .. (h2) -- ++(0.375,0);
      \node (dots) at ($(Z) + (0.5,0.125)$) {\footnotesize$\mathbf\vdots$};
        \node at ($(Z) + (0,-0.375)$) [small H box, draw=none, fill=none, label=below:\phantom{\small$\e^{i\theta}$}] {};
        \node (H) at ($(Z) + (0,0.375)$) [small H box, label=above:\small$\e^{i\theta}$] {};
        \draw (Z) -- (H);
    \end{tikzpicture}
	\end{aligned}
\;\leftrightarrow\!\!\!\!\!
\begin{aligned}
    \begin{tikzpicture}
      \node (Z) at (0,0) [white dot] {};
      \draw (Z) .. controls (-0.3175,0.375) .. ++(-0.75,.375);
      \draw (Z) .. controls (-0.3175,-0.375) .. ++(-0.75,-.375);
      \node (dots) at ($(Z) + (-0.5,0.125)$) {\footnotesize$\mathbf\vdots$};
      \draw (Z) .. controls (0.3175,0.375) ..  ++(0.75,.375);
      \draw (Z) .. controls (0.3175,-0.375) .. ++(0.75,-.375);
      \node (dots) at ($(Z) + (0.5,0.125)$) {\footnotesize$\mathbf\vdots$};
        \node at ($(Z) + (-0.4375,-0.625)$) [small H box, draw=none, fill=none, label=left:\phantom{\small$\e^{i\theta}_{\big.}$}] {};
        \node (H) at ($(Z) + (0.4375,0.625)$) [small H box, label=right:\small$\e^{i\theta}_{\big.}$] {};
        \draw [out=105,in=180] (Z) to node [pos=0.625, not dot, label=left:\small$\neg$] {} (H);
    \end{tikzpicture}
	\end{aligned}
\!\!\!\!\! \leftrightarrow\!\!\!\!\!
\begin{aligned}
    \begin{tikzpicture}
      \node (Z) at (0,0) [white dot] {};
      \draw (Z) .. controls (-0.3175,0.375) .. ++(-0.75,.375);
      \draw (Z) .. controls (-0.3175,-0.375) .. ++(-0.75,-.375);
      \node (dots) at ($(Z) + (-0.5,0.125)$) {\footnotesize$\mathbf\vdots$};
      \draw (Z) .. controls (0.3175,0.375) ..  ++(0.75,.375);
      \draw (Z) .. controls (0.3175,-0.375) .. ++(0.75,-.375);
      \node (dots) at ($(Z) + (0.5,0.125)$) {\footnotesize$\mathbf\vdots$};
        \node at ($(Z) + (-0.4375,-0.625)$) [small H box, draw=none, fill=none, label=left:\phantom{\small$\e^{-i\theta}_{\big.}$}] {};
        \node (H) at ($(Z) + (0.4375,0.625)$) [small H box, label=right:\small$\e^{-i\theta}_{\big.}$] {};
        \draw [out=105,in=180] (Z) to (H);
        \node (H) at ($(Z) + (-0.4375,0.625)$) [small H box, label=left:\small$\e^{i\theta\sigma}_{\Big.}\!\!\!\!\!\!\!$] {};
        \node at ($(Z) + (-0.4375,-0.625)$) [small H box, fill=none, draw=none, label=left:\phantom{\small$\e^{i\theta\sigma}_{\Big.}\!\!\!\!\!\!\!$}] {};
    \end{tikzpicture}
	\end{aligned}
\end{equation}~\\[-3ex]
Applying \textsf{(ZXH-S)} for $\Theta(x) = \e^{i\sigma x}$ for the scalar gadget, and \textsf{(ZXH-GP)} again for the H-box with amplitude $\e^{-i\theta}$, we then have~\\[-3ex]
\begin{equation}
	\begin{aligned}
    \begin{tikzpicture}
      \node (Z) at (0,0) [Z dot, label=above:\small$\theta$] {};
      \node (h1) at (-.5,0.375) [X dot, label=above:\small$\neg$] {};
      \node (h2) at (-.5,-0.375) [X dot, label=below:\small$\neg$] {};
      \draw (Z) .. controls (-0.3175,0.375) .. (h1) -- ++(-0.375,0);
      \draw (Z) .. controls (-0.3175,-0.375) .. (h2) -- ++(-0.375,0);
      \node (dots) at ($(Z) + (-0.5,0.125)$) {\footnotesize$\mathbf\vdots$};
      \node (h1) at (.5,0.375) [X dot, label=above:\small$\neg$] {};
      \node (h2) at (.5,-0.375) [X dot, label=below:\small$\neg$] {};
      \draw (Z) .. controls (0.3175,0.375) .. (h1) -- ++(0.375,0);
      \draw (Z) .. controls (0.3175,-0.375) .. (h2) -- ++(0.375,0);
      \node (dots) at ($(Z) + (0.5,0.125)$) {\footnotesize$\mathbf\vdots$};
    \end{tikzpicture}
	\end{aligned}
\;\leftrightarrow\!\!\!\!\!
\begin{aligned}
    \begin{tikzpicture}
      \node (Z) at (0,0) [white dot] {};
      \draw (Z) .. controls (-0.3175,0.375) .. ++(-0.75,.375);
      \draw (Z) .. controls (-0.3175,-0.375) .. ++(-0.75,-.375);
      \node (dots) at ($(Z) + (-0.5,0.125)$) {\footnotesize$\mathbf\vdots$};
      \draw (Z) .. controls (0.3175,0.375) ..  ++(0.75,.375);
      \draw (Z) .. controls (0.3175,-0.375) .. ++(0.75,-.375);
      \node (dots) at ($(Z) + (0.5,0.125)$) {\footnotesize$\mathbf\vdots$};
        \node at ($(Z) + (-0.4375,-0.625)$) [small H box, draw=none, fill=none, label=left:\phantom{\small$\e^{-i\theta}_{\big.}$}] {};
        \node (H) at ($(Z) + (0.4375,0.625)$) [small H box, label=right:\small$\e^{-i\theta}_{\big.}$] {};
        \draw [out=105,in=180] (Z) to (H);
        \node (H) at ($(Z) + (-0.4375,0.625)$) [small H box, label=left:\small$\e^{i\theta\sigma}_{\Big.}\!\!\!\!\!\!\!$] {};
        \node at ($(Z) + (-0.4375,0.625)$) [small H box, fill=none, draw=none, label=left:\phantom{\small$\e^{i\theta\sigma}_{\Big.}\!\!\!\!\!\!\!$}] {};
    \end{tikzpicture}
	\end{aligned}
\!\!\!\!\!
\leftrightarrow
\begin{aligned}
		\vtikzfig{ZX-green-negated-phase-gadget}
	\end{aligned}
\end{equation}~\\[-1ex]

\medskip
\hrule
\smallskip
\Rule{ZX-RS} Red shear rule.
\\[-3ex] 
\begin{gather*}
  	\biggsem{\vtikzfig{ZX-conjugate-stab-dot}}
\;\;=\;\;
 	\Sem{8ex}{\;\vtikzfig[-4ex]{ZX-shear-gadget}\;}
\end{gather*}

\bigskip
\noindent
\textbf{Proof.}~\\[-7.75ex]
\begin{align*}{}
\mspace{48mu}
	\mspace{48mu}&\mspace{-48mu}
	\mathop{\int\!\!\!\!\int\!\!\!\!\int}\limits_{\mathclap{
		x,y,z \in \D
	}}
		\Bigl(
			\tau^{2cz} \;
				\kket{z}\bbra{z}
		\Bigr)
		\Bigl(
			\tau^{2ax+bx^2} \;
				\kket{\smash{\omega^{-x}}}\bbra{\smash{\;\!\omega^x\;\!}}
		\Bigr)
		\Bigl(
			\tau^{2cy} \; \kket{y}\bbra{y}
		\Bigr)
\\[-1ex]&=\;
	\int\limits_{\mathclap{
		x \in \D
	}}
		\tau^{2ax+bx^2}
	\Biggl[\;\;\;
	\int\limits_{\mathclap{
		z \in \D
	}}
		\tau^{2cz} \;\! \omega^{xz} \;\kket{z}
	\Biggr]	
	\Biggl[\;\;\;
	\int\limits_{\mathclap{
		y \in \D
	}}
		\tau^{2cy} \;\! \omega^{xy} \;\bbra{y}
	\Biggr]
\mspace{-60mu}
\\[1ex]&=\;
	\int\limits_{\mathclap{
		x \in \D
	}}
		\tau^{2ax+bx^2}
	\Biggl[\;\;\;
	\int\limits_{\mathclap{
		z \in \D
	}}
		\omega^{z(x+c)} \;\kket{z}
	\Biggr]	
	\Biggl[\;\;\;
	\int\limits_{\mathclap{
		y \in \D
	}}
		\omega^{y(x+c)} \;\bbra{y}
	\Biggr]
\mspace{-60mu}
\\[1ex]\mspace{-36mu}
\;&=\;
	\int\limits_{\mathclap{
		x \in \D
	}}
		\tau^{2ax+bx^2}
	\;
	\kket{\smash{\omega^{-(x+c)}}}
	\bbra{\smash{\omega^{x+c}}}
\\[1ex]&=\;
	\int\limits_{\mathclap{
		x \in \D
	}}
		\tau^{2ax - 2ac + bx^2 - 2bcx + bc^2}
	\;
	\kket{\smash{\omega^{-x}}}
	\bbra{\smash{\omega^{x}}}
\\[1ex]&=\;
	\tau^{2a(-c)+b(-c)^2}\!
	\int\limits_{\mathclap{
		x \in \D
	}}
		\tau^{2(a-bc)x + bx^2}
	\;
	\kket{\smash{\omega^{-x}}}
	\bbra{\smash{\omega^{x}}}	
\;.
\mspace{-60mu}
\tag*{\qed}
\end{align*}

\medskip
\hrule
\nopagebreak
\smallskip

\Rule{ZH-MH} 
Multiplier-Hadamard rule.
\\[-3ex]
\begin{gather*}
	\Sem{8ex}{\,\vtikzfig[-2ex]{ZX-green-red-multiedge}}
\;\;=\;\;
	\Sem{8ex}{\;\vtikzfig[-2ex]{ZX-multiplier-gadget}\;}
\end{gather*}

\noindent
In this instance, we prove the equality in reverse (by demonstrating that the semantics of the second diagram, is equal to the semantics of the first).
In the proof on the right, we introduce the scalar constant $\gamma_u \!\!\;= \Gamma(0,\!\!\; -u,\!\!\; D) =\!\!\; 
\int_x \!\!\; \tau^{-ux^2}
\!\!\;=
\bigsem{\;%
	\tikz \node (g) at (0,0) [Z dot, label=right:\footnotesize{$[0 \s -u]$}] {};
\;}$.
By Eqn.~\eqref{eqn:quadratic-Gauss-integral-formula}, we have $\lvert \gamma_u \rvert = 1$, and in particular $\gamma_u^{-1} = \gamma_u^\ast = \int_x \tau^{ux^2}$.

\bigskip
\noindent
\textbf{Proof.}~\\[-8.5ex]
\begin{align*}{}
	\mspace{36mu}&\mspace{6mu}
	\gamma_u
	\mathop{\int \!\!\!\! \int \!\!\!\! \int \!\!\!\! \int}\limits_{\mathclap{\substack{
		w,x,y,z \in \D
	}}}
	\tau^{uw^2}	\,
	\tau^{u^{-1}x^2} \,
	\tau^{uy^2} \,
	\omega^{yz}\;
	\kket{z}\,
	\bbracket{y}{\smash{\omega^{-x}}}\,
	\bbracket{\smash{\;\!\omega^x\;\!}}{w} \,
	\bbra{w}
	\mspace{-72mu}
\\[1ex]&=\;
	\gamma_u
	\mathop{\int \!\!\!\! \int \!\!\!\! \int \!\!\!\! \int}\limits_{\mathclap{\substack{
		w,x,y,z \in \D
	}}}
	\tau^{uw^2}	\,
	\tau^{u^{-1}x^2} \,
	\tau^{uy^2} \,
	\omega^{yz}\,
	\omega^{xy}\,
	\omega^{wx}
	\;
	\kket{z}
	\bbra{w}
\\[1ex]&=\;
	\gamma_u
	\mathop{\int \!\!\!\! \int \!\!\!\! \int \!\!\!\! \int}\limits_{\mathclap{\substack{
		w,x,y,z \in \D
	}}}
	\tau^{uw^2}	\,
	\tau^{ux^2} \,
	\tau^{uy^2} \,
	\omega^{yz}\,
	\omega^{uxy}
	\omega^{uwx}
	\;
	\kket{z}
	\bbra{w}
\\[1ex]&=\;
	\gamma_u
	\mathop{\int \!\!\!\! \int \!\!\!\! \int \!\!\!\! \int}\limits_{\mathclap{\substack{
		w,x,y,z \in \D
	}}}
	\tau^{uw^2}	\,
	\tau^{u(x+y)^2} \,
	\omega^{uwx + yz}
	\;
	\kket{z}
	\bbra{w}
\\[1ex]&=\;
	\gamma_u
	\mathop{\int \!\!\!\! \int \!\!\!\! \int \!\!\!\! \int}\limits_{\mathclap{\substack{
		w,x,y,z \in \D
	}}}
	\tau^{uw^2}	\,
	\tau^{ux^2} \,
	\omega^{uwx - uwy + yz}
	\;
	\kket{z}
	\bbra{w}
\\[1ex]&=\;
	\gamma_u
	\mathop{\int \!\!\!\! \int \!\!\!\! \int}\limits_{\mathclap{\substack{
		w,x,z \in \D
	}}}
	\;\;
	\Biggl(\;\;\;
		\int\limits_{\mathclap{y \in \D}}
		\!
		\omega^{y(z-uw)}
		\!
	\Biggr)
	\tau^{u(x+w)^2}
	\;
	\kket{z}
	\bbra{w}
\\[1ex]&=\;
	\gamma_u
	\mathop{\int \!\!\!\! \int \!\!\!\! \int}\limits_{\mathclap{\substack{
		w,x,z \in \D
	}}}
	\bbracket{z-uw}{0} \,
	\tau^{ux^2}
	\;
	\kket{z}
	\bbra{w}
\\[1ex]&=\;
	\Biggl[
		\gamma_u
		\int\limits_{\mathclap{ x \in \D }}
		\tau^{ux^2}
	\Biggr]
	\Biggl[\;\;\;
		\mathop{\int \!\!\!\! \int}\limits_{\mathclap{\substack{
			w,z \in \D
		}}}
		\bbracket{z-uw}{0} \,
		\;
		\kket{z}
		\bbra{w}
	\Biggr]
\\[1ex]&=\;
		\mathop{\int \!\!\!\! \int \!\!\!\! \int}\limits_{\mathclap{\substack{
			w,y,z \in \D
		}}}
		\kket{z}
		\biggl[ \bbracket{z+y}{0} \biggr]
		\biggl[
			\bbracket{\big.y + \!\smash{\textstyle \sum\limits_{j=1}^u w}\big.}{0} 
		\biggr]
		\bbra{w}
\\[1ex]&=\;
		\mathop{\int \!\!\!\! \int \!\!\!\! \int \!\!\!\! \int}\limits_{\mathclap{\substack{
			t \in \D^u \\
			w,y,z \in \D
		}}}
		\kket{z} 
		\biggl[
			\bbracket{z+y}{0} 
		\biggr]
		\biggl[
			\bbracket{\big.y + \!\smash{\textstyle \sum\limits_{j=1}^u t_j}\big.}{0} \bbra{t} 
		\biggr]
		\biggl[
			\kket{w}\sox{u}\bbra{w}
		\biggr]
\;.
\mspace{-36mu}
\tag*{\qed}
\end{align*}

\medskip
\vfill
\hrule
\nopagebreak
\smallskip

\Rule{ZX-ME} 
Multiplier elimination.
\\[-3ex]
\begin{gather*}
	\Sem{8ex}{\,\vtikzfig[-1ex]{ZX-green-red-multiedge-lollipop}}
\;\;=\;\;
	\Bigsem{\;\vtikzfig[-2ex]{ZX-red-quadratic-lollipop}\;}
\end{gather*}

\bigskip
\noindent
\textbf{Proof.}~\\[-8.5ex]
\begin{align*}{}
	\mspace{108mu}&\mspace{-72mu}
		\mathop{\int \!\!\!\! \int \!\!\!\! \int \!\!\!\! \int}\limits_{\mathclap{\substack{
			t \in \D^u \\
			w,y,z \in \D
		}}}
		\;
		\biggl[
			\tau^{2az+bz^2} \bbracket{\smash{\omega^z}}{y}
		\biggr]
		\biggl[
			\bbracket{\big.y + \!\smash{\textstyle \sum\limits_{j=1}^u t_j}\big.}{0} \bbra{t} 
		\biggr]
		\biggl[
			\kket{w}\sox{u}\bbra{w}
		\biggr]
	\mspace{-72mu}
\\[-2ex]&=\;
		\mathop{\int \!\!\!\! \int \!\!\!\! \int}\limits_{\mathclap{\substack{
			w,y,z \in \D
		}}}\;\;
		\biggl[
			\tau^{2az+bz^2} \omega^{zy}
		\biggr]
		\bbracket{y + uw}{0} 
		\bbra{w}
\\[1ex]&=\;
		\mathop{\int \!\!\!\! \int}\limits_{\mathclap{\substack{
			w,z \in \D
		}}}\;\;
		\biggl[
			\tau^{2az+bz^2} \omega^{-zuw}
		\biggr]
		\bbra{w}
\\[1ex]&=\;
		\int\limits_{\mathclap{\substack{
			z \in \D
		}}}
			\tau^{-2au^{\text{--}1}\!\!\;z\,+\,bu^{\text{--}2}\!\!\;z^2}  			
 		\Biggl[\;\;\;\,
			\int\limits_{\mathclap{\substack{
				w \in \D
			}}}
				\omega^{wz}
				\bbra{w}	\,
		\Biggr]
\\[1ex]&=\;
		\int\limits_{\mathclap{\substack{
			z \in \D
		}}}
			\tau^{-2au^{\text{--}1}\!\!\;z\,+\,bu^{\text{--}2}\!\!\;z^2}
			\bbra{\smash{\omega^z}}
\;.
\mspace{-36mu}
\tag*{\qed}
\end{align*}

\bigskip
\hrule
\nopagebreak
\smallskip

\Rule{ZX-MEH} Multi-edge Hopf Law.
$
\begin{aligned}
	\Sem{8ex}{\; %
	\begin{aligned}
	\begin{tikzpicture}
		\node (Z) [Z dot] at (0,0) {};
		\draw (Z) -- ++(-0.375,0);
		\node (G) [X dot] at (1.5,0) {};
		\draw (G) -- ++(0.5,0);
		\draw [out=70,in=110] (Z) to (G);
		\draw [out=45,in=135] (Z) to (G);
		\draw [out=-45,in=-135] (Z) to (G);
		\draw [out=-70,in=-110] (Z) to (G);
		\node at ($(Z)!0.4125!(G) + (0,0.0875)$) {$\vdots$};
		\node at ($(Z)!0.625!(G) + (0,-0.)$) {$\left. \begin{matrix} \\[6ex] \end{matrix}\right\} \! \raisebox{-0.125ex}{$D$}$};
	\end{tikzpicture}
	\end{aligned}\; }
\;=\;
	\Bigsem{
	\begin{aligned}
	\begin{tikzpicture}
		\node (Z) [Z dot] at (-0.1875,0) {};
		\draw (Z) -- ++(-0.375,0);
		\node (Z') [X dot] at (0.1875,0) {};
		\draw (Z') -- ++(0.375,0);
	\end{tikzpicture}
	\end{aligned}}
\end{aligned}
$

\bigskip
\noindent
\textbf{Proof.} This follows from \textsf{(ZH-MH)}, \textsf{(ZXH-GW)}, and \textsf{(ZXH-RG)}.

\medskip
\hrule
\nopagebreak
\smallskip

\Rule{ZX-A} Antipode Law.
$
	\biggsem{%
	\begin{aligned}
	\begin{tikzpicture}
		\node (Z) [Z dot] at (0,0) {};
		\draw (Z) -- ++(-0.375,0);
		\node (G) [X dot] at (.75,0) {};
		\draw (G) -- ++(0.375,0);
		\draw [out=45,in=135] (Z) to (G);
		\draw [out=-45,in=-135] (Z) to node [midway, X dot] {} (G);
	\end{tikzpicture}
	\end{aligned}}
\;=\;
	\Bigsem{
	\begin{aligned}
	\begin{tikzpicture}
		\node (Z) [Z dot] at (0,0) {};
		\draw (Z) -- ++(-0.5,0);
		\node (G) [X dot] at (.5,0) {};
		\draw (G) -- ++(0.5,0);
	\end{tikzpicture}
	\end{aligned}}
$

\bigskip
\noindent
\textbf{Proof.} This follows from \textsf{(ZXH-GW)}, \textsf{(ZXH-RG)}, and \textsf{(ZH-A)}.

\bigskip
\hrule
\bigskip
\noindent
\textit{(End of ZX rewrites.)}
\vfill

\newpage

\section{Constraints on interpretation imposed by discrete integrals}
\label{apx:proofs-rewrites-constrained-Ockhamic}

In this Appendix, we consider in what ways the semantics of the ZX and ZH calculi are constrained, simply by choosing to define them in terms of discrete integrals on a finite set $\D$, but without fixing a specific measure $\mu$ for those integrals beyond requiring $\mu(\{x\})$ to be the same for all $x \in \D$.
We also consider how choosing a different presentation from Eqn.~\eqref{eqn:FT-of-f} for the discrete Fourier transform would impact the measure $\mu$.

In order to limit the scope of this exploration, we limit ourselves to \emph{Ockhamic} semantic maps of ZX and ZH calculi.
This is a constraint on semantic maps which was introduced for the case $\D = \{0,1\}$ in Ref.~\cite[pp.\,31\,\&\,34]{dB-2021}, and which we extend to arbitrary sets $\D = \{L,L{+}1,\ldots,U{-}1, U\}$ below.

\subsection{Ockhamic interpretations of ZX and ZH diagrams}
\label{apx:measure-constraints}

The semantics of ZX and ZH nodes (\emph{e.g.},~as in the original articles~\cite{CD-2011,BK-2019} defining them) are conventionally chosen to be (about) as simple as possible when expressed via summation notation.
However, as Ref.~\cite{dB-2021} noted, there may be practical advantages to considering alternative interpretations of those generators, in the form of simplified rewrite systems.
This raises the question of what a reasonable `search space' of semantic maps may be for these calculi, for the purpose of denoting multilinear operators over $\cH$ for a fixed vector space $\cH \cong \C^D$ (as opposed for instance to \emph{equivalence classes} of such operators).

In Ref.~\cite{dB-2021}, one of us mooted sets of constraints on semantic maps $\sem{\,\cdot\,}$ for ZX- and ZH-diagrams, which is flexible enough to represent a range of possible semantics for the generators without a proliferation of parameters for the dots and boxes.
Those definitions were proposed to guide investigation into semantics for ZX- and ZH-diagrams on qubits, which is a special case which enjoys several special properties which do not readily extend to the general case $D>1$ in general, and which therefore may have been overspecialised in some ways. 
We propose the following revision of the definitions of \cite[pp.\,31\,\&\,34]{dB-2021} for $D > 1$ in general, letting $\D = \{L,L\,{+}\,1,\ldots,U\,{-}\,1,U\}$ for integers such that $U-L+1=D$.%
	\footnote{%
		One way in which these definitions revise those of Ref.~\cite[pp.\,31\,\&\,34]{dB-2021} is in allowing an amplitude different from $1$, for the $\ket{0}\sox{n}\!\bra{0}\sox{m}$ component of green dots, for the $\ket{\smash{\omega^0}}\sox{n}\!\bra{\smash{\omega^0}}\sox{m}$ component of red dots, and the components $\ket{y}\bra{x}$ of H-boxes for which none of the entries $x_1, \ldots, x_m, y_1, \ldots, y_n$ are zero.
		Given the introduction of functions $\Z \to \C$ to represent amplitudes, this seems to us a mild extension.
	}

\begin{itemize}
\item
	A semantic map $\sem{\,\cdot\,}$ of \textbf{ZX\:diagrams} in terms of operators $\cH\sox{m} \to \cH\sox{n}$ for arbitrary $m,n \in \N$, is \emph{Ockhamic} if it satisfies Eqn.~\eqref{eqn:stringGenerators}, and if there exists a sequence $(u_k)_{k \in \N}$ of positive scalars such that $u_2 = 1$,%
		\footnote{%
			\label{fn:u2-constraint-Ockhamic-ZX}%
			The constraint $u_2 = 1$ was not present in the original proposed definitions of `Ockhamic' semantic maps for ZX or ZH~diagrams in Ref.~\cite[pp.\,31\,\&\,34]{dB-2021}.
			We propose this revised definition so that red and green phase-free $1 \to 1$ dots, and the white and gray $1 \to 1$ dots, would all represent unitary transformations.
			This constraint is satisfied in practise for all existing Ockhamic semantic maps for ZX and ZH~diagrams.
		}
	and so that the following equations are satisfied for functions $\Theta: \Z \to \C$\;\!:

\vspace*{-3ex}
 \begin{subequations}%
\label{eqn:ZX-Ockhamic}%
  \begin{align}{}
  \Biggsem{\!\!\!\input{./Figures-Source/ZX-green-phase-dot-arity.tikz}\!\!\!}
  &=\,
    u_{m{\!\!\;+}n} 
    \sum_{x \in \D} \!\!\;
    	\Theta(x)\; \ket{x}^{\!\otimes n}\!\bra{x}^{\!\!\;\otimes m}
	\!,
%
%
\\[1.5ex]
%
%
  \Biggsem{\!\!\!\input{./Figures-Source/ZX-red-phase-dot-arity.tikz}\!\!\!}
  &=\,
    u_{m{\!\!\;+}n} 
    \sum_{k \in \D} \!\!\;
    	\Theta(k)\, \ket{\smash{\omega^{-k}}}^{\!\otimes n}\!\bra{\smash{\!\:\omega^{k}\!\;}}^{\!\!\;\otimes m}
   \!,
%
%
\\[1.5ex]
%
%
  \Bigsem{\input{./Figures-Source/ZX-H-plus-box.tikz}}
  \;\!&=\;\!
  	\text{\small$\dfrac{1}{\sqrt D}$}\mathop{\sum \sum}_{x,k \in \D}
  		\e^{2\pi i k x / D} \ket{k}\bra{x}
  ,
%
%
\\[1.5ex]
%
%
  \Bigsem{\input{./Figures-Source/ZX-H-minus-box.tikz}}
  \;\!&=\;\!
  	\text{\small$\dfrac{1}{\sqrt D}$}\mathop{\sum \sum}_{x,k \in \D}
  		\e^{-2\pi i k x / D} \ket{k}\bra{x}
  .
  \end{align}
  \end{subequations}
  We propose this definition in the interest of having a calculus in which the green dots and the red dots being are related by the two kinds of Hadamard boxes, which in turn relate to the quantum Fourier transform; but also retaining the `Only the Connectivity Matters'~\cite{CD-2011,CW-2018} property.%
		\footnote{%
			It may prove to be preferable, after all, to retain the notational convention seen in most existing presentations of ZX-based calculi~\cite{GW-2017,Wang-2018,Wang-2021,PWSYYC-2023} of having red dots be related to green dots via conjugation by the Fourier transform. 
			One might describe such an interpretation as a `conjugate-Ockhamic' interpretation (or, elect to redefine `Ockhamicity' to favour the  interpretation in which this conjugation relation holds).
		}
	\smallskip

\item
	A semantic map $\sem{\,\cdot\,}$ of \textbf{ZH\:diagrams} in terms of operators $\cH\sox{m} \to \cH\sox{n}$ for arbitrary $m,n \in \N$, is \emph{Ockhamic} if it satisfies Eqn.~\eqref{eqn:stringGenerators}, and if there exist  sequences $(u_k)_{k \in \N}$, $(h_k)_{k \in \N}$, and $(g_k)_{k \in \N}$ of positive scalars such that $u_2 = 1$,%
		\textsuperscript{\ref{fn:u2-constraint-Ockhamic-ZX}}
	$h_0 = 1$, and so that the following equations are satisfied for $c \in \Z_D$ and for functions $\mathrm{A}: \Z \to \C$:~\\[-2.5ex]
\begin{subequations}{}%
\label{eqn:ZH-Ockhamic}%
\begin{align}
  \Biggsem{\!\!\!\input{./Figures-Source/ZH-H-phase-box-arity.tikz}\!\!}
  \,&=\; h_{m{+}n}\! \mathop{\;\sum \sum\;}_{%
      \mathclap{
        {x \in \D^m\!,\, y\in \D^n}
      }}
      \mathrm{A}(x_1 \!\cdot\!\cdot\!\cdot x_m y_1 \!\cdot\!\cdot\!\cdot y_n) \!
      \ket{y}\!\!\bra{x}
%
%
\\[3ex]
%
%
  \Biggsem{\!\!\!\input{./Figures-Source/ZH-gray-dot-arity.tikz}\!\!}
  &=\;
  	g_{m{+}n}\!
    \mathop{\;\sum \sum\;}_{%
      \mathclap{\substack{
        {x \in \Z_D^m \!\!\:,}
        \,
        {y \in \Z_D^n} \\[.5ex]
      	\sum\limits_h x_h + \sum\limits_k y_k \;\!\equiv\, 0
      }}}
      \,
      \ket{y}\!\!\bra{x}
%
%
\\[1.25ex]
%
%
  \Biggsem{\!\!\!\input{./Figures-Source/ZH-white-dot-arity.tikz}\!\!}
  &=
  	u_{m{+}n} \!\sum_{x \in \D}\!
    \ket{x}^{\!\otimes n}\!\bra{x}^{\!\!\;\otimes m}
  	,
%
%
\\[3.5ex]
%
%
  \Bigsem{\input{./Figures-Source/ZH-gen-not-dot.tikz}}
  &=
  	\sum_{x \in \Z_D} \ket{-c{-}x}\bra{x}
  \;.
\end{align}%
\end{subequations}%
~\\[-1.5ex]
	We propose this definition in the interest of emphasising the theme of semantics involving simple constraints on classical indices, 
	having the generalised-not dots represent simple permutations of the standard basis,%
		\footnote{%
			Our attitude is flexible as to whether $c$-not-dots should denote the transformation $\ket{x} \mapsto \ket{-c{-}x}$ as we define it here, or instead $\ket{x} \mapsto \ket{c{-}x}$.
			These two conventions lead to slightly different rewrite systems, either of which may be considered favourable depending on other conventions one might adopt to simplify notation.
			We would in fact prefer the latter convention of representing $\ket{x} \mapsto \ket{c{-}x}$, were it not for the impact on the relationship with ZX red dots, the ZX stabiliser phase notation, and the notation for the Pauli operators.
			A clever notational convention might possibly resolve these issues.
		}
	having degree-0 H-boxes with parameter $\alpha \in \C$ (taken as short-hand for a function $t \mapsto \alpha^t$) stand directly for scalar factors of $\alpha$, and retaining flexsymmetry of the interpretation of the generators.
\end{itemize}
\smallskip

\noindent
As in Ref.~\cite{dB-2021}, these two conditions of `Ockhamicity' (for ZX- and for ZH-diagrams) are proposed as an interpretation of the design principles and features of the ZX- and ZH-calculi~\cite{CD-2011,BK-2019}, but does not necessarily represent some unifying property of these two different families of diagrams.%
	\footnote{%
		Indeed, if one identifies an H-box with degree $2$ and amplitude parameter $\omega = \e^{2\pi i /D} \in \C$ in the ZH~calculus, with the Hadamard plus box of ZX~diagrams, this imposes significant constraints on the parameters $h_2$, $g_k$, and $u_k$.%
	}

We may consider how the constraint of using an `Ockhamic' semantic map, for ZX~diagrams and for ZH~diagrams, interacts with the constraint of imposing a simple semantics for these diagrams via discrete integrals.
Specifically, we determine what constraints are imposed on the parameter $\nu$ governing the measure $\mu$, and on the scalar factors in Eqns.~\eqref{eqn:ZX-Ockhamic} and~\eqref{eqn:ZH-Ockhamic}, by considering Ockhamic semantic maps on the ZX~and ZH~generators for which Eqns.~\eqref{eqn:idealised-ZX-integrals} and~\eqref{eqn:idealised-ZH-integrals} hold.

\subsection{Ockhamic semantics for ZH diagrams}
\label{apx:constraining-Ockhamic-ZH}

Consider what constraints are imposed on an Ockhamic semantic map of ZH~diagrams, by requiring Eqns.~\eqref{eqn:idealised-ZH-integrals} to hold.
(These will in fact be less restrictive in a sense than the comparable constraints for semantics of ZX~diagrams, owing to the explicit role in Eqn.~\eqref{eqn:ZX-Ockhamic} for the quantum Fourier transform.)

We proceed directly by considering how the expressions in Eqn.~\eqref{eqn:ZH-Ockhamic} may be expressed as integrals with respect to the measure $\mu$, and then considering the consequences of requiring Eqns.~\eqref{eqn:idealised-ZH-integrals} to hold.
Note in particular how, for a multi-coefficient vector $x \in \D^m$, we have $\ket{x} = \ket{x_1} {\otimes} \ket{x_2} {\otimes} \cdots {\otimes} \ket{x_m} = \nu^m \kket{x_1} {\!\:\otimes\!\:} \kket{x_2} {\!\:\otimes\!\:} \cdots {\!\:\otimes\!\:} \kket{x_m} = \nu^m \kket{x}$.
Similarly, for every sum over $x \in \D^m$ that we turn into an integral, we introduce a factor of $\nu^{-2m}$ (one factor of $\nu^{-2}$ for each coefficient $x_j \in \D$).
We then have:~\\[-1ex]
\begin{subequations}%
\label{eqn:constrain-Ockhamic-ZH}%
\allowdisplaybreaks
\begin{align}{}
\label{eqn:constrain-Ockhamic-ZH-gen-not-dot}
  \Bigsem{\input{./Figures-Source/ZH-gen-not-dot.tikz}}
\,&=
  	\sum_{x \in \Z_D} \!\ket{-c{-}x}\bra{x}
\;=\;
  	\nu^2 \!\sum_{x \in \Z_D} \!\kket{-c{-}x}\bbra{x}
\;=
  	\int\limits_{\mathclap{x \in \Z_D}} \kket{-c{-}x}\bbra{x} \; 
  	,
\\[3ex]
\label{eqn:constrain-Ockhamic-ZH-white-dot}
  	\Biggsem{\!\!\!\input{./Figures-Source/ZH-white-dot-arity.tikz}\!\!}
&=\;
  	u_{m{+}n} \!\sum_{x \in \D}\!
    \ket{x}^{\!\otimes n}\!\bra{x}^{\!\!\;\otimes m}
\notag\\[2ex]&=\;
  	u_{m{+}n} \cdot \nu^{m{+}n}
  	\!\sum_{x \in \D}\!
    \kket{x}^{\!\otimes n}\!\bbra{x}^{\!\!\;\otimes m}
\notag\\[2ex]&=\;
  	u_{m{+}n} \cdot \nu^{m{+}n{-}2}
  	\!\!\int\limits_{\mathclap{x \in \D}}
  		\kket{x}^{\!\otimes n}\!\bbra{x}^{\!\!\;\otimes m} \;,
\\[3ex]
\label{eqn:constrain-Ockhamic-ZH-H-box}
  \Biggsem{\!\!\!\input{./Figures-Source/ZH-H-phase-box-arity.tikz}\!\!}
  &=\,
  	h_{m{+}n} \cdot
  	\nu^{m{+}n}
	\! \mathop{\;\sum \sum\;}_{%
      \mathclap{
        {x \in \D^m, y\in \D^n}
      }}
      \mathrm{A}(x_1 \!\cdot\!\cdot\!\cdot x_m y_1 \!\cdot\!\cdot\!\cdot y_n)\;
      \kket{y}\bbra{x}
\notag\\[1ex]&=\,
  	h_{m{+}n}\! \cdot
  	\nu^{-(m{+}n)}
    \mathop{\int \!\! \!\! \int}_{%
      \mathclap{
        {x \in \D^m, y\in \D^n}
      }}\!
      \mathrm{A}(x_1 \!\cdot\!\cdot\!\cdot x_m y_1 \!\cdot\!\cdot\!\cdot y_n)\;
      \kket{y}\bbra{x}
    \;,
\\[2ex]
\label{eqn:constrain-Ockhamic-ZH-gray-dot}
\Biggsem{\!\!\!\input{./Figures-Source/ZH-gray-dot-arity.tikz}\!\!}
  &=\,
  	g_{m{+}n}\!
    \mathop{\;\sum \sum\;}_{%
      \mathclap{\substack{
        {x \in \Z_D^m \!\!\:,}
        \,
        {y \in \Z_D^n} \\[.5ex]
      	\sum\limits_h x_h + \sum\limits_k y_k \;\!=\, 0
      }}}
      \,
      \ket{y}\!\!\bra{x}
\notag\\[1ex]&=\;
  	g_{m{+}n}\!
    \mathop{\;\sum \sum\;}_{%
      \mathclap{
        {x \in \Z_D^m \!\!\:,}
        \,
        {y \in \Z_D^n}
    }}
    \bracket{\big.\,
    	\smash{\textstyle \sum\limits_h x_h + \sum\limits_k y_k}
    \,}{\:\!0\:\!}      \,
      \ket{y}\!\!\bra{x}
\notag\\[1ex]&=\,
  	g_{m{+}n} \cdot \nu^{m{+}n{+}2}
    \mathop{\;\sum \sum\;}_{%
      \mathclap{
        {x \in \Z_D^m \!\!\:,}
        \,
        {y \in \Z_D^n}
    }}
    \bbracket{\big.\,
    	\smash{\textstyle \sum\limits_h x_h + \sum\limits_k y_k}
    \,}{\:\!0\:\!}      \,
      \,
      \kket{y}\bbra{x}
\notag\\[1ex]&=\,
  	g_{m{+}n} \cdot
  	\nu^{-m{-}n{+}2}
  	\mathop{\int \!\!\!\! \int}\limits_{%
      \mathclap{
        {x \in \Z_D^m \!\!\:,}
        \,
        {y \in \Z_D^n}
      }}
      \,
      \kket{y}\bbra{x}
    \;\;
    \bbracket{\big.\,
    	\smash{\textstyle \sum\limits_h x_h + \sum\limits_k y_k}
    \,}{0}
   	\;.
\end{align}~\\[-1ex]
In particular, the imposed semantics for the generalised-not dot are automatically satisfied by the definition of the point-mass vectors $\kket{x}$; requiring the leading scalar on the last line of Eqn.~\eqref{eqn:constrain-Ockhamic-ZH-white-dot} to be equal to $1$, imposes the constraint $u_k = \nu^{-k{+}2}$ (which in particular is consistent with $u_2 = 1$); requiring the leasing scalar on the last line of Eqn.~\eqref{eqn:constrain-Ockhamic-ZH-H-box} to be $1$, imposes the constraint $h_k = \nu^k$ (which in particular is consistent with $h_0 = 1$); and requiring the leading scalar of Eqn.~\eqref{eqn:constrain-Ockhamic-ZH-gray-dot} to be equal to $1$, imposes the constraint $g_k = \nu^{k-2}$.
\end{subequations}

Note that none of the above imposes constraints on $\nu$ itself, essentially because the only unitarity constraints happen to involve a single integral over $\D$ (as with the generalised-not dot or the $1 \to 1$ white dot), so that all factors of $\nu$ cancel in those cases.
Thus, the approach of defining semantics for ZH~generators in terms of integrals as in Eqn.~\eqref{eqn:idealised-ZH-integrals} is in itself sufficient to ensure the Ockhamicity of the interpretation, regardless of the normalising factor $\nu$ is used to define the measure $\mu$.

We note in particular that in the case $D=2$, the original semantics for ZH~diagrams presented by Backens and Kissinger~\cite{BK-2019} satisfies these constraints for $\nu = 1$.
For arbitrary $D>1$, such a choice yields a measure for which $\mu(\{\ast\}) = 1$ and $\mu(\D) = D$.
While the rewrite systems that one obtains for this measure are likely to involve more frequent changes to the scalar factors through scalar gadgets, this does seem (both intuitively and in practise~\cite{BKK-2021,LMW-2022}) an appropriate choice to denote and analyse problems in counting complexity.

\subsection{Ockhamic semantics for ZX diagrams}
\label{apx:constraining-Ockhamic-ZX}

By considering Ockhamic semantics for ZX~generators, we necessarily consider an interpretation of the (inverse) discrete Fourier transform which is unitary.

If we impose the constraint that the inverse Fourier transform as expressed in Eqn.~\eqref{eqn:integral-inv-FT} should be equal to the quantum Fourier transform, this is equivalent to imposing unitarity on a Fourier transform as expressed in Eqn.~\eqref{eqn:integral-FT} that preserves the measure.
From this, we would obtain the constraint $\nu = D^{-1/4}$, as described in Section~\ref{sec:normalisation-constraints-unitary-FT}.
We can also see this by directly considering the consequences of imposing this constraint on the meaning of integration over $\D$:
\begin{equation}
\label{eqn:impose-Hadamard-FT-dagger}
\begin{aligned}[b]
	\text{\small$\dfrac{1}{\sqrt D}$}\mathop{\sum \sum}_{x,k \in \D}
  		\e^{2\pi i k x / D} \ket{k}\bra{x}
  \,=\,
  \Bigsem{\input{./Figures-Source/ZX-H-plus-box.tikz}}
  \,:={}&	
	\mathop{\int \!\!\!\! \int}\limits_{\;\mathclap{k,x \in \D}}
		\e^{2\pi i k x / D} \;\kket{k}\bbra{x}
 \\[1ex]{}={}&\;
  \nu^2
	\mathop{\sum \sum}\limits_{\;\mathclap{k,x \in \D}}
		\e^{2\pi i k x / D} \;\ket{k}\bra{x}
\;,
\end{aligned}
\end{equation}~\\[-2ex]
so that $\nu = D^{-1/4}$, and in particular $\mu(\D) = N = D\nu^2 = \sqrt D$.
In this way, unlike the case with Ockhamic semantic map of ZH~generators as we considered in Section~\ref{apx:constraining-Ockhamic-ZH}, relating Ockhamic semantics of ZX~generators to simple integrals necessarily constrains $\mu$.

Following Section~\ref{apx:constraining-Ockhamic-ZH}, to obtain a simple representation of the semantics for green and red dots in an Ockhamic semantic map, we might impose the following constraints:
\begin{small}%
\begin{align}{}
\mspace{-24mu}
\label{eqn:idealised-ZX-integrals-redux-redux}%
 \Biggsem{\!\!\!\input{./Figures-Source/ZX-green-dot-arity.tikz}\!\!}
 &=
  	\int\limits_{\mathclap{x \in \D}}   		
    \kket{x}^{\!\otimes n}\bbra{x}^{\!\!\;\otimes m} 
  	\,,
%
&\quad
%
 \Biggsem{\!\!\!\input{./Figures-Source/ZX-red-dot-arity.tikz}\!\!}
 &=
  	\int\limits_{\mathclap{k \in \D}} 
    \kket{\smash{\omega^{-k}}}^{\!\otimes n}\bbra{\smash{\!\;\omega^{k}\!\;}}^{\!\!\;\otimes m} 
  	\!\;.
\mspace{-18mu}
\end{align}%
\end{small}~\\[-2ex]
As with the white dots in Ockhamic semantic maps of the ZH calculus (which is identical to the analysis of green dots with phase function $\tilde\theta(x) = 1$), we would then obtain the constraint $u_k = \nu^{-k{+}2}$.

\subsection{Relationship to the `well-tempered' calculi for $D=2$}
\label{apx:special-case-qubits}

Notice that the normalisation $\nu = D^{-1/4}$, and the corresponding semantics, essentially reproduces%
	\footnote{%
		Strictly speaking, the calculi of Ref.~\cite{dB-2021} involve red and green dots with parameters $\theta \in \R$, H-boxes with parameters $\alpha \in \C$, only one type of Hadamard box instead of two, and a `nu~box' which is missing from the calculus presented here.
		We may bridge these differences using the short-hand described for phases / amplitudes to parameterise these nodes, identifying the Hadamard plus and minus boxes with the single Hadamard box of Ref.~\cite{dB-2021}, and replacing the nu-boxes with some suitable scalar gadgets (such as H-boxes parameterised by powers of $\nu = D^{-1/4}$).
	}
the `well-tempered' semantics of ZX~and ZH~diagrams~\cite{dB-2021} in the case that $D = 2$.
In this way, Eqns.~\eqref{eqn:idealised-ZX-integrals} and~\eqref{eqn:idealised-ZH-integrals} provide a more intuitive definition of those semantics, and extend them to arbitrary $D>1$.

The process by which we obtain the semantic map in Section~\ref{sec:normalisation-constraints-unitary-FT}
may appear very different from how the well-tempered interpretation was devised in Ref.~\cite{dB-2021}, which solved for a specific Ockhamic semantic map subject to a list of desirable (idealised) rewrites.
However: it is worth noting that the framework of discrete integrals and (what we have here called) `accompanying' point-mass distributions, implicitly impose some of the constraints that were explicitly imposed in Ref.~\cite{dB-2021}.
Specifically: by choosing our notation so that
\begin{equation}
	\int\limits_{\mathclap{x \in \D}} \kket{x}\sox{n}\bbracket{x}{a}		
\;\;=\;\;
	\kket{a}\sox{n}
,
\end{equation}
it is not difficult to show that we automatically ensure the correctness of a number of rewrites including \mbox{\textsf{(ZX-GF)}}, \mbox{\textsf{(ZX-CPY)}}, \mbox{\textsf{(ZX-PU)}},  \mbox{\textsf{(ZH-AI)}}, \mbox{\textsf{(ZH-NS)}}, \mbox{\textsf{(ZH-WGB)}}, and \mbox{\textsf{(ZH-HM)}} for the semantics of Eqns.~\eqref{eqn:idealised-ZX-integrals} and~\eqref{eqn:idealised-ZH-integrals}.
In this way, we may regard the framework of discrete integrals and accompanying point-mass distributions as simplifying both the results and the methodology of Ref.~\cite{dB-2021}.
This provides a clarification and a theoretical justification of the `well-tempered' semantics, and indeed an extension of them to all $D > 1$.

Note that choosing the interpretations of Eqn.~\eqref{eqn:idealised-ZH-integrals} just for the ZH~generators, does not in fact impose any constraints on either the measure $\mu$ on $\D$.
As we see in Section~\ref{apx:constraining-Ockhamic-ZH},  Eqn.~\eqref{eqn:idealised-ZH-integrals} only imposes the constraint on the interpretation that the generators must be related to each other by simple geometric progressions related to the parameter $\nu$, but does not fix what $\nu$ should be.
In particular, if we take ${\nu \!=\! 1}$, Eqn.~\eqref{eqn:idealised-ZH-integrals} reproduces the original semantics provided by Backens and Kissinger~\cite{BK-2019} for the ZH~generators for ${D\!=\!2}$.
In effect, the ZH~calculus prioritises the standard basis to such an extent that it does not impose any strong relationships between that basis and any other, and in so doing leaves $\nu$ unconstrained.
It is the single constraint on the ZX~generators, that $\kket{\smash{\omega^k}} = \int_x \omega^{-kx} \,\kket{x}$ should be unitarily equivalent to $\kket{x}$, which suffices to fix the measure $\mu$ and thus to fix specific semantics for all of the generators through Eqns.~\eqref{eqn:idealised-ZX-integrals} and~\eqref{eqn:idealised-ZH-integrals}.

\subsection{Considering alternative normalisations of the Fourier transform}
\label{apx:alternative-normalisations-FT}

Given the decisive role that fixing the interpretation of the discrete Fourier transform plays in constraining semantic maps for ZX and ZH~diagrams, it may be of interest to consider how choosing a different normalisation factor $\nu$ would affect the presentation of the Fourier transform, or the relationships between the measures of the domain of a function $f: (\Z_D,\mu) \to \C$ and that of $\hat f$.

Above, we described the way in which the measure $\mu$ on $\Z_D$ (induced from the measure on $\D$ defined in Section~\ref{sec:discrete-integrals}) is constrained by various constraints on the discrete Fourier transform, construed as a map $f \mapsto \hat f$ on functions $f: \Z_D \to \C$. 
In particular, if we consider $\Z_D$ as a measure space and take $f: (\Z_D, \mu) \to \C$ and $\hat f: (\Z_D,\mu) \to \C$ to have the same domain considered as measure spaces, and if we also take $F: \kket{f} \mapsto \kket{\smash{\hat f}}$ as defined in Eqn.~\eqref{eqn:integral-FT} to be unitary, we find that $N = \mu(\Z_D) = \sqrt D$ (or equivalently, $\nu = \smash{\bbracket{x}{x}}^{-1/2} = D^{-1/4}$).

If one wished instead to have a more general relation $N = \lambda D$ for some scalar $\lambda \ne D^{-1/2}$, one could instead consider
\begin{itemize}
\item	
	setting $F\herm F = \alpha \mathbf 1$ for a scalar $\alpha > 0$, which is equal to $1$ if and only if $F$ is unitary;
\item
	Defining $\hat f$ to range over the space $(\Z_D,\mu')$ for a measure $\mu' = \beta \mu$ for some scalar $\beta > 0$;
\item
	Defining the Fourier transform $\hat f$ of $f$ by modifying Eqn.~\eqref{eqn:FT-of-f} to incorporate an additional scalar factor $\gamma > 0$ which may differ from $1$.
\end{itemize}
We consider the consequences of all three such modifications.
Let $\mathbf X_{\{0\}} : \Z_D \to \C$ be the function such that $\mathbf X_{\{0\}}(t) = 1$ if and only if $t = 0$, and $\mathbf X_{\{0\}}(t) = 0$ otherwise; then~\\[-1ex]
\begin{equation}
\kket{\smash{\mathbf X}_{\{0\}}}
\;=\;
	\int\limits_{\mathclap{x \in (\Z_{\!\!\;D}\!\!\;,\!\;\mu)}}
		\mathbf X_{\{0\}}(x) \; \kket{x}
\;=\;
	\nu^2 \kket{0}
\;=\;
	\nu \ket{0},
\end{equation}
from which it follows that
\begin{equation}
	\widehat{\mathbf X}_{\{0\}}(k)
\;=\;
	\gamma \int\limits_{\mathclap{x \in (\Z_{\!\!\;D}\!\!\;,\!\;\mu)}}
		\mathbf X_{\{0\}}(x) \; \e^{-2\pi ikx/D}
\;=\;
	\gamma \nu^2
.
\end{equation}
Let $\mathrm J: \Z_D \to \C$ be the constant function $\mathrm J(x) = 1$: then $\mathrm J = \tfrac{1}{\gamma \nu^2} \widehat{\mathbf X}_{\{0\}}$.
For the sake of uniformity, define point-mass functions $\kket{k}$ to accompany the integrals over $k \in (\Z_D,\mu')$ for $\mu' = \beta \mu$, so that
\begin{equation}
	\int\limits_{\mathclap{k \in (\Z_{\!\!\;D}\!\!\;,\!\;\mu')}}
		\bbracket{k}{k} \; \mathrm d\mu'(k)
	\;=\;
		1.
\end{equation}
Using the fact that $\nu^2 = \mu(\{\ast\}) = N/D$, we then have
\begin{equation}
\label{eqn:constraint-on-N-via-FT'}
\begin{aligned}[b]
	N
\;=\;
	\mu(\Z_D)
\;=\;
	\int\limits_{\mathclap{x \in (\Z_{\!\!\;D}\!\!\;,\!\;\mu)}}
		1 \; \mathrm d\mu(x)
\;&=\;
	\beta\!
	\int\limits_{\mathclap{k \in (\Z_{\!\!\;D}\!\!\;,\!\;\mu')}}
		1 \; \mathrm d\mu'(k)
\\[1ex]&=\;
	\beta\!
	\mathop{\int\!\!\!\!\int}\limits_{\mathclap{h,k \in (\Z_{\!\!\;D}\!\!\;,\!\;\mu')}}
		\mathrm J(h)^\ast \mathrm J(k) \; \bbracket{h}{k} \; \mathrm d\mu'(h) \, \mathrm d\mu'(k)
\\[1ex]&=\;
	\beta\,
	\bbracket{\mathrm J}{\mathrm J}
\;=\;
	\frac{\beta}{\gamma^2 \nu^4}\,
	\bbracket{\smash{\widehat{\mathbf X}_{\{0\}}}\big.}{\big.\smash{\widehat{\mathbf X}_{\{0\}}}}
\\[1ex]&=\;
	\frac{\beta}{\gamma^2 \nu^4}\,
	\bbra{\smash{\mathbf X}_{\{0\}}\big.}F\herm F \kket{\big.\smash{\mathbf X}_{\{0\}}}
\\[1ex]&=\;
	\frac{\alpha \beta}{\gamma^2 \nu^4}\,
	\bbracket{\smash{\mathbf X}_{\{0\}}\big.}{\big.\smash{\mathbf X}_{\{0\}}}
\;=\;
	\frac{\alpha \beta}{\gamma^2 \nu^2}
\;=\;
	\frac{\alpha \beta D}{\gamma^2 N},
\end{aligned}
\end{equation}
so that $
	N
\;=\;
    {\sqrt{\alpha \beta D}}/{\gamma}\,.	
$ 
Thus, we may vary any three of $\alpha$, $\beta$, $\gamma$, and $N$ to determine the remaining parameter.

If one wished to fix a different value for $N$ (\emph{e.g.},~$N \!\!\:=\!\!\: 1$ or $N \!\!\:=\!\!\: D$), and separately fix the extent to which $F$ increases or decreases the $\ell_2$ norm of its operands (if one wished to adopt a convention where $F$ might not be a unitary matrix), this would determine the extent to which the difference in the norm of $\smash{\hat f}$ corresponds to a change of units $\mu$ to $\mu'$ as we transform $f \mapsto \hat f$, or to which Eqn.~\eqref{eqn:FT-of-f} must be rescaled, or both.
Instead, we elect to fix $\alpha, \beta, \gamma = 1$, which allows us to take Eqn.~\eqref{eqn:FT-of-f} unmodified, to consider only one measure $\mu$ on $\Z_D$ in the context of the Fourier transform, and to take $F$ to be unitary.
This imposes the constraint $N = \sqrt{D}$, or equivalently $\nu = D^{-1/4}$.

While these remarks stand apart from any analysis over $\R$, it is possible to show how relating the measure $\mu$ on $\Z_D$ to a discrete measure on the compact continuous group $\R_N$ further motivates distinct measures $\mu, \mu'$ for the domains of $f: (\Z_D,\mu) \to \C$ and that of $\hat f: (\Z_D,\mu') \to \C$, if one fixes $N$ to be different from $\sqrt D$.
See Appendix~\ref{apx:FT-and-measure} for more details.

\newpage
\section{Quadratic Gaussian integrals}
\label{apx:quadratic-Gaussian-integrals}

In this Section, we consider how to evaluate the quadratic Gauss integrals described in Eqn.~\eqref{eqn:quadratic-Gauss-integral}.

\subsection{Preliminaries on quadratic Gaussian sums}

The scalars in Eqn.~\eqref{eqn:quadratic-Gauss-integral} may be expressed in terms of quadratic Gaussian sums, which are sums of the form
\begin{equation}
	G(r,s,N)
\;=\;
	\sum_{x=0}^{N{-}1}	\e^{2\pi i (rx^2 + sx) / N} 	
\;=\;
	\sum_{x=0}^{N{-}1}	\omega_N^{rx^2 + sx} 	
\end{equation}
where $\omega_N = \exp(2\pi i /N)$ is a primitive $N\textsuperscript{th}$ root of unity.
Motivated by the analysis of Ref.~\cite{dB-2013}, we consider two separate cases: the case where $N$ is odd, and the case where $N$ is a multiple of $4$.
(The case where $N$ is even but not a multiple of $4$ does not arise in our work, as we consider $N = D$ for $D$ odd, and $N = 2D$ for $D$ even, where $D>1$ is the dimension of a qudit.)

We define the Jacobi symbol $\bigl( \frac{k}{m} \bigr)$ for $m$ and odd integer and $k \in \Z$, as follows.
Let $\uchi_p$ be the quadratic character on $\Z_p$\,:
\begin{equation}
	\uchi_p(k)
\;=\;
	\begin{cases}
		0,	&	\text{if $p$ divides $k$};	\\
		+1, &	\text{if $k \equiv u^2 \pmod{p}$ for some $u \in [p]$};	\\
		-1,	&	\text{otherwise}.
	\end{cases}
\end{equation}
For $m = p_1^{t_1} p_2^{t_2} \cdots p_n^{t_n}$ an odd integer with $p_j$ distinct primes and $t_1 \in \N$, we define the Jacobi symbol as
\begin{equation}
\label{eqn:Jacobi-symbols}
	\Bigl(\frac{h}{m}\Bigr)
\;=\;
	\uchi_{p_1}(k)^{t_1} \, \uchi_{p_2}(k)^{t_2} \,\cdots\, \uchi_{p_n}(k)^{t_n}	\,,
\end{equation}
which is $0$ is $k$ and $m$ have any common prime factors $p_j$, and is $\pm 1$ otherwise.
For odd integers $r$, we also define
\begin{equation}
\label{eqn:number-theory-epsilon-parameter}
	\epsilon_m
\;=\;
	\left\{
	\begin{aligned}
 		1,	\;\;&\quad	\text{if $m \equiv 1 \pmod{4}$}	\\
 		i,	\;\;&\quad	\text{if $m \equiv 3 \pmod{4}$}
	\end{aligned}
	\right\}
\;=\;
	i^{(m-1)^2/4},
\end{equation}
which satisfies $\epsilon_m^2 = \uchi_m(-1)$.
Suppose that $r$ and $N$ have a common divisor other than $1$.
Let $t = \gcd(r,N)$.
It is possible to show that
\begin{equation}
\label{eqn:quadratic-Gauss-sum-common-divisor}
	G(r,s,N)
\;=\;
	\begin{cases}
		t \cdot G\Bigl(\dfrac{r}{t}, \dfrac{s}{t}, \dfrac{N}{t}\Bigr),
			&	\text{if $t$ is a divisor of $s$};
	\\
		0,	&	\text{otherwise}.
	\end{cases}
\end{equation}
We can then reduce the general case, to the special case where $r$ and $N$ have no common divisor $t > 1$, and where $r \ge 0$.
Suppose, then, that $r$ and $N$ have no common divisor $t > 1$, so that $r$ has an inverse $u \in [N]$ modulo $N$ (\emph{i.e.},~$ur \equiv 1 \bmod{N}$), and $N$ has an inverse $q \in [r]$ modulo $r$ (\emph{i.e.},~$qN \equiv q \bmod{r}$).

\begin{itemize}
\allowdisplaybreaks
\item
	If $N$ is odd, let $h \in \D$ be such that $2h \equiv 1 \pmod{N}$, and let $\tilde s = uhs$ so that $2r\tilde s \equiv s \pmod{N}$.
	We then have~\\[-2ex]
	\begin{equation}
	\begin{aligned}[b]
		G(r,s,N)
	\;=\;
		\sum_{x=0}^{N{-}1}
			\omega_N^{r(x-\tilde s)^2 + s(x-\tilde s)} 
	\;&=\;
		\sum_{x=0}^{N{-}1}
			\omega_N^{r(x^2 - 2\tilde sx + \tilde s^2) + sx - s \tilde s} 
	\\&=\;
		\sum_{x=0}^{N{-}1}
			\omega_N^{rx^2  +\;\! r u^2 h^2 s^2  \;\!-\;\! uhs^2} 
	\\[1ex]&=\;
		\omega_N^{ru^2(h-h^2)s^2}\!
		\cdot
		G(r,0,N).
	\end{aligned}
	\end{equation}~\\[-1ex]
	For $N$ odd and $r$ having no factors in common with $N$ greater than $1$,~\\[-2ex]
	\begin{equation}
			G(r,0,N)
		\;=\;
			\epsilon_N \Bigl(\frac{r}{N}\Bigr) \sqrt{N}	
	\end{equation}~\\[-1ex]
	where $\bigl(\tfrac{r}{N}\bigr)$ is the Jacobi symbol.
	Then we have~\\[-2ex]
	\begin{equation}
		G(r,s,N)
	\;\;=\;\;	
		\omega_N^{ru^2(h-h^2)s^2}\!
		\epsilon_N \Bigl(\frac{r}{N}\Bigr) \sqrt{N}	
	\;\;=\;\;
		\e^{2 \pi i ru^2(h-h^2)s^2\!\!\;/N} \,
		i^{(N-1)^2\!\!\:/\!\!\;4}\,
		\Bigl(\frac{r}{N}\Bigr) 
		\sqrt{N}	
	\end{equation}~\\[-1ex]
	In particular, we have $\bigl\lvert G(r,s,N) \bigr\rvert = \sqrt{N}$, and for $s = 0$ we have $G(r,s,N) = i^k \sqrt{N}$ for some $k \in \Z$.

\item
	Suppose that $N$ is even (in which case $r$ must be odd as it would not be divisible by $2$)
	Let $N = 2M$ for some integer $M$.
	We may consider the alternative sum~\\[-2ex]
	\begin{equation}
	\label{eqn:var-Gauss-sum}
		\tilde G(r,s,M)
	\;=\;
		\sum_{x=0}^{M{-}1} \e^{\pi i(rx^2 + sx)/M},
	\end{equation}~\\[-1ex]
	which we may relate to $G(r,s,N)$ as follows:~\\[-2ex]
	\begin{equation}{}
	\label{eqn:relate-var-Gauss-sum}
	\mspace{-18mu}
	\begin{aligned}[b]
		G(r,s,2M)
	\;&=
		\sum_{x=0}^{M{-}1} \e^{2\pi i(rx^2 + sx)/2M}
	+	
		\sum_{x=M}^{2M{-}1} \e^{2\pi i(rx^2 + sx)/2M}
	\\&=
		\sum_{x=0}^{M{-}1} \Bigl[ 
				\e^{\pi i(rx^2 + sx)/M} + \e^{\pi i\bigl(r(M^2 + 2Mx + x^2) + s(M+x)\bigr)/M}
			\Bigr]
	\\&=
		\sum_{x=0}^{M{-}1} \Bigl[
			\e^{\pi i(rx^2 + sx)/M} \,+\,
			\e^{\pi i (Mr + s)} \, \e^{\pi i(rx^2 + sx)/M}
		\Bigr]
	\\&=
		\Bigl(1 + (-1)^{Mr + s} \Bigr) \;\!
		\tilde G(r,s,M)
	\;.
	\end{aligned}
	\end{equation}~\\[-1ex]
	If $M$ is odd and $s$ is even, then we have $(-1)^{Mr+s} = -1$, in which case $G(r,s,N) = 0$; similarly if $M$ is even and $s$ is odd.
	Thus it suffices to consider the cases where $M$ and $s$ are either both even, or both odd, so that $Mr + s$ is even.
	\begin{enumerate}[(a)]
	\allowdisplaybreaks
	\item	
		If $M,s$ are both odd, let $g \in \Z$ be such that $2g = u(M - s)$, where recall $u \equiv r^{-1} \pmod{2M}$.
		\begin{align}{}
		\mspace{-18mu}
			G(r,s,2M)
		\;&=
			\sum_{x=0}^{2M{-}1} \e^{2\pi i(rx^2 + sx)/2M}
		\notag\\&=
			\sum_{x=0}^{2M{-}1} \e^{2\pi i(r(x + g)^2 + s(x + g)/2M}
		\notag\\[1ex]&=
			\sum_{x=0}^{2M{-}1} \e^{2\pi i(rx^2 + 2grx + rg^2 + sx + sg)/2M}
		\notag\\&=\;
			\e^{\pi i (rg^2 + sg)/M}
			\sum_{x=0}^{2M{-}1} (-1)^x \, \e^{\pi i rx^2/M}
		\notag\\&=
			\e^{\pi i (rg^2 + sg)/M}
			\sum_{x=0}^{M{-}1} \Bigl(
				(-1)^x \e^{\pi i rx^2/M} \,+\, (-1)^x \,\e^{\pi i r(M+x)^2/M}
			\Bigr)
		\notag\\&=
			\e^{\pi i (rg^2 + sg)/M}
			\sum_{x=0}^{M{-}1} \Bigl(
				(-1)^x \e^{\pi i rx^2/M} \,-\, (-1)^x \e^{\pi i rx^2/M}
			\Bigr)
		\;=\;
			0
		\;.
		\end{align}
		Together with the fact that $G(r,s,N) = 0$ for $M$ odd and $s$ even, we have in fact $G(r,s,N) = 0$ if $N = 2M$ for $M \in \Z$ odd.
		\medskip

	\item
		If $M,s$ are both even, let $M', s' \in \Z$ be such that $M = 2M'$ and $s = 2s'$.
		Let $\varsigma = +1$ if $r > 0$ and $\varsigma = -1$ if $r < 0$.
		We may show that
	\begin{equation}
	\begin{aligned}[b]
		\tilde G(r,s,M)
	\;&= 	
		\sum_{x=0}^{M{-}1} \e^{\pi i(rx^2 + sx)/M}
	\\&=\, 	
		\sqrt{\lvert M r^{-1}\rvert  \;\!}
		\,
		\e^{\pi i \bigl(\lvert Mr \rvert - s^2\bigr)\!\!\;\big/\!\!\:4Mr} 
		\sum_{k=0}^{\lvert r \rvert {-}1}
			\e^{- \pi i(Mk^2 + sk)/r}
	\\[1ex]&=\, 	
		\sqrt{\lvert M r^{-1} \rvert}
		\,
		\e^{\varsigma  i \pi / 4}\, \e^{-i\pi s^2 \!\!\:/ 4Mr}
		\sum_{k=0}^{\lvert r \rvert {-}1}
			\e^{- \varsigma \;\! i \pi (Mk^2 + sk)\big/\lvert r\rvert }
	\\[1ex]&=\; 	
		\sqrt{ \lvert Mr^{-1} \rvert}
		\,
		\e^{\varsigma i \pi / 4}\, \e^{-i\pi s^2 \!\!\:/ 4Mr}\,
		\tilde G(-  \varsigma M,\,- \varsigma  s,\,\lvert r \rvert )
		,
	\end{aligned}
	\end{equation}
	by quadratic reciprocity.
	By Eqn.~\eqref{eqn:relate-var-Gauss-sum}, we have
	\begin{equation}
	\begin{aligned}[b]
		G\bigl(-\varsigma M,-\varsigma s, \lvert r\rvert\bigr)
	\;&=\;
		\Bigl(1 + (-1)^{Mr+s}\Bigr)\, \tilde G\bigl(-\varsigma M,-\varsigma s, \lvert r\rvert \bigr)
	\\&=\;
		2\,\tilde G\bigl(- \varsigma M, - \varsigma s, \lvert r \rvert\bigr)
	\;=\;
		2\,\tilde G\bigl(\varsigma M, \varsigma s, \lvert r\rvert \bigr)^\ast
	\;.
	\end{aligned}
	\end{equation}
	Recall that $r$ is odd, and has no common divisors $t>1$ with $N$; it then follows that $r$ has no such common divisors with $M$ either, and in particular, we have $(4q)M' = qN \equiv 1 \pmod{r}$.
	Let $\eta \in \Z$ be such that $\eta \equiv 2^{-1} \pmod{r}$: we then have
	\begin{equation}
		\mspace{-36mu}
	\begin{aligned}[b]
		G(r,s,N)
	\;&=\;
		2 \,\tilde G(r,s,M)
	\\[1ex]&=\;
		2\, \sqrt{ \lvert Mr^{-1} \rvert}
		\;
		\e^{\varsigma i \pi / 4}\, \e^{-i\pi s^2 \!\!\:/ 4Mr}\,
		\tilde G\bigl(\varsigma M, \varsigma s, \lvert r\rvert \bigr)^\ast
	\\[2ex]&=\;
		\sqrt{\tfrac{1}{2}\lvert Nr^{-1}\rvert}
		\;
		\e^{\varsigma i \pi /4} \,
		\e^{-\pi i s^2 \!\!\:/ 2Nr}\,
		G\bigl(\varsigma 2 M', \varsigma 2 s', 2 \lvert r\rvert \bigr)^\ast		
	\\[2ex]&=\;
		\sqrt{\tfrac{1}{2}\lvert Nr^{-1}\rvert}
		\;
		\e^{\varsigma i \pi /4} \,
		\e^{-\pi i s^2 \!\!\:/ 2Nr}\,
		\cdot 2\cdot G\bigl(\varsigma M', \varsigma s', \lvert r\rvert \bigr)^\ast	
	\\[2ex]&=\;
		\sqrt{2\lvert Nr^{-1}\rvert }
		\;
		\e^{\varsigma i \pi /4} \,
		\e^{-\varsigma \pi i s^2 \!\!\:\big/ 2N\lvert r\rvert}\,
		\biggl( 
			\e^{2 \pi i \varsigma M' (4q)^2(\eta^2-\eta)s^2\!\!\;/\lvert r \rvert}\;
			i^{-(\lvert r \rvert-1)^2\!\!\:/\!\!\;4}\,
			\Bigl(\frac{M'}{\lvert r \rvert}\Bigr) 
			\sqrt{\lvert r\rvert}	
		\biggr)
		\mspace{-90mu}
	\\[1ex]&=\;
		\e^{\varsigma i \pi /4}  \;
		\e^{-\varsigma \pi i s^2 \!\!\:\big/ 2N\lvert r \rvert}\;
		\e^{-2 \pi i \varsigma q s^2\!\!\;/\lvert r \rvert} \;
		i^{-(\lvert r \rvert-1)^2\!\!\:/\!\!\;4}\,
		\Bigl(\frac{N}{\lvert r \rvert}\Bigr) 
		\cdot \sqrt{2N\;\!}
	\\[1ex]&=\;
		\e^{i \pi \varsigma /4}  \;
		\e^{- \pi i \varsigma s^2 (1 + 4Nq)\!\!\:\big/\!\!\: N\lvert r \rvert}\;
		i^{-(\lvert r \rvert-1)^2\!\!\:/\!\!\;4}\,
		\Bigl(\frac{N}{\lvert r \rvert}\Bigr) 
		\cdot \sqrt{2N\;\!}
		\,.
	\end{aligned}
	\end{equation}
	In particular, we have $\bigl\lvert G(r,s,N) \bigr\rvert = \sqrt{2N}$, and for $s = 0$ we have $G(r,s,N) = \sqrt{i}^{\,k} \!\cdot \sqrt{2N}$ for some odd integer $k \in \Z$.		
	\end{enumerate}
\end{itemize}
In summary, if $r$ and $N$ have no common divisor $t > 1$, we have:
\begin{subequations}%
\begin{align}
\mspace{-12mu}
	G(r,s,N)
\,&=\,
	\left\{
	\begin{aligned}
		\e^{i\varphi(r,s,N)} \sqrt{N},			\quad
		&	\text{if $N$ is odd};
	\\
		\e^{i\varphi(r,s,N)} \sqrt{2N},		\quad
		&	\text{if $N$ is a multiple of $4$ and $s$ is even},
	\\
		0,								\quad
		&	\text{otherwise},
	\end{aligned}
	\right.
\intertext{%
	where $\varphi(r,s,N) \in \R$ is some phase angle satisfying 
}
\label{eqn:quadratic-Gauss-sum-phase}
\mspace{-18mu}
	\e^{i\varphi(r,s,N)}
\,&=\,
	\left\{
	\begin{aligned}
		\e^{2 \pi i ru^2(h-h^2)s^2\!\!\;/N} \,
		i^{(N-1)^2\!\!\:/\!\!\;4}\,
		\Bigl(\frac{r}{N}\Bigr),
			&\quad
			\begin{minipage}[t]{25ex}
				if $N$ is odd	\\
				\small(for
					$u \equiv r^{-1} \!\bmod{N}$ \\
					and
					$h \equiv 2^{-1} \!\bmod{N}$),
			\end{minipage}
	\\[1ex]
		\e^{i \pi \varsigma /4}  \;
		\e^{- \pi i \varsigma s^2 (1 + 4Nq)\!\!\:\big/\!\!\: N\lvert r \rvert}\;
		i^{-(\lvert r \rvert-1)^2\!\!\:/\!\!\;4}\,
		\Bigl(\frac{N}{\lvert r \rvert}\Bigr),
			&\quad
			\begin{minipage}[t]{25ex}
				if $N$ is even \\
				\small(for $q \equiv N^{-1} \!\bmod{r}$),
			\end{minipage}
	\end{aligned}
	\right.
	\mspace{-24mu}
\end{align}%
\end{subequations}
where $\bigl(\tfrac{r}{N}\bigr)$ and $\bigl(\tfrac{N}{r}\bigr)$ are Jacobi symbols as described by Eqn.~\eqref{eqn:Jacobi-symbols}.
In particular, note that if $s = 0$ then  $\e^{i\varphi(r,s,N)}$ is a power of $\e^{i\pi/4}$, and in particular is equal to $\e^{i\pi(N+1)k/4}$ for some integer $k$.
Using Eqn.~\eqref{eqn:quadratic-Gauss-sum-common-divisor} to generalise this to include the case where $t = \gcd(r,N)$ may be greater than $1$, we then have
\begin{equation}
\label{eqn:general-quadratic-Gauss-sum}
\begin{aligned}[b]
	G(r,s,N)
\,&=\,
	\left\{
	\begin{aligned}
		\e^{i\varphi\bigl(\tfrac{r}{t},\, \tfrac{s}{t},\, \tfrac{N}{t}\bigr)} \cdot \sqrt{tN},			\quad
		&	\text{if $N/t$ is odd, and $s$ is a multiple of $t$};
	\\
		\e^{i\varphi\bigl(\tfrac{r}{t},\, \tfrac{s}{t},\, \tfrac{N}{t}\bigr)} \cdot \sqrt{2tN}, \quad
		&	\mathrlap{\text{if $N/t$ is a multiple of $4$, and $s$ is a multiple of $2t$};}
	\\[1ex]
		0,								\quad
		&	\text{otherwise},
	\end{aligned}
	\right.
\\[-1ex]
\end{aligned}
\end{equation}
using the fact that when $G(r,s,N)$ is non-zero, we have $\bigl\lvert G(r,s,N) \bigr\rvert = t \cdot \bigl\lvert G(\tfrac{r}{t}, \tfrac{s}{t}, \tfrac{N}{t}) \bigr\rvert$, which is equal to $t \cdot \sqrt{N/t\,} = \sqrt{tN}$ for $N$ odd and equal to $t \cdot \sqrt{2N/t\,} = \sqrt{2tN}$ for $N$ even.

\subsection{Application to quadratic Gaussian integrals}

In our analysis, we consider two sorts of integrals, depending on the parity of the qudit dimension $D$.
In each case, we are interested in the integral
\begin{equation}
	\Gamma(a,b,D)
\;=\;
	\int\limits_{\mathclap{x \in \D}} \tau^{2ax + bx^2}
\;=\;
	\nu^2 \,\sum\limits_{\mathclap{x \in \D}} \, \tau^{2ax + bx^2}	\;,
\end{equation}
where $\tau = \e^{\pi i (D^2 + 1)/D}$ and $\nu = D^{-1/4}$.
For all integers $D > 1$, we have $\tau^2 = \e^{\pi i (2D^2 + 2)/D} = \e^{2\pi i + 2 \pi i / D} = \e^{2\pi i / D}$.
However, for $D$ odd, $\tau$ is itself also a $D\textsuperscript{th}$ root of unity (specifically: $\tau = \omega^h$ for $h \equiv 2^{-1} \bmod{D}$), whereas for $D$ even we have $\tau = \e^{\pi i / D}$, which is a primitive $2D\textsuperscript{th}$ root of unity.

\paragraph*{For $D$ an odd integer.}
	In the case of $D$ odd, we are interested in the quantity
	\begin{equation}
		\Gamma(a,b,D)
	\;=\;
		\int\limits_{\mathclap{x \in \D}} \tau^{2ax + bx^2}
	\;=\;
		\int\limits_{\mathclap{x \in \D}} \e^{2\pi i h (2ax + bx^2)/D}
			.
	\end{equation}
	Using the definition of the integral and the normalisation $\nu = D^{-1/4}$, we have
	\begin{equation}
		\Gamma(a,b,D)
	\;=\;
		\nu^2 \sum_{x=0}^{D{-}1}
			\e^{2\pi i (2ahx + bhx^2)/D}
	\;=\;
		\tfrac{1}{\sqrt D} G(bh, a, D)
		.
	\end{equation}
	If $t = \gcd(b,D)$, we may apply Eqn.~\eqref{eqn:general-quadratic-Gauss-sum} to obtain
	\begin{equation}
		\Gamma(a,b,D)
	\;=\;
		\left\{
		\begin{aligned}
			\frac{\text{\small$\sqrt{Dt}$}}{\text{\small$\sqrt D$}}
			\cdot \e^{i\varphi\bigl(\tfrac{bh}{t}, \tfrac{a}{t}, \tfrac{D}{t}\bigr)}
			\;&=\;
			\sqrt{t\,}
			\cdot \e^{i\varphi\bigl(\tfrac{bh}{t}, \tfrac{a}{t}, \tfrac{D}{t}\bigr)}\;,
			&&	\text{if $a$ is also a multiple of $t$};
		\\[1ex]
			&\phantom=\;\;\;
				0,
			&&	\text{otherwise};
		\end{aligned}
		\right.
	\end{equation}
	in particular, if $t = 1$, we have $\lvert \Gamma(a,b,D) \rvert = 1$.

\paragraph*{For $D$ an even integer.}
	In the case of $D$ even, we are interested in the quantity
	\begin{equation}
		\Gamma(a,b,D)
	\;=\;
		\int\limits_{\mathclap{x \in \D}} \tau^{2ax + bx^2}
	\;=\;
		\int\limits_{\mathclap{x \in \D}} \e^{\pi i (2ax + bx^2)/D}
			.
	\end{equation}
	Again using the definition of the integral and the normalisation $\nu = D^{-1/4}$, we have
	\begin{equation}
		\Gamma(a,b,D)
	\;=\;
		\nu^2 \sum_{x=0}^{D{-}1}
			\e^{\pi i (2ax + bx^2)/D}
	\;=\;
		\tfrac{1}{\sqrt D} \tilde G(b, 2a, D)
	\;=\;
		\tfrac{1}{2\sqrt D} G(b, 2a, 2D)
		,
	\end{equation}
	where $\tilde G(b,2a,D)$ is as defined in Eqn.~\eqref{eqn:var-Gauss-sum}.
	Note that as $D$ is even, $2D$ is then a multiple of $4$.
	Let $v = \gcd(b,2D)$: we may again apply Eqn.~\eqref{eqn:general-quadratic-Gauss-sum} to obtain
	\begin{gather}{}
	\mspace{-36mu}
		\Gamma(a,b,D)
	\;=\;
		\left\{
		\begin{aligned}
			\frac{\text{\small$\sqrt{4Dv}$}}{\text{\small$2\sqrt D$}}
			\cdot \e^{i\varphi\bigl(\tfrac{b}{v}, \tfrac{2a}{v}, \tfrac{2D}{v}\bigr)}
			\;&=\;
			\sqrt{v\,}
			\cdot \e^{i\varphi\bigl(\tfrac{b}{v}, \tfrac{2a}{v}, \tfrac{2D}{v}\bigr)}\;,
			&&	
				\mathrlap{%
				\begin{minipage}[t]{30ex}
					if $2D/v$ is a multiple of $4$ \\
					and $2a$ is a multiple of $2v$;
				\end{minipage}}
		\\[2ex]
			\frac{\text{\small$\sqrt{2Dv}$}}{\text{\small$2\sqrt D$}}
			\cdot \e^{i\varphi\bigl(\tfrac{b}{v}, \tfrac{2a}{v}, \tfrac{2D}{v}\bigr)}
			\;&=\;
			\sqrt{v\!\!\:/2\,}
			\cdot \e^{i\varphi\bigl(\tfrac{b}{v}, \tfrac{2a}{v}, \tfrac{2D}{v}\bigr)}\;,
			&&	
				\mathrlap{%
				\begin{minipage}[t]{30ex}
					if $2D/v$ is odd and \\
					$2a$ is a multiple of $v$;
				\end{minipage}}
		\\[2ex]
			0\;,
			&&&	\text{otherwise}.
		\end{aligned}
		\right.
		\mspace{-36mu}
	\notag\\[-4ex]
	\label{eqn:cases-Gaussian-integral-a}
	\end{gather}
	Again, let $t = \gcd(b,D)$.
	Furthermore, let $g,g'$ be the largest integers such that $2^g$ divides $D$,and $2^{g'}$ divides $b$.
	For $q = \min\;\{g,g'\}$, this means that $2^q$ is the largest power of $2$ which divides $t$.
	Let $t' = t/2^q$, and $D' = D/2^gt'$ and $b' = b\!\:/\!\:2^{g'}\!t'$: then all of $t'$, $D'$, and $b'$ are odd integers with no common divisors greater than $1$, and $D = 2^g t' D' = 2^{g-q} t D'$ and $b = 2^{g'} t' b' = 2^{g'-q} t b'$.
	Then
	\begin{equation}
			v
		\;=\;
			\gcd(b,2D)
		\;=\;
			\gcd\bigl(2^{g'} t' b', 2^{g+1} t' D' \bigr)
		\;=\;
			 2^{\,\min\,\{g+1, g'\}} \cdot t'
	\end{equation}
	We may then present the cases in Eqn.~\eqref{eqn:cases-Gaussian-integral-a} as follows.
	\begin{itemize}
	\item
		If $2D/v = 2^{g+1-\min\{g+1, g'\}} D'$ is a multiple of $4$, then $g+1-\min\{g{+}1, g'\} > 1$, from which we may infer that $g{+}1 - g' > 1$ and that $g - g' > 0$.
		Then $q = g'$ and $v = t$; and we have $D/t$ even and $b/t$ odd.
		Conversely, if $D/t$ is even, this also implies that $g - g' > 0$, and that $2D/v$ is a multiple of $4$.
		(Notice also that in this case, $2a$ is a multiple of $2v$ if and only if $a$ is a multiple of $t$).
	\item
		If $2D/v = 2^{g+1-\min\{g+1, g'\}} D'$ is odd, then $g{+}1 \le g'$.
		In this case we would have $q = g < g'$ and $v = 2^{q+1} t' = 2t$; and we have $D/t$ odd and $b/t$ even.
		Conversely, if $D/t$ is odd and $b/t$ is even, this implies that $q = g < g'$, which in turn implies that $2D/v$ is odd.
		(In this case as well, $2a$ is a multiple of $v$ if and only if $a$ is a multiple of $t$.)
	\end{itemize}
	From these observations, in the case of $D$ even, we obtain
	\begin{gather}
		\Gamma(a,b,D)
	\;=\;
		\left\{
		\begin{aligned}
			\sqrt{t\,}
			\cdot \e^{i\varphi\bigl(\tfrac{b}{t}, \tfrac{2a}{t}, \tfrac{2D}{t}\bigr)}\;,
			&\qquad	
				\begin{minipage}[t]{55ex}
					if $D$ is even, $D/t$ is even, $b/t$ is odd, \\ and
					$a$ is a multiple of $t$;
				\end{minipage}
		\\[.5ex]
			\sqrt{t\,}
			\cdot \e^{i\varphi\bigl(\tfrac{b}{2t}, \tfrac{a}{t}, \tfrac{D}{t}\bigr)}\;,
			&\qquad
				\begin{minipage}[t]{55ex}
					if $D$ is even, $D/t$ is odd, $b/t$ is even, \\ and
					$a$ is a multiple of $t$;
				\end{minipage}
		\\[1ex]
			0\;,
			&\qquad	\text{otherwise}.
		\end{aligned}
		\right.
		\mspace{-72mu}
	\notag\\[-3.5ex]
	\end{gather}

\bigskip
\paragraph*{Summary.}
From the equations above, if $t = \gcd(b,D)$, we have
	\begin{gather}{}
	\mspace{-36mu}
		\Gamma(a,b,D)
	\,=\,
		\left\{
		\begin{aligned}
			\sqrt{t\,}
			\cdot \e^{i\varphi\bigl(\tfrac{b}{t}, \tfrac{2a}{t}, \tfrac{2D}{t}\bigr)},
			&\quad\!\!
				\text{%
					if $D$ and $D/t$ are even but $b/t$ is odd, and 
					$a$ is a multiple of $t$;
				}
		\\[1ex]
			\sqrt{t\,}
			\cdot \e^{i\varphi\bigl(\tfrac{b}{2t}, \tfrac{a}{t}, \tfrac{D}{t}\bigr)},
			&\quad\!\!
				\text{%
					if $D$ and $b/t$ are even but $D/t$ is odd and 
					$a$ is a multiple of $t$;
				}
		\\[1ex]
			\sqrt{t\,}
			\cdot \e^{i\varphi\bigl(\tfrac{bh}{t}, \tfrac{a}{t}, \tfrac{D}{t}\bigr)},
			&\quad\!\!
				\text{%
					if $D$ is odd, $h \equiv 2^{-1} \!\!\!\!\pmod{D}$, and $a$ is a multiple of $t$;
				}
		\\[1ex]
						0\,,
			&\quad\!\!
				\text{otherwise}.
		\end{aligned}
		\right.
		\mspace{-36mu}
		\notag\\[-3.5ex]
	\label{eqn:quadratic-Gauss-integral-formula}
	\end{gather}
In each case, if we let $t = \gcd(b,D)$, we have $\bigl\lvert \Gamma(a,b,D) \bigr\rvert = \sqrt{t\,}$ if $a$ is a multiple of $t$ and the value of ${(D +\!\!\; Db/t^2)}$ is even, and $\Gamma(a,b,D) = 0$ otherwise.
In particular: for $b$ a multiplicative unit modulo $D$ (so that $t = 1$), we have $\bigl\lvert \Gamma(a,b,D) \bigr\rvert = 1$.
If furthermore we have $a = 0$, we have $\Gamma(a,b,D) = \e^{\pi i (D+1)k/4}$ for some $k \in \Z$, from the formula for $\varphi(b,a,D)$ given in Eqn.~\eqref{eqn:quadratic-Gauss-sum-phase}.

\newpage

\section{Relationships between measures on $\D$, on $\Z_D$, and on $\R$}
\label{apx:discrete-measures-R}

This Appendix consists of entirely supplemental observations, on the way in which discrete integrals over $\D$ (or over $\Z_D$, considered as a topological group) may be related to measures over $\R$.
This motivates some preliminaries on discrete measures over $\R$, such as the Dirac delta.

\subsection{Dirac deltas}
\label{apx:dirac-distributions}

When analysing functions on $\R$, it is not uncommon to consider a Dirac distribution $\delta$ (also known as the `Dirac delta'), which is defined in such a way that for an interval $J \subset \R$,~\\[-1.0ex]
\begin{equation}
\label{eqn:dirac-delta}
	\int\limits_{\mathclap{x \in J}} f(x) \, \delta(x-a) \; \mathrm dx
\;=\;
	\begin{cases}
		f(a),	& \text{if $a \in J$};
	\\[.5ex]
		0,		&	\text{otherwise}.	
	\end{cases}
\end{equation}
One may conceive of $\delta(x)$ as the limit of a family of formally definable functions, such as the Gaussians $\mathcal N_{1\!\!\;/n}(x) = \smash{\tfrac{n}{\sqrt 2\pi}\,\e^{-(n x)^2/2}}$ as $n \to \infty$.
In principle, one may consider it as syntactic sugar for a measure $\mu_\delta$ on $\R$, which for any interval $J \subset \R$ satisfies $\mu(J) = 1$ if $0 \in J$, and $\mu(J) = 0$ otherwise: we call this a `point-mass distribution'.
We may write $\delta_a(x) = \delta(x - a)$ for any $a \in \R$, so that $\delta_a$ describes a point-mass distribution at $a \in \R$: that is, a measure $\mu_a$ such that $\mu_a(S) = 1$ if $a \in S$, and $\mu_a(S) = 0$ otherwise.
(More generally, a point-mass distribution is any distribution of the form $p_a \delta_a$ for $p_a \ne 0$: the `mass' of such a distribution is then $p_a$.)
The purpose of the Dirac distributions would then be to allow us to write $\,\int_J f(x) \,\delta_a(x) \,\mathrm dx = \int_J f(x) \,\delta(x-a) \,\mathrm dx\,$ in place of $\,\int_J f(x\!+\!a) \, \mathrm d\mu_a\,$.
This provides a notational bridge between the discrete measures $\mu_a$ and the more common (Lebesgue) measure, so that we may perform analysis as though consistently working with a single variety of integration.

\subsection{Impulses and Dirac combs}
\label{apx:impluses-combs}

A \emph{discrete measure} $\rho$ on $\R$ is a measure which is a linear combination of a countable (and possibly finite) number of such point-mass distributions.
While such measures are not real-valued functions on $\R$, we may say that $\rho(a) \ne 0$ if for any function $f: \R \to \C$, we have $\int_{(a\!\!\:-\!\!\:\epsilon,a\!\!\:+\!\!\:\epsilon)} f(x)\,\rho(x)\,\mathrm dx \,\longrightarrow\, p_a f(a)$ for some $p_a \ne 0$, as $\epsilon \to 0$. 
We may refer to the contributions of the point-mass distributions as `impulses': for a discrete measure $\rho$, we say that $\rho$ has an impluse at $a$ if $\rho(a) \ne 0$ in this sense.

We define the \emph{Dirac comb} $\Sh$ (see,~\emph{e.g.},~Ref.~\cite{Cordoba-1989}) as a discrete distribution, consisting of a sequence of a sum of unit point-mass distributions on $\Z$:
\begin{equation}
\label{eqn:integer-comb}
	\Sh(x)
\,=\,
	\sum_{t \in \Z}	\,\delta_t(x)
\;.
\end{equation}
The Dirac comb is its own Fourier transform, which allows us to also express it as:
\begin{equation}
\label{eqn:FT-integer-comb}
	\Sh(x)
\,=\,
	\sum_{k \in \Z} \e^{2\pi ikx}
\,=\,
	\sum_{k \in \Z} \e^{-2\pi ikx} .	
\end{equation}
We may use the Dirac comb to express any function $\phi: \Z \to \C$ as a complex linear combination of impulses at the integers: if we let $\phi': \R \to \C$ be any extension of $\phi$ to the real numbers, we may define the complex-valued discrete `distribution' $\mathbf I \:\! \phi$ on $\R$, by
\begin{equation}
	\label{eqn:discrete-complex-distribution-via-impulses}
	\mathbf I \:\! \phi(x)
\;=\;
	\Sh(x) \,\phi'(x)
\;=\;
	\sum_{t \in \Z}	\,\delta_t(x) \,\phi(t).
\end{equation}
This will allow us to express sums on integers, in terms of integrals over $\R$: for instance, for any integers $a < b$, we then have~\\[-2.5ex]
\begin{equation}
	\int\limits_{\mathclap{(a,b]}} \mathbf I \!\: \phi(x) \; \mathrm dx
\;=\;
	\sum_{\mathclap{a < t\le b}}
		\phi(t);
\end{equation}
in particular, we have $\int_{(a,b]} \Sh(x)\,\mathrm dx = b-a$, which is the number of integers in the interval $(a,b]$.
Finally, we may consider normalised versions of the Dirac comb with impulses at integer multiples of any interval length $\ell > 0$:
\begin{equation}
\label{eqn:normalised-comb}
	\Sh_\ell(x)
\,=\,
	\ell \,\sum_{t \in \Z}	\,\delta_{\ell t}(x)
\,=\,
	\ell \,\sum_{t \in \Z}	\,\delta(x - \ell t)
\,=\,
	\ell \,\sum_{k \in \Z} \e^{2\pi ikx/\ell}
	.
\end{equation}
The leading scalar factor of $\ell$ in these sums, ensures that for $a < b$ which are integer multiples of $\ell$, we again have $\int_{(a,b]} \Sh(x)\,\mathrm dx = b-a$.
We may then use this to define a generalisation of $\mathbf I$, to embed functions $\phi: \Z \to \C$ as complex-valued measures in $\R$, but with impulses at intervals of $\ell > 0$: for $\phi'$ again any extension of $\phi$ to $\R$, we define
\begin{equation}
	\label{eqn:scaled-discrete-complex-distribution-via-impulses}
	\mathbf I_\ell \;\! \phi(x)	
\;=\;
	\Sh_\ell(x) \, \phi'(\tfrac{1}{\ell}x)
\;=\;
	\sum_{t \in \Z} \delta_{\ell t}(x) \,\phi(t). 
\end{equation}

\subsection{Motivating measures for $\Z_D$ via continuous groups}
\label{apx:continuous-models}

In principle, the analysis of Section~\ref{sec:discrete-measures-ZD}, on measures on $\Z_D$ and constraints on them motivated by the Fourier transform, stands on its own.
However, the reader may be interested in certain observations of how consistent this is with Fourier analysis on $\R$, or whether our observations of measure could be motivated with reference to continuous cyclic groups.

We discuss how this may be done below, by considering how we may model functions $f: \Z_D \to \C$ in terms of complex-valued discrete distributions with period $N$.
This will require some further preliminaries, in particular on the subject of discrete distributions on $\R$.

\subsubsection{Elementary number theory and periodic functions on $\Z$}
\label{apx:lift-periodic-fns}

We begin with some supplemental background on arithmetic modulo $D$, using this as a framework to consider periodic functions $f: \Z \to \C$.

Elements of $\Z_D$ are often represented by integers drawn from the set of integers $0 \le x < D$, though we may represent them by the set of integers $\D := \{L,L\,{+}\,1,\ldots,U\,{-}\,1,U\}$ for any integers such that $U-L+1=D$.
(In this Appendix, we adopt a convention of $L = -\lfloor \tfrac{1}{2}(D-1) \rfloor$ and $U = \lceil \tfrac{1}{2}(D-1) \rceil$, to simplify a certain description of a relartionship with $\R$.)
On may construct $\Z_D$ as a set of equivalence classes of integers which differ by multiples of $D$,
\begin{equation}
	x + D\Z \,=\, \bigl\{ x + nD \,\big\vert\, n \in \Z \bigr\},
\end{equation}
equipped with addition ${(x\!\!\:+\!\!\:D\Z) + (y\!\!\:+\!\!\:D\Z)} = {(x\!\!\:+\!\!\:y) + D\Z}$ and multiplication ${(x\!\!\:+\!\!\:D\Z) \!\!\:\cdot\!\!\: (y\!\!\:+\!\!\:D\Z)} = {(xy) + D\Z}$.
Representatives $x \in \D$ then serve as short-hand for the corresponding equivalence class $x + D\Z$.
(We may occasionally conflate $\Z_D$ with such sets of representatives.)

There is a ring homomorphism $\pi_D: \Z \to \Z_D$ (the `canonical projection') which may be defined by $\pi_D(x) = x + D\Z$.
We may use this to define a transformation $\mathbf L$ on functions, which carries maps $f: \Z_D \to \C$ to periodic functions $\mathbf Lf: \Z \to \C$, given by $\mathbf Lf = f \circ \pi_D$.
Then $\mathbf L$ `lifts' functions $f: \Z_D \to \C$ to functions $\mathbf Lf$, which are `well-defined' modulo $D$ in the sense that $\mathbf L f(x) = \mathbf L f(x+nD)$ for any $n \in \Z$.
By an abuse of notation, for any integer $x \in \D$, we may write $\mathbf Lf(x) = f(x)$.

\subsubsection{Constructing continuous cyclic groups from $\R$}

In order to consider Fourier analysis over $\Z_D$ in terms of discrete integrals, in which the measure $N = \mu(\Z_D)$ may be a parameter which we do not assign a value in advance, it will be helpful to relate functions $f: \Z \to \C$ with period $D$, to functions $g: \R \to \C$ with period $N$.
We use this to consider a measure on $\Z_D$ in terms of a measure on a compact continuous group.

We define an additive group 
of real numbers modulo $N$, for any $N > 1$, consisting of equivalence classes of real numbers which differ by multiples of $N$:
\begin{equation}
	z + N\Z \,=\, \bigl\{ z + tN \,\big\vert\, t \in \Z \bigr\},
\end{equation}
equipped with addition ${(y\!\!\:+\!\!\:N\Z) + (z\!\!\:+\!\!\:N\Z)} = {(y\!\!\:+\!\!\:z) + N\Z}$.
Just as we may represent $\Z_D$ by representative integers $x \in \D = {(-\tfrac{1}{2}D, \tfrac{1}{2}D] \!\:\cap\!\: \Z}$, we adopt the convention of representing $\R_N$ by an interval $J_N = (-\tfrac{1}{2}N,\tfrac{1}{2}N]$ of representatives.
We may then think of $\R_N$ in terms of the interval $J_N$, with `wrap-around' boundary conditions for addition and subtraction when the result of normal addition over $\R$ yields a result which lies outside of $J_N$.
Note that we may regard $\R_N$ as a topological group, and one with a measure $\mu_N$ inherited from the real numbers, consisting of the restriction of the Lebesgue measure to the interval $J_N$: in particular, we have $\mu_N(\R_N) = N$.

Finally, for a fixed integer $D > 1$, note that we may define an injection $j_N: \Z_D \to \R_N$ from the additive group $\Z_D$ to the group $\R_N$, by simply taking
\begin{equation}
	j_N\bigl(x + D\Z\bigr)
\;=\;
	\tfrac{N}{D}\bigl(x + D\Z\bigr)	
\;=\;
	\tfrac{N}{D}x + N\Z
\end{equation}
for $x \in \Z$; we may abuse notation to describe this more concisely by writing $j_N(x) = \tfrac{N}{D}x$ for $x \in \Z_D$.

\subsubsection{Interpretation of measures on $\Z_D$ in terms of measures on $\R$}
\label{apx:measure-interpretation-continuous}

To consider a notion of measure for $\Z_D$ in terms of a continuous group, we may consider the injection $j_N: \Z_D \to \R_N$ for some $N > 0$.
This motivates considering how we may represent functions $f: \Z_D \to \C$ in terms of `complex valued discrete measures' $\phi: \R_N \to \C$, which we may integrate over $\R_N$ to assign a meaning to an integral of $f$ over $\Z_D$.

We do this by considering the related concept of periodic functions on $\Z$, and their representation by periodic discrete measures on $\R$.
Following Section~\ref{apx:lift-periodic-fns}, we may `lift' any $f: \Z_D \to \C$ to a periodic function $\mathbf Lf: \Z \to \C$ with period $D$.
We may then represent $\mathbf Lf$ as a complex-valued discrete distribution, with impulses at invervals of length $\ell = N/D$, by acting on it with the transform $\mathbf I_\ell = \mathbf I_{N\!\!\:/\!\!\:D}$
as defined in Eqn.~\eqref{eqn:scaled-discrete-complex-distribution-via-impulses}.
Define the composite operator $\mathbf E_N = \mathbf I_{N\!\!\:/\!\!\:D\;\!} \mathbf L$, so that
\begin{equation}
\label{eqn:fun-ZD-into-RN}
	\mathbf E_Nf (z)
\,=\,
	\ell\,\sum_{t \in \Z} \delta_{t\ell}(z) \, \mathbf Lf(t)
\,=\,
	\frac{N}{D}\,\sum_{t \in \Z} \:\!\delta_{tN\!\!\:/\!\!\:D}(z) \; \mathbf Lf(t).
\end{equation}
In particular, representing elements of $\Z_D$ by $x \in \D$, the values of $f: \Z_D \to \C$ at each $x$ is represented by an impulse at $\tfrac{N}{D}x \in (-\tfrac{1}{2}N, \tfrac{1}{2}N] = J_N$, and also at $\tfrac{N}{D}(x + tD) = \tfrac{N}{D}x + tN$ for each integer $t \in \Z$.
Then $\mathbf E_N f$ is a discrete distribution with period $N$, which can be characterised in terms of its behaviour on the interval $J_N$.
In particular, 
\begin{equation}
\label{eqn:interpret-discrete-distribution-reals}
	\int\limits_{J_N} \! \mathrm dz \; \mathbf E_Nf(z)
\;=\;
	\frac{N}{D} \int\limits_{\R_N} \! \mathrm dz \; 
		\sum_{t \in \Z} \,\delta_{t\ell}(z)\, \mathbf Lf(t)
\;=\;
	\frac{N}{D} \sum_{\mathclap{\substack{t \in \D}}}\,
	\mathbf Lf(t)
\;=\;
	\frac{N}{D} \sum_{t \in \Z_D} f(t)
\;=\;
	\int\limits_{\mathclap{x \in \Z_D}} f(x) \; \mathrm d\mu(x)
	\,.
\end{equation}
The fact that $\mu(\Z_D) = N$ then reflects the choice of the period $N$ (or equivalently: the modulus $N$) for the injection $j_N: \Z_D \to \R_N$, and that as a discretisation of the measure $\mu_N$ on $\R_N$ (derived from the Lebesgue measure), we have $\mu(\Z_D) = \mu_N(\R_N) = N$.
In some sense, one may regard the injection $j_N$ as describing a way that $\Z_D$ represents a discretisation of the continuous group $\R_N$, by condensing the usual measure of $\R_N$ at a discrete set of points at intervals of $\tfrac{N}{D}$.

\subsubsection{Relating the Fourier transform on $\R$ to measures on $\Z_D$}
\label{apx:FT-and-measure}

The representation of the Fourier transform in Section~\ref{sec:normalisation-constraints-unitary-FT} can be interpreted, as arising from how the Fourier transform on $\R$ affects the discrete measures $\mathbf E_N f$ arising from functions $f: \Z_D \to \C$.
In particular: the Fourier transform will generally transform such distributions, to other distributions which may be regarded as being defined with respect to a different measure.
Consider the Fourier transform $\widehat{\mathbf E_N f}$:
\begin{equation}
	\widehat{\mathbf E_N f}(y)
\;=\;
	\int\limits_{\R} \mathrm dz \; \mathbf E_N f(z) \; \e^{-2\pi i y z}	\,:
\end{equation}
substituting in the expression for $\widehat f_N$ in Eqn.~\eqref{eqn:fun-ZD-into-RN}, we obtain

\begin{align}
	\widehat{\mathbf E_N f}(y)
\;&=\;
	\frac{N}{D} \int\limits_{\R} \mathrm dz \;
		\sum_{t \in \Z} \,\delta_{tN\!\!\:/\!\!\:D}(z)\,
			\mathbf Lf(t)  \, \e^{-2\pi i y z}
\;=\;
	\frac{N}{D} \sum_{t \in \Z} 
			\mathbf Lf(t)  \, \e^{-2\pi i y (tN\!\!\:/\!\!\:D)}	\;.
\intertext{%
	Separating the summation index $t \in \Z$ into representatives of $\Z_D$ and shifts by an entire period, we may then write
}
	\widehat{\mathbf E_N f}(y)
\;&=\;
	\frac{N}{D}
	\mathop{\sum \sum}_{\mathclap{\substack{x\in \D,\tau \in \Z }}}\, 
		\mathbf Lf(x) \,
			\e^{-2\pi i y N\!\!\;(x+\tau \!\!\: D)\!\!\;/\!\!\;D}
\;=\;
	\frac{N}{D}
	\sum_{\mathclap{x \in \Z_D}}\, f(x) \, \e^{-2\pi i y N x/D}
		\biggl[\,
			\sum_{\tau \in \Z}
				\e^{-2\pi i \tau y N}
		\biggr],
\end{align}
factorising away the dependence on $\tau$ in square brackets.
By Eqn.~\eqref{eqn:normalised-comb}, that bracketed expression is in fact $\tfrac{1}{N} \Sh_N(y)$: rewriting this as a sum of impulses, we then have
\begin{equation}
\begin{aligned}[b]
	\widehat{\mathbf E_N f}(y)
\,&=\,
	\frac{N}{D}
	\sum_{\mathclap{x \in \Z_D}}\, f(x) \, \e^{-2\pi i y N x/D}
		\biggl[
			\frac{1}{N}
			\sum_{k \in \Z}
				\delta_{k\!\!\;/\!\!\:N}(y)
		\biggr]
\;=\;
	\frac{1}{D}
	\mathop{\sum \sum}_{\mathclap{x \in \Z_{\!\!\;D}\!\!\;,\!\: k \in \Z}}\, 
		\delta_{k\!\!\;/\!\!\:N}(y)\,
			f(x) \, \e^{-2\pi i k x/D}	\,.
\end{aligned}
\end{equation}
That is, $\widehat{\mathbf E_N f}$ is itself a complex discrete measure, which is zero for any $y$ which is not an integer multiple of $\tfrac{1}{N}$.
By construction, it repeats after every $D$ impulses, and so has a period of $D/N$.
Then $\widehat{\mathbf E_N f} = \mathbf E_{\raisebox{-0.25ex}{$\scriptstyle D\!\!\:/\!\!\:N$}} g$ for a function $g: \Z_D \to \C$,
\begin{equation}
	\mathbf E_{\raisebox{-0.25ex}{$\scriptstyle D\!\!\:/\!\!\:N$}} g(y)
\;=\;
	\frac{1}{N}\sum_{k \in \Z} \,\delta_{k\!\!\;/\!\!\:N}(y)\, \mathbf Lg(k) 	,
\end{equation}
which we can see holds for
\begin{equation}
	g(k)
\;=\;
	\frac{N}{D} \sum_{\mathclap{x \in \Z_D}} \, f(x) \; \e^{-2\pi i k x/D}
\;=\;
	\hat f(k),	
\end{equation}
for $\hat f$ as defined by Eqn.~\eqref{eqn:FT-of-f}, with the integral over $\Z_D$ explicitly expanded as a sum using Eqn.~\eqref{eqn:integral-over-ZD} and using the measure $\mu$ such that $\mu(\Z_D) = N$.

This demonstrates that $\widehat{\mathbf E_N f} = \smash{\mathbf E_{D\!\!\:/\!\!\:N} \hat f}$\,, subject to this representation of the Fourier transform over $\R$.
However, $\smash{\mathbf E_{D\!\!\:/\!\!\:N} \hat f}$ is a complex-valued discrete measure which has period $D/N$, and where in particular a representative interval would be $J_{D\!\!\:/\!\!\:N}$.
This suggests that $\hat f$ should be interpreted in terms of the group injection $j_{\raisebox{-0.25ex}{$\scriptstyle D\!\!\:/\!\!\:N$}} : \Z_D \to \R_{D\!\!\:/\!\!\:N}$\,, which would motivate a measure $\mu'$ on $\Z_D$ such that $\mu'(\Z_D) = \mu_{\raisebox{-0.25ex}{$\scriptstyle D\!\!\:/\!\!\:N$}} (\R_{D\!\!\:/\!\!\:N}) = D/N$.

In summary: if we take $f: (\Z_D,\mu) \to \C$ as a function on the measure space $(\Z_D,\mu)$ with $\mu(\Z_D) = N$, we effectively interpret $f$ in terms of a measure on the topological space $\R_N$, which has a total measure of $N$.
The Fourier transform $\hat f$ is then most appropriately regarded as a measure on the topological space $\R_{D\!\!\:/\!\!\:N}$, which has a total measure of $D/N$, so that $\hat f: (\Z_D, \mu') \to \C$, where $\mu'(\Z_D) = D/N$.
If we wish for $\mu = \mu'$, so that $f$ and $\smash{\hat f}$ have the same domain when we consider $\Z_D$ as a measure space, we must take $N = \sqrt D$, in agreement with the results of Section~\ref{sec:normalisation-constraints-unitary-FT}.

\end{document}